\documentclass[english]{iopart}

\usepackage{appendix}
\usepackage{graphicx}
\usepackage{url}
\usepackage{iopams} 

\expandafter\let\csname equation*\endcsname\relax
\expandafter\let\csname endequation*\endcsname\relax

\usepackage[colorlinks=true,citecolor=blue]{hyperref}

\usepackage{amssymb}
\usepackage[latin1]{inputenc}
\usepackage{amsmath}
\usepackage{epsfig}
\newcounter{fig}
\usepackage{comment}

\newcommand{\Diag}{\textsf{Diag}}

\begin{document}

\title[Diagonals of rational functions ]
{\Large Ising $\,n$-fold integrals as diagonals of rational functions
and integrality of series expansions: integrality versus modularity}

\vskip .3cm 

{\bf  November 26th, 2012} 

\author{A. Bostan$^\P$, S. Boukraa$||$, G. Christol$^\ddag$, S. Hassani$^\S$, 
J-M. Maillard$^\pounds$}
\address{$^\P$ \ INRIA, B\^atiment Alan Turing, 1 rue Honor\'e d'Estienne d'Orves,
 Campus de l'\'Ecole Polytechnique, 91120 Palaiseau, France}
\address{$||$  \ LPTHIRM and D\'epartement d'A{\'e}ronautique,
 Universit\'e de Blida, Algeria}
\address{$^\ddag$ \ Institut de Math\'ematiques de Jussieu, 
UPMC, Tour 25,
 4\`eme \'etage,   
 4 Place Jussieu, 75252 Paris Cedex 05, France}
\address{\S  \ Centre de Recherche Nucl\'eaire d'Alger, 
2 Bd. Frantz Fanon, B.P. 399, 16000 Alger, Algeria}
\address{$^\pounds$ \ LPTMC, UMR 7600 CNRS, 
Universit\'e de Paris 6, Tour 23, 
 5\`eme \'etage, case 121, 
 4 Place Jussieu, 75252 Paris Cedex 05, France} 

\begin{abstract}

We show that the $\, n$-fold integrals $\, \chi^{(n)}$ of the
magnetic susceptibility of the Ising model, as well as various other 
$\,n$-fold integrals of the ``Ising class'', or $\, n$-fold integrals from
enumerative combinatorics, like lattice Green functions, are actually
\emph{diagonals of rational functions}. As a consequence, the power series
expansions of these solutions of linear differential equations ``Derived From
Geometry'' are {\em globally bounded}, which means that, after just one
rescaling of the expansion variable, they can be cast into series expansions
with \emph{integer coefficients}.
Besides, in a more enumerative combinatorics context, we show that
generating functions whose coefficients are expressed in terms of nested sums
of products of binomial terms can also be shown to be {\em diagonals of
rational functions}.
We give a large set of results illustrating the fact that the 
unique analytical solution
 of  Calabi-Yau ODEs, and more generally of MUM ODEs, is, 
almost always, diagonal of rational functions. We revisit  Christol's conjecture 
that globally bounded series of G-operators are necessarily 
diagonals of rational functions. 
We provide a large set of examples of globally bounded series, or series with 
integer coefficients, associated with modular forms, or Hadamard product of 
modular forms, or associated with Calabi-Yau ODEs, underlying the 
concept of modularity.
 We finally address the question of the relations between
the notion of \emph{integrality\/} (series with integer coefficients, or, more
generally, globally bounded series)
 and the \emph{modularity\/}  (in particular integrality of the Taylor 
coefficients of mirror map), introducing new representations of Yukawa couplings.

\end{abstract}

\vskip .5cm


\noindent {\bf AMS Classification scheme numbers}: 34M55, 
47E05, 81Qxx, 32G34, 34Lxx, 34Mxx, 14Kxx 

\vskip .5cm


\maketitle

\section{Introduction}
\label{introduc}

The series expansions of many magnetic
susceptibilities (or many other quantities, like 
the spontaneous magnetisation)
of the Ising model on various 
lattices in arbitrary dimensions are actually series with 
{\em integer coefficients}~\cite{Butera,Vicari,Fujiwara}. This is 
a consequence of the fact that, in a van der Waerden type expansion 
of the susceptibility, all the contributing graphs are the ones
 with exactly two odd-degree vertices and the number of such graphs 
is an integer.
When series expansions in theoretical physics, or mathematical physics, do not
have such an obvious counting interpretation, the puzzling emergence of series
with {\em integer coefficients} is a strong indication that some fundamental
structure, symmetry, concept have been overlooked, and that a deeper
understanding of the problem remains to be discovered\footnote[2]{The emergence of
{\em positive integer} coefficients corresponds to the existence
of some underlying measure~\cite{Bessis} (see also the concept of Mahler 
measures~\cite{Mahler}).}. 
Algebraic functions are known 
to produce series with {\em integer coefficients}. Eisenstein's 
theorem~\cite{Eisenstein,Heine} states that 
the Taylor series of a (branch of an) algebraic function can be recast
into a series with integer coefficients, up to a rescaling by a constant (Eisenstein
constant).
An intriguing result due to Fatou~\cite{Fatou} (see pp. 368--373)
states that a
power series with \emph{integer\/} coefficients and radius of convergence (at least)
$1$, is either rational, or transcendental. This result also appears in P\'olya
and Szeg\"o's famous Aufgaben book~\cite{Polya} (see Problem VIII-167).
P\'olya~\cite{Polya2} conjectured a stronger result, namely that a power
series with integer coefficients which converges in the open unit disk is
either rational, or  admits the {\em unit circle as a natural boundary}
(i.e. it has no analytic continuation beyond the unit disk). This was
eventually proved\footnote[5]{The P\'olya-Carlson result 
can be used to prove that some integer sequences, such as
 the sequence of prime numbers $\, (p_n)$~\cite{Flajolet},
 do not satisfy any linear recurrence relation with polynomial
 coefficients.} by Carlson~\cite{Carlson,Polya3}.
Along this natural
boundary line, it is worth
recalling~\cite{ze-bo-ha-ma-04,Khi6,ze-bo-ha-ma-05c,bo-ha-ma-ze-07b,mccoy3}
that the series expansions of the full magnetic susceptibility of the 2D Ising
model corresponds to a power series with integer coefficients\footnote[1]{In
some variable $\,w$~\cite{Khi6,ze-bo-ha-ma-04,ze-bo-ha-ma-05c,bo-ha-ma-ze-07b}. 
In the modulus variable $\, k$, one needs to perform a simple rescaling by a
factor $\, 2$ or $\, 4$ according to the type of (high, or low temperature)
expansions.}. For them, a unit circle natural boundary certainly
arises~\cite{bo-gu-ha-je-ma-ni-ze-08} (with respect to the modulus variable
$\, k$), but, unfortunately, this cannot be justified by Carlson's
theorem\footnote[9]{The radius of convergence is $\, 1$ with respect to the
modulus variable $\, k$, in which the series {\em does not have} integer
coefficients, being {\em globally bounded} only (this means that it can be
recast into a series with integer coefficients by one rescaling of the
variable $\, k$). If one considers the series expansion with respect to
another variable (such as~$\, w$) in which the series {\em does have} integer
coefficients, then the radius of convergence is not $\, 1$.}.

A series with natural boundaries {\em cannot be D-finite}, i.e. solution of a
linear differential equation with polynomial 
coefficients~\cite{Stanley80,Lipshitz89}\footnote[8]{D-finite series 
are sometimes called \emph{holonomic}. A priori, these notions differ: 
a function $\, f(x_1,\ldots,x_r)$ is called \emph{D-finite\/} 
if all its partial derivatives $\, D_1^{n_1}\cdots D_r^{n_r} \cdot f $ 
generate a finite dimensional space over $\, \mathbb{Q}(x_1,\ldots,x_r)$, 
and \emph{holonomic\/} if the functions
 $\, x_1^{\alpha_1}\,  \cdots \,
 x_r^{\alpha_r} D_1^{\beta_1}\,  \cdots \, D_r^{\beta_r} \, \cdot\,  f $
 obtained by
multiplying monomials in the variables and higher-order derivatives of $\, f$
 subject to the constraint 
$\, \alpha_1\,  +\,  \cdots \, + \alpha_r \, + \beta_1 \, + \cdots \, + \beta_r\,  \leq\,  N$ 
 span a vector space whose dimension over~$\mathbb{Q}$ grows like $\, O(N^r)$. The
 equivalence of these notions is proved by profound results of 
Bern{\v{s}}te{\u\i}n~\cite{Bernstein} and 
Kashiwara~\cite{Kashiwara,Takayama}.}. For simplicity,
let us restrict to series with {\em integer coefficients} (or series that
have integer coefficients up to a variable rescaling), that are series expansions
of D-finite functions. 
Wu, McCoy, Tracy and Barouch~\cite{wu-mc-tr-ba-76}  
have shown that the previous full magnetic susceptibility of the 2D Ising
model can be expressed (up to a normalisation
factor $\,(1-s)^{1/4}/s$, see~\cite{ze-bo-ha-ma-05c,ze-bo-ha-ma-05b}) as an
infinite sum of $n$-fold integrals, denoted by $\, \tilde{\chi}^{(n)}$, which
are {\em actually D-finite}\footnote[8]{For Ising models on higher dimensional 
lattices~\cite{Butera,Vicari,Fujiwara} no such decomposition
of susceptibilities, as an infinite sum of D-finite functions, 
should be expected at first sight.}.
We found out that the corresponding (minimal order)
differential operators are Fuchsian~\cite{ze-bo-ha-ma-04,ze-bo-ha-ma-05c},
and, in fact, ``special'' Fuchsian operators: the critical exponents for \emph{all\/} their
singularities are {\em rational numbers}, and their Wronskians are $\, N$-th
roots of {\em rational functions}~\cite{High}. Furthermore, it has been shown
later that these $\, \tilde{\chi}^{(n)}$'s are, in fact, solutions of 
{\em globally nilpotent} operators~\cite{bo-bo-ha-ma-we-ze-09}, or
 $\, G$-operators~\cite{Andre5,Andre6}. It is worth noting that the series
expansions, at the origin, of the $\, \tilde{\chi}^{(n)}$'s, 
in a well-suited variable~\cite{ze-bo-ha-ma-05c,ze-bo-ha-ma-05b} 
$\,w$, actually have {\em integer coefficients}, even if this  result
does not have an immediate proof\footnote[1]{We are interested in this paper
in the emergence of integers as coefficients of D-finite series. In general,
this emergence is not obvious: it cannot be simply explained at the level of
the linear recurrence satisfied by the coefficients, as illustrated by the
case of Ap\'ery's calculations (see also \ref{proof} below).} for all integers
$\, n$ (in contrast with the full susceptibility).

{}From the first truncated series expansions of $\, \tilde{\chi}^{(n)}$, the
coefficients for generic $\, n$ can be inferred~\cite{bo-bo-ha-ma-we-ze-09}
\begin{eqnarray} 
\label{closedn} 
\hspace{-0.9in}&&\tilde{\chi}^{(n)}(w) \, \, \,
= \, \, \, \, \, \, 2^n \cdot w^{n^2} \cdot 
\Bigl(1 \,\, + \, 4 \, n^2 \cdot w^2 \,\, + \,
2 \cdot (4\, n^4 \, +13\, n^2 \, +1)\cdot w^4 \, \, 
\nonumber \\
\hspace{-0.9in}&& \quad \quad + \, {{8} \over {3}} \cdot 
(n^2+4)\, (4\, n^4\,+23\, n^2+3) \cdot w^6 \\ 
\hspace{-0.9in}&& \quad \quad + \,{{1} \over {3}}
\cdot \, (32\,n^8\,+624\,n^6\,+4006\,n^4\,+8643\,n^2\,+1404)\cdot w^8
\nonumber \\ \hspace{-0.9in}&& \quad \quad + \,{{4} \over {15}} \cdot \,
(n^2+8)\cdot \, (32\, n^8 \, +784\, n^6 \, +6238\, n^4 \, +16271\, n^2 \,
+3180)\cdot w^{10} \,\, + \, \cdots \, \Bigr).
\nonumber 
\end{eqnarray} 
Note that the coefficients of the expansion of $\, \tilde{\chi}^{(n)}(w)/2^n$
 depend on $\, n^2$. 
Note that these coefficients are {\em integer coefficients} when 
$\, n$ is {\em any integer}, this integrality property~\cite{Kratten} of 
the coefficients becoming  straightforward to see when one remarks that
$(4\, n^4\,+23\, n^2+3)$ and  
$\, (32\,n^8\,+624\,n^6\,+4006\,n^4\,+8643\,n^2\,+1404)$
are of the form $\, n \cdot \, (n^2-1)\, f(n)+ \, 3 \,g(n)$
(respectively $ f(n)=\, 4\, n$ and $\, g(n) \, = \, 9\,n^2\,+1$, and
 $ f(n)=\,2\,n \cdot \, (16\,n^4\, +328\,n^2\,+2331)$ and
 $\, g(n) \, = \,4435\,n^2\, +468 $),
 and, hence, are always divisible by $\, 3$, that
 $\, (32\, n^8 \, +784\, n^6 \, +6238\, n^4 \, +16271\, n^2 \,+3180) $
is of the form 
$\, n\cdot \, (n^2-1)\, (n^2-4)\cdot \, f(n)\, +3 \cdot \, 5 \cdot g(n)$,
with $\, f(n) \, = \, 16\, n \cdot \, (2\, n^2+59)$ and 
$g(n) \, = \, 722\,n^4 \, +833\,n^2\, +212$, 
hence, always divisible by $\, 3$ and $\, 5$. 

These coefficients are valid up to $\, w^2$ for $\, n\, \ge \, 3$, $\, w^4$ for
 $\,n\, \ge \, 5$, $\, w^6$ for $\, n\, \ge \, 7$, $w^8$ for $n\ge 9$, and $w^{10}$
for $n\ge 11$ (in particular it should be noted that $\tilde{\chi}^{(n)}$ is
an even function of $\, w$ only for even $n$). Further studies on these 
$\,\tilde{\chi}^{(n)}$'s showed the fundamental role played by the theory of
elliptic functions\footnote[5]{Which is not a surprise for Yang-Baxter
integrability specialists.} (elliptic integrals, {\em modular forms}) and, much more
unexpectedly, \emph{Calabi-Yau ODEs}~\cite{CalabiYauIsing1,CalabiYauIsing}. These
recent structure results thus suggest to see the occurrence of series with
integer coefficients as a consequence of {\em modularity}~\cite{SP4} (modular
forms, mirror maps~\cite{CalabiYauIsing1,CalabiYauIsing,SP4,LianYau}, etc)
 in the Ising model. 

Along this line, many other examples of series
with {\em integer coefficients} emerged in mathematical physics (differential
geometry, lattice statistical physics, enumerative combinatorics, 
{\em replicable functions}\footnote[8]{The concept 
of replicable functions~\cite{Ford}
 is closely related to {\em  modular 
functions}~\cite{McKay,Basraoui}, (see the replicability
 of Hauptmoduls~\cite{Basraoui}), 
Calabi-Yau threefolds, and more generally the concept
 of {\em modularity}~\cite{SP4,Livne,Schutt,Gouvea,Fontaine,Saito}
(the third \'etale cohomology of a rigid Calabi-Yau 
threefold comes from
a modular form of weight 4, ...).} \ldots). One
must, of course, also recall Ap\'ery's results~\cite{Apery}. We give, in
\ref{modular}, a list of {\em modular forms}, and their associated series with
integer coefficients, corresponding to various lattice Green
functions~\cite{GlasserGuttmann,Prell,Zucker,GoodGuttmann}.This integrality 
is also seen in the {\em nome } and in other quantities
like the {\em Yukawa coupling}~\cite{CalabiYauIsing1}. 

 We restrict to series with integer
coefficients, or, more generally, {\em globally bounded}~\cite{Christol}
series of {\em one complex variable}, but it is clear that this integrality
property does also occur in physics with {\em several complex variables}: they
can, for instance, be seen for the 
previous (D-finite\footnote[3]{For several complex variables the ODEs of the paper are
replaced by Picard-Fuchs systems.}) $\, n$-fold integrals $\,
\tilde{\chi}^{(n)}$ for the anisotropic Ising model~\cite{WuJPA} (or for the
Ising model on the checkerboard lattice), or on the example of the lattice
Ising models with a magnetic field\footnote[2]{Along this line original
alternative representations of the partition function of the Ising model in a
magnetic field are also worth recalling~\cite{Barouch}.} (see for instance,
Bessis et al.~\cite{Bessis}).

One purpose of
this paper is to ``disentangle'' the notion of series with integer
coefficients ({\em integrality}) and the notion of 
{\em modularity}~\cite{SP4,Livne,Schutt,Gouvea,Fontaine,Saito}.
In this down-to-earth paper we will use the wording of 
``modularity'', not to refer to the {\em modularity conjecture} 
and other Serre's results that certain Geometric Galois  representations 
are modular, but as a quick proxy word to say that a series solution of a 
linear differential operator {\em as well} as the nome, and hopefully other series
(Yukawa coupling, ...) {\em are all series with integer coefficients}. 

We will show that the
$\,\tilde{\chi}^{(n)}$'s are {\em globally 
bounded} series,  as a
consequence of the fact that they actually are {\em diagonals of rational
functions for any value of the integer} $\, n$. We will generalise 
these ideas, and show that an extremely large class of problems of 
mathematical physics can be interpreted in terms of {\em diagonal 
of rational functions}: $\, n$-fold integrals with algebraic integrand 
of a certain type that we will characterise, 
Calabi-Yau ODEs, MUM linear ODEs~\cite{Guttmann}, series whose coefficients are
{\em nested sums of binomials}, etc. We take, here,
 a learn-by-example approach: on such questions one gets a much deeper 
understanding from highly
non-trivial examples than from general mathematical
demonstrations~\cite{Arnold,Arnold2}. 

\section{Series integrality}
\label{locally}

\subsection{Globally bounded series}
\label{locally}

Let us recall the definition of being 
{\em globally bounded}~\cite{Christol} for a series.
Consider a series expansion with rational coefficients,
with non-zero radius of convergence\footnote[1]{A 
series like the Euler-series 
$\, \sum_{n=0}^{\infty} \, n! \cdot \, x^n$ which has 
integer coefficients is excluded.}. The series is said to be 
globally bounded if there exists an integer $\, N$ such that the
series can be recast into a series with integer coefficients 
with just one rescaling $\, x \, \rightarrow \, N\, x$.

A necessary condition for being globally bounded is that
only a finite number of primes occurs for the factors
of the denominators of the rational number series coefficients.  
There is also a condition on the growth of these denominators, 
that must be bounded exponentially~\cite{Christol},
in such a way that the series has a non-zero $p$-adic radius 
of convergence for all primes~$p$.

When this is the case, it is easy to see that these series
can be recast, with just one rescaling, into series with 
{\em integer coefficients}\footnote[5]{For a first set of series with 
integer coefficients, see  \ref{modular}, where a set of such
series with integer coefficients corresponding to \emph{modular forms\/}
is displayed.}.

\vskip .1cm 

It will be seen, in a forthcoming section (see (\ref{minimal}) below), 
that the series expansion of 
{\em diagonals of rational 
functions}~\cite{Pochekutov,Purdue,Lipshitz} {\em are 
necessarily globally bounded}.

\subsection{Globally logarithmically bounded series}
\label{logarithbounded}
There is another notion, weaker than being globally bounded,
namely the notion of being {\em globally logarithmically\footnote[3]{For
a series $\,\sum  a_n  \,x^n$ being ``bounded'' means  bounded by $\, 1$,  
p-adically:   $\, |a_n|_p\, \leq\, 1\, $, i.e. $\,a_n$ has no $\,p$ 
factor at the denominator.)
Being logarithmically bounded  means "with logarithmic grows", 
 i.e.  $ \, |a_n|_p \,\leq\, n$.}  bounded}. 

As an example consider the series expansion of 
 $_2F_1([1/4, 1/2],[5/4], 4\, x)$. This series
 is {\em not globally bounded}
\begin{eqnarray}
\label{logarithbounded2F1}
\hspace{-0.6in}&&_2F_1\left(\left[{{1} \over {4}},
 {{1} \over {2}}\right],\left[{{5} \over {4}}\right], \, 4\, x\right) 
\,\,\,  = \,\,\, \, \, \, \,\,1 \,\,\, 
 + {{2} \over {5}}\, x\, \, + {{2} \over {3}}\, x^2\, 
+ {{20} \over {13}} \, x^3\, + {{70} \over {17}}\, x^4\, +12\, x^5\,
 \nonumber \\
\hspace{-0.6in}&&\qquad \quad \quad \quad \, \, + {{924} \over {25}}\, x^6\,
 + {{3432} \over {29}}\, x^7\, +390\, x^8\, 
+ {{48620} \over {37}}\, x^9\,\,\, \, \,  + \, \, \cdots 
\end{eqnarray}
When looking at the denominators of the series coefficients, one finds that
\emph{almost all primes\/} of the form $\, 4 \, \ell\, +1$ occur. There is no
way to recast this series into a series with integer coefficients with one
rescaling.

Let us denote $\, \theta \, = \, \, x \cdot \, d/dx$. 
The hypergeometric function 
$_2F_1([1/4,\, 1/2],[5/4],x)$ is solution of the
operator 
\begin{eqnarray}
\omega \,\,\, = \,\,\,\,\,\,\,
  \theta \cdot \Bigl(\theta\, + \,{{1} \over {4}}\Bigr) 
\, \,\,\,
 - \, x \cdot \, \Bigl(\theta\, + \,{{1} \over {4}}\Bigr) \cdot 
\, \Bigl(\theta\, + \,{{1} \over {2}}\Bigr),
\end{eqnarray}
which clearly factors\footnote[1]{This result
can, of course, straightforwardly be generalised to 
$_2F_1([a, \,b],[1+a],x)$.} $\, 4\,\theta\, + \, 1$ at the right:
\begin{eqnarray}
 \Bigl( 2\, \theta  - x \, \cdot \,  (2\,\theta\, + \, 1)    \Bigr) 
  \cdot \, (4\,\theta\, + \, 1).
\end{eqnarray}
Consequently, the action of   $\, 4\,\theta\, + \, 1\,$ 
on  $_2F_1([1/4, 1/2],[5/4],x)$ 
becomes the solution of the order-one globally nilpotent operator
 $\,\,2\, \theta \, - x \, \cdot \,  (2\,\theta\, + \, 1)$, 
and is, thus, an {\em algebraic function}
(rational or $\, N$-th root of rational), namely $(1-x)^{-1/2}$.
The hypergeometric function  $_2F_1([1/4, 1/2],[5/4],x)$ is
 \emph{not\/} globally bounded.
We will see below that, consequently,
 it {\em cannot be the diagonal of a rational function}, however 
it is not a ``wild'' series, the denominators 
do not grow ``too fast'': it is actually such that
a simple order-one operator, namely $4\,\theta\, + \, 1$, acting on this series, changes it 
into a {\em diagonal of rational function}.
Other examples of globally logarithmically  bounded hypergeometric series 
are given in \ref{locally}.

\section{Minimal recalls on diagonals of rational functions}
\label{minimal}

Let us recall here the concept of  {\em diagonal} 
of a ``function''\footnote[5]{This
 is an abuse of language: the ``functions'' are in
 fact defined by {\em series} of several 
complex variables: they have to be analytical (no Puiseux series).}, 
and some of its most important properties.

\subsection{Definition of the diagonal of a rational function}
\label{def}

Assume that 
$\,{\cal F}(z_1, \ldots, z_n)\, = \,\, P(z_1, \ldots, z_n)/Q(z_1, \ldots, z_n)$ 
is a rational function, where $P$ and $Q$ are
 polynomials with {\em rational coefficients} 
such that $Q(0, \ldots, 0) \neq 0$. This assumption implies that ${\cal F}$ 
can be expanded as a Taylor series at the origin
\begin{eqnarray}
\label{defdiag}
\hspace{-0.9in}&&{\cal F}\Bigl(z_1, \, z_2, \, \ldots, \, z_n \Bigr)
\, \, \, = \, \, \, \\
\hspace{-0.9in}&& \quad \quad 
\sum_{m_1 \, = \, 0}^{\infty} \, \sum_{m_2\, = \, 0}^{\infty} 
 \, \cdots \, \sum_{m_n\, = \, 0}^{\infty} 
 \,F_{m_1, \, m_2, \, \ldots, \, m_n}
\cdot  \, z_1^{m_1} \, z_2^{m_2} \,\,  \cdots \,\, z_n^{m_n} \quad
 \in \,\, \,\mathbb{Q}[[z_1, \ldots, z_n]].
\nonumber
\end{eqnarray}
The \emph{diagonal of $\, {\cal F}$\/} is defined as the series
of {\em one variable}
\begin{eqnarray}
\label{defdiag2}
\hspace{-0.6in}&&\Diag\Bigl({\cal F}\Bigl(z_1, \, z_2, \,
 \ldots, \, z_n \Bigr)\Bigr)
\, \, \, = \, \,  \quad \sum_{m \, = \, 0}^{\infty}
 \,F_{m, \, m, \, \ldots, \, m} \cdot \, z^{m}\quad \,\, \in  \,\,\, \mathbb{Q}[[z]].
\end{eqnarray}

More generally, one can define, in a similar way, the diagonal of any
multivariate power series $ \, {\cal F} \in\, K[[z_1, \ldots, z_n]]$, with
coefficients in an arbitrary field $K$ (possibly of positive
characteristic)\footnote[2]{The definition even extends to multivariate
Laurent power series, see e.g.~\cite{BA-JPB}.}.

\subsection{Main properties of diagonals}
\label{propr-diag}

The concept of diagonal of a function has a lot of interesting properties (see
for instance~\cite{legacy2}). Let us recall, through examples, some of the
most important ones.

The study of diagonals goes back, at least, to P\'olya~\cite{Polya21}, in a
combinatorial context, and to Cameron and Martin~\cite{CaMa38} in an
analytical context {\em related to Hadamard products}~\cite{Hadamard}.  
P\'olya showed that the diagonal of a rational function
in {\em two variables\/} is always an {\em algebraic function}. The most
basic example is $\, {\cal F} =\,  1/(1\,-z_1\,-z_2)$, for which  
\begin{eqnarray}
\hspace{-0.3in}&&\Diag ({\cal F})\, \, \, = \, \, \, \,\,
\Diag \left(\sum_{m_1 = 0}^\infty \sum_{m_2 = 0}^\infty  \binom{m_1+m_2}{m_1} 
\cdot \, z_1 ^{m_1} z_2 ^{m_2}\right) 
\nonumber \\ 
\hspace{-0.3in}&& \qquad \quad \quad  \quad \,  \, =\, \,  \, \, \,
 \sum_{m = 0}^\infty \binom{2m}{m} \cdot \, z^m 
\, \,\, \, = \,  \, \, \, \, \,  \frac{1}{\sqrt{1-4z}}.
\end{eqnarray}

The proof of P\'olya's result is based on the simple observation that the
diagonal  $\textsf{Diag} ({\cal F})$ is equal to the coefficient of $\, z_1^0$
in the expansion of ${\cal F}(z_1,z/z_1)$. Therefore, by Cauchy's integral
theorem, $\, \textsf{Diag} ({\cal F})$ is given by the contour integral
\begin{eqnarray}
\hspace{-0.3in}\textsf{Diag} ({\cal F})  \, \,\,  = \,  \, \, \,\,
 [z_1^{-1}] \, {\cal F}(z_1,z/z_1)/z_1  \, \,\,  =  \,\,  \,\,\,\,
\frac{1}{2\pi i}\, \oint_\gamma {\cal F}(z_1,z/z_1) \, \frac{dz_1}{z_1}, 
\end{eqnarray}
where $\, [z_1^{n}]$ means\footnote[1]{This is a convenient notation, 
very often used in combinatorics~\cite{FlSe09}.} 
extracting the $\, n$-th coefficient of a power series, 
and where the contour $\,\gamma$ is a small circle 
around the origin. Therefore, by
Cauchy's residue theorem, $\,\textsf{Diag} ({\cal F})$ is the sum of the
residues of the rational function ${\cal G} = \,{\cal F}(z_1,z/z_1)/z_1$ at all
its singularities $\,s(z)$ with zero limit at $z=\, 0$. Since the residues of a
rational function of two variables are algebraic functions,
 $\, \textsf{Diag}({\cal F})$ is itself an algebraic function.

For instance, when $\,{\cal F} = \, 1/(1-z_1-z_2)$, then 
${\cal G} =\, {\cal F}(z_1,z/z_1)/z_1$ has two poles at
 $\,s\, =\, \frac12 (1 \pm \sqrt{1-4z})$. The
only one approaching zero when $\, z\,\, \rightarrow\,\, 0$ is
 $\,s_0 = \,\frac12 (1 -\sqrt{1-4z})$. If $\, p(s)/q(s)$ has 
a simple pole at $s_0$, then its residue at
$\,s_0$ is $\, p(s_0)/q'(s_0)$. Therefore 
\begin{eqnarray}
\hspace{-0.4in}&&  \Diag ({\cal F})
\, \,\,  = \, \,\, \,\, \, \frac{1}{2\pi i}
\oint_\gamma \frac{dz_1}{z_1-z_1^2-z} 
\nonumber \\
\hspace{-0.4in}&& \qquad  \quad \quad 
 \,  =\,\,\,\, \,  \textsf{Res}_{s_0}
\, \frac{dz_1}{z_1-z_1^2-z}\,\, \,  =\, \,\,\, \,
  \frac{1}{1-2s_0}
\, \,\,\,  = \,\,\, \, \, \frac{1}{\sqrt{1-4z}}.
\end{eqnarray}

When passing from two to more variables, diagonalisation may still be
interpreted using contour integration of a multiple complex integral over a
so-called {\em vanishing cycle}~\cite{Deligne84}. However, the result {\em is not}
an algebraic function anymore. A simple example is 
$\,{\cal F}\, = \,1/(1-z_2-z_3-z_1 z_2-z_1 z_3)$, for which  
\begin{eqnarray}
\hspace{-0.7in}&& \Diag ({\cal F})\,\, \, = \, \,\,\,\,\, 
 1 \,\,\,\, +4 z\, \,+36 z^2 \,\, +400 z^3\, \,+4900 z^4 \,
 +63504 z^5 \,\,+853776 z^6 \,\, 
 \nonumber \\
\hspace{-0.7in}&& \quad \quad \quad \quad   \quad   \quad   \quad   \quad  
 +11778624 z^7 \,\,\,\,  + \, \cdots 
\end{eqnarray}
is equal to the complete elliptic integral of the first kind 
\begin{eqnarray}
\hspace{-0.7in}&&
\Diag ({\cal F})\,\, \,  =\, \, \,\,\, \sum_{m \geq 0} \binom{2m}{m}^2 \cdot \, z^m
\,\,\,\,  
\nonumber \\
\hspace{-0.7in}&& \quad \quad  \quad \,  =\, \,\,\,
  \frac{2}{\pi} \cdot \int_{0}^{{\pi}/{2}} 
\,  \frac{d\vartheta}{\sqrt{1 - 16z \sin^2 (\vartheta)}}
\, \,\, \, = \,\, \, \, \, 
_2F_1\Bigl([{{1} \over {2}}, \,{{1} \over {2}}], \,  [1], \, 16 \, z \Bigr),
\end{eqnarray}
which is a \emph{transcendental\/} function. 

Less obvious examples (see~\cite{Pech}) are
\begin{eqnarray}
\hspace{-0.4in}&& \Diag \left( \frac{1}{1-z_1-z_2-z_3-z_1z_2-z_2z_3-z_3z_1-z_1z_2z_3} \right)
\\
\hspace{-0.4in}&& \qquad  \qquad   \quad   \quad \,  = \,\,\, \,\,\, 
 {{1} \over { 1 \, -z}} \, \cdot \, _2F_1\Bigl(
[{{1} \over {3}}, \, {{2} \over {3}}], \, [1], \, {{54 \, z } \over { (1\, -z)^3}}, 
\Bigr), \nonumber 
\end{eqnarray}
and
\begin{eqnarray}
\hspace{-0.8in}&& \Diag \left( \frac{(1-z_1)(1-z_2)(1-z_3)}{1 \, \,
 -2 \, \, (z_1+z_2+z_3)\, \,  +3 \, \, (z_1z_2+z_2z_3+z_3z_1) \, -4 \, z_1z_2z_3} \right) 
  \\
\hspace{-0.8in}&& \, \,  \quad  \quad \,\,  = \, \, \,\,\, \, \,
  1 \, \, \, \, + \, 6 \cdot \int_0^z \, 
 _2F_1\Bigl(
[{{1} \over {3}}, \, {{2} \over {3}}], \, [2], \, 
{{27 \, w \cdot (2\,-3 \, w)  } \over { (1\, -4 \, w)^3}}\Bigr)
\cdot \, {{dw} \over { (1\,-4w) \, (1\,-64w)}}. \nonumber  
\end{eqnarray}

It was shown by Christol~\cite{Christol84,Christol85,Christol369} that the
diagonal $\, \Diag({\cal F})$ of \emph{any\/} rational function $\, {\cal F}$ is
\emph{D-finite}, in the sense that it satisfies a linear differential equation
with polynomial coefficients\footnote[5]{A more general result was proved by
Lipshitz~\cite{Lipshitz}: {\em the diagonal of any D-finite series is
D-finite}, see also~\cite{Purdue}.}. Moreover, the diagonal of any algebraic
power series in $ \, \mathbb{Q}[[z_1,\, \ldots,\, z_n]]$ is a G-function {\em
coming from geometry}, i.e. it satisfies the Picard-Fuchs type
differential equation associated with some one-parameter family of algebraic
varieties. Diagonals of algebraic power series thus appear to be a {\em
distinguished class} of G-functions\footnote[8]{Such diagonals are solutions
of G-operators. They are functions that are always algebraic mod. a prime $\,
p$. They fill the gap between algebraic functions and $\, G$-series: they can
be seen as {\em generalisations of algebraic functions}.}.
It will be seen below (see (\ref{recall})) that algebraic functions
with $\, n$ variables can be seen as diagonals of rational functions with 
 $\, 2\, n$ variables. Thus diagonals of rational functions also 
 appear to be a {\em distinguished class} of $G$-functions. It is worth 
noting that this  distinguished class is stable by the Hadamard product: 
the {\em Hadamard product of two diagonals of rational functions
is the diagonal of rational function}.

An immediate, but important property of diagonals of rational functions in
$\mathbb{Q}[[z_1,\,\ldots, \,z_n]]$ is that they are {\em globally bounded},
which means that they have {\em integer coefficients} up to a simple change
of variable $\,z \,\, \rightarrow \,\,N \, z$, where 
$ \,N \, \in \, \,\mathbb{Z}$. 

Furstenberg~\cite{Fu} showed that if $\, K$ has positive
characteristic, then the diagonal of any rational power series in
 $ \,K[[z_1,\ldots, z_n]]$ is algebraic. Deligne~\cite{Deligne84,BA-JPB} 
extended this result to
diagonals of algebraic functions.
For instance, when $\,{\cal F}\, =\, \,$
$  1/(1\,-z_2-z_3\,-z_1 z_2\,-z_1 z_3)$, one gets 
 modulo 3,  modulo 5 and  modulo 7
respectively
\vskip .1cm 
\begin{eqnarray}
\hspace{-0.7in}&& \Diag ({\cal F}) \quad   \bmod 3
  \, \,\, =  \,\, \,\, \,
 1 \,\, \,+z \,+z^3 \,+z^4 \, +z^9 \, +z^{10} \, +z^{12} \, +z^{13}
 \, \, \,+ \,\, \cdots
\nonumber \\ 
\hspace{-0.7in}&& \qquad \qquad \qquad  \, =\, \,\,\, 
 \frac{1}{\sqrt{1\, +z}}  \quad  \bmod 3,
\nonumber 
\end{eqnarray}
\begin{eqnarray}
\hspace{-0.9in}&&  \Diag ({\cal F})  \quad   \bmod 5
\,\, \,= \,\,\,\,  \, 1\,\,\, +4\,z\,\, +z^2\, +4z^5\,
 +z^6\,+4z^7\,+z^{10}\, +4z^{11}\, +z^{12} \, \, \, + \,\, \,\cdots
\nonumber \\ 
\hspace{-0.9in}&& \qquad \qquad \qquad 
  \,  \,  =\, \,  \,  \, \, \,  \frac{1}{\sqrt[4]{1\,-z\,+z^2}}  \quad  \bmod 5,
\nonumber 
\end{eqnarray}
\begin{eqnarray}
\hspace{-0.7in}&&  \Diag ({\cal F})  \quad   \bmod 7
 \, \,\, = \, \, \,\, \, \, \,
1 \,\, \, \, +4\,z \,\,  +z^2 \,  +z^3 \,  +4z^7 \,  +2z^8 \,  +4z^9
 \, \,\,  \, + \, \cdots  \, \, 
\nonumber \\ 
\hspace{-0.7in}&& \qquad \qquad \qquad 
=  \, \, \, \,\,  \frac{1}{\sqrt[6]{1\, +4z\,+z^2\, +z^3}}  \quad \, \bmod 7.
\nonumber 
\end{eqnarray}
More generally, for any prime $\, p$, one has
\begin{eqnarray}
\hspace{-0.7in}&&\qquad \quad \quad  \Diag ({\cal F}) \quad \,  \bmod p
 \, \,\, \,\,\, = \, \, \,\,\,\, \,\,  P(z)^{1/(1-p)} \quad  \, \bmod p 
\end{eqnarray}
where the polynomial $\, P(z)$ is nothing
 but~\cite{Ihara,Honda,Koike99}
\begin{eqnarray}
\label{polP}
\hspace{-0.9in}&&\quad \quad  P(z) \, \, \,  = \, \, \,  \, 
_2F_1\Bigl([{{1} \over {2}}, \,{{1} \over {2}}], \,  [1],
 \, 16 \, z \Bigr)^{1-p} \quad \, \,  \,   \bmod \, p 
\nonumber \\
\hspace{-0.7in}&&\quad \quad  \quad  \qquad  \quad \, \,\, = \, \,\, \, \,\,\, 
\,\, \, \sum_{n=0}^{(p-1)/2} \, {p \, -1/2 \choose n}^2 \cdot \, (16 \, z)^n 
\quad \quad  \bmod \, p. 
\end{eqnarray}
For instance, modulo 11,  the polynomial (\ref{polP}) reads:
\begin{eqnarray}
\hspace{-0.9in}&& _2F_1\Bigl([{{1} \over {2}}, \,{{1} \over {2}}], 
\,  [1], \, 16 \, z \Bigr)^{-10} \quad  \bmod 11
\, \, \,  \, = \,\, \, \,  \, 
1\, +4\,z\,+3\,z^2\,+4\,z^3 +5\,z^4\, +z^5 \quad   \bmod 11.
\nonumber 
\end{eqnarray}

Interestingly enough, the polynomial modulo $p$ 
\begin{eqnarray}
\hspace{-0.9in}&&\qquad \qquad \qquad \quad  \quad  P\Bigl({{\lambda} \over {16}} \Bigr)
\, \, \, \, \, = \,\, \,\, \,\,
 \sum_{n=0}^{(p-1)/2} \,  {p \, -1/2 \choose n}^2 \cdot \, \lambda^n, 
\end{eqnarray} 
is~\cite{Deuring,Igusa}, up to a sign $(-1)^{(p-1)/2}$,  the 
{\em Hasse invariant}\footnote[1]{Note Igusa's sentence
``Hence the elliptic differential of the first kind has only one period, 
and that is $\, A(\lambda)$,  up to an arbitrary differential constant.
 This version of Hasse 
invariant has not yet been explicitly remarked.''} of 
 $\,\, y^2 \,=\, x \cdot \,(1-x) \cdot \, (\lambda - x)$.

Note, however, that the Furstensberg-Deligne result~\cite{Fu,Deligne84},
 that we illustrate, here, 
with $\,{\cal F}\, =\, \,$ $  1/(1\,-z_2-z_3\,-z_1 z_2\,-z_1 z_3)$, 
goes far beyond the case of hypergeometric functions for which 
simple closed formulae can be displayed. 

\vskip .4cm 

\subsection{Hadamard, and other products}
\label{propr-diag}

Let us also recall the notion of {\em Hadamard product}~\cite{Hadamard,Necer}
 of two series, that 
we will denote by a star.
\begin{eqnarray}
\hspace{-0.7in}&&\hbox{If} \qquad 
\, \, f(x) \, = \, \, \, \sum_{n=0}^{\infty} \, a_n \cdot x^n,
 \qquad \, \,   
g(x) \, = \, \, \, \sum_{n=0}^{\infty} \, b_n \cdot x^n, 
\qquad \quad \, \, \hbox{then:} 
\nonumber \\
\hspace{-0.7in}&&\qquad  \qquad  f(x) \,\star\, g(x)\, \, \, = \, \,\, \,  \, 
\sum_{n=0}^{\infty} \, a_n \cdot b_n \cdot x^n.
\end{eqnarray}

The notion of diagonal of a function
and the notion of Hadamard product are obviously 
related: 
\begin{eqnarray}
\hspace{-0.6in}\Diag\Bigl( f_1(x_1) \cdot
 f_2(x_2)\,\, \cdots \, \, f_n(x_n)\Bigr)
 \, \,  \,\, = \, \, \, \,  \, \,  
f_1(x) \, \star\, f_2(x) \, \star \, \cdots \, \star \,f_n(x).
\end{eqnarray}
In other words, the diagonal of a product of functions with separate variables is
equal to the Hadamard product of these functions in a common variable. In
particular, the Hadamard product of $n$ rational (or algebraic, or even
D-finite) power series is D-finite.

\vskip .1cm The Hadamard product of two series with integer coefficients is
straightforwardly a series with integer coefficients. Furthermore, the {\em
Hadamard product of two operators}, annihilating two series, defined as the
(minimal order, monic) linear differential operator annihilating the Hadamard
product of these two series, is a {\em product compatible with a large number
of structures and concepts} that naturally occur in lattice statistical
mechanics. For instance, the Hadamard product of two globally
nilpotent~\cite{bo-bo-ha-ma-we-ze-09} operators is {\em also globally
nilpotent}.

\vskip .1cm 
Let us introduce another product, namely the {\em Hurwitz} (shuffle)
{\em product}\footnote[1]{The Hurwitz product of two algebraic functions is 
not algebraic in general, e.g. the Hurwitz square of
 $(1-4z)^{-1/2}$ is equal to
 $\, _2F_1([1/2, 1/2], [1], 16z \, (1-4z)) $. However, 
the Hurwitz product of two algebraic functions is actually 
an algebraic function \emph{modulo a prime} $p$, cf. Prop. 8 in~\cite{Fliess74}. 
The Hurwitz product of a rational and of an algebraic power series with
coefficients in $\mathbb{Q}$ is algebraic, cf. Prop. 3 \& 7
of~\cite{Fliess74}.}
of two series which is defined
as~\cite{Hurwitz1898,Hurwitz1899,Fliess74}:
\begin{eqnarray}
\label{Hurwitz}
\hspace{-0.9in}HurwitzProd\Bigl( \sum_n \, \alpha_n \cdot \, x^n,
 \,  \,  \, \sum_n \, \beta_n \cdot \, x^n \Bigr)
 \,  \, \,  =  \,  \, \, \,  \, \,  
\sum_n\,  \sum_m \, {n +m \choose n} \cdot \, 
\alpha_n \,\beta_m \cdot \, x^{n+m}. 
\nonumber 
\end{eqnarray}
A very simple example is, for instance,
\begin{eqnarray}
\hspace{-0.6in}HurwitzProd\Bigl( {{1} \over {1\, - a \cdot \, x}}, \,\, 
 {{1} \over {1\, - b \cdot \, x}} \Bigr)
 \,  \, \,\,   =  \,  \, \,   \, \,  {{1} \over {1\, - (a\,+b) \cdot \,  x}}.
\end{eqnarray}
Again, we have a remarkable compatibility property between the diagonal
and the Hurwitz product. 
The Hurwitz product of two series that are diagonals of two 
power series $\, A(x_1, \, x_2, \, \, \ldots, \, x_n)$ 
and  $\, B(y_1, \, y_2, \, \, \ldots, \, y_m)$ can itself be seen as 
the diagonal of a power series~\cite{Christol369}, that is very close to the product 
 of these two power series\footnote[2]{Note a misprint 
in the second equation of page 68
of~\cite{Christol369}. }:
\begin{eqnarray}
\hspace{-0.7in}&&HurwitzProd\Bigl(\Diag( A(x_1, \, x_2, \, \, \ldots, \, x_n)), \, 
\, \Diag( B(y_1, \, y_2, \, \, \ldots, \, y_m))\Bigr) \\
\hspace{-0.7in}&& \qquad  \qquad \quad  \, \, \,  \,  = \, \,  \, \,  \, \, \,
 \Diag\Bigl( {{ A(x_1, \, x_2, \, \, \ldots, \, x_n) \cdot \, 
B(y_1, \, y_2, \, \, \ldots, \, y_m)} \over { 
1 \,\, - \,\, t \cdot \,  x_1 \, x_2 \, \, \cdots \, x_n \, 
y_1 \, y_2 \, \, \cdots \, y_m}} \Bigr). \nonumber 
\end{eqnarray}
where $\, t$ is an additional variable. 
In fact there exists an infinite number of products that enjoy a similar property of
compatibility with the diagonal.
The most general products of series compatible with the diagonal are, beyond 
the Hadamard and Hurwitz products: 
\begin{eqnarray}
\hspace{-0.9in}GeneralProd\Bigl( \sum_n \, \alpha_n \cdot \, x^n,
 \, \sum_n \, \beta_n \cdot \, x^n \Bigr)
 \,  \, \, \,  =  \,  \, \,  \, \,  
\sum_n\,  \sum_m \, p(n , \, m) \cdot \, 
\alpha_n \, \beta_m \cdot \, x^{n+m},  \nonumber 
\end{eqnarray} 
where the $\, p(n, \, m)$'s  are coefficients of {\em any rational function}
of two variables $\, R(x, \, y)$:
\begin{eqnarray}
R(x, \, y)  \,  \,\,   =  \,  \, \,  \, \,\, 
 \sum_n\,\sum_m \, p(n , \, m)  \cdot  \,  x^n \, y^m.  
\end{eqnarray} 
Special cases are {\em Lamperti's product}~\cite{Lamperti,Fliess74} 
and {\em Trjitzinsky's product}~\cite{Trjitzinsky}.

\vskip .1cm 

\subsection{Chiral Potts examples}
\label{kit}

Let us consider, for instance, the Hadamard cube of a simple algebraic function 
\begin{eqnarray}
\hspace{-0.8in}&&_3F_2\left(\left[{{1} \over {3}},  \,{{1} \over {3}},
  \, {{1} \over {3}}\right], \, \left[1,\, 1\right]; \, z\right) 
\,  \, \, =    \, \, \,\, \,  \Diag\Bigl(  (1\, -\, x)^{-1/3}  
 \, \star \,(1\, -\, y)^{-1/3}   \, \star \,(1\, -\, z)^{-1/3}  \Bigr) 
\nonumber \\
\hspace{-0.8in}&&\qquad  \qquad  \quad \,  \,  =  \,  \, \,  \,  \, 
_1F_0([{{1} \over {3}}], \, []; \, z)  \, \star \,
_1F_0([{{1} \over {3}}], \, []; \, z)  \, \star \, 
_1F_0([{{1} \over {3}}], \, []; \, z) 
\nonumber \\
\hspace{-0.8in}&& 
\qquad \qquad  \quad\,  \,  =    \, \, \, \,  \,  \,  (1\, -\, z)^{-1/3}  
 \, \star \,(1\, -\, z)^{-1/3}   \, \star \,(1\, -\, z)^{-1/3}.
\end{eqnarray}
It is globally bounded:
\begin{eqnarray}
\hspace{-0.8in}&&
_3F_2\left(\left[{{1} \over {3}},  \,{{1} \over {3}},  \, {{1} \over {3}}\right],
 \, \left[1,\, 1\right]; \, 3^5 \, x\right) 
\,  \, \,  \,  =   \, \,   \, \,  \, \, \,
 1 \,\,\,\, +9\, x\,\,  +648\, x^2\,\,  + 74088\, x^3\,\,  +10418625\, x^4\,
\nonumber \\
\hspace{-0.8in}&& \qquad \quad \quad  \quad   +1648059777\, x^5\,\, 
 +281268868608\, x^6\, +50621016116736\, x^7\, \,  \, \,  + \,\, \cdots 
\nonumber
\end{eqnarray}

Other examples, related to the chiral Potts model
 and its associated Fermat
curves~\cite{legacy2} are  $\, _2F_1([1/2,\, N/3 ], \, [1], 36 \, t)$ and  
$\, _2F_1([1/2,\, N/5 ], \, [1], 100 \, t)$
which have series expansions {\em with integer coefficients}, or, more generally:
\begin{eqnarray}
\hspace{-0.85in}&&
_3F_2\left(\left[{{t} \over {N}},  \,{{q} \over {N}}, 
 \, {{s} \over {N}}\right], \, \left[1,\, 1\right]; \,\, x\right) 
 \, \, \,\, = \,\, \,\,\, (1\, -x)^{-t/N} \,
 \star  \,(1\, -x)^{-q/N} \, \star  \,(1\, -x)^{-s/N}.
 \nonumber 
\end{eqnarray}

\subsection{Furstenberg's result on algebraic functions}
\label{recall}

It was shown by Furstenberg~\cite{Fu} that \emph{any algebraic series\/} in one
variable can be written as the \emph{diagonal of a rational function of two
variables\/} (however, this representation is, by no means unique).  For instance,
\begin{eqnarray}
\hspace{-0.9in}&&f\,\, =\,\,\,\,\, \frac{x}{\sqrt{1-x}}\,\,\, = \,\,\,
\,\,\, x\,\,\,\, +\frac12\,x^2\,\,
+\frac38\,x^3\,\, +\frac{5}{16}\,x^4\,\,
+\frac{35}{128}\,x^5\, \, +\frac{63}{256}x^6\,\,\,\,\, + \, \, \cdots 
\end{eqnarray}
is the diagonal of $\,\,{(2\,x\,y - cx + cy)}/(x\,+y\,+2)\,$ 
for {\em any rational number} $c$.

The basis of Furstenberg's result is the fact that if $\, f(x)$ is a power series
without constant term, and is a root of a polynomial $ \, P(x,y)$ such that
$P_y(0,0)\,\neq\, 0$, then
\begin{eqnarray}
\hspace{-0.7in} f(x) \,\,\,\,  = \,\, \, \,\, \Diag \left( y^2 \cdot \, 
\frac{P_y(xy,\, y)}{P(xy,\, y)} \right)
 \qquad \, \hbox{where:} \quad \quad \quad 
P_y  \, \, = \, \, \, {{\partial P} \over {\partial y}}.
\end{eqnarray}

When $\, P_y(0,0) =\,  0$, this formula is not
true anymore. For instance, it does not apply to the algebraic function 
$\, f\,  =\, {x}/{\sqrt{1-x}}$, annihilated by $\, P = \, (x-1)\,y^2 \, + x^2$,
 since the diagonal of
 $\,\, y^2 \, {P_y(xy,y)}/{P(xy,y)} \,  =\, \,$
$   2y \, (xy-1)/(x^2 \, +xy \, -1)\,$
 is zero.
 However, Furstenberg's
result still holds. A way of seeing this on our example is to observe that 
$\, g\, =\, \,f\, - x\, - \frac12 x^2$ is an algebraic series
 annihilated by a polynomial $Q$
such that $\, Q(0,0)=\, 0$ and $\,Q_y(0,0)\,\neq\, 0$. This
 reasoning, which extends to
the general situation, yields the following rational function whose diagonal
equals $\, f$:
\begin{eqnarray} \label{eq:Furstenberg}
\hspace{-0.6in}&& x \, y \, \cdot \, 
{{ {\cal P}(x, \, y) } \over {{\cal Q}(x, \, y)}}
\qquad \qquad \quad  \hbox{where:} \\ 
\hspace{-0.6in}&&\quad  {\cal P}(x, \, y)
 \,\, \,=\, \,  \, \,\,\,
16 \, x^3y^5\, \,+4 \cdot \, (3x-4)\cdot \,x^2 y^4\,\,
 +4 \,\cdot \, (3+x) \cdot  \, x^2y^3\,\nonumber \\ 
\hspace{-0.6in}&& \qquad  \qquad \quad \quad 
 +\, (12x-24+x^2)\,\, x y^2\,\, +5\,yx^2\,\, +6\,x\,\,-16, 
\nonumber \\ 
\hspace{-0.9in}&&\quad   {\cal Q}(x, \, y)
 \,\,\, \,=\, \,\,  \, \,\,\,
8\, x^2y^3 \,\, \,\,+8 \cdot\, (x-1) \cdot \, x y^2\,\,\,
 +2 \cdot\,(x+4) \cdot \, x y\,\,\, +6 \,x\,\,\,\, -16.
\nonumber 
\end{eqnarray}
When compared to $\,{(2\,x\,y - cx + cy)}/(x\,+y\,+2)$, whose diagonal is also
$f$, this shows that Furstenberg's proof does not necessarily produce
the easiest rational function.

Furstenberg's result has been generalised to algebraic power series 
in an {\em arbitrary number of variables}: 
any  power series \footnote[5]{In the one-variable case, Puiseux 
 series could be considered 
but only after ramifying the variable.}  
 in $\mathbb{Q}[[x_1,...,x_n]]$ algebraic over $\mathbb{Q}(x_1,...,x_n)$ is 
the diagonal of a rational function with $\,2n$ variables 
(see Denef and Lipshitz~\cite{Denef}). 

\section{Selected $n$-fold integrals are
 diagonals of rational functions}
\label{nfoldas}

\newcommand{\iii}{\iota}

\vskip .1cm

Among many multiple integrals that are important in various domains of mathematical 
physics, let us consider the $n$-particle contribution to the magnetic 
susceptibility of the Ising model which we 
denote $\,\tilde{\chi}^{(n)}(w)$. They are given by $\,(n-1)$-dimensional 
integrals~\cite{ze-bo-ha-ma-04,bo-ha-ma-ze-07}:
\begin{eqnarray}
\label{chin}
\hspace{-0.6in}\tilde{\chi}^{(n)}(w) \,\, \,\, = \, \, \,\,\,\, 
\, \frac {(2w)^n}{n!}\ \Big(\prod_{j=1}^{n-1}\,
  \int_0^{2\pi}\, \frac{d\Phi_j}{2\pi} \Big)\ 
 \cdot  \,  Y\,\cdot  \, \frac{1+X}{1-X}\,\cdot\, X^{n-1} \, \cdot  \, G\ ,
\end{eqnarray}
where, defining  $\Phi_0$ by $ \, \, \sum_{i = 0}^{n-1}\,  \Phi_i \,=\, 0$, we set 
\begin{eqnarray}
\hspace{-0.9in}&&X \,=\, \,  \prod_{i = 0}^{n-1}\, x_i , \,\,  \,
\quad 
x_i\, = \,\, \frac{2w}{A_i  + \sqrt{A_i^2-4 w^2} },\,\, \, 
\quad \, Y \, = \,   \prod_{i = 0}^{n-1}\, y_i, \, 
\, \,\,  \quad \,
y_i\, =\,\, \frac{1}{\sqrt{A_i^2 \,-4 w^2} },
\nonumber \\
\hspace{-0.9in}&& G \, = \,\,
 \prod_{0\le i < j\le n-1}  \, \frac{2 \,-2 \cos{(\Phi_i-\Phi_j)}}{(1\,-x_i\, x_j)^2}, 
\quad \,  \hbox{where:} \qquad
A_i \,=\,\, \, 1 \, \, -2 \,w \cos(\Phi_i).
\end{eqnarray}
The integrality property (\ref{closedn}) had been checked~\cite{Khi6} 
for the first $ \tilde{\chi}^{(n)}$  
 and inferred~\cite{bo-bo-ha-ma-we-ze-09} for generic $n$.
We are going to {\em prove it\footnote[2]{Actually we only prove global 
boundedness for the Taylor expansion
 $\tilde{\chi}^{(n)}(w)\,=\,\,\sum\, a_k\, w^k $. However, looking 
et the process more carefully, and, in particular, adding corresponding 
properties on the set  $\,\mathcal{T}_n$ below, one can find out that 
 only powers
of 2  appear in the denominators of the $\, a_k$: the rescaling 
factor (``Eisenstein constant'') is $\,2$ or $\,4$ according to the fact that
one considers high or low temperature series~\cite{Khi6,High},
and that $\sum a_k w^k$  do converge for $|w|<1/4$.}
 for any integer $n$, showing a much fundamental result, namely that
all the $(n-1)$-fold integrals  $\, \tilde{\chi}^{(n)}$'s are very special}:
they are actually {\em diagonals of rational functions}.

\vskip .1cm

\subsection{ $ \, \tilde{\chi}^{(3)}$'s as a toy example}
\label{toy}

At first sight the $ \, \tilde{\chi}^{(n)}$'s are involved transcendental holonomic 
functions. Could it be possible that they correspond to the
 {\em distinguished class}~\cite{BA-JPB} of $G$-functions, 
generalising algebraic functions, which have an interpretation as
 diagonals of multivariate algebraic functions (and consequently diagonals 
of rational functions with twice more variables)? If this is the case, 
then the series of the $ \, \tilde{\chi}^{(n)}$'s
must {\em necessarily reduce modulo any prime to an algebraic function}
 (see (\ref{diagprime})). 
The $ \, \tilde{\chi}^{(1)}$ and $ \, \tilde{\chi}^{(2)}$ contributions
being too degenerate
(a rational function and a too simple elliptic function), 
let us consider the first non-trivial case, namely $ \, \tilde{\chi}^{(3)}$. Its 
series expansion has already been displayed in~\cite{ze-bo-ha-ma-04}. It reads
$\, \, {\tilde{\chi}}^{(3)}/8   \,\,=\,\,\, \, w^9 \cdot \, F(w)\, $ with: 
\begin{eqnarray}
\hspace{-0.9in}&& F(w)\,\, \,=\,\,\, \, 
1 \, \, + \, 36 \, w^2 \, + \, 4 \, w^3 \, 
+ \, 884 \, w^{13} \, + \, 196 \, w^5 \, + \, 18532 \, w^6 \,
 + \, 6084 \, w^7 \, \, \, \, + \, \, \cdots 
\nonumber 
\end{eqnarray}
Since we have obtained the exact ODE satisfied by $\, \tilde{\chi}^{(3)}$
we can produce as many coefficients as we want in its series expansion.
Let us consider this series modulo the prime $\, p \, = \, \, 2$. It now reads
the lacunary series
\begin{eqnarray}
\hspace{-0.9in}&&F(w) \quad  \, \bmod 2  \,\, \,\, = \,\,\, \,\,
1 \, \, + w^8\, +w^{24} \, +w^{56} \, +w^{120} \, +w^{248} \, +w^{504} \, \,
 +w^{1016} \,   \, + \cdots,   \nonumber
\end{eqnarray}
solution of the functional equations on $\, F(w)$ or, with
 $\, z \, = \, \, w^8$, on
 $\, G(z) \, = \, \, 1 \, + \, w^8 \cdot \,  F(w)$ 
\begin{eqnarray}
\hspace{-0.9in}&&\quad \quad \quad \quad  F(w)\,\,=\,\,\, 
1 \, \, + \, \, \, w^8 \cdot \, F(w^2), \qquad \quad 
G(z) \, \,  \, = \, \,  \, z \, \, + \, \, G(z^2),   
\end{eqnarray}
where one recognises, with equation $\, G(z) \,  = \,  \, z  \, + \,  G(z^2)$,
Furstenberg's example~\cite{Fu}
 of the simplest algebraic function in 
characteristic~2\footnote[3]{Modulo the prime $\, p\, = \, 2$,
 the previous functional equation becomes 
$ \, G(z) \, \,  \, = \, \,  \, z \, \, + \, \, G(z)^2$.}.
In fact $ \, H(w) \, = \, \, \, w^9 \, F(w)\, $ is solution of the 
 quadratic equation:
\begin{eqnarray}
\label{degree2}
\hspace{-0.9in}&&\qquad \qquad \quad 
H(w)^2\, \,  + w \cdot \, H(w) \,\,\,   + w^{10}
  \,   \, \, \,  \, = \, \, \, \, \,  \,   0 \quad \mod 2.  
\end{eqnarray}
The calculations are more involved modulo $\, p  = \, 3$.
Indeed, $H(w)=\, \, {\tilde{\chi}}^{(3)}(w)/8$ satisfies, modulo 3,
 the polynomial equation of degree nine
\begin{eqnarray}
\label{degree9}
\hspace{-0.95in}&&\quad \quad \quad \quad p_9 \cdot \, H(w)^9 \,\, \, 
 + \, \,  w^{6} \cdot \, p_3 \cdot \, H(w)^3  \, \, \, 
+ \, \,  w^{10} \cdot \, p_1 \cdot \, H(w) \,\, \,
 \nonumber   \\
\hspace{-0.95in}&& \qquad \qquad \qquad \quad 
 \qquad \qquad \qquad \qquad 
+ \, \, w^{19} \cdot \, p_0^{(1)}\cdot \, p_0^{(2)}\, 
 \, \,\,\, \,  = \, \,\, \,\,  \, 0, 
\end{eqnarray}
where:
\begin{eqnarray}
\hspace{-0.95in}&&\quad p_9 \, \, = \, \, \,\, 
  \, (w+1) ^{3} \, ({w}^{2}+1)^{18} \, (w-1)^{24}, 
\nonumber \\
\hspace{-0.95in}&&\quad p_3 \, \, = \, \, \,
 ({w}^{2}+1)^{18} \, (1-w)^{15} \, ({w}^{4}-{w}^{2}-1),
\quad \, \, p_1 \, \, = \, \, \,
 ({w}^{2}+1)^{20} \, (1-w)^{13}, \
\nonumber \\
\hspace{-0.95in}&&\quad p_0^{(1)} \, \, = \, \, \,
{w}^{6}+{w}^{5}+{w}^{4}-{w}^{2}-w+1,
  \\
\hspace{-0.95in}&&\quad p_0^{(2)} \,  = \,  \,{w}^{37} \, - {w}^{36} \, 
+{w}^{35} \, - {w}^{33} \, +{w}^{31} \, -{w}^{30} \, +{w}^{28} \,
 +{w}^{27} \, +{w}^{24} - {w}^{23} +{w}^{22}
 \nonumber  \\
\hspace{-0.95in}&&\qquad \quad   \, 
- {w}^{21} - {w}^{18} - {w}^{16}+{w}^{14} \, - {w}^{12} \, 
-{w}^{11} \, -{w}^{10}\, +{w}^{7} \, -{w}^{5} \, -{w}^{3}\, -1.
 \nonumber 
\end{eqnarray}

The calculations are even more involved modulo larger primes.
The series for $ \, \tilde{\chi}^{(3)}$  mod. $5$ reads:
\begin{eqnarray}
\label{chi35}
\hspace{-0.95in}&&\quad \quad  \tilde{\chi}^{(3)}
 \,\,  \, = \, \, \,\, \,  w^9 \, \, + \, {w}^{11} \, 
+4\,{w}^{12} \,+4\,{w}^{13} \,+{w}^{14} 
\,+2\,{w}^{15} \,+4\,{w}^{16} 
\, \,+ \,\, \cdots 
\end{eqnarray}
The (minimal order) linear differential
 operator annihilating
the  $ \, \tilde{\chi}^{(3)}$ series mod. $5$,
 reads\footnote[1]{This operator is of zero 
$\, 5$-curvature~\cite{bo-bo-ha-ma-we-ze-09}.
}:
\begin{eqnarray}
\hspace{-0.95in}&&\quad (x+1)  \,  ({x}^{2}+x+1)
  \,  (x+2) \cdot \,  {x}^{4} \cdot \,  D_x^{4}
\, \,\,  +2\,{x}^{3} \cdot \, ({x}^{3}+2\,{x}^{2}+4\,x+4) 
 \, (x+4) \cdot \,  D_x^{3} \, 
\nonumber \\
\hspace{-0.95in}&&\quad \quad \quad \quad \quad 
  +{x}^{2} \cdot \, ({x}^{4}+3\,{x}^{3}+4) \cdot \,   D_x^{2}
\,\,   +4  \cdot  \, ({x}^{4}+3) \cdot \,   x \cdot \,   D_x \,\,   +3
\end{eqnarray}
If one can easily get this  linear differential
 operator, finding the minimal polynomial of $\, \tilde{\chi}^{(3)}$ 
modulo $\,5$, generalising
 (\ref{degree2}) or (\ref{degree9}),
such that $\, P(\tilde{\chi}^{(3)}(w), \, w) \, = \, \, 0 \, \, $ 
mod. $5$, requires a {\em very large} 
number of coefficients. 
 Since the series (\ref{chi35})
 starts with $\, w^9$, it is more convenient to consider 
the polynomial $\, \tilde{P}(\kappa, \, w)$, relating
 $\, \kappa \, = \, \, \tilde{\chi}^{(3)}(w)/w^9$
and $\, w$.  This (minimal) polynomial\footnote[2]{This
 polynomial has been checked 
with a series (\ref{chi35}) of 
380000 coefficients.} is a polynomial of degree 50 in $\, \kappa$, 
and degree 832 in $\, w$, sum of 4058 monomials. This 
(minimal) polynomial of the form:
\begin{eqnarray}
\hspace{-0.95in}&&\quad \quad \tilde{P}(\kappa, \, w) \, \,\, = \, \, \,\,
 P_{50}^{(832)}(w) \cdot \, \kappa^{50} \, + \, P_{30}^{(652)}(w) \cdot \, \kappa^{30} 
\, + \, P_{26}^{(612)}(w) \cdot \, \kappa^{26}
\nonumber \\
\hspace{-0.95in}&&\quad \quad \quad  \quad \, + \, P_{25}^{(601)}(w) \cdot \, \kappa^{25}
 \, + \, P_{10}^{(472)}(w) \cdot \, \kappa^{10} 
 \, + \, P_{6}^{(432)}(w) \cdot \, \kappa^{6} 
\, + \, P_{5}^{(421)}(w) \cdot \, \kappa^{5} 
\nonumber \\
\hspace{-0.95in}&& \quad \quad \quad  \quad 
 \, + \, P_{2}^{(392)}(w) \cdot \, \kappa^{2} 
 \, + \, P_{1}^{(381)}(w) \cdot \, \kappa\, + \, P_{0}^{(369)}(w),
\end{eqnarray}
where the $\, P_{n}^{(m)}(w)$'s are polynomials of degree $\, m$ in $\, w$,
and where the head polynomial reads:
\begin{eqnarray}
\hspace{-0.95in}&& \quad \quad \quad  P_{50}^{(832)}(w)\, \, = \, \, \,
{w}^{382} \cdot \, (w+2)^{20} \, ({w}^{2}+2\,w+4)^{75} \, (w+1)^{70} \,  (w+4)^{20} \,
\nonumber \\
\hspace{-0.95in}&& \quad  \quad \quad \quad  \quad \quad  \quad 
\times \,   ({w}^{2}+3\,w+4)^{75}
 \, ({w}^{4}+4\,{w}^{3}+w+1)^{10}.
\end{eqnarray}
This (minimal) polynomial is a factor of a much larger polynomial 
(in  $\,1$,  $\,\kappa$, $\,\kappa^5$, $\,\kappa^{5^2}$, and
$\,\kappa^{5^3}$)
 of a more ``$p$-adic nature'',
 which is of degree 125 
in $\,\kappa$,  sum of 
3559 monomials, and of degree 1941 in $\, w$. 

One can imagine, in a first step that the
 $ \, \tilde{\chi}^{(3)}$ series mod. {\em any prime} $\, p$
are {\em also algebraic functions}, and, 
in a second step, that $ \, \tilde{\chi}^{(3)}$ 
may be the diagonal of a rational function. In fact we are going to show, 
in the next section, a stronger result: 
the $ \, \tilde{\chi}^{(n)}$'s are 
{\em actually diagonals of rational functions, for any integer $\, n$}. 

\subsection{The $ \, \tilde{\chi}^{(n)}$'s are diagonals of rational functions}
\label{calcula}
Let us, now, consider the general case where $\, n$ is an arbitrary 
integer.

With the change of variable $\,z_i\,=\,\exp(\iii \Phi_i)$ (where $\iii^2 =\, -1$),
 one clearly gets
\begin{eqnarray}
\hspace{-0.9in}&&\qquad \qquad \qquad \quad  \prod_{i = 0}^{n-1} z_i \,\,=\,\,\, 1\ , 
\quad \quad \quad  \, \, \frac{dz_j}{z_j}\,  = \, \, \iii \,d \Phi_j\ ,
\\ 
\hspace{-0.9in}&& \qquad \qquad 2 \cos(\Phi_i)\,\,\, =\,\,\, \, z_i\, +\frac1{z_i}\ , 
\quad \quad \quad 
  2 \cos( \Phi_i - \Phi_j )\,\, =\,\, \,  \frac{z_i}{ z_j} + \frac{z_j}{ z_i} \ ,
\nonumber 
\end{eqnarray}
and \eqref{chin} becomes
\begin{eqnarray}
\label{prop1}
\hspace{-0.4in}&&\tilde{\chi}^{(n)}(w) \,\, \, = \, \,\,\, \, \frac {(2w)^n}{n!}\ 
 \Big( \prod_{j=1}^{n-1} \, \frac{1}{2\iii \pi}\,  \oint_{{\cal C}} \,  \frac{dz_j}{z_j} \Big) \cdot 
\, F(w,z_1,\dots, z_{n-1} )\ ,
\end{eqnarray}
where $\, {\cal C}$ is the path ``turning once counterclockwise around the unit circle''
and where $\, F$ is algebraic over $\, \mathbb{Q} (w, z_1,\,\dots,\, z_{n-1})$ and reads:
\begin{eqnarray} 
\label{integrantF}
 F(w,\, z_1,\dots,\,  z_{n-1}) \, \,\,\, = \, \,\, \,\,\,  \, 
 Y \cdot \, X^{n-1}  \cdot \,{{1 \, + \, X} \over {1 \, - \, X}}  \cdot \, G. 
\end{eqnarray}

\vskip .1cm

Now, let us suppose that $ \,F\, $  {\em is analytic\footnote[1]{One could  
consider Laurent, instead of Taylor, expansions, but this is a slight 
generalisation. The rational function $\,1/(x+xy)$ would become allowed but 
rational functions like  $\, 1/(x+y)$ would remain forbidden. For
 similar purposes, B. Adamczewski and
 J. P. Bell~\cite{BA-JPB} recently used a  generalised notion of Laurent 
expansions in the several variables case due to Sathaye~\cite{Sathaye}, 
see also~\cite{Kauers} for various other generalisations.}
at the origin}, namely that it has a Taylor expansion \eqref{defdiag}.
Then applying $(n-1)$ times the residue formula, one finds 
\begin{eqnarray}
\label{prop2}
\hspace{-0.6in}\tilde{\chi}^{(n)}(w) 
\, \,\,\,  = \, \, \, \, \,\,
 \Diag\Big(\frac {(2\,z_1\cdots z_n)^n}{n!}\ 
 \cdot 
\, F(z_1\cdots z_{n-1} z_n,\,\,  z_1,\, \dots,\,  z_{n-1})\Big).
\end{eqnarray}
To check that this is actually true,
 we introduce an auxiliary set, namely
 ${\cal T}_n$  
 the subset of
  $\mathbb{Q} [ z_1,\, \dots,\,  z_{n-1}, z_1^{-1}, \dots , z_n^{-1} ] [[w]]$, 
consisting of series
\begin{eqnarray}
\hspace{-0.2in}f(w,\, z_1,\, \dots,\,  z_{n-1} )
\,\,\, \,\,  = \, \, \,\, \, \,
 \sum_{m=0}^\infty\,  P_m \cdot \,  w^m,
\nonumber
\end{eqnarray}
where 
$\, P_m$ belongs to $\,\,  \mathbb{Q} [ z_1,\dots, z_{n-1}, z_1^{-1}, \dots , z_n^{-1} ]$
and is such that $\,\,\, \widetilde{ f}(z_1, \, \cdots \, ,z_n) \, \,  $ belongs to
$\,\,\,\,\mathbb{Q} [ z_1,..., z_{n-1} ] [[z_n]]\,\,\, \subset \,\,\,
 \mathbb{Q} [[ z_1,...,z_n]]$, 
where $\, \widetilde{ f}$ is defined by: 
\begin{eqnarray}
\label{defwilde}
\hspace{-0.2in}\widetilde{ f}(  z_1, \, \cdots \, ,z_n)
\,\,\,\, \, = \, \,\, \, \,\,
f(z_1 \cdots z_n, \, \,   z_1, \, \cdots \, ,z_{n-1} )
\end{eqnarray} 

In other words, we ask the degree of $\, P_m$, in each of the $\, z_i^ {-1}$,
to be at most~$\, m$.
Then to prove that~$\widetilde{F}$ has a Taylor expansion, we only have
 to  verify that $\, F$ belongs to~$\, {\cal T}_n$.
Checking this is a straightforward step-by-step computation on auxiliary functions:
\begin{eqnarray}
\label{G}
\hspace{-0.8in}&&  A_i\,\,  =\, \,\,\,    1 \,\, \,   -w \cdot (z_i \, +\frac{1}{z_i}), 
\quad \quad \quad \quad \quad \quad \quad
 \quad \quad \quad \quad  \quad  \,  
\hbox{for:} \quad \quad
 1\, \le\, i \,\le \, n-1, \nonumber \\
\hspace{-0.8in}&&  A_i\,\,  =\, \,\,   \, \,   
1 \,\, \,   -w \cdot \, \big(\frac{1}{z_1 \cdots \, z_{n-1}} \, + z_1\cdots \, z_{n-1}\big),  
\quad \quad \quad \quad \hbox{for:} \quad \quad  i\,= \,\, 0,  \\
\hspace{-0.8in}&& \tilde{A}_i\, \, = \, \,\, \,  \, \,  
 1 \, \,\,   -z_1 \,\, \cdots \, \, z_{i-1} \cdot \, z_{i+1} \,  z_n
 \, \cdot\,   (z_i^2 \, +1),
\quad \quad \quad\quad  \hbox{for:} \quad \quad  1\, \le \, i\, \le \, n-1 , 
\nonumber  \\
\hspace{-0.8in}&& \tilde{A}_i\, \, = \, \,\,\,\,    
 1 \,\, \,   - z_n \,\cdot \,\big(1 \, +z_1^2 \,  \cdots \, z_{n-1}^2\big),  
\quad  \quad \quad \quad \quad  \quad \quad \quad
\hbox{for:}  \quad  \quad  \quad  i\,= \,\, 0.
\nonumber 
\end{eqnarray}

Hence $\, A_i \, \in \, {\cal T}_n$. 
The set $\, {\cal T}_n$ being clearly a $\mathbb{Q}[w]$-algebra, 
$A_i^2\, -4w^2 \, \in \, {\cal T}_n$. 
But ${\cal T}_n$ is complete for the $w$-adic valuation.
In particular, it is stable by the 
operations\footnote{More generally it is stable by the operations
$\, \,  f \, \rightarrow \, (1\, +\,w\, f)^{\delta} \, $ for 
$\, \delta\,  \in\,  \mathbb{Q}$.}
\begin{eqnarray}
\hspace{-0.7in}f(w) \, \, \,\,\, \,  \longrightarrow  \, \, \, \, \,\quad 
\frac1{1\, + w  \cdot \,f(w)}\,\, \,  \, = \,\,\, \, \,\, 
 1\, \, \,\,  -w \cdot \, f(w)\, \, \,\,  \,  +\,\, \cdots,
\qquad \quad \hbox{and:} \\
\hspace{-0.7in} f(w)  \, \,\,\,  \,\,  \longrightarrow \, \,\, \,  \,\quad 
 \sqrt{1\,+ \, w  \cdot \, f(w)}
\, \, \,\,   = \, \,\,   \, \,  \, 1 \, \,\,  \,  +\frac12 \,  w \cdot \, f(w)
\, \,  \,\,  \, +\,\, \cdots 
\end{eqnarray}
So to be sure that the inverse or the square root of some function
 in ${\cal T}_n$ is also in ${\cal T}_n$
 we have only to check that its first Taylor coefficient is actually $1$:
\begin{eqnarray}
\hspace{-0.6in}&&A_i^2\,-4\,w^2\,\,\,\, =
\,\,\,\,\, \, \,\, \,  1\,\, \,  \,\, \,
-2\,w \cdot \,(z_i\,+\frac{1}{z_i})\,\,  \, + w^2 \cdot \,(z_i\,-\frac{1}{z_i})^2, 
 \\
\hspace{-0.6in}&&\quad \quad \hbox{hence} \quad  \quad  \quad  \quad \quad
 \sqrt{A_i^2\,-4 w^2}\,\,\,  =\,\,\,\, \, \,
1\, \, \,\, +\,\,  \cdots \ \,\,\, \, \in \,\,\, {\cal T}_n, 
\nonumber
\end{eqnarray}
\begin{eqnarray}
\hspace{-0.9in}&&y_i\, \,  =\, \, \,\,   \frac{1}{\sqrt{A_i^2\, -4w^2}}
 \,\,  =\,\, \,\, \,  \,
  1\, \,\,   \, +\, \cdots  \quad \,\, \in \, \,\, {\cal T}_n,  
\qquad \quad Y\,\,  =\,\,  \,  1\,\,  \, +\, \cdots  \quad\,  \in \,\,  \,\, {\cal T}_n,
 \nonumber \\
\hspace{-0.9in}&& x_i\,\,  =\, \,\,   \frac{2\,w}{A_i\,+\sqrt{A_i^2\,-4 w^2}}
 \, =\, \,\,  \, w\,\,\,  \, + \,\, \cdots \,\, \,  \in \,\,  \,\, {\cal T}_n, 
\quad \, \,  \, \,  \,   x_i \, x_j \,\, =\,\,\,
 w^2\, \, +\,  \, \cdots\,\,\,\, \, \in  \,\,\, {\cal T}_n,
 \nonumber 
\end{eqnarray}
\begin{eqnarray}
\hspace{-0.8in}&& X\,= \,\,\, \, 
w^n\,  \,\, +\,\,  \cdots \, \,\, \in \,\,   \,\,{\cal T}_n, 
\quad \quad \quad \quad \quad 
\frac{1+X}{1-X}\,\,\,  = \, \,\, \,  1\,\,\,   +\, \cdots \,\,  \,\,\,  \in \,\,   
 \,\,\, {\cal T}_n, 
\nonumber 
\end{eqnarray}
\begin{eqnarray}
\hspace{-0.8in}&&
 G\,\,=\,\,\,
\prod_{0\le i<j\le n-1} \frac{(z_i\,-z_j)^2}{(1\,-x_i x_j)^2 \cdot \,z_i\,z_j }
\,\, \,  \,\,= \,\,\,
\prod_{0\le i<j\le n-1} {{(z_i-z_j)^2} \over {z_i\,z_j }}
\,\,\, \, \, 
 +\,\, \cdots \,\, \,\,\,   \in \,\,\,   {\cal T}_n.
\nonumber
\end{eqnarray}
From the definition of $\, \Phi_0$ which implies  
$\,\, \prod_{i=0}^{n-1} z_i\, = \, 1\,$ 
we also have 
\begin{eqnarray}  
\hspace{-0.3in}\quad  \prod_{0\leq i<j\leq n-1}\,  z_i z_j
\,\, \, \,= \, \,\, \,\Big(\prod_{i=0}^{n-1} z_i\Big)^{n-1}\,\,\, =\,\,\, \, 1, 
\end{eqnarray}
enabling to rewrite $\, G$ in other ways.

Thus, $\, F$ belongs to $\,{\cal T}_n$ and it makes sense to take its
diagonal. As~$\, F$ is algebraic, $\, \tilde{\chi}^{(n)}$ {\em is the diagonal
of an algebraic function} of $\, n$ variables and, consequently, {\em the
diagonal of a rational function} of $ \, 2\, n$ variables.

\vskip .1cm 

We thus see that we can actually {\em find explicitly}
the algebraic function such that its diagonal is 
the $\, n$-fold integrals $\, \widetilde{\chi}^{(n)}$: it is
 {\em nothing but the integrand} of the $\, n$-fold integral, 
up to trivial transformations, namely (\ref{integrantF}).

\vskip .2cm

{\bf Remark }: $\tilde{\chi}^{(n)}$ is a solution of a linear 
differential equation, and has a radius of convergence equal to $\, 1/4$ 
in  $w$. Among the other solutions 
of this equation, there is the function obtained by changing the radical 
appearing in $\, x_i$ into its opposite. A priori there are 
$\, 2^n$ ways to do this, hence
$\, 2^n$ new solutions but, not all distinct. At first sight, for
these new solutions, the $\, x_i$'s are no longer in $\, {\cal T}_n$. 

In fact, we find some quite interesting structure. Let us consider, for
instance, the case of $\tilde{\chi}^{(3)}$. If one considers other choices of
sign in front of the nested square roots in the integrand,
the series expansions of the corresponding $\, n$-fold integrals read:
\begin{eqnarray}
\label{thesetwo}
\fl \qquad w +6\,{w}^{2} +28\,{w}^{3} +124\,{w}^{4} +536\,{w}^{5} +2280\,{w}^{6}
+9604\,{w}^{7}+40164\,{w}^{8}
 \nonumber \\
\fl \qquad \qquad \quad  + 167066\,{w}^{9} +692060\,{w}^{10} +2857148\,{w}^{11}
 \,  \, + \cdots 
\nonumber \\
\fl \qquad {w}^{2} +6\,{w}^{3} +30\,{w}^{4} +140\,{w}^{5} +628\,{w}^{6} +2754\,{w}^{7}
+11890\,{w}^{8} +50765\,{w}^{9}
 \nonumber \\
\fl \qquad \qquad \quad +214958\,{w}^{10} +904286\,{w}^{11} \,  \, + \cdots 
\end{eqnarray}
These two series expansions (\ref{thesetwo}) are solutions
of the same order-seven operator~\cite{ze-bo-ha-ma-04}
 $\, L_7$ as  $\tilde{\chi}^{(3)}$.

We know that other forms (equivalent for integration purposes) of $G$ exist
(see~\cite{nickel-99,nickel-00,2005-chi3-method}). For these forms, other
choices of sign in front of the nested square roots give
\begin{eqnarray} 
\hspace{-0.8in}&&S(+,-,+) \, \, = \, \, \, w +6\,{w}^{2}+28\,{w}^{3}+126\,{w}^{4}
+552\,{w}^{5}+2388\,{w}^{6}
+10192\,{w}^{7}
 \nonumber \\
\hspace{-0.8in}&& \qquad \qquad +43238\,{w}^{8} +181936\,{w}^{9}
+762836\,{w}^{10} +3180800\,{w}^{11} 
 \, \,\,  + \cdots 
\nonumber 
\end{eqnarray}
\begin{eqnarray}
\hspace{-0.8in}&&
S(+,+,-) \, \, = \, \, \, w-2\,{w}^{2}-20\,{w}^{3}-110\,{w}^{4}
-552\,{w}^{5}-2536\,{w}^{6}-11428\,{w}^{7}
\nonumber \\
\hspace{-0.8in}&& \qquad \qquad -49898\,{w}^{8} -216016\,{w}^{9}-920776\,{w}^{10} 
-3905764\,{w}^{11}
 \, \, \,   + \cdots 
\nonumber 
\end{eqnarray}
\begin{eqnarray}
\hspace{-0.8in}&&
S(-,-,+) \, \, = \, \, \,w \, +6\,{w}^{2} +8\,{w}^{3} +14\,{w}^{4}
 -84\,{w}^{5} -596\,{w}^{6}-4004\,{w}^{7}
-19610\,{w}^{8}
 \nonumber \\
\hspace{-0.8in}&& \qquad \qquad -99148\,{w}^{9}-447332\,{w}^{10}
-2068492\,{w}^{11} 
\,  \, \,  + \cdots 
\nonumber \\
\hspace{-0.8in}&&
S(-,+,-) \, \, = \, \, \, w \, +10\,{w}^{2}+44\,{w}^{3}+202\,{w}^{4}
 +848\,{w}^{5} +3672\,{w}^{6}
+15200\,{w}^{7}
 \nonumber \\
\hspace{-0.8in}&& \qquad \qquad +64310\,{w}^{8} +264424\,{w}^{9} +1104872\,{w}^{10} 
+4523656\,{w}^{11} 
\,\, + \cdots 
\nonumber 
\end{eqnarray}

\begin{eqnarray}
\hspace{-0.8in}&&
S(+,-,-) \, \, = \, \, \, {w}^{2} \, +3\,{w}^{3} +13\,{w}^{4} 
+47\,{w}^{5} +189\,{w}^{6} +707\,{w}^{7}
+2800\,{w}^{8}
 \nonumber \\
\hspace{-0.8in}&& \qquad \qquad +10637\,{w}^{9} +41865\,{w}^{10}
 +160535\,{w}^{11}\,\,   + \cdots 
\nonumber 
\end{eqnarray}
\begin{eqnarray}
\hspace{-0.8in}&&
S(-,+,+) \, \, = \, \, \, {w}^{4} \, +4\,{w}^{5}+25\,{w}^{6}
+103\,{w}^{7} +496\,{w}^{8} +2042\,{w}^{9}
+9013\,{w}^{10} \nonumber \\
\hspace{-0.8in}&& \qquad \qquad +36931\,{w}^{11} \,\,  + \cdots 
\nonumber
\end{eqnarray}
to be compared with $\tilde{\chi}^{(3)} \, = \, \, S(+,+,+) \, = \, S(-,-,-)$.
These alternative series are not solutions of 
the order-seven operator~\cite{ze-bo-ha-ma-04}
 $\, L_7$ annihilating  $\tilde{\chi}^{(3)}$, however, some linear 
combination of these series are solutions of $\, L_7$. 
For instance the  linear 
combination $\,S(+,+,-)-S(-,+,+)-S(+,-,+)$,  
which is actually equal to $\,S(-,+,-)+4\, S(+,-,-)-S(-,-,+)$, 
is solution of $\, L_7$. 

Another form  of $G$ (again equivalent for integration purposes) 
gives:
\begin{eqnarray}
\hspace{-0.8in}&&
\tilde{S}(+,-,+) \, \, = \, \, \, w \, +8\,{w}^{2}+36\,{w}^{3}+164\,{w}^{4}
+704\,{w}^{5}+3041\,{w}^{6} +12786 \,{w}^{7} 
 \nonumber \\
\hspace{-0.8in}&& \qquad \qquad +54067\,{w}^{8} +224864\,{w}^{9}
+939709\,{w}^{10}+3881708\,{w}^{11}
 \,  \, \, + \cdots 
\nonumber 
\end{eqnarray}
\begin{eqnarray}
\hspace{-0.8in}&&\tilde{S}(+,+,-) \, \, = \, \, \,w \, +2\,{w}^{2}-6\,{w}^{3}
-48\,{w}^{4}-314\,{w}^{5}-1555\,{w}^{6}
-7626\,{w}^{7} 
 \nonumber \\
\hspace{-0.8in}&& \qquad \qquad -34461\,{w}^{8}-155898\,{w}^{9} -678199\,{w}^{10}
-2957648\,{w}^{11} \, \, \, + \cdots 
\nonumber \\
\hspace{-0.8in}&&
\tilde{S}(-,+,+) \, \, = \, \, \, {w}^{2}\, +3\,{w}^{3} +14\,{w}^{4} +51\,{w}^{5}
 +214\,{w}^{6} +810\,{w}^{7} +3296\,{w}^{8} 
\nonumber \\
\hspace{-0.8in}&& \qquad \qquad +12679\,{w}^{9}  +50878\,{w}^{10} 
+197466\,{w}^{11} \, + \cdots 
\nonumber 
\end{eqnarray}
Again, these other alternative series are not solutions of 
 $\, L_7$, but  the  linear 
combination $\,2 \, \tilde{S}(-,+,+)+\tilde{S}(+,-,+)-\tilde{S}(+,+,-)$
is solution of $\, L_7$. 

All these alternative series are, in fact, solution of a higher order 
linear differential operator that $\, L_7$ rightdivides.

One does remark that {\em all these alternative series} are,
as $\tilde{\chi}^{(3)}$, series with {\em integer coefficients}.

\vskip .1cm

\subsection{More $\, n$-fold integrals of the Ising class 
and a simple integral  of the Ising class}
\label{Isingclass}

It is clear that the demonstration we have performed on the $\, \chi^{(n)}$'s
can also be performed straightforwardly, mutatis mutandis, with other $\, n$-fold 
integrals of the ``Ising class'' like the $\, n$-fold integrals 
$\, \Phi_H$ in~\cite{bo-ha-ma-ze-07}, which amounts to getting
 rid of the fermionic term 
$\, G$ (see (\ref{G})), the  $\, \chi_d^{(n)}$'s corresponding to  $\, n$-fold 
integrals associated with the diagonal\footnote[1]{Of course this 
 ``diagonal wording'' should not be confused with the notion of diagonal
of a function.} susceptibility~\cite{mccoy3,CalabiYauIsing}
(the magnetic field is located on a diagonal of the square lattice), 
 the $\, \Phi_D^{(n)}$'s
 in~\cite{bo-ha-ma-ze-07b} which are simple integrals, and also 
for all the lattice Green functions displayed 
in~\cite{GoodGuttmann,Guttmann}, and the list is
far from being exhaustive. For instance,  the simple integral 
 $\, \Phi_D^{(n)}$ is the
 diagonal of the algebraic function:
\vskip .1cm 

\begin{eqnarray}
\label{good}
\hspace{-0.6in}\frac{2}{n!} \cdot (1-t^2)^{-1/2} \cdot { G_n \, {F_n^{n-1} } \over {
G_n \, F_n^{n-1}  \, -\, (2 \, w \,t)^n}} \, \, \, \,\,   - \, \frac{1}{n!} \,,
 \qquad   \quad  \hbox{where:} 
\end{eqnarray}
\begin{eqnarray}
\hspace{-0.6in}&&F_n \, \, = \, \, \, \,\, \,
1\,\,\,  \, \,-2\, w \,\, \, \, +(1-4\,w\,+4\,w^2\,- 4\,w^2\,t^2)^{1/2},  \, \\
\hspace{-0.6in}&&G_n \, \, = \, \, \, \,\, \,
1 \, \,\,\,   \,-2\,w \, t\cdot \,T_{n-1}\Bigl({{1} \over {t}}\Bigr)\, \,\, \,
+\Bigl(\big(1 \, \,\,  \,-2\,w \, t\cdot \,T_{n-1}\Bigl({{1} \over {t}}\Bigr) \big)^2
 \, - 4\,w^2 \cdot \,t^2\Bigr)^{1/2},
 \nonumber 
\end{eqnarray}
and where $\, T_{n-1}(t)$ is the $\, (n-1)$-th Chebyshev polynomial of the first kind.
The way we have obtained these Chebyshev results (\ref{good}) is displayed in \ref{Chebi}.

\vskip .2cm 

The integral $\, \Phi_D^{(n)}(w) $ is the diagonal
 of an algebraic function of {\em two} variables, and also the 
diagonal of a rational function of \emph{four\/} variables, and this 
{\em independently of the actual value of $n$}.

\vskip .2cm 

\subsection{More general $\, n$-fold integrals as diagonals}
\label{gener}

More generally   the demonstration we have performed
 on the  $\, \tilde{\chi}^{(n)}$'s
can be performed for {\em any} $\, n$-fold integral that can be recast
in the following form:
\begin{eqnarray}
\label{form}
\hspace{-0.5in}\quad \int_{C}  \, \int_{C} \,  \cdots \, \int_{C} \,
 {{dz_1} \over {z_1}}  \, {{dz_2} \over {z_2}} \,\, \, \cdots \,\,\,  {{dz_n} \over {z_n}}
\,\cdot \,  {\cal A}\Bigl(x, \, z_1, \, z_2, \, \cdots, \, z_n \Bigr), 
\end{eqnarray}
where the subscript $\, C$ denotes the unit circle, and where $\,{\cal A}$ denotes 
an algebraic function of the $\,n$ variables, which (this is the crucial ingredient),
as a function of several variables $x$ and the $\, z_i$'s, has 
an {\em analytical} expansion at $(x , \, z_1, \, z_2, \, \cdots \, z_n) \, = \, \,$
$ (0 , \, 0, \, 0, \, \cdots,  \, 0)$:
\begin{eqnarray}
\label{form2}
\hspace{-0.6in}&&{\cal A}\Bigl(x, \, z_1, \, z_2, \, \cdots, \, z_n \Bigr)
\, \,\,  \,\, = \\
\hspace{-0.6in}&& \quad \quad 
\sum_{m \, = \, 0}^{\infty} \, \sum_{m_1 \, = \, 0}^{\infty} \, \sum_{m_2\, = \, 0}^{\infty} 
 \,\, \cdots \, \sum_{m_n\, = \, 0}^{\infty} 
 \,A_{m, \, m_1, \, m_2, \, \cdots, \, m_n}
\cdot  \, z_1^{m_1} \, z_2^{m_2} \,\,  \cdots \, \,z_n^{m_n} \, \cdot x^m.
\nonumber
\end{eqnarray}

Consequently, an extremely large set of $\, n$-fold integrals occurring in
theoretical physics (lattice statistical mechanics, enumerative combinatorics,
number theory, differential geometry, ...) can actually be seen to be
\emph{diagonal of rational functions}. Consequently these $\, n$-fold
integrals correspond to series expansions (in the variation parameter $\,x$)
that are {\em globally bounded} (can be written after one rescaling into
series with integer coefficients), and are solutions of {\em globally
nilpotent}~\cite{bo-bo-ha-ma-we-ze-09} linear differential operators.

Such a general $\, n$-fold integral is, thus, the diagonal of an algebraic
function (or of a rational function with twice more variables~\cite{Denef})
which is essentially the integrand of such $\, n$-fold integral. Furthermore,
such a general $\, n$-fold integral is solution of a (globally nilpotent)
linear differential operator, that can be obtained exactly from the integrand,
using the creative telescoping method (see \ref{telesc}). \vskip .1cm

Finally, in the case of Calabi-Yau ODEs (see below), these functions
can be interpreted as periods of Calabi-Yau varieties, these 
algebraic varieties being essentially the
 integrand of such $\, n$-fold integrals.
The integrand is thus the key ingredient to wrap, in the same bag,
algebraic geometry viewpoint, differential geometry viewpoint and 
analytic and arithmetic approaches (series with integer coefficients). 

\section{Calabi-Yau ODEs}
\label{plan}

Calabi-Yau ODEs have been defined in~\cite{Almkvist} as order-four linear
differential ODEs that satisfy the following conditions: they are maximal
unipotent monodromy~\cite{Morrison,Batyrev} (MUM), they satisfy a ``Calabi-Yau
condition'' which amounts to imposing that the exterior squares of these
order-four operators are of order {\em five} (instead of the order six one
expects in the generic case), the series solution, analytic at $\, x=\, 0$, is
globally bounded (can be reduced to integer coefficients), the series of their
nome and Yukawa coupling are globally bounded\footnote[2]{The 
instantons numbers are integers.}. In the literature, one finds 
also a cyclotomic condition on the monodromy at the point at
 $\, \infty$, $\, x \, = \, \,  \infty$, and/or
 the conifold\footnote[1]{The local exponents
are $0, 1, 1, 2$. For the cyclotomic condition on the
monodromy at $\, \infty$ see Proposition 3 in~\cite{TablesCalabi}.} 
character of one of the singularities~\cite{chen-yang-yui-08}.

Let us recall that a linear ODE has MUM (maximal unipotent
monodromy~\cite{CalabiYauIsing1,TablesCalabi}) if all the
exponents at (for instance) $x= \, 0$ are zero.
 
In a hypergeometric framework the MUM condition amounts to restricting to
hypergeometric functions of the type
$\, _{n+1}F_n([a_1,\,a_2,\, \cdots \, a_n], \, [1, \, 1, \, \cdots \, 1]; x)$,
since the indicial exponents at $\, x \, = 0$ 
 are the solutions of
 $\rho \,(\rho+b_1-1) \cdots (\rho+b_n-1) \, = \, \,$
$\rho^{n+1} \, = \, \, 0$,
where the $b_j$ are the lower parameters which 
are here all equal to $\, 1$.

Let us consider a MUM order-four linear differential operator.
The four solutions $y_0, \,y_1, \,y_2, \,y_3$ of this order-four 
linear differential operator read:
\begin{eqnarray}
\hspace{-0.8in}&&y_0, \quad \quad \, \, \,\,\,\,
y_1 \, = \,\,\, \,\, y_0 \cdot \ln(x)\,\,   \,  +\,   \tilde{y}_1,
 \quad \quad  \, \, \,\,\,\,
y_2 \, = \,\,\, \, y_0 \cdot  {{\ln(x)^2} \over {2}} \,
  \,  +\, \tilde{y}_1\cdot \ln(x)
 \,\,  \,  + \,\,\, \,   \tilde{y}_2,
\nonumber \\
\hspace{-0.8in}&&y_3 \, \, = \,\,\, \,\, \, 
 y_0 \cdot  {{\ln(x)^3} \over {6}} \, \,\, \, 
 + \, \, \tilde{y}_1 \cdot  {{\ln(x)^2} \over {2}} \, \, \, \, 
+ \, \, \tilde{y}_2 \cdot  \ln(x) \, \, \, \, 
+ \,  \, \tilde{y}_3, 
 \nonumber
\end{eqnarray}
where $\, y_0$, $\,\tilde{y}_1$,
 $\,\tilde{y}_2$, $\,\tilde{y}_3$
 are  analytical at $\, x \, = \, \, 0$
(with also $\,  \tilde{y}_1(0) \, = \,
   \tilde{y}_2(0) \, = \, \tilde{y}_3(0) \, = \,  0$).

The nome of this linear differential operator
reads:
\begin{eqnarray}
\label{nome}
q(x) \, \, \, = \,  \,\, \, \, 
\exp\Bigl({{y_1 } \over {y_0 }}  \Bigr)
 \, \, \, \,  = \, \,  \, \, \, \,
x \cdot \, \exp\Bigl({{\tilde{y}_1 } \over {y_0 }}  \Bigr). 
\end{eqnarray}

\vskip .1cm

Calabi-Yau ODEs have been defined as being MUM, thus having 
one solution analytical at $\, x \, = \, \, 0$. 
As far as Calabi-Yau ODEs are concerned, the fact that 
this solution analytical at $\, x \, = \, \, 0$ has an integral representation,
and, furthermore, an integral representation of the form (\ref{form})
together with (\ref{form2}),
is far from clear, even if one may have a ``Geometry-prejudice'' that 
this solution, analytical at $\, x \, = \, \, 0$, can be interpreted 
as a ``Period'' and ``Derived From Geometry''~\cite{Andre5,Andre6,Andre7}. 

Large tables of Calabi-Yau ODEs have been obtained by 
Almkvist et al.~\cite{TablesCalabi,Almkvist1,Almkvist2}. 
It is worth noting that the coefficients $\, A_n$ of the 
series corresponding to the solution analytical at $\, x \, = \, \, 0$,
are, most of the time, {\em nested sums of product of binomials}, less frequently 
 nested sums of product of binomials and of harmonic
 numbers\footnote[5]{The generating function of Harmonic numbers is 
$\, H(x) \, = \, \,  \sum \, H_n \cdot x^n$
$ \, \, = \, \, \, -\, \ln(1-x)/(1-x)$.} $\, H_n$,
and, in rare cases, no ``closed formula'' is known for these coefficients.

Let us show, in the case of  $\, A_n$
coefficients being nested sums of product of binomials, that the 
solution of the Calabi-Yau ODE, analytical at $\, x \, = \, \, 0$,
which is by construction a series with integer coefficients, 
is actually a diagonal of rational function, and furthermore,
that this rational function can actually be easily built. 

\vskip .1cm 

\subsection{Calculating the rational function
 for nested product of binomials}
\label{theorem}

For pedagogical  reasons we will just consider, here, 
a very simple 
example\footnote[3]{See Proposition 7.3.2 in~\cite{Batyrev}.} 
 of a series $\, {\cal S}(x)$, with integer coefficients, 
given by a sum of product of binomials
\begin{eqnarray}
\label{112}
\hspace{-0.7in}&&{\cal S}(x) 
\,\, \,\, =\, \,\, \, \,\,
\sum_{n=0}^{\infty} \, \sum_{k=0}^{n}
 \, {n\choose k}^3 \, \cdot x^n 
\nonumber \\
\hspace{-0.7in}&&\quad \,\, \, = \, \,\,  \,\,\,
HeunG(-8, -2, 1, 1, 1, 1, \, \, 8 \, x) \,\, \,\,  = \, \,\,  \,\,\,
 HeunG(-1/8, 1/4, 1, 1, 1, 1,\, -x)
\nonumber \\
\hspace{-0.7in}&&\quad \,\, \, = \, \,\,  \,\,\,
1\,\, \,\, +2\,x \,\,\, +10\,x^2 \,\, +56\,x^3 \, +346 \,x^4\,\,
 +2252\, x^5 \,\, +15184\, x^6 \,\, +104960\, x^7\, 
 \nonumber  \\
\hspace{-0.7in}&& \qquad \quad  \quad \qquad +739162\,x^8 \, \, 
+5280932\,x^9\,\, +38165260\,x^{10}
 \,\, \,\,  + \, \,\, \cdots 
\end{eqnarray}
This is the generating function of sequence {\bf A} in Zagier's tables
 of binomial coefficients sums (see p.~354 in~\cite{Zagier}).

The reader can easily get convinced
that the calculations of this section can
straightforwardly (sometimes tediously) be generalised
to more complicated~\cite{Egorychev} nested sums of product of
 binomials\footnote[1]{Not necessarily corresponding to modular forms
as can be seen on (\ref{sol1}), (\ref{modcurve}).}. 
 
The diagonal of a rational function $\, P/Q$ 
is written using Deligne's trick
\begin{eqnarray}
\label{4056}
\hspace{-0.5in} \quad \Diag\Bigl( {{P} \over {Q}}   \Bigr)  
\, \, \, \, \,  = \, \, \, \,  \, \, 
\Bigl({{1} \over {2\, i \, \pi}}\Bigr)^m \cdot 
\int_{C} \,  {{P} \over {Q}} \cdot    {{dz_1} \over {z_1}} 
 \cdot    {{dz_2} \over {z_2}}  \,  \, \cdots   \, \,  {{dz_m} \over {z_m}},  
\end{eqnarray}
where $\, C$ a vanishing\footnote[1]{Cycle \'evanescent in french.}
 cycle~\cite{Letterto}, 
which is, with everyday words, the $\, n$-variables residue formula.  
Finding that a series is a diagonal of a rational function
amounts to framing it into a residue form like (\ref{form}). In order to
achieve this, we write the binomial  $\, {n\choose k}$ as the residue
\begin{eqnarray}
{n\choose k} \,\, \,  = \, \, \, \,\, {{1} \over {2 \, i \, \pi}} \cdot \, 
\int_{C} {{(1\, +z)^n} \over {z^k}} \cdot {{dz } \over {z}},
\end{eqnarray}
and, thus, we can rewrite $\, {\cal S}(x)$ as
\begin{eqnarray}
\label{112bis}
\hspace{-0.95in}&& (2\, i\, \pi)^3 \cdot \, {\cal S}(x)
\, \, \,    =  \,\, \, \, 
\nonumber \\
\hspace{-0.95in}&& \, \,   =  \,\, \,  \, \,\sum_{n=0}^{\infty} \int \int  \int  \,
\sum_{k=0}^{n} \, {{1} \over {(z_1\, z_2 \, z_3)^k}}
 \,  \cdot \, \Bigl(
 (1+z_1)\,  (1+z_2)\,  (1+z_3)\cdot \, x\Bigr)^n 
\,  \cdot \, {{dz_1\, dz_2 \, dz_3 } \over { z_1\, z_2 \, z_3}} 
\nonumber 
\end{eqnarray}
\begin{eqnarray}
\hspace{-0.95in}&& 
=  \,  \int \int \int  
\sum_{n=0}^{\infty} \, {{1 \, -\Bigl(1/(z_1\, z_2 \, z_3)\Bigr)^{(n+1)}} \over
  {1 \, -\Bigl(1/(z_1\, z_2 \, z_3)\Bigr)}}
\cdot \Bigl(
 (1+z_1)\,  (1+z_2)\,  (1+z_3)\cdot \, x\Bigr)^n 
\,  \cdot \, {{dz_1\, dz_2 \, dz_3 } \over { z_1\, z_2 \, z_3}}
\nonumber 
\end{eqnarray}
\begin{eqnarray}
\hspace{-0.95in}&& \quad 
 =  \,\, \,\,  \,  -\, \int \int \int  \,  
\sum_{n=0}^{\infty} \, {{z_1\, z_2 \, z_3 } \over
  {1 \, -\, z_1\, z_2 \, z_3}}
\cdot \Bigl(
 (1+z_1)\,  (1+z_2)\,  (1+z_3)\cdot \, x\Bigr)^n 
\,  \cdot \, {{dz_1\, dz_2 \, dz_3 } \over { z_1\, z_2 \, z_3}}
\nonumber \\
\hspace{-0.95in}&& \quad \quad 
\, \,\, \,\,  \,  + \, \, \int \int \int  
\sum_{n=0}^{\infty} \, {{1 } \over
  { 1 \, -\, z_1\, z_2 \, z_3}}
\cdot \Bigl(
 {{(1+z_1)\,  (1+z_2)\,  (1+z_3)\cdot  x} \over {z_1\, z_2 \, z_3 }}\Bigr)^n 
\,  \cdot \, {{dz_1\, dz_2 \, dz_3 } \over { z_1\, z_2 \, z_3}}
\nonumber 
\end{eqnarray}
\begin{eqnarray}
\hspace{-0.95in}&& \quad 
=  \,\, \,\,  \,  \, \int \int \int  \,  R(x; \, z_1, \, z_2,  \, z_3) \,
\cdot \, {{dz_1\, dz_2 \, dz_3 } \over { z_1\, z_2 \, z_3}},
 \qquad  \qquad  \qquad
\end{eqnarray}
where $\, R(x; \, z_1, \, z_2,  \, z_3)$ reads: 
\begin{eqnarray}
\hspace{-0.65in}{{z_1\, z_2 \, z_3 } \over { \big(1\, - \, x \cdot \,(1+z_1)(1+z_2)(1+z_3)\big)
\, \big( z_1\, z_2 \, z_3  \, - \, x \cdot \,(1+z_1)(1+z_2)(1+z_3)\big)}}.
 \nonumber 
\end{eqnarray}
{}From this last result one deduces immediately that (\ref{112})
is actually the diagonal of:
\begin{eqnarray}
\label{gooddiag}
\hspace{-0.9in}&&\quad {{ 1} \over {
 \big(1\,-z_0 \cdot \,(1+z_1)(1+z_2)(1+z_3)\big) \cdot 
\,\big(1\,-z_0\,z_1\, z_2\,z_3\,(1+z_1)(1+z_2)(1+z_3)\big)
}}.
\nonumber 
\end{eqnarray}

Note that, as a consequence of a combinatorial identity due 
to Strehl and Schmidt~\cite{Strehl,Schmidt,Zudilin},
  $\, {\cal S}(x)$ can also be written as 
\begin{eqnarray}
\label{112sec}
\hspace{-0.95in}&&\quad  {\cal S}(x) 
\,\, \, =\, \,\, \, \,
\sum_{n=0}^{\infty} \, \sum_{k=0}^{n}
 \, {n\choose k}^2 \, {2\, k\choose n} \, \cdot x^n 
\,\, \, =\, \,\, \, 
\sum_{n=0}^{\infty} \, \sum_{k=[n/2]-1}^{n}
 \, {n\choose k}^2 \, {2\, k\choose n} \, \cdot x^n.  
\end{eqnarray}
Calculations similar to  (\ref{112bis}) on this other binomial
representation (\ref{112sec}), enable to express
 (\ref{112}) as the diagonal of 
an alternative rational function:
\begin{eqnarray}
\hspace{-0.9in}&&\quad {{ 1} \over {
 \big(1\,\,  -z_0 \cdot \,(1+z_1)(1+z_2)(1+z_3)^2\big) \cdot
 \,\big(1\, \, -z_0\,z_1\, z_2 \cdot \,(1+z_1)(1+z_2) \big)
}}.
\end{eqnarray}

\vskip .1cm 

\vskip .2cm 

We thus see that we can actually {\em get explicitly},
from straightforward calculations, the rational function (\ref{gooddiag})
for the Calabi-Yau-like ODEs (occurring from
{\em differential geometry} or {\em enumerative combinatorics}) 
when series with nested sums of binomials take place, and, more generally,
for enumerative combinatorics problems (related or not to Calabi-Yau manifolds)
 where series with {\em nested sums of binomials} take place. 

\vskip .1cm 

{\bf Remark:} These straightforward effective 
calculations guarantee to obtain
 an {\em explicit expression} for 
the rational function (\ref{gooddiag}), however the rational function is 
far from being unique, and worse, the number of variables the rational function
depends on is far from being the smallest possible number. Finding the 
``minimal''  rational function (whatever the meaning of ``minimal'' may be) is 
a very difficult problem. Recalling the 
well-known Ap\'ery series $\, \mathcal{A}(x)$, 
and its rewriting due to Strehl and Schmidt~\cite{Strehl,Schmidt,Zudilin},
\begin{eqnarray}
\label{Aper}
\hspace{-0.9in}&&\mathcal{A}(x) \, =  \, \, \sum_{n=0}^{\infty} \, \sum_{k=0}^{n}
 \, \, {n\choose k}^2 \,  {n+\, k \choose k}^2 \cdot x^n 
\, \, \, = \,\,  \,\,  \,  \,
\sum_{n=0}^{\infty} \, \sum_{k=0}^{n} \, \sum_{j=0}^{k}
 \, \, {n\choose k} \,  {n+\, k \choose k} {k\choose j}^3 \,  \cdot x^n 
\nonumber  \\
\hspace{-0.9in}&& \quad  \quad \quad  \quad  \qquad 
\, \, = \, \, \, \,  \,\,  \,\,
1 \, \, \,\, \,  + 5\, x \, \,  \, + \, 73 \, x^2 \, \,  \, +1445 \, x^3 
\, \,  \, + \, 33001 \, x^4 
\,\,  \,  \, \, + \, \cdots,  
\end{eqnarray}
$\mathcal{A}(x)$ is known to be 
the diagonal of the rational function in five variables 
$\, 1/R_1/R_2$
where $\, R_1, \, R_2$ read~\cite{Christol84}:
\begin{eqnarray}
\hspace{-0.9in}R_1 \, \, = \, \, \, \,  \,
 1 \, - \, z_0, \quad \quad \quad \, \,\, 
R_2 \, \, = \, \, \, \,  \,
 (1 \, - \, z_1)(1 \, - \, z_2) (1 \, - \, z_3)(1 \, - \, z_4) \, \, 
- \, z_0 z_1 z_2, \nonumber
\end{eqnarray}
as well as the diagonal of the rational function in five variables 
$\, 1/Q_1/Q_2$
where $\, Q_1, \, Q_2$ read~\cite{Christol85,Christol}:
\begin{eqnarray}
\hspace{-0.9in}\, \,\, Q_1 \, \, = \, \, \, \,  \,
 1 \, - \, z_1\,  z_2\,  z_3\, z_4, \quad \quad \, \,\,  
Q_2 \, \, = \, \, \, \,  \, (1 \, - \, z_3)(1 \, - \, z_4) \, \, \,  
- \, z_0 \cdot \, (1 \, + \, z_1)(1 \, + \, z_2), \nonumber
\end{eqnarray}
and {\em also}  the diagonal of the rational function 
in six variables $\, 1/P_1/P_2/P_3$
where $\, P_1, \, P_2, \, P_3$ read~\cite{Christol84}:
\begin{eqnarray}
\hspace{-0.9in} P_1  \, = \, 
 1\,\, -z_0 \, z_1 ,\quad \, \,\,  P_2  \, = \, 
1\,\, -z_2 \, - z_3 \, - z_0 \, z_2 \, z_3,  \quad \, \,\,  P_3  \, = \, 
1\,\, -z_4 \, - z_5 \, - z_1 \, z_4 \, z_5. \nonumber
\end{eqnarray}

A yet different diagonal representation for the Ap\'ery series, due 
to Delaygue\footnote[1]{Private communication.}, is provided by 
the diagonal of the rational function in eight variables:
\begin{eqnarray}
\hspace{-0.95in}{{1} \over {
(1\; \, - \; z_4 z_5 z_6 z_7) \cdot \,  (1\; - \; z_0 \cdot \,(1+z_4))
 \cdot(1 \, -z_1 \cdot \,(1+z_5)) \cdot (1-z_2-z_6) \cdot (1-z_3-z_7)}}.
\nonumber 
\end{eqnarray}

Calculations similar to  (\ref{112bis}) on these new binomial
expressions 
provides two new rational functions such that (\ref{Aper}) 
can be written as the diagonal of one of these two rational functions.
One is a rational function of five variables, of the form $\, 1/Q^{(5)}_1/Q^{(5)}_2$ 
\begin{eqnarray}
\hspace{-0.6in}&&Q^{(5)}_1 \, \,\, = \, \,  \, \, \,\,
 1\,\,\, \, -z_0 \, z_1 \, z_2 \, z_3 \, z_4  \, \cdot \,(1+z_1)\, (1+z_2)\, (1+z_3)\, (1+z_4),
\nonumber  \\
\hspace{-0.6in}&&Q^{(5)}_2 \, \,\, = \, \,  \, \, \,\,
1\,\,\, \, -z_0\,  \cdot \,(1+z_1)\, (1+z_2)\, (1+z_3)^2\, (1+z_4)^2,
\end{eqnarray}
and the other one, is a rational function of six variables, 
 of the form $\, 1/Q^{(6)}_1/Q^{(6)}_2/Q^{(6)}_3$ 
\begin{eqnarray}
\hspace{-0.6in}&&Q^{(6)}_1 \, \, = \, \,  \,  \,  \,\,
 1\,\, \, \,  -z_0 \, z_3 \, z_4 \, z_5 \, \cdot
 \,(1+z_1)\, (1+z_2)^2\, (1+z_3)\, (1+z_4)\, (1+z_5),
\nonumber  \\
\hspace{-0.6in}&&Q^{(6)}_2 \, \, = \, \,  \, \, \, \,
1\,\, \,  \, -z_0\,z_1\, z_2 \, z_3 \, z_4 \, z_5 \cdot \,(1+z_1)\, (1+z_2),
\nonumber  \\
\hspace{-0.6in}&&Q^{(6)}_3 \, \, = \, \,  \, \,  \,\,
1\,\, \,  \, -z_0 \,\cdot  \,(1+z_1)\, (1+z_2)^2\, (1+z_3)\, (1+z_4)\, (1+z_5).
\end{eqnarray}
We thus see that, when a given function is a diagonal of a rational function,
the rational function is far from being unique, the ``simplest'' representation 
(minimal number of variables, lowest degree polynomials, ...) being hard to find.

Similar computations show that the generating function of
 sequence {\bf E} in Zagier's list~\cite{Zagier}:
\begin{eqnarray}
\label{Zagier-E}
\hspace{-0.9in}&&\,{\cal E}(x) 
\,\, \,\,\, =\, \,\, \, \,\,\,
\sum_{n=0}^{\infty} \, \sum_{k=0}^{\lfloor n/2 \rfloor}
 \, \,  4^{n-2k} \, \cdot \,  \binom{n}{2k} \, \binom{2k}{k}^2 \cdot \, x^n
\nonumber \\
\hspace{-0.9in}&&\quad\, \,\, \, = \, \,\,  \,\,\,\,
1\,\, \,\,\, +4\,x \,\,\, +20\,x^2 \,\, +112\,x^3 \, +676 \,x^4\,\,
 +4304\, x^5 \,\, +28496\, x^6 \,\, +194240\, x^7\, 
 \nonumber  \\
\hspace{-0.9in}&& \qquad \quad \qquad \, +1353508\,x^8\, \, 
+9593104\,x^9\,\, +68906320\,x^{10}
 \,\, \,\,  + \, \,\, \cdots 
\end{eqnarray}
is the diagonal of the rational function in four variables
\[ \frac{1}{(1-4 z_0 z_1 z_2 z_3 \cdot (1+z_1))
 \cdot (1-z_0^2 z_2 z_3 \cdot (1+z_2)^2 (1+z_3)^2 (1+z_1)^2)}, \]
while the generating function of Zagier's sequence {\bf B}
\begin{eqnarray}
\label{Zagier-B}
\hspace{-0.7in}&&{\cal B}(x) 
\,\, \,\, =\, \,\, \, \,\,
\sum_{n=0}^{\infty} \, \sum_{k=0}^{\lfloor n/3 \rfloor}
 \, (-1)^{k} \, \cdot  \, 3^{n-3k} \cdot \, \binom{n}{3k} \,
 \binom{3k}{k} \, \binom{2k}{k} \cdot \, x^n
\nonumber \\
\hspace{-0.7in}&&\quad \quad \,\, \, = \, \,\,  \,\,\,\,
1\,\,\, \,\, +3\,x \,\,\, +9\,x^2 \,\, +21\,x^3 \, +9 \,x^4\,\,
 -297\, x^5 \,\, -2421\, x^6 \,\, -12933\, x^7\, 
 \nonumber  \\
\hspace{-0.7in}&& \quad \qquad \quad \qquad -52407\,x^8\, \, 
-145293\,x^9\,\, -35091\,x^{10}
 \,\, \,\,  + \, \,\, \cdots 
\end{eqnarray}
is the diagonal of the rational function in four variables
\begin{eqnarray} 
\hspace{-0.7in} \frac{1}{(1 \, \, -3 \, z_0 z_1 z_2 z_3 \cdot (1+z_1))
 \cdot (1\, \, +z_0^3 z_2^2 z_3^2 \cdot (1+z_1)^3 (1+z_2)^3 (1+z_3)^2)}.
\end{eqnarray}

\vskip .1cm 

\vskip .1cm 

Such calculations can systematically  be performed on any series defined
by nested  sums of product of binomials. We have performed such calculations 
on a large number of the series corresponding to the list 
of Almkvist et al~\cite{TablesCalabi}, that are given by such 
nested  sums of product of binomials. 

In fact, any $s$-nested sum of products of binomials raised to powers 
$\ell_1, \, \dots, \,  \ell_t$ can be written as the diagonal 
of a rational function in
 $\,\,  \ell_1 \, + \, \cdots\,  + \ell_t\,  + 1\, $ variables, 
of the form $\, \, \left((1-Q_0) (1-Q_1) \, \cdots \, (1-Q_s)\right)^{-1}$,
 where the $Q_i$'s are products of powers of the
variables $z_i$ and of the linear forms $1+z_i$.

For instance, when $s=\,1$, it is easy to prove using the same
 technique that the power series
\begin{eqnarray} 
 \sum_{n=0}^\infty \,  \sum_{k=0}^n \, \,
 \prod_{i=1}^p \, c^{an+bk} \cdot \binom{\alpha_i n \,
 + \beta_i k}{\gamma_i n + \delta_i k} \cdot x^n,
\end{eqnarray}
is the diagonal of the rational function in $ \,p+1 \,$ variables
\begin{eqnarray} 
\hspace{-0.1in} {{1} \over { (1 \,\,  - c^a u \,\cdot
 \, z_0\,  z_1\,  \cdots\,  z_p) \cdot (1 \, \, -c^{a+b} \,\cdot \,
  u\, v \,\, z_0\,  z_1 \, \cdots \, z_p)}}, 
\end{eqnarray}
where $\, u$ and $\, b$ read:
\begin{eqnarray} 
\hspace{-0.3in}\quad \quad u \, = \, \,  \,
 \prod_{i=1}^p \, \frac{(1+z_i)^{\alpha_i}}{z_i^{\gamma_i}}, 
\qquad \quad 
 v \, = \, \, \,  \prod_{i=1}^p \,  \frac{(1+z_i)^{\beta_i}}{z_i^{\delta_i}}.
\end{eqnarray}

The same machinery provides in some cases diagonal representations for
algebraic power series (as diagonals of bivariate rational functions) that are
much simpler than those produced by Furstenberg's result sketched
 in Section~\ref{recall}. For instance, using the fact that
\begin{eqnarray}
\hspace{-0.1in}\quad {2n-2\choose n-1} \,\, \,  = \, \, \, \,\,\, 
 {{1} \over {2 \, i \, \pi}} \cdot \, 
\int_{C} {{(1\, +z)^{2n-2}} \over {z^{n-1}}} \cdot {{dz } \over {z}},
\end{eqnarray}
the algebraic function
\begin{eqnarray}
\hspace{-0.1in}\quad f\,\,\,\, =\,\,\,\,\,\,  \frac{z}{\sqrt{1-z}}
\,\, \,\,  = \,\,\,  \,\,\,
4 \cdot \sum_{n=0}^\infty \, \binom{2n-2}{n-1}\, \Big(\frac{z}{4} \Big)^n,
\end{eqnarray}
is readily seen to be the diagonal of the rational function
\begin{eqnarray} 
\hspace{-0.1in}\quad \quad \quad \frac{z_0 \, z_1}{
1\,\, -\, \; z_0 \cdot \left(1 \, + \, z_1 \right)^2/4}, 
\end{eqnarray}
which is much simpler than (\ref{eq:Furstenberg}).

Similarly, the power series
\begin{eqnarray}
\hspace{-0.1in}\qquad \quad \quad \quad \sum_{n=0}^\infty \, \binom{sn}{n} \cdot x^n,
\end{eqnarray}
is seen to be the diagonal of the rational function
\begin{eqnarray}
\hspace{-0.1in}\qquad \quad \quad \quad  \frac{1}{1\,\,  -z_0 \cdot \, (1+z_1)^s}.
\end{eqnarray}

\section{Comments and speculations}
\label{commspec}

\subsection{Christol's theorem}
\label{subtheorem}

\vskip .1cm 

In~\cite{Christol} (page 61 Theorem 12, see also Proposition 7 
in page 50 of~\cite{Christol369} ) it is proved 
that any power series with an {\em integral representation} 
(as defined in (\ref{repint1}) see below)
 and of {\em maximal weight} for  the corresponding
 {\em Picard-Fuchs linear differential equation}
 (denoted by $\, L_V$ below) is the {\em diagonal of a rational function}
 and, in particular, is {\em globally bounded}.

The technical nature of the original papers is such that
the result itself is difficult to find. This paragraph is devoted to explain,
 in down-to-earth terms, the somewhat esoteric expressions used
 in its wording, and to explain what it means on explicit examples.
As the original proof is very obfuscated its principle is 
 sketched  in \ref{onfuscation}.

A function  $f$, analytic near $0$, is said 
to have an ``integral representation''
 if it can be written in the following form\footnote[1]{Following~\cite{KZ}
 or, better,  its version with parameter~\cite{JA}, 
one could define integral representations
by integrating in (\ref{repint1}) on  domains $C$  defined
 by polynomial (in $\, x_1, \, \ldots, \,  x_n$)
 equalities or inequalities. Actually the case
 $ \,C = \, \{ x_i\in[0,1]\,;\, 1 \,\leq \, i \,\leq \, n\} $ 
would be enough. To connect this definition with ours, one needs 
to use the Stokes theorem. However, 
the dimension $\, n$ of the underlying
complex manifold is basic for our next definitions and it is unfortunately
 not preserved by Stokes theorem.}:
\begin{eqnarray}
\label{repint1}
\hspace{-0.2in} \quad f(x)\,\,\,\,= \,\,\,\,\,
 \int_C \,F(x;\,x_1,\,\ldots,\,x_n)\cdot \,   dx_1\, \,\cdots \, \,dx_n,
\end{eqnarray}
where $F$ is an {\em algebraic function}, hence living on 
some (projective) {\em complex} $(n+1)$-fold  $V$,  and 
$C$ is a ``cycle'', namely an (oriented)  compact 
(i.e. without boundary) {\em real} $n$-fold contained in $\,V$.
In (\ref{repint1}) $x$ must be seen as a parameter. Then one integrates the
 $n$-differential $ \,\, F(x;\ldots) \,dx_1\cdots dx_n$, that
 depends on $x$, on a ``constant'' cycle\footnote[2]{One could, more
 generally, integrate an  $m$-differential for $\, m < \, n$. But Lefschetz
 theorems assert that, up to  taking hyperplane sections, one
 can reduce to the case $n= \, m$.} $C$. If it exists, 
the integral representation is far from unique. Formula (\ref{form})  
shows that any diagonal
 of a rational function has an integral representation for which $C$ is 
the so-called ``vanishing~\cite{evanescent} cycle''. This 
is straightforwardly extended to diagonals of algebraic functions.
 
In practical examples, to obtain a cycle, one often needs to complete the
 integration domain by means of symmetries in the usual way 
when dealing with the method of residues.
That happens for instance for the hypergeometric function
 $ \, f(x) = \, B(a_2,\, b_1 \, -a_2) \cdot \, _2F_1([a_1,a_2],\,\,[b_1]; \, x)$ 
($B$ is the beta function, $ \, \Re(b_1) \, > \, \Re(a_2) \, > \, 0$), which has
 the following Euler integral representation:
\begin{eqnarray}
\hspace{-0.4in}f(x)\,\,\, \, = \, \, \,\, 
\int_0^1 \,  x_1^{a_2-1}\cdot \, (1-x_1)^{b_1-a_2-1} \, \cdot \,
 (1 \,\, -x \,x_1)^{-a_1} \cdot \, dx_1.
\end{eqnarray}
More generally the only hypergeometric functions $ \, _pF_q$ having 
an (Euler) integral representation are the 
$_{n+1}F_n$ with rational\footnote[8]{The parameters have to be rational for the
 $n$-form in the integral representation (\ref{repint1})
 be algebraic. More deeply, when globally bounded, the hypergeometric 
function is a $G$-function and  the
corresponding linear differential operator is  globally
nilpotent hence it has only regular singularities with rational exponents.  
The parameters $\,a_i$ and $ \, b_j$ are directly linked
 to exponents at $ \, 0$ and $ \, \infty$.}
 parameters~\cite{Hattori,Driver}  $ \, a_i,\,b_j$:
\begin{eqnarray}
\label{nplusunFn} 
\hspace{-0.95in}&&\qquad _{n+1}F_n([a_1, \, a_2,  \, \,  \cdots,  \, \, a_n, 
\,  a_{n+1}],\, [b_1, \, b_2,  \, \cdots,  \, b_n], \,  x) 
 \nonumber  \\
\hspace{-0.95in}&&\quad \quad \quad \quad \,\, \,\, \, = \, \, \, \, \,  \,
\rho \cdot \int_0^1 \, \cdots \, \int_0^1 \, x_1^{a_1}   \, \, \cdots \, x_n^{a_n}
 \cdot \, 
(1\, -x_1)^{b_1-a_1-1}\, \, \cdots \, \, (1\, -x_n)^{b_n-a_n-1} 
\nonumber \\
\hspace{-0.95in}&& \quad \, \, \quad \qquad \quad  \qquad \,\,
\,\times \, (1\, -x_1 \, x_2 \,\cdots \, x_n \cdot  x)^{-a_{n+1}} \cdot 
{{dx_1} \over {x_1}}  \, \,  \cdots \, \,   {{dx_n} \over {x_n}}.
\end{eqnarray}

Moreover, the integration cycle $C$ of (\ref{repint1}) can involve 
points at infinity (we are dealing with projective geometry).

As $\, C$ is a cycle, adding an exact differential to $\, F\,dx_1\cdots dx_n$ clearly 
do not change $f$. But a famous Grothendieck's theorem \cite{GR1} asserts that 
  $n$-differentials on $V$ up to exact ones built up a finite 
 $\mathbb{C}(x)$-space. Moreover\footnote[5]{See~\cite{Hodge}  specially
 \S\kern.15em1 and \S\kern.15em 4 for instance but notice that the weight filtration
 used there and the monodromy weight we will use are quite distinct even
 if both constructions could look similar.} differentiation under the integral
 endows this space with a connection, namely the Gauss-Manin one. In other words,
 $\, f$ is solution of an ordinary linear differential equation (ODE) $L_V$,
 namely the {\em Picard-Fuchs differential equation}. This ODE depends only on $\, V$ 
and do not involve any cycle but choosing a particular cycle $\, C$ amount
to choosing a particular solution $\, f$ of $ \, L_V$.

The Picard-Fuchs ODE $L_V$ is known to have only regular
 singularities with rational exponents.
Now $ \, L$ being a linear ODE for which $\, 0$
 is a regular singularity, solutions near $ \,0$
 of $\, L$ are endowed with a ``monodromy\footnote[2]{The monodromy operator  $T$,
 ``turning once counterclockwise around $0$'', 
acts on the space  $\mathit{S}$  of solutions of $ \, L$ by 
 $ \,T(x)= \, x\,e^{2i\pi}= \, x$ and 
$T(\log(x)) \,= \, \log(x) \,+2i\pi$. The solution $ \,f$ is of weight
 $W$ if and only if it is
 in the image  $(T-1)^W(\mathit{S})$ but not 
in  $(T-1)^{W+1}(\mathit{S})$.} weight'':
 $\, f$ is of weight $W$ if $L$ has $\, W+1$ solutions  
that are built in the following way:
\begin{eqnarray}
\hspace{-0.75in}&&\quad \quad   f(x)\ ,\quad f(x) \cdot \,\log(x)\, +f_1(x)\ ,
 \quad \cdots\ ,\quad  \\ 
\hspace{-0.75in}&&\qquad \quad \quad \quad \quad 
 \,  \,  f(x) \cdot \,\frac{\log^W(x)}{\small{W!}} \, \,  \, 
+ f_1(x)\cdot \, \frac{\log^{W-1}(x)}{\small{(W-1)!}}
\, \, \,  \,   +\, \, \cdots \, \,  \,  \,  +\, f_W(x), 
\nonumber 
\end{eqnarray}
where the $f_i$ are analytic near $0$, 
and no solution involving $\, f(x) \cdot \log^{W+1}(x)$. For instance, a MUM 
 (maximal unipotent monodromy) ODE $L$ of order $\mu$ has a unique 
(up to a multiplicative constant) analytic solution  near $ \, 0$ and this solution
 is of logarithm weight $\mu \, -1$.

Geometric considerations imply that, for solutions of $\, L_V$, the 
maximum monodromy weight  is  $n-1$. So we will say that $f$ is of 
{\em maximal weight} for $ \, L_V$ if it is of weight $\, n-1$.
 
The Picard-Fuchs linear ODE is difficult to  determine and, moreover, 
depends on the particular integral representation.
What is well defined is the minimal linear ODE  $L_f$ of which $f$, with
 an integral representation, is solution. Then its Picard Fuchs ODE $\,L_V\,=\, M\,L_f$ 
is a left multiple of $\,L_f$ and the monodromy weight of $\,f$ for $\,L_V$ is at least 
the monodromy weight of $\,f$ for $\,L_f$ (it is likely that the two monodromy weights
 are actually the same). The order of $\,L_f$ is  
smaller, and often much smaller, than the order of $L_V$.

There is absolutely no reason for the order of $L_f$ to be the number $n$ of variables
 of the integral representation,  but this effectively do happen in examples, 
notably for hypergeometric functions $_{n+1}F_n$ and
 for certain Calabi-Yau  linear ODE. Under
 these circumstances, if $L_f$ is MUM then $f$ is of maximum weight for $L_V$
 and the theorem asserts that $f$  is the diagonal of 
a rational function. Let us remark that in that case
$f$  is  the unique (up to a multiplicative constant)
 analytic solution  near $\, 0$ of $\, L_f$. 

Disappointingly,  when this result 
can be applied to $\, _{n+1}F_n$,  it
becomes somewhat trivial. More precisely, the hypergeometric function is of 
maximal weight if and only if $\, b_j \, = \, 1$ for all $j$
 (there is only $n!$'s in the denominator of coefficients). In that case it is
obviously the 
Hadamard product of algebraic functions:
\begin{eqnarray}
\label{hyperdiag}
\hspace{-0.6in}&&\quad _{n}F_{n-1}([\alpha_1, \,\alpha_2, \, \cdots,  \, \alpha_n], 
\,[1, \, 1, \, \cdots \, 1], \, x) \\
\hspace{-0.6in}&& \quad \quad  \quad  \quad  \quad \quad 
\, \,   = \, \,  \,    \,   \,   \,
  (1\, -x)^{-\alpha_1} \, \star \, 
(1\, -x)^{-\alpha_2} \,  \, \cdots \, \,  \star \,  (1\, -x)^{-\alpha_n}.
 \nonumber 
\end{eqnarray}

Therefore, we now have (at least) three sets
 of problems yielding diagonal of rational 
functions: the $\, n$-fold integrals of the form 
(\ref{form}) with (\ref{form2}),
the Picard-Fuchs linear ODEs with solution of maximal monodromy weight
 and, finally, the problems of
enumerative combinatorics where nested sums
 of products of binomials take place. 
Diagonal of rational functions, thus, occur in a quite large set of problems
of theoretical physics.

\subsection{Christol's conjecture}
\label{conjec}

The diagonal of a rational function is 
 globally bounded  (i.e. it has non zero radius
 of convergence and
 integer coefficients up to one rescaling) and  {\em D-finite}
 (i.e. solution of a linear differential equation with polynomial 
coefficients)\footnote[2]{The series expansion of the susceptibility
of the isotropic 2-D Ising model can be recast into a series with integer
coefficients (see~\cite{Khi6,High,bernie2010,bo-gu-ha-je-ma-ni-ze-08}),
 but it cannot be the diagonal of rational functions
since the full susceptibility is
 {\em not a D-finite function}~\cite{bernie2010}.}.

The reciprocal statement is the ``Christol's conjecture''~\cite{Christol}
 saying that any D-finite, globally bounded 
series  is necessarily the diagonal 
of a rational function. 

A fantastic Chudnovski theorem (\cite{DGS} page 267) asserts that the minimal
linear differential operator of a $G$ function (and in particular of a
D-finite globally bounded series) is a $\, G$-operator (i.e. at least
conjecturally, a globally nilpotent
operator)~\cite{bo-bo-ha-ma-we-ze-09,Andre5,Andre6}. ``Christol's conjecture''
amounts to saying something more: if the solution of this globally nilpotent
linear differential operator is, not only a $\, G$-series, but a {\em globally
bounded series, then it is the diagonal of a rational function}.

Conversely the solution, analytical at $0$, of a globally nilpotent linear
differential operator is necessarily a $\, G$-function~\cite{Andre5,Andre6}.
Moreover, a ``classical'' conjecture, with numerous avatars, claims that any
$G$-function comes from geometry i.e. roughly speaking, it has an integral
representation\footnote[3]{Bombieri-Dwork conjecture
 see for instance~\cite{Andre6}.}.

To test the validity of Christol's conjecture we look for counter-examples not
contradicting classical conjectures. Then we search D-finite power series with
integer coefficients which are not algebraic but have an integral
representation and are not of maximal weight for the corresponding
Picard-Fuchs linear ODE.

As a first step let us limit ourself to hypergeometric functions
 $\, _{n+1}F_n$. The monodromy weight $\, W$ is exactly the number of $ \, 1$ 
among the $\, b_i$.

When $\, _{n+1} F_n\, $ is globally bounded and has no integer parameters
 $ \,b_i$ ($ W= \, 0$), its minimal ODE has a $p$-curvature zero 
for almost all primes $\, p$. However, a Grothendieck conjecture,
 proved for $\, _3F_2\, $  in~\cite{Hodge}, and 
generalised to $ \, _{n+1} F_n\, $ in~\cite{BeHe89},
asserts that, under these circumstances, the hypergeometric function is {\em
algebraic}.

So we are looking for  {\em  globally bounded} 
hypergeometric functions satisfying $\quad$
$ \, 1 \, \leq  \, W \, \leq \,  n-1$.
In general such hypergeometric functions are $\, G$-series but are very far 
from being globally bounded. The hypergeometric world extends largely 
outside the  world of diagonal of rational functions.

Such an example in the first case $\, n=\, 2$, $\, W =\, 1 \, $ 
was given in~\cite{Christol}:
\begin{eqnarray}
\label{contre1}
\hspace{-0.8in}&&\quad 
_3F_2\left(\left[{{1} \over {9}},\, {{4} \over {9}},\,{{5} \over {9}}\right],
\, \left[{{1} \over {3}},\, 1\right], \,  3^6\, x\right)
\, \,\, \,  \, = \,\,  \, \, \, \,\, 
1\,\,\,\, +60\,x\, \, +20475\,x^2\, \, +9373650\,x^3 
\nonumber \\
\hspace{-0.8in}&&\,\quad \, \, \,  +4881796920\,x^4\,\, +2734407111744\,x^5\,\, 
+1605040007778900\,x^6 \,  \,\,+ \,\, \,  \cdots  
\end{eqnarray}
The integer coefficients read with the rising factorial
 (or Pochhammer) symbol
\begin{eqnarray}
{{ (1/9)_n \cdot (4/9)_n \cdot (5/9)_n } \over {
(1/3)_n \cdot (1)_n \cdot n!}} \cdot \,  3^{6n}
\, \,  \,= \, \, \,\, \, {{\rho(n)} \over {\rho(0)}}, 
\end{eqnarray}
where:
\begin{eqnarray}
\label{far1}
\rho(n)\,\,\, = \, \, \,  \, \, 
 {{ \Gamma(1/9+n)\,\Gamma(4/9+n) \, \Gamma(5/9+n)} \over {
\Gamma(1/3+n)\, \Gamma(1+n) \, \Gamma(1+n)}} \cdot \,  3^{6n}.
\end{eqnarray}
Note that, at first sight, it is 
{\em far from clear}\footnote[3]{In contrast with cases 
where binomial (and thus integers) 
expressions take place.} on (\ref{far1}), or 
 on the simple recursion
on the $\, \rho(n)$ coefficients (with the initial 
value $\, \rho(0)\, = \, \, 1$)
\begin{eqnarray}
\label{ratio1}
{{\rho(n+1)} \over {\rho(n)}}\, \, \,= \, \, \, \,  \,
3 \cdot \, 
{{(1 + 9 n) \, (4 + 9 n) \, (5 + 9 n)  } \over {(1 + 3 n) (1 + n)^2 }}, 
\end{eqnarray}
to see that the $\, \rho(n)$'s are actually integers. A 
sketch of the (quite arithmetic) proof that the  $\, \rho(n)$'s 
are actually integers, is given in \ref{proof}. 

Because of the $\, 1/3$ in the right (lower) parameters of (\ref{contre1}),
the hypergeometric function (\ref{contre1}) is not an obvious
 Hadamard product 
of algebraic functions (and thus a diagonal of a rational function), and 
one can see that it is not an algebraic hypergeometric function
either by calculating its $\, p$-curvature and finding that
 it is not zero~\cite{Andre7}, or using~\cite{JAW}. 
 Proving that an algebraic function  is
the diagonal of a rational function and proving that a solution 
of maximal weight for a Picard-Fuchs equation is
the diagonal of a rational function use two entirely 
distinct ways. The hope is to  combine 
both techniques to conclude in the intermediate situation.

This example remained for twenty years, the only ``blind spot''
on Christol's conjecture. We have recently found many other 
 $\, _3F_2 \, $ examples\footnote[2]{$_2F_1$ cases 
are straightforward, and cannot provide counterexamples 
to  Christol's conjecture.}, 
such that their series expansion have {\em integer coefficients} 
but are not obviously diagonal of rational functions.
These new hypergeometric examples are displayed 
in \ref{black}. Unfortunately these hypergeometric 
examples are on the same ``frustrating 
footing'' as Christol's example (\ref{contre1}): we are not able to show that 
one of them is actually a diagonal of a rational function, or, 
conversely, to show that one of them cannot be the 
diagonal of a rational function.

\section{Integrality versus modularity: learning by examples}
\label{versus}

A large number of examples of integrality of  series-solutions
comes from modular forms. Let us just display two such modular 
forms associated with HeunG functions of the form 
$\, HeunG(a, q, 1, 1, 1, 1; x)$. 

\subsection{First modular form example}
\label{ex1}

 One can, for instance, rewrite the example (\ref{112}) 
of subsection (\ref{theorem}),  
namely $\, HeunG(-1/8, 1/4, 1, 1, 1, 1, -x)$, 
as a hypergeometric function\footnote[1]{The relation between modular forms
and hypergeometric functions can be simply seen in the identity 
$\, _2F_1([1/12,5/12],[1], \, 1728/j(\tau))^4 \, = \, \, E_4(\tau)$, 
where $\, E_4$ is an Eisenstein series.} 
 with {\em two rational pullbacks}:
\begin{eqnarray}
\label{sol1}
\hspace{-0.6in}&&HeunG(-1/8, 1/4, 1, 1, 1, 1, -x)\
\, \, \, \, \, = \, \, \, \,  \,
\, \,  \sum_{n=0}^{\infty} \, \sum_{k=0}^{n} \, {n\choose k}^3 \, x^n
  \, \, \, \, \,
 \\
\hspace{-0.6in}&&\quad \quad \,  
 = \,\, \,  \,\Bigl((1\, +4\, x) \cdot
 (1\, +228\, x \,+48\, x^2\, +\, 64\, x^3 ) \Bigr)^{-1/4} 
\nonumber \\
\hspace{-0.6in}&&\quad \quad \quad \quad \quad \quad \times \, 
 _2F_1\Bigl([{{1} \over {12}}, \, {{5} \over {12}}], \, [1]; \, {{
1728 \cdot  (1\, -8 \, x)^6 \cdot (1\, +x)^3 \cdot x } \over {
(1\, +228\, x \,+48\, x^2\, +\, 64\, x^3)^3 \cdot 
(1\, +4\, x)^3 }}   \Bigr)
 \nonumber 
\end{eqnarray}
\begin{eqnarray}
\hspace{-0.6in}&&\quad \quad 
\,  = \, \, \,  \, 
\Bigl((1\, -2 \, x) \cdot (1\, -6 \,x\,+228 \,x^2\,-8\,x^3 ) \Bigr)^{-1/4} 
\nonumber \\
\hspace{-0.6in}&&\quad \quad \quad \quad \quad \quad   \times \, 
 _2F_1\Bigl([{{1} \over {12}}, \, {{5} \over {12}}], \, [1]; \, {{
1728 \cdot  (1\, -8 \, x)^3 \cdot (1\, +x)^6 \cdot x^2 } \over {
(1\, -2 \, x)^3 \cdot (1\, -6 \,x\,+228 \,x^2\,-8\,x^3)^3 }}   \Bigr).
\nonumber 
\end{eqnarray}
The relation between the two pullbacks,
that are related by the ``Atkin\footnote[5]{In previous
 papers~\cite{Renorm,CalabiYauIsing1}, with some abuse of language, we
called such an involution 
an \emph{Atkin-Lehner involution}. In fact this terminology is commonly used 
in the mathematical community for an involution $ \, \tau \, \rightarrow \, -N/\tau$,
 on~$\tau$, the ratio of periods,
 and {\em not} for our $x$-involution. However,{\em  when
 the modular curve is of  genus zero}
one has a parametrisation in term of the variable 
 $\, x_N(\tau) \, = \, \,  (\eta(\tau)/\eta(N \tau))^{24/(N-1)}$ 
(see eq. (27) in~\cite{AtkinMorain}), which actually transforms 
{\em as an involution} ($x_N \, \rightarrow \, A/x_N$) 
under the Atkin-Lehner involution. 
This is why we switch to the wording 
\emph{``Atkin" involution}.}'' involution\footnote[3]{The 
relevance of the ``Atkin'' involution 
 $\, x \, \leftrightarrow \, -1/8/x$ is also clear 
on the operator:
note that operator $\, \Omega$ is invariant
by changing $\, \theta \, \leftrightarrow \,  -1 \, -\theta$, 
and  $\, x\, \leftrightarrow \, -1/8/x$.
Also note that changing $\, \Omega$ by a pullback 
$\, x\, \leftrightarrow \, -1/8/x$, amounts to
changing  $\, \Omega$ into  $\,\, x \cdot \Omega \cdot x^{-1}$.
}
 $\, x \, \leftrightarrow \, -1/8/x$,
 gives the modular curve:
\begin{eqnarray}
\label{modcurve}
\hspace{-0.8in}&&1953125\, y^3 \, z^3 \, \,  \,  \, 
 -187500 \, y^2 \, z^2 \cdot (y+z) \,
  \,  +375 \, y \, z \cdot (16\, z^2\, -4027\, z\, y\, +16\, y^2)
\nonumber \\
\hspace{-0.8in}&& \qquad \quad \quad    -64 \cdot \, (z+y) \cdot
 (y^2+z^2\, +1487 \, z\, y)
\,   \, \,  +110592 \cdot \, z\, y 
\, \, \,\, \,  = \, \,\, \,  \, \, 0. 
\end{eqnarray}

Series (\ref{112}) is solution of the  (exactly)
 {\em self adjoint} linear differential operator
$\, \Omega$ where ($\theta \, = \, \, x \cdot D_x$):
\begin{eqnarray}
\label{aux1}
\hspace{-0.2in}&&x \cdot \Omega\,\, \,\,\, = \, \, \, \, \, \,
\theta^2 \,\, \,  \,  -\, x \cdot 
(7\, \theta^2 \, +7 \,\theta \, +2) 
\,  \, \, \,  -8 \, x^2 \cdot   (\theta \, +\, 1)^2.
\end{eqnarray}

The relevance of the ``Atkin'' involution 
 $\, x \, \leftrightarrow \, -1/8/x$ is also clear 
on the operator:
note that operator $\, \Omega$ is invariant
by changing $\, \theta \, \leftrightarrow \,  -1 \, -\theta$, 
and  $\, x\, \leftrightarrow \, -1/8/x$.
Also note that changing $\, \Omega$ by a pullback 
$\, x\, \leftrightarrow \, -1/8/x$, amounts to
changing  $\, \Omega$ into  $\,\, x \cdot \Omega \cdot x^{-1}$. 

\subsubsection{Modular invariance  \newline \newline }
\label{modinv}

Do note that these pullbacks are respectively of the
 form ($\circ$ denotes the composition of functions):
\begin{eqnarray}
\hspace{-0.9in} {\cal M}_2\, = \, \, {{1728 \, x} \over {(x\, + \, 16)^3 }} \, 
\circ \, {{(1 \, -8 \, x)^3} \over {x  \, (1+x)^3 }} 
\quad \hbox{ and:} \quad \, \,  
{\tilde {\cal M}}_2\, = \, \, {{1728 \, x} \over {(x\, + \, 256)^3 }} \, 
\circ \, {{(1 \, -8 \, x)^3} \over {x  \, (1+x)^3 }}, 
\end{eqnarray}
where one recognises the two Hauptmoduls of the modular curve
corresponding to $\, \tau \, \rightarrow \, 2 \, \tau$.
Introducing the Dedekind-like parametrisation:
\begin{eqnarray}
\hspace{-0.8in} \quad x(q) \, = \, \,  \, 
q \cdot \, \prod_{n=1}^{\infty} (1 \, -q^n)^{a(n)}
 \qquad  \, \,  \hbox{where:}\,    \quad  \quad 
\sum_{n=1}^{\infty} \, a(n) \cdot \, t^n \,\, = \, \,  \,\, 
{{3 \, t \cdot \, (1\, -t^2)} \over { 1 \, - \, t^6}},  
\nonumber 
\end{eqnarray}
one can rewrite the Hauptmoduls as:
\begin{eqnarray}
\hspace{-0.9in}\qquad \quad \quad   {\tilde {\cal M}}_2\Bigl(
 {{(1 \, -8 \,x(q))^3 } \over {x(q) \cdot \, (1 \, + \,x(q))^3 }}
 \Bigr)\,\,\,\, = \,\, \,\, \,\,
{\cal M}_2\Bigl( 
{{(1 \, -8 \,x(q^2))^3 } \over {x(q^2) \cdot \, (1 \, + \,x(q^2))^3 }}
 \Bigr).
\end{eqnarray}

\subsubsection{Other representations \newline \newline }
\label{7.2}

In fact, using Kummer's relation and other relations on
$\, _2F_1$'s: 
\begin{eqnarray}
\hspace{-0.7in}&&_2F_1\Bigl([{{1} \over {6}}, \, {{1} \over {3}}],
 \, [1],\, 4 \, x \, (1-x))
 \,\,\,  = \, \, \, \, \,
 _2F_1\Bigl([{{1} \over {3}}, \, {{2} \over {3}}], \, [1],\, x\Bigr)
  \nonumber \\
\hspace{-0.7in}&&\,\qquad \quad \quad \quad   \,\,  = \, \, \, \, \, 
(1-x)^{-1/3} \cdot \,  _2F_1\Bigl([{{1} \over {3}}, \,
 {{1} \over {3}}], \, [1],\,
 -{{x} \over  {1\, -x}}\Bigr) 
  \nonumber \\
\hspace{-0.7in}&&\,\qquad \quad \quad \quad  \,\,  = \, \, \, \, \, 
(1-x)^{-2/3} \cdot \,  _2F_1\Bigl([{{2} \over {3}}, \,
 {{2} \over {3}}], \, [1],\,
 -{{x} \over  {1\, -x}}\Bigr)
 \\  
\hspace{-0.7in}&&\,\qquad \quad \quad \quad  \,\,  = \, \, \, \, \, 
\Bigl({{9} \over { 9 \, - \, 8 \, x}} \Bigr)^{1/4} \cdot \, 
 _2F_1\Bigl([{{1} \over {12}}, \,
 {{5} \over {12}}], \, [1],\,
 {{64 \, x^3 \cdot \, (1-x)} \over  {(9 \, - \, 8 \, x)^3}}\Bigr).
 \nonumber
\end{eqnarray}
 $\, HeunG(-1/8, 1/4, 1, 1, 1, 1, -x)$
 can be written in many different
ways as hypergeometric $_2F_1$ with {\em two} pullbacks:
\begin{eqnarray}
\label{sol1other}
\hspace{-0.7in}&&HeunG(-1/8, 1/4, 1, 1, 1, 1, -x)
\, \, \, \,  = \, \,  \, \, \,
{{1} \over {1\, +\, 4 \, x}} \cdot \,
 _2F_1\Bigl([{{1} \over {3}}, \, {{2} \over {3}}], \, [1],\,
  {{27 \, x} \over {(1\, + \, 4 \, x)^3}}  \Bigr)
\nonumber \\
\hspace{-0.7in}&&\quad \quad \, \, \, = \, \,  \,  \, 
{{1} \over {1\, -\, 2 \, x}} \cdot \,
 _2F_1\Bigl([{{1} \over {3}}, \, {{2} \over {3}}], \, [1],\,
  {{27 \, x^2} \over {(1\, - \, 2 \, x)^3}}  \Bigr) \\
\hspace{-0.7in}&&\quad \quad \, \, \, = \, \, \, \,  \, 
 (1\,+x)^{-1/3}\cdot (1\,-8\, x)^{-2/3} \cdot \, 
 _2F_1\Bigl([{{1} \over {3}}, \, {{1} \over {3}}], \, [1],\, 
 -\, {{27 \, x} \over {(1\, +  \, x) 
\cdot  (1\,-8\, x)^2 }}  \Bigr)
 \nonumber \\
\hspace{-0.7in}&&\quad \quad \, \, \, = \, \, \, \,   \, 
{{1} \over {1\, +\, 4 \, x}} \cdot \,
 _2F_1\Bigl([{{1} \over {6}}, \, {{1} \over {3}}], \, [1],\, 
 {{ 108 \, x \, (1+x) \, (1\, -8 \, x)^2 } \over {
(1\, + \, 4 \, x)^6 }}  \Bigr) \, \,   \, \,
 \, \,  = \,\,  \,  \, \, \,  \, \cdots
\nonumber 
\end{eqnarray}

\vskip .1cm 

The  relation between the two pullbacks in (\ref{sol1other}), namely 
$\, u \, = \, \, 27\, x/(1+\, 4\, x)^3$, and 
$ v \, = \, \,  27\, x^2/(1-\, 2\, x)^3)$, 
is the genus-zero modular curve:
\begin{eqnarray}
\label{moducurve1}
\hspace{-0.4in}&&8\,{u}^{3}{v}^{3} \, \, 
-12\,{u}^{2}{v}^{2} \cdot \,  \left( u+v \right)\, \, 
 +3\,uv \cdot \,  (2\,{u}^{2}+2\,{v}^{2}+13\,uv) \\
\hspace{-0.4in}&& \qquad \qquad \, \,
  - \, (u+v)  \cdot \, ({v}^{2}+29\,uv+{u}^{2}) \, \, +27\,uv 
\, \,\,\,\,   = \, \,\, \,\, \,  0. 
 \nonumber 
\end{eqnarray}
Let us consider the modular curve (\ref{moducurve1})
for $\, u \, \, = x$, seeing $\, v$ as a algebraic
 function of $\, x$.
One of the three root-solutions 
expands as:
\begin{eqnarray}
\label{tildev0}
\hspace{-0.9in}&& \tilde{v}_0(x) \,  \, = \, \, \,\, \, \,
 {{1} \over {27}} \,{x}^{2} \, \, \,  +{\frac {10}{243}}\,{x}^{3} \,
 +{\frac {256}{6561}}\,{x}^{4} \, 
+{\frac {18928}{531441}}\,{x}^{5} \, 
+{\frac {154000}{4782969}}\,{x}^{6} \,
 \, \,  \, + \, \, \cdots,  
\end{eqnarray}
the two other ones being Puiseux series (here $\, t$ denotes $\, x^{1/2}$):
\begin{eqnarray}
\label{tildev1}
\hspace{-0.9in}&& \tilde{v}_1(x) \, \, \, = \, \,   \, \, 
3\,\sqrt {3}\,\,  t\,\,  \, -15\,\, {t}^{2} \, +{\frac {119}{6}}\,\sqrt {3}\,\,  {t}^{3}\, 
-{\frac {1904}{27}}\,\, {t}^{4}\, 
\, +{\frac {50701}{648}}\,\sqrt {3}\,\,  {t}^{5}\, 
\, \,  \, \, + \, \,\,  \cdots, 
\end{eqnarray}
the third root, $\, \tilde{v}_2(x)$ corresponding to change  $\, t$ into  $\,- t$. 
These series expansions compose nicely: $\, \tilde{v}_0(\tilde{v}_1(x)) \, \, = \, \, 
 \tilde{v}_0(\tilde{v}_2(x)) \, \, = \, \,   \, x$.

\vskip .1cm 

\vskip .1cm 

{\bf Remark: Ramanujan's cubic transformation}. 
For the hypergeometric function $\,_2F_1([1/3,2/3],[1],x)$, the existence
of {\em two} pullbacks is also reminiscent of 
relation (28) in~\cite{CalabiYauIsing}, or to 
the known Ramanujan's cubic transformation formula~\cite{Chan}:
\begin{eqnarray}
\label{cubic}
\hspace{-0.9in}&&(1\, + \, 6 \, x)\cdot \, 
 _2F_1\Bigl([{{1} \over {3}}, \, {{2} \over {3}}], \, [1],\, 27\, x^3 \Bigr)
 \,\, \, = \,  \,  \, \, \, \, 
 _2F_1\Bigl([{{1} \over {3}}, \, {{2} \over {3}}], \, [1],\, \, 
 1 \, - \, \Bigl({{1 \, -3 \, x } \over {1 \, + \, 6 \, x }}\Bigr)^3 \Bigr)
\nonumber \\
\hspace{-0.9in}&&\quad \quad \,  \,\,\,\, = \, \, \, \,\,  \, \,  \,  \,
1\,\,\,  \, \,  +6\,x\, \, \,  +6\,x^3\, \,  +36\,x^4\, \,  +90\,x^6\, \, 
+540\,x^7\,+1680\,x^9\,\,  \,  \, \,+ \, \, \cdots 
\end{eqnarray}
Noting that $\,  _2F_1\Bigl([1/3, \, 2/3], \, [1], \, x)$ and 
$\,  _2F_1\Bigl([1/3, \, 2/3], \, [1], \, 1\, -x)$ are solutions of the
{\em same} second order linear ODE, one can change one of the two pullbacks, 
$\, P_1\, = \, 27\, x^3$ and
 $\, P_2\, = \, 1 \, -(1 \, -3 \, x)^3/(1 \, + \, 6 \, x)^3$
  in (\ref{cubic}), into $\, 1\, - \, P_i$, $\, i \,= \,  1, \, 2$. The relation
between the two pullbacks 
$\, u \, =  \, \, 1 \, - \, P_1$ and $\, v \, =  \,  P_2$, 
is a simple (genus zero, $(u,v)$-symmetric) 
modular curve:
\begin{eqnarray}
\label{moducurve}
\hspace{-0.9in}&& 512\cdot \, u^3\cdot \, v^3\,\,
  -1728\cdot \, u^2\cdot \, v^2 \cdot \, (u+v)\,\,\,\,
 +216\cdot \, u \, v\cdot \,
 (17\cdot \, u \, v +9\cdot \, (u^2+v^2)) 
 \nonumber \\
\hspace{-0.9in}&& \, \, \,\, \,-243\cdot \, (v+u) \cdot \,
 (3\cdot \, (u^2+v^2) \,\, \, +8 \cdot \, u\, \, v)\, \,  \, 
+729\cdot \, (u^2+u \, v+v^2) \, \,\, \, = \,\,  \, \, \, \, 0. 
\end{eqnarray}

\vskip .1cm 

\subsubsection{Schwarzian condition \newline \newline }
\label{schwarcond}

A necessary condition for {\em two different} rational (resp. algebraic) 
pullbacks to exist for a hypergeometric
function like $\,  _2F_1([1/3, \, 2/3], \, [1], \, x)$, 
i.e. a necessary condition  
for a relation 
\begin{eqnarray}
\hspace{-0.5in} _2F_1\Bigl([{{1} \over {3}}, \, {{2} \over {3}}], \, [1], \, p_1(x)\Bigr)
 \, \,\,\, = \, \,\, \, \, \, r_{1,2}(x) \cdot  \,
  _2F_1\Bigl([{{1} \over {3}}, \, {{2} \over {3}}], \, [1], \, p_2(x)\Bigr), 
\end{eqnarray}
for some algebraic functions $\, r_{1,2}(x)$, is the (symmetric) condition
\begin{eqnarray}
\label{Schwarz}
\hspace{-0.7in}&&\{p_1(x), \,x\}\, \,  \,
+ \, \, {{1} \over {18}} \cdot  {{8 \,p_1(x)^2\, -8 \,p_1(x) \, +9 
} \over {p_1(x)^2 \cdot (p_1(x)-1)^2}}
  \cdot  \Bigl( { { d p_1(x) } \over { d x }}   \Bigr)^2
 \nonumber \\
\hspace{-0.7in}&&\qquad \quad  \quad = \, \,\,  \, \, \, \{p_2(x), \,x\}
\, \, \,  \,
+ \, \, {{1} \over {18}} \cdot  {{8 \,p_2(x)^2\, -8 \,p_2(x) \, +9
 } \over {p_2(x)^2 \cdot (p_2(x)-1)^2}} 
 \cdot  \Bigl( { { d p_2(x) } \over { d x }}   \Bigr)^2, 
\end{eqnarray}
where $\{p_1(x), \,x\}$ denotes the {\em Schwarzian derivative}. 
This condition
is invariant by the simple transformations 
$\, p_i \, \rightarrow \, 1\, -\, p_i$, $\, i \, =\, 1,\,2$.
One immediately verifies that the {\em Schwarzian condition}
 (\ref{Schwarz})
 is actually verified for the pullbacks
$\, (p_1(x), \, p_2(x))  \, \, $ occurring in
 (\ref{cubic}), namely 
$(27\, x^3, \,\, 1 \, -(1 \, -3 \, x)^3/(1 \, + \, 6 \, x)^3)$,
or in (\ref{sol1other}), namely
 $\, (27\, x/(1+\, 4\, x)^3, \, 27\, x^2/(1-\, 2\, x)^3)$. 

Let us consider the modular curve (\ref{moducurve})
for $\, u \, \, = x$, seeing $\, v$ as a algebraic
 function of $\, x$. The three root-solutions 
expand as:
\begin{eqnarray}
\hspace{-0.95in}&& v_0(x) \, = \, \, \,\, \, \,  1  \, \, \,\, 
 - {{1} \over {729 }}\, x^3  \, \, 
\, - {{5} \over {2187 }} \, x^4 \,  \, - {{56} \over {19683 }} \, x^5 \, \,
 - {{1691} \over {531441 }} \, x^6 \, \,  
 - {{5390 } \over {1594323 }} \, x^7 \,\,  + \, \, \, \cdots, 
\nonumber 
\end{eqnarray}
\begin{eqnarray}
\label{v1}
\hspace{-0.95in}&& v_1(x) \, = \, \, \,
 \left({{i\sqrt {3}} \over {2}}  -{\frac {1}{2}} \right) \cdot \, x 
\, +\,{{ 5 \, i\sqrt {3}} \over {9}} \, {x}^{2}\,\,
+ \left(  {{i\sqrt {3}} \over {2}} +{\frac {19}{54}} \right) \cdot \, {x}^{3}
 \,+ \left( {\frac {65}{162}}
\,i\sqrt {3}+{\frac {95}{162}} \right) \cdot \, {x}^{4}
\nonumber \\
\hspace{-0.95in}&& \quad \quad \quad \quad  \,
\,+ \left( {\frac {70}{243}}\,i\sqrt {3}+{\frac {532}{729}} \right) \cdot \, {x}^{5}\,
\,+ \left( {\frac {1171}{1458}} +{\frac {2297}{13122}}\,i\sqrt {3} \right) \cdot \, {x}^{6}\,
\, \,\,  + \, \, \, \cdots, 
\end{eqnarray}
the third series expansion $\, v_2(x)$ having its coefficients complex conjugate 
of the ones in the series expansion $\, v_1$. 
These series expansions compose nicely:
\begin{eqnarray}
\hspace{-0.9in}&& v_1(v_2(x)) \,\, = \, \,\, v_2(v_1(x)) \,\, = \, \,\, x,  
 \qquad \, \,\,
v_0(v_1(x)) \,\, = \, \,\, v_0(v_2(x)) \,\, = \, \,\, v_0(x), 
 \\ 
\hspace{-0.9in}&&  v_1(v_1(x)) \,\, = \, \,\,  v_2(x),   \quad 
v_2(v_2(x)) \,\, = \, \,\, v_1(x),  \quad 
 v_1(v_1(v_1(x))) \,\, = \, \,\,v_2(v_2(v_2(x))) \,\, = \, \,\,  x. \nonumber 
\end{eqnarray}

One verifies on these three series expansions (\ref{v1}) the
condition corresponding to
the $\, p_1(x) \, = \,  x \, \, $ subcase in (\ref{Schwarz}): 
\begin{eqnarray}
\label{Schwarz2}
\hspace{-0.9in}&& \{v(x), \,x\}\, \, 
+ \, \, {{1} \over {18}} \cdot  {{8 \,v(x)^2\, -8 \,v(x) \, +9 
} \over {v(x)^2 \cdot (v(x)-1)^2}}
  \cdot  \Bigl( { { d v(x) } \over { d x }}   \Bigr)^2 \,  \,  \, 
= \, \, \, \, 
 \, \, {{1} \over {18}} \cdot  {{8 \,x^2\, -8 \,x \, +9 } \over {x^2 \cdot (x-1)^2 }}.
\end{eqnarray}
One also verifies that the three series expansions (\ref{tildev0}) and 
 (\ref{tildev1}) also satisfy, as they should,  the previous Schwarzian 
condition (\ref{Schwarz2}). One also verifies, as it should, that
the composition of all these series satisfy (\ref{Schwarz2}).
For instance 
\begin{eqnarray}
\label{tildev0square}
\hspace{-0.7in}&& \tilde{v}_0(\tilde{v}_0(x)) \,  \, = \, \, \,\, \,
{\frac {1}{19683}}\,{x}^{4} \, +{\frac {20}{177147}}\,{x}^{5} \, 
+{\frac {274}{1594323}}\,{x}^{6} \, +{\frac {86636}{387420489}}\,{x}^{7} \, 
 \, \, \, + \, \, \cdots, 
\nonumber  \\
\label{tildev0square}
\hspace{-0.7in}&& v_0(\tilde{v}_0(x)) \,  \, = \, \, \,\, \,
1 \,\,  -{\frac {1}{14348907}}\,{x}^{6} \, -{\frac {10}{43046721}}\,{x}^{7} \, 
-{\frac {187}{387420489}}\,{x}^{8}  \,\, \,\, 
+ \,\, \, \cdots, 
\end{eqnarray}
satisfies (\ref{Schwarz2}).

\vskip .2cm

In other words, the {\em Schwarzian condition} (\ref{Schwarz}) 
is another way to encode 
the modular curves (\ref{moducurve}) or (\ref{moducurve1}),
 and in fact, an infinite 
number of  modular curves. The emergence of Schwarzian derivatives
 should not be seen as a surprise. A relation like
$_2F_1([1/3, \, 2/3], [1], \, v(x))
  =  \, R_{1,2}(x) \cdot  \,  _2F_1([1/3,  2/3],  [1], \, x)$
is obviously stable by the composition\footnote[1]{
The Schwarzian derivative $\, S(f)  = \, \{f(x), \,x\}$
is the well-suited derivative to take into account when
 composition of functions occurs. This  
can, for instance,  be seen on the chain rule:
$\,\,(S(f \circ g))(z) $
$\, = \, \, \, S(f)(g(z)) \cdot g'(z)^2 $ 
$ \, + \, \, S(g)$. } 
of the pullback functions $\, v(x)$.

\vskip .1cm 

\subsection{Second modular form example}

The integrality of  series-solutions
 can be quite non-trivial like 
the solution of the 
 Ap\'ery-like operator
\begin{eqnarray}
\hspace{-0.7in}&&\Omega \,\,  \, = \, \, \, \,\, \, 
x \cdot (1 \, -11 \, x \, -x^2) \cdot D_x^2 \,\,  \,\,\,
 + (1 \, -22 \, x \, -3 \, x^2) \cdot D_x\,\,
\, \, -(x+3), \\
\hspace{-0.7in}&& \hbox{or:} \qquad \quad x \cdot \, \Omega
 \,\,  \, = \, \, \,\, \,\,  \,  
\theta ^2 \,\,\,  \,  -x \cdot \, (11 \, \theta^2 \, +11 \, \theta\, +\, 3)
\,\, \,    -x^2 \cdot \, (\theta\, +1)^2, \nonumber 
\end{eqnarray}
which can be written  as a HeunG
function. Introducing 
$\, \alpha \, = \, \, 11/2 \, - 5  \cdot 5^{1/2}/2, \, $
this (at first sight involved) HeunG
function reads: 
\begin{eqnarray}
\label{11253}
\hspace{-0.9in}&&HeunG\Bigl(-\, {{123} \over {2}} \,
 +{{55} \over {2}} \cdot 5^{1/2},\,
 -\, {{33} \over {2}} \, +{{15} \over {2}} \cdot 5^{1/2},1,1,1,1,\,
 \alpha \cdot x\Bigr) 
\nonumber \\
\hspace{-0.9in}&& \quad \, \, = \, \, \,  \,  \, \, 
{{1} \over {1\, -\alpha \, x}} \cdot \, HeunG\Bigl({{1} \over {2}} \,
 -{{11} \over {50}} \cdot 5^{1/2},\,
 {{1} \over {2}} \, -{{1} \over {10}} \cdot 5^{1/2},1,1,1,1,\,
 -\, {{ \alpha \, x} \over{1 \, -\, \alpha \, x }} \Bigr) 
\nonumber \\
\hspace{-0.9in}&& \quad \, \, = \, \, \,  \,  \, \, 
\sum_{n\, = \, 0}^{\infty} \,
 \sum_{k\, = \, 0}^{n} \, {n\choose k}^2 \, {n+k\choose k} \cdot \, x^n 
\nonumber 
\end{eqnarray}
\begin{eqnarray}
\hspace{-0.9in}&& \quad \, \, = \, \, \, \, \, \, \,\, 
1 \,\,\,\,  \, +3 \cdot x \, \,+19 \cdot x^2 \,\, +147 \cdot x^3  \,\,
+1251 \cdot x^4\,\,
 +11253 \cdot x^5 \,  \, +104959 \cdot x^6 \,
\nonumber \\
\hspace{-0.9in}&& \quad\quad \quad \,
 + 1004307 \cdot x^7 +9793891 \cdot x^8\,
 +96918753 \cdot x^9\, +970336269 \cdot x^{10} \,
 \nonumber \\
\hspace{-0.9in}&& \quad \quad\quad 
  +9807518757 \cdot x^{11}\, \, +99912156111 \cdot x^{12}\,
 +1024622952993 \cdot x^{13}
\nonumber \\
\hspace{-0.9in}&& \quad \quad \quad  +10567623342519 \cdot x^{14}\,
+ 109527728400147 \cdot x^{15}\, + 1140076177397091 \cdot x^{16}
\nonumber \\
\hspace{-0.9in}&& \quad \quad\quad 
 + 11911997404064793 \cdot x^{17} \, +124879633548031009 \cdot x^{18} \,
 \nonumber \\
\hspace{-0.9in}&& \quad \quad \quad 
+1313106114867738897 \cdot x^{19} \, \,\,\, \,  + \,\,  \cdots 
\end{eqnarray}
but  {\em actually corresponds to a modular form}, which
 can be written in
two different ways using {\em two pullbacks}:
\begin{eqnarray}
\label{cinqun}
\hspace{-0.6in}&&(x^4\,+12\,x^3+14\, x^2\,-12\,x\,+1)^{-1/4} \,
\nonumber \\
\hspace{-0.6in}&&\qquad \quad \quad \, \,\times  \,
   _2F_1\Bigl([{{1} \over {12}}, \,{{5} \over {12}}], \, [1]; \,
 {{1728 \cdot x^5 \cdot (1 \, -11\, x \, -x^2) } \over {
(x^4\,+12\, x^3\,+14\,x^2\,-12\,x\,+1)^3 }}\Bigr)
\nonumber \\
\hspace{-0.6in}&&\,\quad  = \, \, \, \, \,\,\,  
(1 \, + \, 228 \, x \, +\, 494 \, x^2 \, -228 \, x^3 \,
 + \, x^4)^{-1/4} \,  \\
\hspace{-0.6in}&&\qquad \quad \quad \quad  \, 
\times  \,  _2F_1\Bigl([{{1} \over {12}}, \,{{5} \over {12}}], \, [1]; \,
 {{1728 \cdot x \cdot (1 \, -11\, x \, -x^2)^5 } \over {
(1 \, + \, 228 \, x \, +\, 494 \, x^2 \, -228 \, x^3 \, + \, x^4)^3 }}\Bigr).
 \nonumber 
\end{eqnarray}

Do note that these two pullbacks are respectively of the form:
\begin{eqnarray}
\hspace{-0.5in}&& {{1728 \cdot x^5 \cdot (1 \, -11\, x \, -x^2) } \over {
(x^4\,+12\, x^3\,+14\,x^2\,-12\,x\,+1)^3 }} \\
\hspace{-0.5in}&& \qquad \qquad \quad \quad  \, \, = \, \, \, \, \, \,  \,  
{{1728 \cdot x } \over {(x^2\,+10\, x\, +5)^3 }} \, 
\circ \, {{1\,-11\,x\,-x^2 } \over {x }},
\nonumber 
\end{eqnarray}
and
\begin{eqnarray}
\hspace{-0.5in}&& {{1728 \cdot x \cdot (1 \, -11\, x \, -x^2)^5 } \over {
(1 \, + \, 228 \, x \, +\, 494 \, x^2 \, -228 \, x^3 \, + \, x^4)^3 }} \\
\hspace{-0.5in}&& \qquad \qquad \quad \quad   \, \, = \,\, \, \, \, \, \,   
{{1728 \cdot x^5 } \over {(x^2\,+250\, x\, +3125)^3 }} \,  \, 
\circ \, \,   \, {{1\,-11\,x\,-x^2 } \over {x }},
\nonumber 
\end{eqnarray}
where one recognises~\cite{Maier} the two Hauptmoduls
of the modular curve corresponding to
 $\, \tau \, \rightarrow \, 5 \cdot \tau$:
\begin{eqnarray}
\label{totalrecall}
\hspace{-0.4in}&&{\cal M}_5(x) \, \, = \, \,  \,
{{1728 \cdot x } \over {(x^2\,+10\, x\, +5)^3 }}, 
 \,  \,  \, \\ 
\hspace{-0.4in}&&{\tilde {\cal M}}_5(x) \, \, = \,  \,  \,  \,
{{1728 \cdot x^5 } \over {(x^2\,+250\, x\, +3125)^3 }} 
\, \,  \, = \,\, \, \, \,
{\cal M}_5 \left({{5^3} \over {x}} \right).
\nonumber 
\end{eqnarray}

\vskip .1cm

Also note that the fact that the relation between the two Hauptmoduls
 (\ref{totalrecall}) corresponds to 
 $\, \tau \, \rightarrow \, 5 \cdot \tau$
can be seen straightforwardly if one introduces the quite non-trivial
Dedekind-like parametrisation 
\begin{eqnarray}
\hspace{-0.9in}&&x(q) \, \, =  \, \, 
q \cdot \, \prod_{n=1}^{\infty} \, (1\, -\, q^n)^{a(n)},
 \, \quad  \hbox{where:}  \, \,  \quad  
\sum_{n=\, 1}^{\infty} \,  a(n) \cdot \, t^n \, \, =  \, \,  \, 
\, {{5\cdot \, t \cdot \, (1\, -t) \, (1\, -t^2)
} \over {1 \, -t^5}}, 
\nonumber
\end{eqnarray}
yielding:
\begin{eqnarray}
\hspace{-0.5in} {\tilde {\cal M}}_5\Bigl( {{1 \, -11 \,x(q^5) \, -x^2(q^5)
 } \over {x(q^5) }}
 \Bigr)\,\,\,\,  = \,\, \, \,\, \, \,
{\cal M}_5\Bigl( {{1 \, -11 \,x(q) \, -x^2(q) } \over {x(q) }} \Bigr).
\end{eqnarray}

\vskip .1cm

More modular form examples of series with integer coefficients 
are given in \ref{Goly}.
The modular form examples displayed in \ref{modular}, 
 corresponded to lattice Green functions~\cite{Guttmann}.
Therefore, they have {\em $\, n$-fold integral representations},
and, {\em after} section (\ref{gener}), can be seen 
to {\em be diagonals of rational functions}.
In contrast the modular form examples displayed  in \ref{Goly} correspond to 
differential geometry examples discovered by 
Golyshev and Stienstra~\cite{Golyshev}, 
where no $\, n$-fold integral representation is available at first sight.

\subsection{Remark}
\label{7.3}

\vskip .1cm

At first sight one might think, and this is almost suggested in the 
literature, that the integrality of the series-solutions of 
the globally nilpotent operators, corresponds to some deep 
arithmetic property and, in the same time, that these 
integer coefficients have a deep ``physical'' meaning 
(instantons, ...). There is often some
confusion in the literature between the
 concept of {\em integrality} of series (globally bounded
series) and the concept of 
{\em modularity}~\cite{SP4} which suggests a connection with selected 
algebraic varieties (modular forms and elliptic functions~\cite{Huse}, 
mirror maps, Calabi-Yau manifolds). 

Note that, along this ``selected algebraic varieties'' line, 
it is also worth recalling  Krammer-Deitweiler's
 counter-example~\cite{Bouw,Dettweiler}
 to Dwork's conjecture of a globally nilpotent operator
that cannot be reduced to an operator having hypergeometric 
solutions up to a pullback, which corresponds 
to the two following HeunG
functions: 
\begin{eqnarray}
\label{HeunG81}
&&  HeunG(81, \, 1/2,\,  1/6,\,  1/3, \, 1/2,\,  1/2,\, 81\, x),  \\
&& x^{1/2}  \cdot
  HeunG(81,\,  21,\,  2/3, \, 5/6,\,  3/2,\,  1/2,\, 81\, x). 
\nonumber 
\end{eqnarray}
These two HeunG are solutions of a globally nilpotent 
linear differential operator\footnote[1]{Generically 
HeunG functions with rational 
parameters are not globally nilpotent, which indicates that they do not 
have an integral representation as an integral of some
algebraic integrand.}
but {\em are not globally bounded}: they are $\, G$-series that 
{\em cannot} be reduced to series with {\em integer coefficients}.
However, this  special example corresponds to a special 
{\em arithmetic} situation:  we are dealing, here,
 with very special HeunG functions
 associated with periods of a family of
 {\em abelian surfaces over a Shimura curve}~\cite{Dettweiler},
and  very special ODEs corresponding to 
uniformising linear differential equations
of {\em arithmetic Fuchsian lattices}. 

The interest of demonstration of sections  (\ref{calcula}), (\ref{gener})
is to show that this integrality can be, in fact, 
the straight consequence of a quite 
simple form of the $\, n$-fold integral namely (\ref{form}) with the analyticity
condition (\ref{form2}), that can be expected 
to be seen in a quite general framework 
(for instance in enumerative combinatorics), far from the theory
of elliptic curves, or the theory of Calabi-Yau manifolds 
and other Frobenius manifolds~\cite{Manin}.
Let us now try to ``disentangle'' the concept of  {\em integrality} of series 
(globally bounded series) and the concept of 
{\em modularity}. 

\section{Integrality versus modularity}
\label{learning}
\vskip .1cm

\subsection{Diffeomorphisms of unity pullbacks}
\label{diffeofunit}

Let us consider a first simple example of a hypergeometric function
which is solution of a Calabi-Yau ODE, and which occurred, at least
two times in the study of the Ising 
susceptibility $\, n$-fold integrals~\cite{CalabiYauIsing1,CalabiYauIsing}
$\, \chi^{(n)}$ and $\, \chi_d^{(n)}$, namely 
$_4F_3([1/2,\,1/2,\,1/2,\,1/2 ], \, [1, \, 1, \, 1], 256 \,x)$, where 
we perform a (diffeomorphism of unity) pullback:
\begin{eqnarray}
\label{diffeo}
\hspace{-0.9in}&&_4F_3\Bigl([{{1} \over {2}},\,{{1} \over {2}},
\,{{1} \over {2}},\,{{1} \over {2}}], 
\, [1, \, 1, \, 1], \, \,  \,{{256 \,\, x} \over {
1 \,+c_1\,x \, +c_2\, x^2+ \, \cdots }}\Bigr)
\, \, \, \,= \,\,  \,\,\,\, \,   1 \,\,\,\,  +16\cdot \, x \,
    \\
\hspace{-0.9in}&&\quad \quad  \quad \,  \, \, \, +(1296-16\,c_1)\cdot x^2 
\,\, \,  +(160000 \,+16\,c_1^2 \, -16\,c_2 \, -2592\, c_1)\cdot x^3
\,\,\,\,  \,  + \,\,  \cdots 
\nonumber 
\end{eqnarray}
If the pull-back in (\ref{diffeo}) is such that the
coefficients $\, c_n$, at its denominator, are integers, one finds that the 
series expansion is actually a series with integer coefficients, for {\em every
such pullback} (i.e. for every integer coefficients $\, c_n$).
Furthermore, a straightforward calculation of the corresponding nome $\, q(x)$ 
and its compositional inverse (mirror map) $\, x(q)$, 
{\em also yields series with integer coefficients}:
\begin{eqnarray}
\label{q}
\hspace{-0.7in}&& q(x) \, \,= \,\,\, \,  \,\,\,x\, \,  \,  \,
 +(64-c_1)\cdot x^2 \, \, 
+(c_1^2\,+7072 \, -c_2\,-128\,c_1)\cdot x^3 \, \, \, \, + \, \cdots, \\
\hspace{-0.7in}&&x(q) \, \,  = \,\, \,  \, \,\,   q \,\, \,\, \,  
 +(c_1 \, -64)\cdot q^2 \,\, 
 +(c_1^2 \, +1120 \, +c_2 \, -128\, c_1)\cdot q^3 \, \, \, \,  + \, \cdots, 
\end{eqnarray}
when its Yukawa coupling~\cite{CalabiYauIsing1}, seen
 as a function of the nome $\, q$,
 $\, K(q)$ is also a series with integer coefficients
and is {\em independent of the pullback}:
\begin{eqnarray}
\label{Yuku}
\hspace{-0.4in}K(q) \, \,= \,\, \, \,\, \,1 \,\, \,\, \, + \, 32 \cdot \,q\,  \,
 +  4896 \cdot \, q^2 \,\, +702464 \cdot \, q^3 \, \,\,\, \,   + \, \cdots 
\end{eqnarray}

This independence of the Yukawa coupling with regards to  pullbacks,
is a known property, and has been proven in~\cite{Almkvist}, for
any pullbacks of the diffeomorphism of unity 
form $\,\, p(x) \, = \, \, \, x \, \, + \, \cdots$

\vskip .1cm

{\bf Remark:} Seeking for Calabi-Yau ODEs, Almkvist et al. 
have obtained~\cite{TablesCalabi} a quite large list
 of fourth order ODEs, which are MUM
by definition and have, by construction, the {\em integrality} for 
the solution-series\footnote[2]{Hence, these operators are $\, G$-operators. One
 can also, calculating their $\, p$-curvatures,
 check, directly, that the Calabi-Yau operators 
in Almkvist et al.~\cite{TablesCalabi} tables 
are actually globally nilpotent~\cite{bo-bo-ha-ma-we-ze-09}, 
thus yielding automatically the rationality
of the exponents for all the singularities}
 analytic at $\, x \, = \, \, 0$.
Looking at the Yukawa coupling of these ODEs is a way to define 
{\em equivalence classes up to pullbacks} of ODEs sharing the 
same Yukawa coupling. This ``wraps in the same bag'' all the linear ODEs that 
are the same {\em up to pullbacks}. Let us recall how difficult it is to see
if a given Calabi-Yau ODE has, up to operator equivalence, and 
up to pullback, a hypergeometric function 
solution~\cite{CalabiYauIsing1,CalabiYauIsing}, because
finding the pullback is extremely
 difficult~\cite{CalabiYauIsing1,CalabiYauIsing}. We may
 have, for the Ising model,
some $\, _{n+1}F_n$ hypergeometric function
 prejudice~\cite{CalabiYauIsing1,CalabiYauIsing}:
it is, then, important to have an invariant that is independent of 
this pullback (we cannot find most of the time).

\vskip .1cm 

{\bf Remark:} The Yukawa coupling is not preserved by 
the operator equivalence. Two linear differential operators, that are 
 homomorphic, do not necessarily have the same Yukawa
 coupling (see \ref{Yukawaratio}).

\vskip .1cm

\subsection{Yukawa couplings in terms of determinants}
\label{Yukdet}

Another way to understand this fundamental 
{\em pullback invariance}, amounts to rewriting
 the Yukawa coupling~\cite{mirror,Almkvist},
not from the definition usually given 
in the literature (second derivative with
respect to the ratio of periods), but in terms 
of determinants of solutions (Wronskians, ...) that naturally 
present nice covariance properties with respect 
to pullback transformations (see \ref{Yukawaratio}).

We have the alternative definition for the 
{\em Yukawa coupling} given in \ref{Yukawaratio}: 
\begin{eqnarray}
\label{Yukawa}
K(q) \,\,\, = \, \, \,\,\, 
 \Bigl( q \cdot {{d} \over {dq }} \Bigr)^2
 \Bigl(  {{y_2} \over {y_0}}\Bigr)
 \, \,\,\, = \, \,\,\,\, \, 
 {{W_1^3 \cdot W_3 } \over {W_2^3 }}, 
\end{eqnarray}
where the determinantal variables $\, W_m$'s 
are the determinants built from the four solutions 
of the MUM differential operator. 
This alternative definition, in terms of these $\, W_m$'s, 
enables to understand the {\em remarkable invariance 
of the Yukawa coupling by pullback 
transformations}~\cite{CalabiYauIsing}. 
These determinantal variables $\, W_m$ 
quite naturally, and canonically, yield to introduce another 
 ``Yukawa coupling'' (which, in fact, 
{\em corresponds to the Yukawa coupling
of the adjoint operator} (see \ref{Kstar})). This ``adjoint Yukawa coupling''
 is {\em also invariant by pullbacks}. It 
has, for the previous example, the following series 
expansion with integer coefficients:
\begin{eqnarray}
\label{Yukstar}
\hspace{-0.4in}K^{\star}(q) \,\,\, = \, \,\, \, \,\, \,
1 \,\,\,\,  + \, 32 \cdot \,q\,  
\, +  4896 \cdot \, q^2 \,\, +702464 \cdot \, q^3 \, \,\,\, \,\, + \,\, \cdots 
\end{eqnarray}
which actually identifies with (\ref{Yuku}). 
The equality of the Yukawa coupling for this order-four
 operator, and for its 
(formal) adjoint  operator, is a straightforward 
consequence of the fact
that the order-four operator annihilating 
$ \, _4F_3\Bigl([{{1} \over {2}},\,{{1} \over {2}},
\,{{1} \over {2}},\,{{1} \over {2}}], 
\, [1, \, 1, \, 1], \, \, 256 \,x\Bigr)$
is exactly {\em self-adjoint}, and, more generally, 
 of the fact that the order-four operator, 
annihilating (\ref{diffeo}),
is conjugated to its adjoint by a simple function (which is nothing
but the denominator of the pullback).

\vskip .1cm

\subsection{Modularity}
\label{modu}

This example, with its corresponding relations 
(\ref{diffeo}), (\ref{q}), (\ref{Yuku}), (\ref{Yukstar})
 may suggest a quite wrong prejudice 
that the {\em integrality of the solution} of an 
order-four linear differential operator
automatically yields to the integrality of the
 nome, mirror map and Yukawa
coupling, that we will call,
 for short, ``{\em modularity}''. This is {\em far from being
the case}, as can be seen, for instance, in the following interesting
example, where the nome and Yukawa coupling $\, K(q)$ 
{\em do not correspond to globally bounded series}, when the $\, _4F_3$
solution of the order-four operator as well as the 
Yukawa coupling {\em seen as a function of $\, x$},  $\, K(x)$,
are, actually, both {\em series with integer coefficients}. 

\vskip .1cm

Let us consider the following $\, _4F_3$ hypergeometric function
which is clearly a Hadamard product of algebraic functions
 and, thus, the diagonal of a rational function:
\begin{eqnarray}
\label{saoud}
\hspace{-0.8in}&&\quad _4F_3\Bigl([{{1} \over {2}},\, {{1} \over {3}},
 \,{{1} \over {4}},\, {{3} \over {4}}],
\, [1,\,1,\,1 ], \, x\Bigr) 
\,\, \,  \, 
\nonumber   \\
\hspace{-0.8in}&&\quad \qquad \quad \,  \,= \, \, \,\,  \, 
(1\,-x)^{-1/3} \,\star \,  (1 \,-x)^{-1/2} \,
 \star \, (1 \,-x)^{-1/4} \,\star \,  (1 \,-x)^{-3/4}
\nonumber  \\
\hspace{-0.8in}&&\quad \qquad \quad  \,  \,= \, \, \,  \,\, \, 
\Diag\Bigl( 
(1-z_1)^{-1/3} \, (1-z_2)^{-1/2} \, (1-z_3)^{-1/4} \, (1-z_4)^{-3/4}
\Bigr) , \nonumber
\end{eqnarray}
It is therefore globally bounded:
\begin{eqnarray}
\label{saoud}
\hspace{-0.9in}&&_4F_3([{{1} \over {2}},\, {{1} \over {3}}, \,
{{1} \over {4}},\, {{3} \over {4}}],
\, [1,\,1,\,1 ], \, 2304\, x) 
\,   \, \,  \,  \,= \,\, \, \,\, \,  \,1 \, \,\,  \, +72\, x\,\, 
+45360\, x^2\,\, +46569600\, x^3\,
 \nonumber \\
\hspace{-0.9in}&&\qquad \quad +59594535000\, x^4 \,\, +86482063571904\, x^5\,\,
 + 136141986298526208\, x^6\, 
 \nonumber \\
\hspace{-0.9in}&& \qquad  \quad +226888189910421811200\, x^7\, \,
+394399917777684601926000\, x^8\,  
\nonumber \\
\hspace{-0.9in}&& \qquad \quad 
+708188604075430924446000000\, x^9 \,\,\,\, \,  +  \,\cdots 
\end{eqnarray}

Its Yukawa coupling, seen as a function of $\, x$,  is actually a  
{\em series with integer coefficient} in $\, x$:
\begin{eqnarray}
\label{Ksaoudx}
\hspace{-0.9in}&&K(x) \,\, = \, \,\, \,\,\,\, \,  
1\,\, \,\, +480\,x\,\, +872496\,{x}^{2}\, \,
+1728211968 \,{x}^{3}\,\, +3566216754432\,{x}^{4}\, 
\nonumber  \\
\hspace{-0.9in}&& \quad \quad  \quad \, +7536580798814208 \,{x}^{5}\,\,
 +16177041308360579328 \,{x}^{6}\, \\
\hspace{-0.9in}&& \quad \quad  \quad  \, 
 + 35105183794659521064960 \,{x}^{7}
\,  +76799014669577085362391024\,{x}^{8}\, 
\nonumber \\
\hspace{-0.9in}&& \quad \quad  \quad \, 
+169059790576811511759706311168 \,{x}^{9}\,\,\, \, \,  + \, \, \cdots 
\nonumber 
\end{eqnarray}
However, do note that the series, in term of the nome, is 
{\em not globally bounded}:
\begin{eqnarray}
\label{Ksaoudq}
\hspace{-0.9in}&&K(q) \,\,\, = \,\, \,\,\,\,\, 
 \, 1\,\,\, \, +480\,\, q\,\,\, +653616\,  \,{q}^{2}\,\,
 +942915456\,  \,{q}^{3}\,\, +1408019875200 \, \,{q}^{4}\,
\nonumber \\
\hspace{-0.9in}&& \quad \quad \quad \quad  +2146833138536640\,  \,{q}^{5}
\, \,\,\, + \,\,  \, \cdots  \\
\hspace{-0.9in}&&  \quad \quad \quad \quad
+ \, 571436303929319146711343817202689132288 \,  \,{{ \, \, q^{12}} \over {11}}
\, \,\, \,+\, \, \,   \cdots \nonumber
\end{eqnarray}

In fact, the nome  $ \, q(x)$, and the mirror map $\, x(q)$, 
 are {\em also not globally bounded}.
Note that in this example, the non integrality appears at order twelve 
(for $ \, x(q)$, $ \, q(x)$ and $ \, K(q)$).
If the prime 11 in the denominator in (\ref{Ksaoudq})
was the only one, one could recast the series into 
a series with integer coefficients introducing
 another rescaling
 $2304 \,  x \, \rightarrow \,11 \times 2304 \, \,x$.
But, in fact, we do see the appearance of an {\em infinite number of 
other primes} at higher 
orders  denominators in $\, x(q)$, $ \, q(x)$ and $\, K(q)$.

\vskip .1cm

\subsection{Hadamard products of $\, \omega_n$'s}
\label{Hadomegan}

Let us consider the two order-two operators 
\begin{eqnarray}
\label{tau2}
\hspace{-0.3in}&& \omega_2 \,\, = \,\, \,\, \,  \,\,\,
D_x^{2} \, \,\,\, \,
 +{\frac { (96\,x+1) }{(64\,x+1)  \cdot \,x }} \, \cdot D_x \, \,\,
+ \,{\frac {4}{(64\,x+1)\, x }},
\\
\label{tau3}
\hspace{-0.3in}&&\omega_3 \,\, = \,\, \, \,\, \,\,\,
D_x^{2} \, \,\, \, \,
+{\frac { (45\,x+1) }{ (27\,x+1) \cdot \, x}} \cdot D_x
\, \,\,\, +\,{\frac {3}{ \left( 27\,x+1 \right) x}},
\end{eqnarray}
which are associated with two modular forms corresponding, on their associated
nomes $\,q$, to the transformations $\, q \, \rightarrow \, q^2$ and 
$\, q \, \rightarrow \, q^3$ respectively (multiplication 
of $\, \tau$, the ratio of their periods by $\, 2$ and $\, 3$),
as can be seen on their respective solutions:
\begin{eqnarray}
\label{stau2}
\hspace{-0.9in}&&_2F_1\Bigl([{{1} \over {4}}, \,{{1} \over {4}} ],
 \, [1], \, -64 \, x)
\, \,\,  = \, \, \,\,  \,
(1 \, +256\, x)^{-1/4} 
\cdot \, _2F_1\Bigl([{{1} \over {12}}, \,{{5} \over {12}} ],
 \, [1], \, {{1728 \, x } \over {(1 \, +256\, x)^3 }}   \Bigr) 
 \nonumber \\
\hspace{-0.9in}&&\, \,  \qquad \, \, \,\,  \, \,
\, \,\,  = \, \, \,\,  \,
(1 \, +16\, x)^{-1/4} 
\cdot \, _2F_1\Bigl([{{1} \over {12}}, \,{{5} \over {12}} ],
 \, [1], \, {{1728 \, x^2} \over {(1 \, +16\, x)^3 }}   \Bigr)
 \\
\hspace{-0.9in}&&\, \,   = \, \, \,\,  \, \, \,
1 \,\,\,  \,  \, -4\,x\,\,  +100\,{x}^{2}\,\,  -3600\,{x}^{3}\,\,  +152100\,{x}^{4}\,\, 
-7033104\,{x}^{5}\,\,  +344622096\,{x}^{6} 
\nonumber \\
\hspace{-0.9in}&&\quad \quad  \quad  \quad  \quad 
\,-17582760000\,{x}^{7} \,+924193822500\,{x}^{8}\,\, 
-49701090010000\,{x}^{9}\, \, \, \, \, \,  + \, \,  \,\cdots 
\nonumber  
\end{eqnarray}
\begin{eqnarray}
\label{stau3}
\hspace{-0.9in}&&\Bigl((1 \, +27\, x) \, (1 \, +243\, x)^3\Bigr)^{-1/12} 
\cdot \, _2F_1\Bigl([{{1} \over {12}}, \,{{5} \over {12}} ],
 \, [1], \, {{1728 \, x } \over {
(1 \, +243\, x)^3\, (1\, +27 \, x) }}  \Bigr) \nonumber \\
\hspace{-0.9in}&&\,    = \, \, \Bigl((1 \, +27\, x) \, (1 \, +3\, x)^3\Bigr)^{-1/12} 
\cdot \, _2F_1\Bigl([{{1} \over {12}}, \,{{5} \over {12}} ],
 \, [1], \, {{1728 \, x^3 } \over {
(1 \, +3\, x)^3\, (1\, +27 \, x) }}  \Bigr) 
 \\
\hspace{-0.9in}&&\,    = \, \, 
\,  \,_2F_1\Bigl([{{1} \over {3}}, \,{{1} \over {3}} ],
 \, [1], \, -27 \, x) \, \,\,  = \, \, \, \,\,\, 
1 \,\,\,   -3\,x\, +36\,{x}^{2}\, 
 -588\,{x}^{3}\,  +11025\,{x}^{4}\,
-223587\,{x}^{5}\, 
\nonumber   \\
\hspace{-0.9in}&&\quad \quad \,\, \, \,  \, \,    \, \,\,  
 +4769856\,{x}^{6}\,\, -105423552\,{x}^{7}\, +2391796836\,{x}^{8}\,\, 
-55365667500\,{x}^{9} \,\,\,  \, + \,\, \, \cdots \nonumber
\end{eqnarray}

The relation between the two Hauptmodul pullbacks in (\ref{stau2}) 
\begin{eqnarray}
\hspace{-0.7in}&&\quad \quad \quad  u \,\, \,   = \, \, \,\,
 {{1728 \, x } \over {(1 \, +256\, x)^3 }}, \qquad \quad  \, \, \, 
v \,\, \,   = \, \, \,\, {{1728 \, x^2} \over {(1 \, +16\, x)^3 }}, 
\end{eqnarray}
corresponds to the (genus-zero) fundamental modular curve:
\begin{eqnarray}
\hspace{-0.9in}&&\quad  \, \,   1953125\,\,  {u}^{3}{v}^{3} \, \, \,
-187500\,\,  {u}^{2}{v}^{2} \cdot \, \left( v+u \right)\,\,\,  
 +375\,uv \cdot \, \left( 16\,{u}^{2}+16\,{v}^{2}\,  -4027\,uv \right)\,
\nonumber \\ 
\hspace{-0.9in}&&\quad \quad \quad \quad  \quad  -64\, \, (u\,  +v) \cdot \, 
 \left( {v}^{2}+1487\,uv+{u}^{2} \right)\,\, 
 +110592\,uv \,\,  \, = \, \,\,  \, 0, 
\end{eqnarray}
The relation between the two Hauptmodul pullbacks in (\ref{stau3}) 
\begin{eqnarray}
\hspace{-0.9in}&&\quad \quad  \quad  u \,\, \,   = \, \, \,\,
{{1728 \, x} \over {(1 \, +243\, x)^3 \, (1 \, +27\, x) }},
 \quad \, \quad 
v \,\, \,   = \, \, \,\, {{1728 \, x^3 } \over {(1 \, +3\, x)^3 \, (1 \, +27\, x)  }}, 
\end{eqnarray}
corresponds to the (genus-zero) modular curve:
\begin{eqnarray}
\hspace{-0.95in}&& 262144000000000\,\,{u}^{3}{v}^{3}  \cdot \, (u \, +v) \, 
 \, +4096000000\,\, {u}^{2}{v}^{2} \cdot \, (27\,{v}^{2}+27\,{u}^{2}-45946\,uv) \, 
\nonumber \\ 
\hspace{-0.95in}&&\quad \quad
 +15552000 \,uv \cdot \,  (u \,  + v) \cdot \,  \left( {v}^{2}+241433\,uv+{u}^{2} \right) \,  \, 
 \\ 
\hspace{-0.95in}&&\quad \quad \quad \quad \quad 
+729\,({u}^{4}\, +{v}^{4}) \,\,\,\, -779997924\,({u}^{3}v+\,u{v}^{3})\,\,\,
 +1886592284694\,\,{u}^{2}{v}^{2}
\nonumber \\ 
\hspace{-0.95in}&&\quad \quad \quad \quad \quad \quad \quad 
+2811677184\,uv \cdot \, (u\, + v)
 \, \,\, \, -2176782336\,uv \, \,\,\,  = \,\, \,\,  \, 0.
 \nonumber 
\end{eqnarray}

Similarly, one can consider the order-two operators
$\,\omega_n$ associated with other modular forms corresponding to 
$\, \tau \, \rightarrow \, n \cdot \tau$.
The $\, \omega_{n}$'s can be simply deduced from Maier~\cite{Maier1},
 for modular forms corresponding to genus-zero curves i.e. for 
$\, n \, = \, 2, \, 3,  $
$4, \, 5, \, $ $6, \, 7, \, 8, \, $ $9, \, 10, \, 12, \, $
 $ 13, \, 16, \, 18, \, 25$.
After a simple rescaling,
one gets series with integer coefficients. For instance 
considering the linear differential operator $\, {\cal L}_7$
(annihilating the modular form $\, h_7$)
in Table 13 of~\cite{Maier}
\begin{eqnarray}
\label{Maier7}
\hspace{-0.8in}\, \, {\cal L}_7 \, \, \,\,  = \, \, \,  \,\, \,\,  D_x^2 \,  \,\,\, 
+ \, \, \,{\frac {7\,{x}^{2}+65\,x+147}{ 3 \, ({x}^{2}+13\,x+49)\,  x}} \, \cdot \, D_x 
\,\, +\, \, \,\,{\frac {4\,x+21}{ 9 \, ({x}^{2} +13\,x +49) \, \cdot \,  x}},
\end{eqnarray}
one has the modular form solution
\begin{eqnarray}
\label{modformMaier7}
\hspace{-0.6in}&&D_7(x)^{1/12} \cdot \,
 _2F_1\Bigl([{{1} \over{12}}, \,{{5} \over{12}}], \, [1]; \, {{1728} \over {j_7(x)}}\Bigr)
\nonumber \\ 
\hspace{-0.6in}&&\quad \quad \, \, \, = \, \, \,  \,\, \,  \, 
 {\frac {7^{7/6}}{{x}^{2/3}}} \, \cdot \, 
D'_7(x)^{1/12} \cdot \,
 _2F_1\Bigl([{{1} \over{12}}, \,{{5} \over{12}}], \, [1]; \, {{1728} \over {j'_7(x)}}\Bigr)
\\ 
\hspace{-0.6in}&&\quad \quad \, \, \, = \, \, \,  \,\, \,  \,\,  1\,\,  \,
-{{1} \over {21}}\,x \,\,\, +{\frac {11}{3087}}\,{x}^{2} \, \,
-{\frac {380}{1361367}}\,{x}^{3} \, \, 
+{\frac {3887}{200120949}}\,{x}^{4} \,\, \,\,\,\, +\, \cdots 
\nonumber
\end{eqnarray}
where
\begin{eqnarray}
\hspace{-0.9in}&&j_7(x) \, \, \, = \, \, \,  \, 
{\frac { \left( {x}^{2}+13\,x+49 \right)  \left( {x}^{2}+5\,x+1 \right)^{3}}{x}}, 
\qquad \quad \, j'_7(x) \, \, \, = \, \, \, \, \,j_7\Bigl( {{49} \over {x}} \Bigr), 
\nonumber \\
\hspace{-0.9in}&& 
D_7(x)  \, \, \, = \, \, \,  \,  \,
{\frac {49}{ ({x}^{2}+13\,x+49)  \cdot ({x}^{2}+5\,x+1)^{3}}}, 
\quad  \quad \,
D'_7(x) \, \, \, = \, \, \,\,  \, D_7\Bigl( {{49} \over {x}} \Bigr), 
\end{eqnarray}
The series (\ref{modformMaier7}) is globally bounded. 
Rescaling the $\, x$ into $\,\, 3^2 \, \, 7^2 \cdot   \, x$
$ \, = \, \, 441 \cdot \, x$, 
Maier's linear differential operator (\ref{Maier7}) becomes 
\begin{eqnarray}
\hspace{-0.95in}\omega_{7}  \, \, \, = \, \, \,  \, \, D_x^2 \,  \,\, + \, \,
 {{1 \, +195\, x + \, 9261\, x^2} \over {
(1 \, +117\, x + \, 3969\, x^2) \cdot \, x }}
 \cdot \, D_x  
\, \,\, + \, \, {{ 21 \cdot \, (1\, + \, 84\, x)} \over {
(1 \, +117\, x + \, 3969\, x^2) \cdot \, x  }},
\end{eqnarray}
which has the following series with {\em integer} coefficients:
\begin{eqnarray}
\hspace{-0.6in}&&D_7(441 \, x)^{1/12} \cdot \,
 _2F_1\Bigl([{{1} \over{12}}, \,{{5} \over{12}}], \, [1]; \, {{1728} \over {j_7(441 \,x)}}\Bigr)
 \\ 
\hspace{-0.6in}&&\quad \, \, \,\, = \, \, \,  \, \,\,
1 \,\, \, -21\,x\, \, +693\,{x}^{2}\, \, -23940\,{x}^{3}\,
+734643\,{x}^{4}\,-13697019\,{x}^{5}\,
\nonumber \\ 
\hspace{-0.6in}&&\quad \quad \quad \quad \quad \quad
-494620749\,{x}^{6}\,+83079255420\,{x}^{7}\,-6814815765975\,{x}^{8}
\,\, \,  \,\, + \, \, \cdots 
\nonumber 
\end{eqnarray}

\vskip .1cm 

The two operators  $\, \omega_{2}$ and $\, \omega_{3}$ 
have a ``modularity'' property: their series expansions 
analytic at $\, x=\, 0$, 
(\ref{stau2}) and (\ref{stau3}), {\em as well as}
 the corresponding nomes, mirror maps
are series with integer coefficients. The Hadamard product 
is a quite natural transformation to introduce 
because {\em it preserves the global nilpotence
of the operators},  {\em it preserves the integrality of series-solutions}, 
and it is a {\em natural transformation to introduce when seeking for
diagonals of rational functions}\footnote[5]{And,
 consequently, has been heavily used
 to build Calabi-Yau-like ODEs 
(see Almkvist et al.~\cite{Almkvist}).}.  Let us perform the 
 Hadamard product of these two operators.
With some abuse of language~\cite{CalabiYauIsing}, the Hadamard product 
of the two order-two operators
 (\ref{tau2}) and  (\ref{tau3}) 
\begin{eqnarray}
\label{H23}
\hspace{-0.5in}&&H_{2,3}\,  \, \, = \, \, \, \,\, \, D_x^{4} \, \, \, \,\,
+6\,{\frac { (2064\,x-1)}{ (1728\,x-1) \cdot \, x }} \cdot D_x^3 \, \, \,
\,+{\frac { (19020\,x-7) }{ (1728\,x-1) \cdot \, x^2 }} 
 \cdot D_x^2 \, \,
\nonumber \\
\hspace{-0.5in}&& \quad \quad  \quad \quad  \quad \quad 
\, \,+{\frac { (4788\,x-1) }{(1728\,x-1) \cdot \, x^3 }}
\cdot D_x \,\,\,
\,+ \,{\frac {12}{(1728\,x-1) \cdot \, x^3  }}, 
\end{eqnarray}
is defined as the  (minimal order) linear differential 
operator having, as a solution,
the Hadamard product of the solution-series (\ref{stau2}) and (\ref{stau3}),
which is, by construction, a series with integer coefficients.  
This series is,  of course, nothing but the expansion of
the hypergeometric function:
\begin{eqnarray}
\label{1728}
\hspace{-0.3in}&& _4F_3([{{1} \over {4}}, \, {{1} \over {4}}, 
\, {{1} \over {3}}, \, {{1} \over {3}}],
 \, [1, \, 1, \, 1], 1728 \, x) \,  \\
\hspace{-0.3in}&& \qquad \qquad \, \, = \, \, \, \,\,\, 
 _2F_1([{{1} \over {4}}, \, {{1} \over {4}}], \, [1], \, -64 \, x) \, \star \,
 _2F_1([{{1} \over {3}}, \, {{1} \over {3}}], \, [1], \, -27 \, x). 
\nonumber    
\end{eqnarray}
\vskip .1cm
\vskip .1cm 
The Hadamard product of the order-two operator (\ref{tau2}) with itself
 (Hadamard square) 
\begin{eqnarray}
\label{H22}
\hspace{-0.6in}&&H_{2,2}\, \,  \, = \, \,\, \,  \,  \, D_x^{4} \,\,\, \,  \,  
 +2\,{\frac {(14336\,x-3)}{ (4096\,x-1) \cdot \, x }} \cdot D_x^3
\,\,\, \,  +{\frac { (42496\,x-7) }{(4096\,x-1) \cdot \, x^2 }} \cdot D_x^2
\nonumber \\
\hspace{-0.4in}&& \quad \quad \quad \quad \quad  \quad 
\, +{\frac { (9984\,x-1)}{ \left( 4096\,x-1 \right) \cdot \, x^3  }}\cdot D_x
\,\,\, \, + \,{\frac {16}{(4096\,x-1) \cdot \, x^3  }}, 
\end{eqnarray}
is defined as the  (minimal order) linear differential 
operator having the series-solution
\begin{eqnarray}
\hspace{-0.9in}&&1 \,  \, +16\,x \, +10000\,{x}^{2} \, +12960000\,{x}^{3} \, 
+23134410000\,{x}^{4} \, +49464551874816\,{x}^{5}\,  \, \, +  \,\cdots, 
\nonumber
\end{eqnarray}
which is the  Hadamard product of the solution-series (\ref{stau2}) 
with itself. This series is,  of course, nothing but the expansion of:
\begin{eqnarray}
\label{4096}
\hspace{-0.5in}&& _4F_3([{{1} \over {4}}, \, {{1} \over {4}}, \,
 {{1} \over {4}}, \, {{1} \over {4}}],
 \, [1, \, 1, \, 1],  \, 4096 \, x) \, 
 \\
\hspace{-0.5in}&& \qquad  \qquad  \qquad \, \, = \, \, \, \, \,\,
  _2F_1([{{1} \over {4}}, \, {{1} \over {4}}], \, [1], \, -64 \, x) \, \star \,
 _2F_1([{{1} \over {4}}, \, {{1} \over {4}}], \, [1], \, -64  \,  x). 
\nonumber    
\end{eqnarray}
This operator $\, H_{2,2}$ is a MUM operator.  We can, therefore,
 define, without any ambiguity,
the nome (and mirror map) and Yukawa coupling 
of this order-four operator~\cite{CalabiYauIsing}.
One finds out that the nome\footnote[1]{The nome of the Hadamard product
of two operators has no simple relation with the nome
of these two linear differential operators.}, and the mirror map (and the Yukawa coupling 
as a function of the $\, x$ variable), are {\em not globally bounded}:
they {\em cannot} be reduced, by one rescaling, to series with integer coefficients.

\vskip .1cm 

Similarly, one can also introduce the Hadamard square of (\ref{tau3})
\begin{eqnarray}
\label{H33}
\hspace{-0.4in}&&H_{3,3}\, \, \, = \, \,\,  \, \, \, \, \,
 D_x^{4} \,\, \,  \, \,
 +6\,{\frac { (891\,x-1)^{3}}{(729\,x-1) \cdot \, x }} \cdot D_x^3
\,\,\,  \,  +7\,{\frac { \left( 
1215\,x-1 \right) }{(729\,x-1)\cdot \, x^2 }} \cdot D_x^2\, 
\nonumber \\
\hspace{-0.4in}&& \quad \quad \quad \quad  \quad  \quad  \quad 
\, +{\frac { (2295\,x-1) }{(729\,x-1) \cdot \, x^3 }} \cdot D_x
\,\,\,\, +\,{\frac {9}{(729\,x-1)\cdot \, x^3 }} ,
\end{eqnarray}
which has the hypergeometric solution: 
\begin{eqnarray}
\label{729}
\hspace{-0.5in}&& _4F_3([{{1} \over {3}}, \, 
{{1} \over {3}}, \, {{1} \over {3}}, \, {{1} \over {3}}],
 \, [1, \, 1, \, 1], 729 \, x) \, 
 \\
\hspace{-0.5in}&& \qquad \qquad \qquad \, \, = \, \, \, \,\,\, 
 _2F_1([{{1} \over {3}}, \, {{1} \over {3}}], \, [1], \, -27 \, x) \, \star \,
 _2F_1([{{1} \over {3}}, \, {{1} \over {3}}], \, [1], \, -27 \, x). 
\nonumber 
\end{eqnarray}

Let us remark that the three linear differential operators (\ref{H23}),
 (\ref{H22}) and (\ref{H33}), are MUM and of order four. However, 
they are not of the Calabi-Yau type.

\subsection{Hadamard products versus Calabi-Yau ODEs}
\label{hadamard}

This is not the case for other values of $\, n$ and $\, m$.
 For instance
 one can introduce\footnote[2]{To get the Hadamard product
of two linear differential operators use, for instance,
 Maple's command \textsf{gfun[hadamardproduct]}.} 
 $\, H_{4,4} \, = \, \,\omega_4 \, \star \, \omega_4$,
the Hadamard square of $\, \omega_4$, which is an irreducible
order-four linear differential operator, and has 
the hypergeometric solution already encountered 
for some $\, n$-fold integrals of the decomposition of the full
 magnetic susceptibility of
the Ising model~\cite{CalabiYauIsing1,CalabiYauIsing}:
\begin{eqnarray}
\label{256}
\hspace{-0.5in}&& _4F_3([{{1} \over {2}}, \, {{1} \over {2}},
 \, {{1} \over {2}}, \, {{1} \over {2}}],
 \, [1, \, 1, \, 1], \, 256 \, x) \, 
 \\
\hspace{-0.5in}&& \qquad \qquad \qquad \, \, = \, \,  \, \, \,\, 
 _2F_1([{{1} \over {2}}, \, {{1} \over {2}}], \, [1], \, -16 \, x) \, \star \,
 _2F_1([{{1} \over {2}}, \, {{1} \over {2}}], \, [1], \, -16 \, x). 
\nonumber 
\end{eqnarray}

The associated operator having (\ref{256}) as a solution, 
obeys the ``Calabi-Yau condition'' that its exterior square is of {\em order five}.

Let us give in a table the orders (which go from $\, 4$ to $\, 20$)
 of the various $\, H_{m,n}\, = \,\, H_{n,m}\,  $ Hadamard products
of the order-two operators associated with the (genus-zero) modular forms 
 operators $\, \omega_n$ and $\, \omega_m$: 

\vskip .1cm 

\vskip .2cm 
\hspace{-0.2in}
\begin{tabular}{|l|p{.4cm}|p{.4cm}|p{.4cm}|p{.4cm}|p{.4cm}|p{.4cm}|p{.4cm}|p{.4cm}|p{.4cm}|p{.4cm}|p{.4cm}|p{.4cm}|p{.4cm}|p{.4cm}|}
\hline
n\textbackslash m     & $2$ & $3$ & $4$ & $5$ & $6$ & $7$ & $8$ & $9$ & $10$ & $12$ & $13$ & $16$ & $18$  & $25$  \\ \hline
2   & $4$ & $4$ & $4$ & $6$ & $4$ & $6$ & $4$ & $4$ & $10$ & $8$  & $10$  & $8$ & $12$ & $14$  \\ \hline
3   &     & $4$ & $4$ & $6$ & $4$ & $6$ & $4$ & $4$ & $10$ & $8$  & $10$  & $8$ & $12$ & $14$  \\ \hline
4   &     &     & $4\,\, *$ & $6$ & $4\,\, *$ & $6$ & $4\,\, *$ & $4\,\, *$ & $10$ & $8$  & $10$  & $8$ & $12$ & $14$  \\ \hline
5   &     &     &     & $6$ & $6$ & $8$ & $6$ & $6$ & $12$ & $10$ & $12$  & $10$ & $14$ & $16$  \\ \hline
6   &     &     &     &     & $4\,\, *$ & $6$ & $4\,\, *$ & $4\,\, *$ & $10$ & $8$  & $10$  & $8$ & $12$ & $14$  \\ \hline
7   &     &     &     &     &     & $6$ & $6$ & $6$ & $12$ & $10$ & $12$   & $10$ & $14$ & $16$ \\ \hline
8   &     &     &     &     &     &     & $4\,\, *$ & $4\,\, *$ & $10$ & $8$  & $10$   & $8$ & $12$ & $14$ \\ \hline
9   &     &     &     &     &     &     &     & $4\,\, *$ & $10$ & $8$  & $10$   & $8$ & $12$ & $14$ \\ \hline
10  &     &     &     &     &     &     &     &     & $10$ & $14$ & $16$   & $14$ & $18$ & $20$\\ \hline
12  &     &     &     &     &     &     &     &     &      & $8$  & $14$   & $12$ & $16$ & $18$ \\ \hline
13  &     &     &     &     &     &     &     &     &      &      & $10$   & $14$ & $18$ & $20$ \\ \hline
16  &     &     &     &     &     &     &     &     &      &      &        & $8$ & $16$ & $18$ \\ \hline
18  &     &     &     &     &     &     &     &     &      &      &        &     & $12$ & $20$ \\ \hline
25  &     &     &     &     &     &     &     &     &      &      &        &     &      & $14$ \\ \hline
\hline
\end{tabular}

\vskip .1cm 

\vskip .1cm 

\hskip -.7cm 
where the star $\, *$ denotes Calabi-Yau 
ODEs\footnote[8]{Recall that Calabi-Yau ODEs are defined by a list of 
constraints~\cite{Almkvist}, the most important ones being, besides
being MUM, that their {\em exterior square
 are of order five}. There are more exotic conditions like 
the cyclotomic condition on the monodromy at $\, \infty$,
 see Proposition 3 in~\cite{TablesCalabi}.}. 

\vskip .1cm 

\vskip .1cm 

The following operators are of order four:
$\, H_{2,2}$,  $\, H_{2,3}$,   $\, H_{2,4}$, 
 $\, H_{2,6}$, $\, H_{2,8}$, $\, H_{2,9}$,   
$\, H_{3,3}$,  $\, H_{3,4}$,  $\, H_{3,6}$, $\, H_{3,8}$,  $\, H_{3,9}$,  ...
Their exterior squares, which are of order six, do not have 
rational solutions\footnote[5]{They cannot be 
homomorphic to Calabi-Yau ODEs.}. 

The following operators are of order six:
$\, H_{2,5}$,  $\, H_{2,7}$,    
$\, H_{3,5}$,  $\, H_{3,7}$,  $\, H_{4,5}$, $\, H_{4,7}$,  $\, H_{5,5}$, 
 $\, H_{5,6}$,  $\, H_{5,8}$,  $\, H_{5,9}$, 
 $\, H_{6,7}$, $\, H_{7,7}$, $\, H_{7,8}$,$\, H_{7,9}$, ...
Their  exterior square, which are of order fifteen, do not have 
rational solutions (and cannot be homomorphic to
 higher order Calabi-Yau linear ODEs). 

\vskip .1cm 

Remarkably the following ten order-four operators 
$\, H_{4,4}$,  $\, H_{4,6}$,   $\, H_{4,8}$,  $\, H_{4,9}$, $\, H_{6,6}$, 
$\, H_{6,8}$, $\, H_{6,9}$, $\, H_{8,8}$, $\, H_{8,9}$, 
$\, H_{9,9}$ (with a star in the previous table)
are all MUM, and {\em are such that their exterior squares
 are of order five}\footnote[3]{They
are conjugated to their (formal) adjoint by a function.}:
they are  {\em Calabi-Yau ODEs}. 
Actually the nome, mirror map and Yukawa coupling series 
are {\em series with integer coefficients for all these 
order four Calabi-Yau operators}.
The Yukawa coupling series of these Calabi-Yau operators 
are respectively, for $\, H_{4,4}$
\begin{eqnarray}
\label{YukuM44}
\hspace{-0.9in}&&K(q) \, \,= \,\, \,K^{\star}(q) 
\,\, \,  \,= \,\, \, \,\,\,  \,  \,1 \,\,\, \, \,  + \, 32 \cdot \,q\,  \, 
 +  4896 \cdot \, q^2 \,\,  +702464 \cdot \, q^3 \,\, \, + 102820640 \cdot \, q^4
  \,\,  \nonumber \\
\hspace{-0.9in}&&\qquad \,\,  + 15296748032 \cdot \, q^5
\, + \,  2302235670528\cdot \, q^6\, \,  + \,349438855544832 \cdot \, q^7 
  \nonumber \\
\hspace{-0.9in}&&\qquad \quad \quad  + \,53378019187206944 \cdot \, q^8  
\,\,  + \,8194222260681725696 \cdot \, q^9 \,\,  \,    \, + \,\, \,\cdots,  
\end{eqnarray}
which is $\, Number  3 \, $ in Almkvist et al. large
 tables of Calabi-Yau ODEs~\cite{TablesCalabi}, 
and is the well-known one for\footnote[9]{Actually $\, H_{4,4}$
is exactly the operator for $\, _4F_3([1/2,1/2,1/2,1/2], \, [1,1,1], 256 \, x)$.}
 $\, _4F_3([1/2,1/2,1/2,1/2], \, [1,1,1],\, 256 \,x)$,
for $\, H_{4,6}$: 
\begin{eqnarray}
\label{YukuM46}
\hspace{-0.9in}&&K(q) \, \,\, \,= \,\, \,K^{\star}(q) \, \,= \,\,\,
 \, \, \, \,  \,1 \,\, \,\, \,  + \, 20 \cdot \,q\,  
 +  36 \cdot \, q^2 \, +15176 \cdot \, q^3 \,\,  + \, 486564 \cdot \, q^4 \,\,  
\,\,    \nonumber \\
\hspace{-0.9in}&&\qquad \quad  \quad + \,21684020 \cdot \, q^5 \,\, 
+ \,1209684456 \cdot \, q^6 \,\,+ \,58513394904 \cdot \, q^7 \,\,  
 \,\, \nonumber \\
\hspace{-0.9in}&& \qquad \quad  \quad \quad \quad  + \,2921860726948 \cdot \, q^8 \,\, 
+ \,141376772107064 \cdot \, q^9 
\,\, \, \, \,    + \,\, \, \cdots 
\end{eqnarray}
which is $\, Number  137 \, $ in tables~\cite{TablesCalabi}.

We give, in \ref{YukawaHmn}, the expansion of the Yukawa coupling 
for a set of other $\, H_{m,n}$ that are Calabi-Yau:
in particular their exterior square is order {\em five} (not 
six as one could expect for a generic irreducible 
order-four operator).
It will be shown, in a forthcoming publication, that
the fact that the order five exterior power
property occurs means that these operators are necessarily {\em conjugated} 
(by an algebraic function)
to their adjoints. Thus, the ``adjoint Yukawa coupling'' $\, K^{\star}(q)$
is necessarily equal to the  Yukawa coupling 
 $\, K(q)$ for these operators.

\vskip .1cm 

{\bf Remark:}  The operator having the Hadamard product of the
two HeunG functions $\, HeunG(a, \, q, \, 1, \, 1, \,  1, \, 1; \, x)$
and $\, HeunG(A, \, Q, \, 1, \, 1, \,  1, \, 1; \, x)$ as a solution
reads:
\begin{eqnarray}
\label{abAB}
\hspace{-0.7in}&&(x-1)  \, (x-a) 
 \, (x\,-A)  \, (x\,-A\,a)  \, ( A\,a \, -{x}^{2})^{2}
\cdot \,x^3 \cdot \,  D_x^{4}  \\
\hspace{-0.7in}&& \qquad \quad\, +2\, \, (x ^2 \,-A\,a) \cdot U_3 
 \cdot \,{x}^{2}\cdot \,  D_x^{3} \, 
\, \,\,\,  - U_2 \cdot \, x \cdot \,  D_x^{2}\, \, \,\,  - U_1 \cdot \,  D_x \, \,\,\,  +U_0, 
 \nonumber
\end{eqnarray}
where the polynomials $\, U_n$ are given in \ref{Un}.

The exterior square of this order-four operator (\ref{abAB}) 
is of order {\em five} for {\em any value} of the 
parameters $a, \, q, \, A, \, Q$ (instead of the order-six one expects for the  
exterior square of a generic irreducible order-four operator).

The HeunG functions 
solutions  of the form 
$\, HeunG(a, \, q, \, 1, \, 1, \,  1, \, 1; \, x)$
are an interesting set of HeunG functions. They verify the following 
(six M\"obius) identities~\cite{Maier7}:
\begin{eqnarray}
\label{HeunGidentity}
\hspace{-0.7in}&&HeunG(a, \, q, \, 1, \, 1, \,  1, \, 1;\,  \, x)
 \, \,\, \,  = \, \, \,  \,   \,\,    \, 
HeunG\Bigl({{1} \over { a }}, \, {{ q} \over { a}}, 
\, 1, \, 1, \,  1, \, 1;\,  \, {{ x } \over { a }}\Bigr)
\nonumber \\
\hspace{-0.7in}&&\quad \, \,  \, = \, \,  \,  \, \, 
{{1} \over { 1 \, -x}} \cdot \,   \, 
HeunG\Bigl({{a} \over { a\, -1}},\,  \, {{ a\, -q } \over { a\, -1 }},\, 
 \, 1, \, 1, \,  1, \, 1; \,\,  - \,{{x} \over {1\, -x  }}\Bigr)
\nonumber \\ 
\hspace{-0.7in}&&\quad \, \,  \, = \, \,  \,  \, \, 
{{1 } \over { 1 \, -x/a }} \cdot \,  \,  
HeunG\Bigl({{ 1} \over { 1\, -a }},\,\,   
 \, {{ q\, -1} \over { a \, -1}}, \, 1, \, 1, \,  1, \, 1; 
\,\,  -\,{{ x} \over { a-\, x }}\Bigr)
\end{eqnarray}
\begin{eqnarray}
\hspace{-0.7in}&&\quad \, \,  \, = \, \,  \,  \, \, {{a} \over {a\, -x}} \cdot \, 
HeunG\Bigl(1\, -a,\,\,   
 \, 1 \, -q, \, 1, \, 1, \,  1, \, 1;
 \,\,  \,{{ (a\, -1) \cdot \, x} \over { a-\, x }}\Bigr)
\nonumber \\ 
\hspace{-0.7in}&&\quad \, \,  \, = \, \,  \,  \, \, 
{{1} \over { 1 \, -x}} \cdot \,   \, 
HeunG\Bigl({{a \, -1} \over { a}},\,  \, {{ a\, -q } \over { a}},\, 
 \, 1, \, 1, \,  1, \, 1;
 \,\,  - \,{{x} \over {1\, -x  }} \cdot {{a \, -1} \over {a }}\Bigr).
\nonumber 
\end{eqnarray}

\vskip .1cm

\vskip .1cm

{\em The ten linear differential operators denoted by a star $\, *$ 
in the previous table are all 
of this form}: they have the Hadamard product of two HeunG functions 
solutions  of the form 
$\, HeunG(a, \, q, \, 1, \, 1, \,  1, \, 1; \, x)$
as a solution.  Note, however, that this  HeunG-viewpoint of 
the most interesting $\, H_{m,n}$'s
does not really help. Even inside this restricted set of  HeunG functions 
solutions  of the form $\, HeunG(a, \, q, \, 1, \, 1, \,  1, \, 1; \, x)$
it is hard to find exhaustively the values of the two parameters
$\, a$ and of the accessory parameter\footnote[1]{The accessory parameter
appears in many applications as a spectral parameter~\cite{Maier7}.} $\, q$
such the series $\, HeunG(a, \, q, \, 1, \, 1, \,  1, \, 1; \, x)$
is globally bounded\footnote[2]{Along this line
see the paper by Zagier~\cite{Zagier}
on integral solutions of Ap\'ery-like equations.}, or, just, such
that the order-two operator having 
$\, HeunG(a, \, q, \, 1, \, 1, \,  1, \, 1; \, x)$ 
as a solution is globally nilpotent (see \ref{Un}). 

\vskip .1cm 

The order-four operators $\, H_{3,3}$, $\, H_{3,4}$, 
are all MUM operators\footnote[5]{Note that the 
Hadamard product of two MUM ODEs is not necessarily a MUM ODE:
the order-six operator $\, H_{3,7}$ is {\em not} MUM.},  
but, similarly to the situation
encountered with  $\, H_{2,2}$, their nome, mirror map and Yukawa 
couplings are {\em not globally bounded}.

\vskip .1cm 

Many $\,H _{m,n}$ are not MUM, for instance the
 order-eight operator $\, \, H_{12,12}$, or 
the order-six operator $\, H_{3,7}$, are {\em not MUM}. Concerning $\,H_{3,7}$
and  as far as its
six solutions are concerned, it is ``like'' the four solutions of an
order-four MUM operator, together with the two solutions 
of another order-two MUM operator, but the  order-six operator $\, H_{3,7}$
is not a direct-sum of an  order-four and order-two operator.
We have two solutions analytical at $\, x\, = \, \, 0$ (no $\ln(x)$),
 two solutions with a $\ln(x)$. A linear combination of
these two solutions analytical at $\, x\, = \, \, 0$ is, 
by construction a series with 
{\em integer coefficients} (the Hadamard product of the 
two series with integer coefficients which are the initial
ingredients in this calculation), when the other linear combinations 
are {\em not globally bounded}.

\vskip .1cm 

\subsection{``Atkin'' transformations}
\label{Attransf}

It is worth noting that the globally bounded character
 of some of these $\, H_{m,n}$,
at $\, x=\, 0$, is simply related to the globally
 bounded character at $\, x=\, \infty$ of a conjugate operator, 
as a consequence of some simple homomorphism relation with 
their transformed by ``Atkin'' pullbacks:
\begin{eqnarray}
\label{proper}
\hspace{-0.9in}&&\qquad \quad \, \, \,  \qquad  x^{1/4} \cdot \, H_{2,2}(x)  \, \,  = \, \,\,   
 H_{2,2}\Bigl( x \rightarrow \, {{1} \over {2^{24}  \, x}} \Bigr) \cdot \, x^{1/4}, 
\nonumber \\
\hspace{-0.9in}&&\qquad \quad \, \, \,  \qquad x^{1/3} \cdot \, H_{3,3}(x)   \,\,  = \, \, \, 
 H_{3,3}\Bigl( x \rightarrow \, {{1} \over {3^{12} \, x}} \Bigr) \cdot \, x^{1/3},   
 \\
\hspace{-0.9in}&&x^{1/2} \cdot \, H_{4,4}(x)  \,  = \, \,  
 H_{4,4}\Bigl( x \rightarrow \, {{1} \over {2^{16} \, x}} \Bigr) \cdot \, x^{1/2}, 
\quad  x \cdot \, H_{6,6}(x)  \,  = \, \, 
 H_{6,6}\Bigl( x \rightarrow \, {{1} \over {2^6  \,3^4 \,  x}} \Bigr) \cdot \, x,    
\nonumber \\
\hspace{-0.9in}&&x \cdot \, H_{8,8}(x)  \, \, = \, \, \, 
 H_{8,8}\Bigl( x \rightarrow \, {{1} \over {2^{10} \, x}} \Bigr) \cdot \, x, 
\quad \quad \quad \quad  x \cdot \, H_{9,9}(x)  \, \, = \, \, \, 
 H_{9,9}\Bigl( x \rightarrow \, {{1} \over {3^6 \, x}} \Bigr) \cdot \, x, 
\nonumber
\end{eqnarray}
but also for the order-six operators
\begin{eqnarray}
\label{h55}
&&x^{1/2} \cdot \, H_{5,5}(x)  \, \, = \, \, \, \, 
 H_{5,5}\Bigl( x \rightarrow \, {{1} \over {2^{12}\, 5^6 \, \, x}} \Bigr) \cdot \, x^{1/2}, \\
&&x^{2/3} \cdot \, H_{7,7}(x)  \, \, = \, \, \, \, 
 H_{7,7}\Bigl( x \rightarrow \, {{1} \over {3^{8}\, 7^4 \, \, x}} \Bigr) \cdot \, x^{2/3},
\end{eqnarray}
and the order-eight  operators
\begin{eqnarray}
&&x^{2} \cdot \, H_{12,12}(x)  \, \, = \, \, \, \, 
 H_{12,12}\Bigl( x \rightarrow \, {{1} \over {2^{4}\, 3^2 \, \, x}} \Bigr) \cdot \, x^{2}, 
\\
&&x^{2} \cdot \, H_{16,16}(x)  \, \, = \, \, \, \, 
 H_{16,16}\Bigl( x \rightarrow \, {{1} \over {2^{6}\,  \, \, x}} \Bigr) \cdot \, x^{2},
\end{eqnarray}
or the order-ten  operators
\begin{eqnarray}
&&x^{3/2} \cdot \, H_{10,10}(x)  \, \, = \, \, \, \, 
 H_{10,10}\Bigl( x \rightarrow \, {{1} \over {2^{8}\, 5^2 \, \, x}} \Bigr) \cdot \, x^{3/2}, 
\\
&&x^{7/6} \cdot \, H_{13,13}(x)  \, \, = \, \, \, \, 
 H_{13,13}\Bigl( x \rightarrow \, {{1} \over {2^{8}\, 3^3\, 13^2 \, \, x}} \Bigr) \cdot \, x^{7/6}, 
\end{eqnarray}
or the order-twelve operator
\begin{eqnarray}
&&x^{3} \cdot \, H_{18,18}(x)  \, \, = \, \, \, \, 
 H_{18,18}\Bigl( x \rightarrow \, {{1} \over {2^{2}\, 3^2 \, \, x}} \Bigr) \cdot \, x^{3}, 
\end{eqnarray}
or the order-fourteen operator
\begin{eqnarray}
\hspace{-0.3in}&&x^{5/2} \cdot \, H_{25,25}(x)  \,  \, \, = \, \, \, \, 
 H_{25,25}\Bigl( x \rightarrow \, {{1} \over {2^{8}\, 5^2 \, \, x}} \Bigr) \cdot \, x^{5/2}. 
\end{eqnarray}
to be compared with
\begin{eqnarray}
\hspace{-0.9in}&&x^{1/4} \cdot \, \omega_{2}(x)  \, \, = \, \, \, 
 \omega_{2}\Bigl( x \rightarrow \, {{1} \over {2^{12}\, x}} \Bigr) \cdot \, x^{1/4}, 
\quad \quad  x^{1/3} \cdot \, \omega_{3}(x)  \, \, = \, \, \, 
 \omega_{3}\Bigl( x \rightarrow \, {{1} \over {3^6 \, x}} \Bigr) \cdot \, x^{1/3},
\nonumber \\
\hspace{-0.9in}&&x^{1/2} \cdot \, \omega_{4}(x)  \, \, = \, \, \, 
 \omega_{4}\Bigl( x \rightarrow \, {{1} \over {2^{8} \, x}} \Bigr) \cdot \, x^{1/2}, 
\quad \quad \quad  \, \,  \,    x \cdot \, \omega_{6}(x)  \, \, = \, \, \, 
 \omega_{6}\Bigl( x \rightarrow \, {{1} \over {2^3\, 3^2\, x}} \Bigr) \cdot \, x,
\nonumber \\
\hspace{-0.9in}&&x \cdot \, \omega_{8}(x)  \, \, = \, \, \, 
 \omega_{8}\Bigl( x \rightarrow \, {{1} \over {2^{5} \, x}} \Bigr) \cdot \, x, 
\quad \quad \quad \quad  \quad  \quad  \, \,  \,  
 x \cdot \, \omega_{9}(x)  \, \, = \, \, \, 
 \omega_{9}\Bigl( x \rightarrow \, {{1} \over {3^3 \, x}} \Bigr) \cdot \, x,  
\nonumber
\end{eqnarray}
and:
\begin{eqnarray}
\label{omeg5}
&&x^{1/2} \cdot \, \omega_{5}(x)   \,  \, \, = \,\, \, \,  \,\,  
 \omega_{5}\Bigl( x \rightarrow \, {{1} \over {2^{6}\, 5^3 \, x}} \Bigr) \cdot \, x^{1/2}, 
\\
&& x^{2/3} \cdot \, \omega_{7}(x)   \,  \, \, = \,\, \, \,  \, \, 
  \omega_{7}\Bigl( x \rightarrow \, {{1} \over {3^{4}\, 7^2 \, \, x}} \Bigr) \cdot \, x^{2/3}, 
\quad \quad \,\,  \, \,\,\cdots \nonumber 
\nonumber 
\end{eqnarray}

Note, however, that these properties (\ref{proper}) are no longer valid for the
off-diagonal operators $\, H_{m,n}$, $\, m \, \ne \, n$.
For instance, the other order-four Calabi-Yau
 operators $\, H_{4,6}$, $\, H_{4,8}$, $\, H_{4,9}$
are {\em not globally bounded} at $ \, x \, = \, \, \infty$: if one changes these 
operators by a $\, x \rightarrow \, 1/x$ pullback, the new order-four operators
after a well-suited conjugation by $\, x^r$ ($r$ rational number)
 are {\em not globally bounded}. For instance $\, H_{4,6}(x  \rightarrow 1/x)$
requires to be conjugated by $\, x^{1/2}$ (Puiseux series): after conjugation
 by $\, x^{1/2}$,  the series analytical at $\, x \, = \, \, 0$
 is not globally bounded.

\vskip .1cm 
\vskip .1cm 

{\bf Remark}:
Let us denote $\, A_n$ the constant in the ``Atkin'' involution 
$\, x \rightarrow A_n/x \, $ for  modular form of order $n$
 (i.e. $\tau \, \rightarrow \, n \, \tau$).
We denote by $\omega_{n}(x)$, the order-two operator associated
 with a modular form of order $\, n$.

One has
\begin{eqnarray}
\label{omegaAnBn}
 x^{r_n} \cdot \, \omega_{n}(x)   \,  \,  \, \,\, = \, \, \, \,  \, \, \, 
 \omega_{n}\Bigl( x \rightarrow \, 
{{1} \over {A_n \, B_n^2 \, x}} \Bigr) \cdot \, x^{r_n},  
\end{eqnarray}
where the $\, B_n$'s are integers, 
and, quite remarkably\footnote[5]{If one forgets, for a second, that sum 
and multiplication do not commute, 
it is tempting to see naively these conjugation 
results (\ref{HnnAnBn}), on the $\, H_{n,n}$'s, as a straight consequence 
of the relation, like (\ref{omegaAnBn}), on the $\, \omega_n$'s,
 since the $\, H_{n,n}$'s 
are Hadamard squares of the $\, \omega_n$'s, the constant in the pullback 
involution in the right-hand side of (\ref{HnnAnBn}) being the square 
of the constant in (\ref{omegaAnBn}). 
This is not the case.}, one also has for $ \, H_{n,n}(x)$ the
 Hadamard product of two $\, \omega_{n}(x)$'s
\begin{eqnarray}
\label{HnnAnBn}
 x^{r_n} \cdot \, H_{n,n}(x)   \,  \, \,  \, = \, \, \,   \, \, \, 
H_{n,n}\Bigl( x \rightarrow \,
 {{1} \over {A_n^2 \, B_n^4 \, x}} \Bigr) \cdot \, x^{r_n}.
\end{eqnarray}
The values of $\, A_n$, $\, B_n$ and $\, r_n$ are given in the following table:

\vskip .2cm 

\vskip .1cm 

\hspace{-0.3in}\begin{tabular}{|l|l|l|l|l|l|l|l|l|l|l|l|l|l|l|}
\hline
$n$    	        & 2 & 3 & 4 & 5 & 6 & 7 & 8 & 9 & 10 & 12 & 13& 16 & 18 & 25  	\\ 
\hline
$A_n$    	& $2^{12}$ & $3^6$ & $2^8$ & $5^3$ & $2^3\, 3^2$ & $7^2$ & $2^5$ & $3^3$ & $2^2 \, 5$ & $2^2 \, 3$ & $13$ & $2^3$ & $2 \cdot \, 3$ & $5$
	\\ \hline
$B_n$    	& $1$ & $1$ & $1$ & $2^2$ & $1$ & $3^2$ & $1$ & $2$ & $1$ & $1$ & $2^2 \, 3^2$ & $1$ & $1$ & $2^2$ \\ \hline
$r_n$   	& $\frac{1}{4}$ & $\frac{1}{3}$ & $\frac{1}{2}$ & $\frac{1}{2}$ & $1$ & $\frac{2}{3}$ & $1$ & $1$ & $\frac{3}{2}$ & $2$ & $\frac{7}{6}$ & $2$	& $3$ & $\frac{5}{2}$ \\ 
\hline
\end{tabular}

\vskip .1cm 

\vskip .1cm 

\subsection{$\, _2F_1([1/N, \,1/N], \, [1], \,  x)$ hypergeometric functions }
\label{Attransf}

Modular forms can always be written as the hypergeometric function $\, _2F_1$,  
up to an algebraic pre-factor, and {\em up to a pullback} (see Maier~\cite{Maier1}).
Relations (\ref{1728}), (\ref{4096}), (\ref{729}) and  (\ref{256})
 underline the special role
of modular forms that can be written as  $\, _2F_1$ with no algebraic pre-factor,
and no pullbacks (see Table 15 in~\cite{Maier1}), namely 
$\, _2F_1([1/4,\, 1/4],\, [1], \, -64 \, x)$, 
$\, _2F_1([1/3,\, 1/3],\, [1], \, -27 \, x)$, $\, _2F_1([1/2,\, 1/2],\, [1], \,-16 \, x)$.
Along this line it is worth considering the hypergeometric functions 
$\, _2F_1([1/N,\, 1/N],\, [1], \, -N^3 \, x)$ 
which always yield globally bounded series,
together with their associated linear differential 
operators. The nome for these operators
{\em does not} (generically) correspond to globally bounded series. One notes, however, 
that $\, N\, = \, 6$ is ``special'', yielding series with integer coefficients for
the hypergeometric function\footnote[1]{We use identity $_2F_1([1/6,1/6],[1],x) $
$\, = \, (1-x)^{-1/6} \cdot \, _2F_1([1/12,5/12],[1], -4\, x/(1-x)^2)$, which 
singles out the known~\cite{Renorm} pullback
 $\, x \, \rightarrow \,  -4\, x/(1-x)^2$.} 
\begin{eqnarray}
\label{forminv}
\hspace{-0.7in}&&_2F_1\Bigl([{{1} \over {6}},\, {{1} \over {6}}],
\, [1], \, -432 \, x \Bigr) \\
\hspace{-0.7in}&&\quad \quad \, \, \, = \, \,\,  \, \,  \,  (1\, + \, 432 \, x)^{-1/6}  \cdot \, 
_2F_1\Bigl([{{1} \over {12}},\, {{5} \over {12}}],\, [1], \, 
{{ 1728\, x} \over  {(1\, + \, 432 \, x)^2}}\Bigr) 
\nonumber 
\end{eqnarray}
\begin{eqnarray}
\hspace{-0.7in}&&\quad \quad \, \, \, = \, \,\,  \,  \, \, 
1\, \, \,  \,  -12\,x \, \, +1764\,x^2\, \, -397488\,x^3\, \,
 +107619876\,x^4 \, \, -32285962800 \,x^5 
\nonumber \\
\hspace{-0.7in}&& \quad \quad \quad \quad \quad \quad \quad \, 
\, +10342270083600 \,x^6
\, \, -3467404345579200 \,x^7 \,\, \, \, + \, \,\, \cdots
\nonumber 
\end{eqnarray}
but also for the corresponding nome
\begin{eqnarray}
\hspace{-0.9in}&&q \,\, \, \, = \, \,  \, \, \, \, \,x \, \, \, \,  \,  -120\, x^2 \,
 \, +24660\, x^3 \, \, -6322720 \, x^4 \,
\,  +1828573410 \, x^5\,  \, -570359919024 \, x^6 
\nonumber \\
\hspace{-0.9in}&& \quad \quad \quad \quad \quad \, +187363061411720\, x^7
\,  \, -63912709875600960 \, x^8\,\,\, \, \,  \, 
+ \, \,\, \cdots 
\end{eqnarray}
as well as the mirror map:
\begin{eqnarray}
\hspace{-0.9in}&&x(q) \,\, = \, \, \, \, \,\,  q\, \,\, \, \,
 +120\, q^2 \, \, +4140\, q^3\,  \, +166720\, q^4\, \,  -6012210 \, q^5\,  \,
 +1165528224 \, q^6 \nonumber \\
\hspace{-0.9in}&&\quad \quad \quad \quad  \quad \quad \, -178811454280 \, q^7\,  
+29512658112000 \, q^8 \,\,\,\,  \,  \, + \,\, \, \cdots 
\end{eqnarray}
In fact  the hypergeometric function 
$\, _2F_1([1/6,\, 1/6],\, [1], \,  \, x)$
 (as well as the hypergeometric function 
 $\, _2F_1([1/6,\, 5/6],\, [1], \, \, 1\, -x)$)
is actually ``special'' (as was first
seen by Ramanujan, see for instance Cooper~\cite{Cooper}).
In \ref{Special} the modular form character of (\ref{forminv})
is made crystal clear in a quite heuristic way.

\vskip .1cm 
{\bf Remark:} 
Using the $\, _2F_1([a,b],[1],x)$ hypergeometric functions 
associated to  modular forms\footnote[2]{For $\, [1/2,1/2], \,  [1/3,1/3], 
 \, [1/4,1/4], \, [1/6,1/6]$,
see the Ramanujan's 
theories to alternative basis~\cite{Cooper} for 
other ``signatures''. For $\, [1/8,3/8]$
see the third example in \ref{modular}.} 
(namely $[a,b] \, = \, [1/2,1/2], \,  [1/3,1/3],  \, [1/4,1/4], \, [1/6,1/6],$ 
 $\,  [1/3,2/3],\, [1/3,1/6], \, [1/6,5/6],$ $ \,[1/4,3/4], \,[1/8,3/8],
 \,  [1/12,5/12],$ ...), 
 one can build $_4F_3$ globally bounded examples
 by simple Hadamard products
 of these selected $\, _2F_1$. We give in \ref{Miscell} a miscellaneous set
of identities expressing HeunG functions, or modular forms,
as the previously selected $\, _2F_1$ hypergeometric functions
with {\em two pullbacks}.

\vskip .1cm

\subsection{Modularity and Hypergeometric series 
with coefficients ratio of factorials}
\label{ratiooffac}

As a consequence of the classification by Beukers and Heckman~\cite{BeHe89}
 of all algebraic $_{n}F_{n-1}$'s, 
the $\, \, _8F_7$ hypergeometric series
\begin{eqnarray}
\hspace{-0.75in}_8F_7\biggl(
\left[\frac{1}{30}, \, \frac{7}{30},\,\frac{11}{30},\,
\frac{13}{30},\,\frac{17}{30},\,\frac{19}{30},\,\frac{23}{30},\,
\frac{29}{30}\right],\,
\left[\frac{1}{5},\,\frac{2}{5},\,\frac{3}{5},\,\frac{4}{5},\,
\frac{1}{2},\,\frac{2}{3},\,\frac{1}{3}\right],
\,\, \, 
2^{14} \, \, 3^9 \, \, 5^5   \, \, x \biggr),  
\nonumber 
\end{eqnarray}
has {\em integer coefficients}, and is an {\em algebraic function}. 
The Galois group belonging to this function is the Weyl group $\, W(E_8)$
which has 696729600 elements~\cite{Varilly}.
It is an {\em algebraic series} of degree 483840.
More precisely, it was noticed~\cite{Villegas}
 by Rodriguez-Villegas that the previous 
power series reads: 
\begin{eqnarray}
\sum_{n =0}^{\infty} \,\,
 {{(30 \, n)! \, \, n!} \over {(15 \,n)! \,  \, (10 \,n)! \, \, (6 \,n)!}} \cdot \, x^n,
\end{eqnarray}
which is precisely the series introduced by Chebyshev 
during his work~\cite{Chebyshev} on the
 distribution of prime numbers to establish the estimate 
\begin{eqnarray}
\hspace{-0.1in}\quad 0.92 \,\, \, \,  \frac{x}{\log x} \,\,\, \,\, \, \,\,
 \leq\, \, \, \,  \, \, \pi (x) \,\,\, \,\,\,   \leq\,\, \,\, \, \, \, 
  1.11 \,\,\, \,\,   \frac{x}{\log x}, 
\end{eqnarray}
on the prime counting function $\, \pi(x)$.

\vskip .1cm  

Considering hypergeometric series such that their coefficients are ratio
of factorials, a paper by Rodrigues and Villegas~\cite{Villegas} 
 gives the conditions of these factorials for the hypergeometric series to be 
algebraic (all the coefficients are thus integers). A simple example is, 
for instance the algebraic function:
\begin{eqnarray}
\hspace{-0.4in}\quad _3F_2\Bigl([{{1} \over {4}}, \,{{1} \over {2}}, \, {{3} \over {4}}], \, 
[{{1} \over {3}}, \,{{2} \over {3}}]; \, \,  {{256} \over{27}} \cdot \, x\Bigr) 
\, \, \, \, = \,\,\, \, \,  \,  \, \sum_{n=0}^{\infty} \, \, {4 \, n \choose n} \cdot \, x^n. 
\end{eqnarray}

\vskip .1cm

Along this line it is worth recalling Delaygue's Thesis~\cite{Delaygue} 
(see also Bober~\cite{Bober})
which gives some results\footnote[8]{Necessary and sufficient conditions for
the integrality of the mirror maps series.} 
for series expansions\footnote[5]{These series are not algebraic functions.}
 such that their coefficients are
{\em ratio of factorials}:
\begin{eqnarray}
\hspace{-0.4in}&&_2F_1\Bigl([{{1} \over {3}}, \,{{2} \over {3}}], [1]; \,\, 27 \, x\Bigr)
\, \,\,\,  = \, \,\, \, \,   \, \sum_{n=0}^{\infty} \, {{(3n)!} \over {(n!)^3}}   \cdot \, x^n,
\\
\hspace{-0.4in}&&_4F_3 \Bigl([{{1} \over {2}}, {{1} \over {2}}, 
{{1} \over {2}}, {{1} \over {2}}],[1,1,1]; \, \, 256 \cdot \, x\Bigr)
\, \,\,\,  = \, \,\, \, \, \, \, \,
 \sum_{n=0}^{\infty} \, {{((2n)!)^4 } \over { (n!)^8}} \cdot \, x^n,
\end{eqnarray}

\begin{eqnarray}
\label{last}
\hspace{-0.4in}&&_4F_3 \Bigl([{{1} \over {2}}, {{1} \over {2}}, 
{{1} \over {6}}, {{5} \over {6}}],[1,1,1]; \,\,\, 2^8 \, \, 3^3 \cdot \,  x\Bigr)
\, \,\,  \,= \, \, \, \, \,  \, \,
\sum_{n=0}^{\infty} \, {{(6n)! \, \, (2n)! } \over {(3n)! \, (n!)^{5}}} \cdot \, x^n.
\end{eqnarray}
These ratio of factorials are integer numbers. The series expansion of (\ref{last}) 
reads:
\begin{eqnarray}
\label{lastser}
\hspace{-0.9in}&& \qquad 1 \,\, +240\,x \,\,+498960\,{x}^{2} \,\,
+1633632000\,{x}^{3}\, \,+6558930378000\,{x}^{4} \,
 \nonumber \\
\hspace{-0.9in}&&\qquad\qquad  \quad +29581300719210240\,{x}^{5}
\,+143836335737833939200\,{x}^{6} \,  \, \, \,  \, + \,  \, \, \, \cdots 
 \nonumber
\end{eqnarray}

\subsection{More Hadamard products: Batyrev and van Straten examples~\cite{Batyrev}}
\label{hadamardmore}

\subsubsection{A first auto-adjoint Calabi-Yau ODE \newline \newline }
\label{B1}

\vskip .1cm 

An order-four operator has been found by Batyrev and van Straten~\cite{Batyrev} 
\begin{eqnarray}
\label{Batyrev1first}
\hspace{-0.5in}&& B_1\,\,  \,  \, = \,\,  \, \, \,  \, \,
\theta^4 \, \, \,\,  -3 \, x \cdot 
(7\, \theta^2 \, +7 \,\theta \, +2) \cdot 
  (3\, \theta \, +\, 1) \cdot   (3\, \theta \, +\,2)
\, \\
\hspace{-0.5in}&&\quad \quad \quad \quad \quad \,  \,  \,  \, 
 -72 \, x^2 \cdot  
   (3\, \theta \, +\, 5) \cdot   (3\, \theta \, +\,4) \cdot
   (3\, \theta \, +\, 2)   \cdot  (3\, \theta \, +\, 1), 
  \nonumber
\end{eqnarray}
which is conjugated to its adjoint:
$\, B_1 \cdot \, x\, = \, \, x \cdot \, adjoint(B_1)$.

Operator (\ref{Batyrev1first}) is a Calabi-Yau operator: it is MUM, and it is such
 that its exterior square is of {\em order five}.
Its has a solution analytical at $\, x\, = \, 0$ which 
is actually the Hadamard product
of the previous selected hypergeometric $\, _2F_1$: 
 \begin{eqnarray}
\label{sumup}
\hspace{-0.7in}\quad  _2F_1\Bigl([{{1} \over {3}}, \,{{2} \over {3}}], [1]; \,
\, 27 \, x\Bigr) \, \star \, 
\Bigl( {{1 } \over {1\, +4 \, x}}\,  \cdot \,
 _2F_1\Bigl([{{1} \over {3}}, \, {{2} \over {3}}], \, [1]; \, 
{{27 \cdot x } \over {(1\, +4 \, x )^3 }}  \Bigr) \Bigr).
\end{eqnarray}

\subsubsection{A second auto-adjoint Calabi-Yau ODE \newline \newline }
\label{B2}
\vskip .1cm 

A second example~\cite{Batyrev} (see\footnote[1]{There is a 
small misprint in~\cite{Batyrev} page 34:
 $(2\, \theta \, +\, 1)$ must be replaced 
by $(2\, \theta \, +\, 1)^2$ in the $\, 4\, x$ term.} page 34)
 of order-four operator
 corresponds to  Calabi-Yau 3-folds 
in $\, P_1 \times  P_1 \times  P_1 \times  P_1$:
\begin{eqnarray}
\label{Batyrev2}
\hspace{-0.5in}&&B_2\,\,  \, \, = \,\,  \, \, \, \,  \, \,  
\theta^4 \, \, \, \, \,   -4 \, x \cdot 
(5\, \theta^2 \, +5 \,\theta \, +2) \cdot   (2\, \theta \, +\, 1)^2
\,  \nonumber \\
\hspace{-0.5in}&&\quad \quad \quad \quad \quad \quad \quad \quad
 +64 \, x^2 \cdot  
   (2\, \theta \, +\, 3) \cdot   (2\, \theta \, +\,1) \cdot
   (2\, \theta \, +\, 2)^2,   
\end{eqnarray}
corresponding to the series-solution with coefficients:
\begin{eqnarray}
\label{integB2}
\hspace{-0.5in}&&{2n \choose n} \cdot \sum_{k=0}^{n} \,
 {n\choose k}^2 \cdot {2k \choose k} \cdot {2n-2k \choose n-k} \\
\hspace{-0.5in}&&\qquad \quad \, \, = \, \,  \, \,  \, 
{2n \choose n}^2  \cdot 
\,  _2F_1\Bigl([{{1 } \over {2}}, \, -n,\,  -n,\,  -n],
\,  \,[1,\,  1,\,  -\, {{2\, n \, -1 } \over {2}}]; \, 1\Bigr).
 \nonumber 
\end{eqnarray}
Its Wronskian reads:
\begin{eqnarray}
\label{wronskB2}
\hspace{-0.95in}W \, = \, \, 
{{1} \over { (1 \, -64 \, x)^2 \cdot \, (1\, -16\, x)^2 \cdot \, x^6}},
\, \quad
x^3 \cdot \, W^{1/2} \, = \, \,
 {{1} \over { (1 \, -64 \, x) \, (1\, -16\, x)}}.
\end{eqnarray}
This operator is also a Calabi-Yau operator: it is MUM, and it is such
 that its exterior square is {\em order five}.  This order five property 
is a consequence of  $\, B_2$ being conjugated to its adjoint:
$\, B_2 \cdot \, x\, = \, \, x \cdot \, adjoint(B_2)$.

The series-solution of (\ref{Batyrev2}) 
 can be written as an Hadamard product 
\begin{eqnarray}
\label{serB2}
\hspace{-0.9in}&&{\cal S}  \,  \,   \, = \,\, \,\, \, 
 (1\, -4\, x)^{-1/2} 
\, \star \, HeunG(4, 1/2, 1/2, 1/2, 1, 1/2; \,  16\, x)^2 \\
\hspace{-0.9in}&&\, \,  \,   \, = \,\, \,\, \,
 1\,+8\,x\,+168\,x^2\,
+5120\,x^3\,+190120\,x^4\,+7939008\,x^5\,+357713664\,x^6 
\, \, \,  + \, \cdots, 
\nonumber 
\end{eqnarray}
the modular form character of 
$\, HeunG(4, 1/2, 1/2, 1/2, 1, 1/2; \,  16\, x)$
being illustrated with identities (\ref{this}) in \ref{modular}.
Its nome reads:
\begin{eqnarray}
\label{nomeB2}
\hspace{-0.9in}&&q \, \, = \, \, \,  \, \,
x \,\, \,  \, +20\,x^2 \,+578\,x^3 \,+20504\,x^4 \,+826239\,x^5 \,
+36224028\,x^6 \,+1684499774\,x^7 \,
\nonumber  \\
\hspace{-0.9in}&& \quad +81788693064\,x^8 \,
 \,+4104050140803\,x^9 \,+211343780948764\,x^{10}
 \, \,\, \,  \,+\,\, \,\cdots 
\end{eqnarray}
Its {\em mirror map} reads:
\begin{eqnarray}
\label{mirrorB2}
\hspace{-0.9in}&&x(q) \, \, = \, \, \,  \, \, \,\,
q \, \, \, \, \,  -20\,q^2\, \,  +222\,q^3\, \,
  -2704\,q^4\, \,  +21293\,q^5\,  \, 
-307224\,q^6\, \,  +80402\,q^7\,   \nonumber  \\
\hspace{-0.9in}&& \qquad  \qquad  \, \, \,
 -67101504\,q^8\, \,  -1187407098\,q^9\, \, 
 -37993761412\,q^{10}\,\, 
 \, \, \,\,   + \, \, \cdots 
\end{eqnarray}

The Yukawa coupling of (\ref{Batyrev2}) reads: 
\begin{eqnarray}
\hspace{-0.9in}&&K(q) \, \, = \, \, \, K^{\star}(q) 
\, \,\, = \, \, \, \, \,  \, \, \,
1\, \,\, \,  +4\, q\, \,  +164\, q^2\, \,  +5800\, q^3\, \,  +196772\, q^4\, \, 
 +6564004\, q^5 \, 
\nonumber  \\
\hspace{-0.9in}&&\qquad \quad  \quad \quad +222025448\, q^6 \, \,  
 +7574684408\, q^7 \, +259866960036\, q^8 
\,\,\,  \,  \,   + \, \, \, \cdots 
\end{eqnarray}
The equality of the Yukawa coupling with the
``adjoint'' Yukawa coupling, $\,K(q) \, \, = \, \, \, K^{\star}(q)$, is a straight
consequence of relation  $\,\, B_2 \cdot \, x\, = \, \, x \cdot \, adjoint(B_2)$.

\vskip .1cm 
Do note that recalling Batyrev and van Straten~\cite{Batyrev},
(see step2 page 496), and following Morrison~\cite{Morrison},
 one can also write the Yukawa coupling as:
\begin{eqnarray}
\label{otherYuk}
\hspace{-0.65in}K(q) \, \, \, \, = \, \, \, \, \, \,
  {{ x(q)^3 \, \cdot W_4^{1/2}} \over { y_0^2}} \cdot
 \, \Bigl({{q} \over {x(q)}} \cdot {{d x(q)} \over {dq}}\Bigr)^3
\, \, = \, \, \, \, \,
  {{ W_4^{1/2}} \over { y_0^2}} \cdot
 \, \Bigl( q \cdot {{d x(q)} \over {dq}}\Bigr)^3,
\end{eqnarray}
where $\, W_4$ is the Wronskian (\ref{wronskB2}).
{}From this alternative expression 
for the Yukawa coupling it is obvious that if 
the analytic series $\, y_0(x)$, {\em as well as} the nome (\ref{nomeB2}) 
are series with integer coefficients, then, the mirror map (\ref{mirrorB2}) 
is also a series with integer coefficients, and, therefore,
  $\, y_0$ seen as a function
of the nome $\, q$, as well as $\,x^3 \, W_4^{1/2}$, and, consequently, 
the Yukawa coupling  is a series with integer coefficients 
(as a series in $\, q$ or in $\, x$).

{\em The globally bounded character of the analytic series $\, y_0(x)$
together with the nome, thus yields the globally bounded character of 
the mirror map, Yukawa coupling, that we associate with the modularity}.
Similar results can be found in Delaygue's 
thesis~\cite{Delaygue}, in a framework when the coefficients of
hypergeometric series are ratio of factorials. 

\vskip .1cm 

In contrast the globally bounded character of the analytic 
series $\, y_0(x)$, together with the globally bounded character of 
the Yukawa coupling (seen for instance as a series in $\, x$) 
does not imply that the nome, or the mirror map, are globally bounded
as can be seen on example (\ref{saoud}) (see (\ref{Ksaoudx}) and (\ref{Ksaoudq})).

\vskip .1cm 

\subsubsection{An operator non trivially homomorphic to $\, B_2$ \newline \newline }
\label{operatornontriv}

\vskip .1cm 

Let us, now, introduce the order-four operator 
\begin{eqnarray}
\label{calB2}
\hspace{-0.6in}&&{\cal B}_2 \, \, = \, \, \,\, \,
 256\, x^2 \cdot \, \theta^2\, (2\, \theta+3)\, (2\, \theta+1)\,
 \nonumber  \\
\hspace{-0.6in}&&  \quad  \quad  \qquad \qquad  
-4\, x \cdot \, (2\, \theta+1)\, (2\, \theta-1)\, (5\, \theta^2-5\, \theta+2)
\, \, \,  + \, (\theta-1)^4. 
\end{eqnarray}
This operator is non-trivially\footnote[2]{The
 intertwiners between $\,{\cal B}_2 $
 and $\,B_2 $ are operators not simple functions.} homomorphic to
 the Calabi-Yau operator (\ref{Batyrev2}):
\begin{eqnarray}
{\cal B}_2 \cdot \, x \cdot (2\, \theta\, +1)  \, 
\, \, = \, \, \,\, \, x \cdot (2\, \theta\, +1) \cdot \, B_2. 
\end{eqnarray}
As a consequence of the previous intertwining relation,
one immediately finds that the series-solution analytic at $\, x\, = \, 0$
of this new MUM operator (\ref{calB2}) is nothing but the action of the
order-one operator $\,\, x \cdot (2\, \theta\, +1)\, $
 on the series (\ref{serB2}), 
and reads:
\begin{eqnarray}
\hspace{-0.9in}&&x \cdot (2\, \theta\, +1)[{\cal S}]
 \, \, \, = \,\, \, \, \, \,  \, 
x \,\, \,  \,  +24\, x^2 \, \,  +840\, x^3 \, \,  +35840\, x^4 \, \, 
 +1711080\, x^5 \, \,  +87329088\, x^6 \, 
\nonumber \\
\hspace{-0.9in}&& \qquad \qquad \quad   +4650277632\, x^7 \, \,  
+254905896960\, x^8 \,\, \,  \,\, + \,\,\, \, \cdots 
\end{eqnarray}
It is obviously also a series with {\em integer coefficients} (the action of 
 $\, x \cdot (2\, \theta\, +1) $ on the series with integer coefficient
is straightforwardly a series with integer coefficients). More generally, the
globally bounded series remain globally bounded series by operator equivalence
(non trivial homomorphisms between operators: generically the intertwiner 
operators are not simple functions).

The exterior square of the order-four operator 
(\ref{calB2}) is an order-six operator which is, in fact 
the LCLM of an order-five operator $\,{\cal E}_5$ and an order-one operator:
\begin{eqnarray}
\hspace{-0.95in}&& Ext^2({\cal B}_2) \,  \, = \, \, \,  \, 
  {\cal E}_5 \, \oplus \Bigl(D_x \, -{{d \ln(\rho(x)} \over {dx}} \Bigr),
\quad \hbox{where:} \quad \quad 
\rho(x) \,  \, = \, \, \,  \, {{x} \over {(1\, -16\, x) \,  (1\, -64\, x)}}.
\nonumber 
\end{eqnarray}

Operator $\,{\cal B}_2$  is non-trivially homomorphic to its adjoint:
\begin{eqnarray}
\hspace{-0.7in}{\cal B}_2 \cdot \, x^3 \cdot
 (2\, \theta\, +3) \cdot (2\, \theta\, +5)  \, 
\, \, = \, \, \,\,\, \, x^3 \cdot (2\, \theta\, +3)
 \cdot (2\, \theta\, +5)  \cdot \, adjoint({\cal B}_2). 
\end{eqnarray}
The Yukawa coupling of this order-four operator (\ref{calB2}),  
non-trivially homomorphic to (\ref{Batyrev2}), reads: 
\begin{eqnarray}
\label{YucalB2}
\hspace{-0.9in}&&\quad K(q) \, \, \, \, \, = \, \, \, \, \, \, \,
 1\,\, \, \, \,-4\, q\, \, \, -140\, q^2\, \, \, -4040\, q^3\, \, \,
 - 64436  \, \,{{ q^4} \over {3}} \, \,
\, + 1889332 \,  \, {{q^5 } \over {3 }}
\nonumber  \\
\hspace{-0.9in}&&\quad \quad \quad \quad \quad  
+ 88331368 \, \, {{q^6} \over {5 }}\, \, + 1652707624 \,\, {{q^7} \over { 9}}\, \, 
- 69295027684 \, \, {{q^8} \over {63 }} \, 
\,\,\,   + \, \, \, \cdots 
\end{eqnarray}
The Yukawa coupling series (\ref{YucalB2}) is {\em not globally bounded}.

The ``adjoint Yukawa coupling'' of this order-four operator (\ref{calB2})
reads:
\begin{eqnarray}
\label{adjYucalB2}
\hspace{-0.9in}&&K^{\star}(q) \, \, \, \, = \, \, \, \, \, \, \, \,
1\,\, \, \, \, +12\, q\, \, \, +564\, q^2\, \, \, +20440\, q^3\, \, \, 
+865732\, q^4\, \, \, +37162444\, q^5
\nonumber  \\
\hspace{-0.9in}&& \quad  \, \, \, \, \, \,  
 +8255346664 \, \, {{q^6} \over {5}} \, \, 
 + 1121762648248 \, \, {{q^7} \over {15}}\, \, 
+72336859374772 \, \, {{q^8} \over {21}} \,  \,  \,  \,  +  \,  \,     \, \cdots 
\end{eqnarray}
Again, the  adjoint Yukawa coupling series (\ref{adjYucalB2}) 
is {\em not globally bounded}.

On this example one sees that the Yukawa coupling of two 
non-trivially homomorphic operators
are {\em not necessarily equal}. The Yukawa couplings of two  homomorphic 
operators are equal when the two operators are {\em conjugated by a function}
(trivial homomorphism). The modularity property is {\em not preserved}
 by (non-trivial) operator equivalence: it can depend on a condition 
that the exterior square of the order-four operators are of order {\em five}.
The Calabi-Yau property is not preserved by operator equivalence. 

\vskip .1cm 

\vskip .1cm

{\bf To sum-up:} All these examples show that the 
{\em integrality} (globally bounded series) 
is {\em far from identifying with modularity}. All these examples have to be 
taken into account if one has in mind to build new conjectures
combining these  globally boundedness of various series with the 
concept of  diagonal of rational functions: 
for instance, can we imagine that being a diagonal of rational 
functions automatically yields that the nome or the Yukawa coupling
are globally bounded series in $\, q$ or $\, x$, etc ? 

\section{Conclusion}
\label{concl}

Seeking for the linear differential operators for the  $\, \chi^{(n)}$'s,
we first discovered that they were Fuchsian
 operators~\cite{ze-bo-ha-ma-04,ze-bo-ha-ma-05c}, and,
in fact, ``special'' Fuchsian operators, namely 
Fuchsian operators with rational exponents for all their singularities,
and with Wronskians that are $\, N$-th roots of rational functions.
Then we discovered that they were $\, G$-operators
(or equivalently globally nilpotent~\cite{bo-bo-ha-ma-we-ze-09}), and 
more recently, we accumulated 
results~\cite{CalabiYauIsing} indicating that 
they are ``special'' $\, G$-operators.
 There is, in fact, {\em two quite different 
kinds of ``special character''} of these $\, G$-operators. On one side,
we have the fact that one of their solutions is not only $\, G$-series, 
but is a {\em globally bounded} series. This special character 
has been addressed in this very paper,
 and we have seen that, in fact, this ``{\em integrality} 
property~\cite{Kratten}'' is a consequence of
 {\em quite general mathematical assumptions}
 often satisfied in physics (the  integrand is not only algebraic but analytic 
in all the variables~\eqref{form2}). However, we have also seen another special property
of these $\, G$-operators, namely the fact that they seem to be quite systematically
{\em homomorphic to their adjoints}~\cite{CalabiYauIsing}. We will 
show, in a forthcoming publication, 
that this last property amounts, on the associated linear differential systems,
to having {\em special differential Galois groups}, and  
that their exterior or symmetric square, have {\em rational solutions}. This 
last property is a  property of a more ``physical'' nature than the previous one,
related to an underlying {\em Hamiltonian structure}~\cite{Manin},
 or as this is the case,
for instance in the Ising model,
related to the underlying isomonodromic structure in the problem, which
yields the occurrence of some underlying Hamiltonian structure~\cite{Manin}. 
In general the {\em integrality} of $\, G$-operators 
{\em does not} imply the operator to be
homomorphic to its adjoint, and conversely being homomorphic to its adjoint
{\em does not} imply\footnote[2]{See \ref{CalabiYaucond}
which gives an example of a (hypergeometric) family 
of order-four operators satisfying the Calabi-Yau condition that their
exterior square is of order five, and, even, a  family 
of self-adjoint order-four operator, the 
corresponding hypergeometric solution-series
being {\em not globally bounded}. See also \ref{appendO}.}
 integrality (and even does not
 imply\footnote[1]{For 
instance the operator $\,\, D_x^n \, -x\, D_x\, -1/2\,$
 (see page 74 of~\cite{Katz})
 with an irregular singularity is self-adjoint.}
 the operator to be Fuchsian). Interestingly,
 the $\,\chi^{(n)}$'s, as well as many important 
problems of theoretical physics, correspond to $\, G$-operators that present 
these two complementary ``special characters'' (integrality and, up to homomorphisms,
self-adjointness), and, quite often, this is seen in the framework of the
emergence of ``modularity''. 

Nomes, mirror maps, and Yukawa couplings are {\em not D-finite} functions:
they are solutions of quite involved {\em non-linear} (higher order
Schwarzian) ODEs (see for instance Appendix D in~\cite{bo-bo-ha-ma-we-ze-09}).
Therefore, the question of the series integrality of the nomes, mirror maps,
Yukawa couplings, and other pullback-invariants (see \ref{Yukawaratio})
requires to address the very difficult question of series-integrality for
(involved) {\em non-linear} ODEs, or, equivalently, the problem of
non-linear recursions with integer sequence solutions. 
 Note, however, as seen in
Section~\ref{B2}, in particular in~\eqref{otherYuk}, that the integrality of the
series $\,y_0(x)$ and of the nome $\, q(x)$ are
 {\em sufficient to ensure}, {\em provided the operator is conjugated to its
adjoint} (see (\ref{condmagic})), 
the integrality of the other quantities such as the
Yukawa coupling, mirror maps. However the integrality of the nome remains 
an involved problem. These questions will
certainly remain open for some time.

In contrast, and more modestly, we have shown that a {\em very large sets of
problems in mathematical physics} (see sections  (\ref{gener}), (\ref{plan})
 and (\ref{subtheorem})) 
{\em actually correspond to diagonals of rational
functions}. In particular,  we have been able to show
that the $\, \chi^{(n)}$'s $\, n$-fold integrals of the
susceptibility of the two-dimensional Ising model
 are actually {\em diagonals of rational
functions for any value of the integer} $\, n$, thus proving that the
 $\,\chi^{(n)}$'s are 
{\em globally bounded for any value of the integer} $\, n$.
As can be seen in the ``ingredients'' of our simple 
demonstration (see (\ref{gener})),
no elliptic curves, and their modular forms~\cite{Kean}, no Calabi-Yau~\cite{Huse}, 
or Frobenius manifolds~\cite{Manin}, 
or Shimura curves, or arithmetic lattice
 assumption~\cite{Bouw,Dettweiler} is required to prove the result. 
We just need to have a $\, n$-fold integral such that its integrand is not only
algebraic, but {\em analytic in all the variables}. 

\vskip .3cm 

\vskip .2cm 

{\bf Acknowledgments} 
We would like to thank A. Enge and F. Morain for interesting and detailed 
discussions on Fricke and Atkin-Lehner involutions.
S. B. would like to thank the LPTMC and the CNRS for kind support. 
A. B. was supported in part by the Microsoft Research--Inria Joint Centre.
As far as physicists authors 
are concerned, this work has been performed without
 any support of the ANR, the ERC or the MAE. 

\vskip .1cm 

\appendix

\section{Modular forms and series integrality}
\label{modular}

{\bf First example}:
The generating function of the integers 
\begin{eqnarray}
\label{hadaprod}
\hspace{-0.6in}&&\quad \sum_{k=0}^{n} \, {n\choose k}^2 \cdot {2k \choose k}
 \cdot {2n-2k \choose n-k}  \\
\hspace{-0.6in}&&\quad \quad \quad \quad  \quad \, \, = \, \, \, \, \,  {2n \choose n} \cdot
\,  _2F_1\Bigl([{{1} \over {2}}, \, -n,\,  -n,\,  -n],
\,  \,[1,\,  1,\,  -\, {{2\, n \, -1} \over {2}}]; \, 1\Bigr),  
\nonumber 
\end{eqnarray}
is nothing else but the expansion of the square of a HeunG function
\begin{eqnarray}
\label{this}
\hspace{-0.7in}&& HeunG\Bigl(4,\, {{1} \over {2}},\, {{1} \over {2}},
 \, {{1} \over {2}},\, 1,\, {{1} \over {2}}; 
\, 16 \cdot x\Bigr)
\,  \, = \, \, \,  \, \, \,\, \, \,
1\, \,\,  \,\,\,+2\,x\,\, \, \,+12\,x^2\,\,+104\,x^3\,
 \nonumber \\
\hspace{-0.7in}&&\qquad \quad \quad  \, \, +1078\,x^4\,\,+12348\,x^5\,\,+150528\,x^6\,\,
+1914432\,x^7\,\,\,\, \, + \, \, \cdots  
\end{eqnarray}
solution of the  order-two operator
\begin{eqnarray}
\label{GoodHeunDiam}
\hspace{-0.4in}H_{diam} \,\, = \, \, \,\,\,\, \, \,
 \theta^2 \, \, \, \, \, -2 \cdot x \cdot (10 \, \theta^2\, +5\, \theta\, +1)
 \, \,\, \,
 +\,  16 \, x^2 \cdot  (2\, \theta\, +1)^2.
\end{eqnarray}
which corresponds to the {\em diamond lattice}~\cite{GoodGuttmann}. 
This HeunG function (\ref{this}) is {\em actually 
a modular form}\footnote[1]{Generically HeunG functions are far from being
modular forms. They are even far from being solutions of globally nilpotent 
operators (they generically have no integral 
representations~\cite{Vidunas,Valent}). There is a relation
between these operators being finite-gap~\cite{Verdier}
and their globally nilpotence.} which can be written in two different ways:
\begin{eqnarray}
\label{twopull}
\hspace{-0.4in}&&HeunG(4, {{1} \over {2}}, {{1} \over {2}}, 
{{1} \over {2}}, 1, {{1} \over {2}}; \,  16\, x)
 \,  \, \,\, 
\nonumber \\
\hspace{-0.4in}&&\qquad 
\,  \, = \, \,\,  \, \, (1\, -4 \, x)^{-1/2} \, \cdot \,
 _2F_1\Bigl([{{1} \over {6}},\, {{1} \over {3}}], \, [1]; \,
 {{108 \, x^2} \over {(1\, -4 \, x)^3 }} \Bigr)  \\
\hspace{-0.4in}&&\qquad 
\,  \, = \, \, \, \, \, (1\, -16 \, x)^{-1/2} \, \cdot \,
 _2F_1\Bigl([{{1} \over {6}},\, {{1} \over {3}}], \, [1]; \,
 -\, {{108 \, x } \over {(1\, -16 \, x)^3 }} \Bigr). 
\nonumber
\end{eqnarray}

These two pullbacks are {\em related by an ``Atkin'' involution}
 $\, x \, \leftrightarrow \, 1/64/x$. 
The associated modular curve, relating these
 two pullbacks (\ref{twopull}) 
yielding {\em  the modular curve}:
\begin{eqnarray}
\label{encore2}
\hspace{-0.5in}&&4\cdot \, y^3 \,z^3 \,\,\,\, -12 \,y^2\, z^2\, \cdot (y+z)\,\,\,\,
+3  \,y \, z \cdot(4\, y^2\,+ \, 4\, z^2 \, -127 \,y \, z)\,\,
\nonumber \\
\hspace{-0.5in}&&\quad \quad  \quad  \quad 
 -4 \cdot (y \, +z) \cdot (y^2\, + z^2 \, +83 \,y \,z)
\,\,\,\, +432  \,y \, z
\,\,\,\, \,\,\, = \,\,\,\,\,  \,\, 0, 
\end{eqnarray}
which is $(y, \, z)$-symmetric and is 
{\em exactly the rational modular
 curve} in eq. (27) already found
for the order-three operator $\, F_3$ in~\cite{CalabiYauIsing}
for the five-particle contribution $\, \tilde{\chi}^{(5)}$ 
of the magnetic susceptibility
of the Ising model. 
\vskip .3cm 
This result in~\cite{Prell,GoodGuttmann} 
can be rephrased as follows. 
One introduces
the order-three operator which has the following $\, _3F_2$
solution
\begin{eqnarray}
{{1} \over {(4\, -\, x^2)^3}} \, \cdot \, \, \, 
_3F_2\Bigl([{{1 } \over {3}}, \, {{1 } \over {2}}, \,{{2 } \over {3}}],\, 
[1, \, 1], \, \, {{27 \, x^4 } \over { (4\, -\, x^2)^3}}   \Bigr), 
\end{eqnarray}
associated with the {\em Green function of the diamond lattice}.
Along a {\em modular form line} lets us note that
this hypergeometric function actually has {\em two} pullbacks:
\begin{eqnarray}
\hspace{-0.5in}&&_3F_2\Bigl([{{1} \over {3}}, \,
 {{1} \over {2}}, \,{{2 } \over {3}}],\, 
[1, \, 1], \, \, {{27 \, x^4 } \over { (4\, -\, x^2)^3}}\Bigr) 
\,\,\,\,   \\
\hspace{-0.5in}&& \quad \quad \quad \, = \, \, \,  \,  \,\, 
 {{x ^2 \, -4  } \over {4  \cdot (x^2-1) }} \cdot \, \, 
_3F_2\Bigl([{{1 } \over {3}}, \, \, {{1 } \over {2}}, \,{{2 } \over {3}}],\, 
[1, \, 1],\,\, {{27 \, x^2 } \over {4 \cdot \, (x^2 -1)^3}}\Bigr). 
\nonumber  
\end{eqnarray}
These two pullbacks related by the ``Atkin'' involution 
$ \, \,  x \, \rightarrow \, 2/x$:
\begin{eqnarray}
\label{pull}
\hspace{-0.3in}u(x)   \, = \, \, {{27 \, x^4} \over { (4\, -\, x^2)^3}}, 
\qquad \quad 
v(x) \,\, = \, \,\, \,  u({{2} \over {x}}) \,\, \,  = \,  \, \, \,
{{27 \, x^2} \over {4 \cdot \, (x^2 -1)^3}},
\end{eqnarray}
corresponding, again, to the modular curve (\ref{encore2}).

\vskip .1cm 

{\bf Second example}. The HeunG function 
\begin{eqnarray}
\label{HeunGminus3}
\hspace{-0.6in}&&HeunG(-3, 0,1/2,1,1,1/2; \, 12\cdot x)
\, \, \, \, \, \,  \\
\hspace{-0.6in}&& \quad \, \,\, = \, \, \, \, (1\, +4 \, x)^{-1/4} \cdot \, 
HeunG\Bigl(4, \, {{1} \over {2}}, \, {{1} \over {2}}, \, 
{{1} \over {2}}, \, 1, \, {{1} \over {2}}, \, {{16 \, x} \over {1\, +4 \, x}}\Bigr)
\, \,\,  
 \nonumber \\
\hspace{-0.6in}&& \quad \quad \quad  \, \,\, = \, \, \, \,  \,\,\,\,
1\,\,\, \,\, +6\, x^2\,\, +24\,x^3\, \,  +252\,x^4\, \, +2016\,x^5  \, \,
+19320\,x^6\, \,  +183456\,x^7\,
 \nonumber \\
\hspace{-0.6in}&& \quad\qquad \qquad \quad  \quad 
  \,+1823094\,x^8 \,\, \, +18406752\,x^9 \,\, \, +189532980\,x^{10}
\, \, \,\,\, + \,\,\cdots 
\nonumber 
\end{eqnarray}
is solution of 
\begin{eqnarray}
\label{Heunfcc}
\hspace{-0.8in}Heun_{fcc}\,\, \, = \,\,\, \, \, \, \, 
\theta^2 \,\, \,\,-2 \, x \cdot \theta \cdot (4\, \theta \, +\, 1)
 \, \, \,  \, 
-24 \cdot x^2 \cdot  (2\, \theta \, +\, 1) \cdot  (\theta \, +\, 1),
\end{eqnarray}
The  square of (\ref{HeunGminus3})
is actually the solution of an order-three operator
 (see equation (19) in~\cite{GoodGuttmann})
emerging for lattice Green functions of the 
face-centred cubic (fcc) lattice
which is thus the symmetric square of (\ref{Heunfcc}).
This  hypergeometric function with a polynomial
pull-back can also be written:
\begin{eqnarray}
\label{identitybis}
\hspace{-0.5in}&&HeunG(-3,\,  0,\, 1/2,\, 1,\, 1,\, 1/2; \, 12 \cdot x)
\, \,\, \, \, 
  \nonumber \\
\hspace{-0.5in}&&\qquad \, \, = \, \, \, \,  \, 
 _2F_1\Bigl([{{1} \over {6}}, \, {{1} \over {3}}],[1];\, 
 108\cdot x^2 \cdot(1+4\, x)\Bigr) \\
\hspace{-0.5in}&& \qquad \, \, = \, \, \,  \, 
(1-12\, x)^{-1/2} \, \cdot \,
 _2F_1\Bigl([{{1} \over {6}}, \, {{1} \over {3}}],[1];\, 
 -\, {{ 108 \cdot x \cdot(1+4\, x)^2} \over {(1-12\, x)^3}} \Bigr),
 \nonumber 
\end{eqnarray}
where the involution 
$\, x  \, \leftrightarrow \, -1/4\cdot (1+4\, x)/(1-12\, x)\, \, $
takes place.
The modular curve relating these two pullbacks reads {\em exactly 
the rational curve} (\ref{encore2}) already 
obtained in~\cite{CalabiYauIsing}.

\vskip .1cm 

{\bf Third example}. The HeunG function 
$\, HeunG(1/9, 1/12, 1/4, 3/4, 1, 1/2; \,  4\,x)$
is solution of the order-two operator corresponding to the 
simple cubic lattice Green function 
\begin{eqnarray}
\label{364bis}
\hspace{-0.6in}&&H_{sc} \, \, \, = \, \, \, \, \, \, \,\,
 \theta^2 \, \,  \, \, \,  \, -x \cdot (40\, \theta^2+20\, \theta\, +3)
\, \,  \,  \,
 +9 \cdot x^2  \cdot (4\, \theta\, +3) \cdot (4\, \theta\, +1). 
\nonumber 
\end{eqnarray}
The square of this HeunG function is a series with
integer coefficients which identifies with the
 Hadamard product of  $\, (1\, -4\, x)^{-1/2}$
with a modular form :
\begin{eqnarray}
\hspace{-0.9in}&& HeunG(1/9, 1/12, 1/4, 3/4, 1, 1/2; \,  4\,x)^2 
\, \,\,    \\
\hspace{-0.9in}&& \qquad \quad = \, \, \, \,\, \,
 (1-4\, x)^{-1/2}\, \star \, HeunG(1/9, 1/3, 1, 1, 1, 1; \, x)  
\nonumber \\
\hspace{-0.9in}&& \qquad \quad= \, \, \, \, \, \,\,
1\,  \,\,\, \,+6\, x\, \, + 90\, x^2 \,  \,+ 1860\, x^3 \, \,
 +44730\, x^4\, \,
+ 1172556\, x^5\, \,+ 32496156\, x^6\, \nonumber \\
\hspace{-0.9in}&& \qquad \quad \quad \quad \qquad  + 936369720\, x^7 \, \,
 + 27770358330\, x^8\, \, +842090474940\, x^9\,\, \, \,\,
\, + \, \, \cdots 
\nonumber
\end{eqnarray}
The HeunG function 
$\, HeunG(1/9, 1/12, 1/4, 3/4, 1, 1/2; \,  4\,x)$ is globally bounded:
 the series of 
$\, HeunG(1/9, 1/12, 1/4, 3/4, 1, 1/2; \,  8\,x)$ 
 is a series with integer coefficients. 
One can also write this HeunG function
in terms of a $\, _2F_1([1/6,1/3],\, [1],x)$ hypergeometric function 
up to a simple algebraic pullback (with a square root), or in terms 
 of a $\, _2F_1([1/8,3/8],\, [1],x)$ hypergeometric function:
\begin{eqnarray}
\label{even}
\hspace{-0.9in}&&HeunG(1/9, 1/12, 1/4, 3/4, 1, 1/2; \,  4\,x) 
\, \,  =  \, \,  \, \, \, C_2^{1/4} \, \cdot  \,
 _2F_1\Bigl([1/8,\,  3/8],\,  [1]; \, P_2   \Bigr), \quad \hbox{with:} 
\nonumber 
\end{eqnarray}
\begin{eqnarray}
\hspace{-0.9in}&&C_2\, \,  =  \, \, \,
 {{1} \over {9 \cdot \, (1\, +12\, x)^2}} \cdot
 \Bigl(5\, -36\, x \,+4\cdot (1-36\,x)^{1/2}\Bigr), \quad\quad
P_2\, \,  =  \, \, \, {{ 128 \cdot x} \over { (1\, +12 \, x)^4}} \cdot \, p_2, 
 \nonumber \\
\hspace{-0.9in}&&p_2\, \,  =  \, \, \, \, 
(1\, -42\, x \, +352\, x^2\, -288\, x^3) \, \, \,
 +\, (1\, -4\, x) \cdot (1\, -20\, x) \cdot \, (1\, -36 \, x)^{1/2}.
\nonumber
\end{eqnarray}
Do note that taking the Galois conjugate (changing $(1-36\,x)^{1/2}$ into 
$-(1-36\,x)^{1/2}$) gives the series expansion
of  $\,\,\, 3^{-1/2} \cdot \, HeunG(1/9, 1/12, 1/4, 3/4, 1, 1/2; \,  4\,x)$. 
This shows that there exists an identity for $\, _2F_1([1/8,3/8],\, [1],x)$ 
with {\em two different pullbacks}, namely the previous $\, P_2$  
and its Galois conjugate, these two pullbacks being related by a
(symmetric genus zero) modular curve:
\begin{eqnarray}
\label{A13}
\hspace{-0.9in}&&5308416 \cdot  y^4\, z^4\, \,  \,
+442368 \cdot y^3\, z^3 \cdot (y+z)\,\,  \,
+512\, y^2 \,z^2\cdot (27\,y^2\, +27\,z^2\, -27374\,x\, y)\,
\nonumber \\
\hspace{-0.9in}&&\quad \, \, \, \,   +192\, y\, z \cdot (y\,+z) 
\cdot (y^2\,+z^2\,+10718 \,y\, z)
\, \, \,\, +y^4\, +z^4\,\,  +3622662\, y^2\, z^2
\nonumber \\
\hspace{-0.9in}&&\, \, \, \, \,   \quad  \, 
-19332 \cdot y\, z \cdot \, (y^2\, +  \,z^2) \, \,
 +79872 \cdot y \, z \cdot (y\, +z)\, 
 -65536  \cdot y \, z\, \,\, \, = \,\,\, \, \,\, 0. 
\end{eqnarray}

\vskip .1cm 

{\bf Revisiting the examples}. In a recent paper~\cite{spanning} corresponding to 
spanning tree generating functions and Mahler measures, a result from Rogers 
(equation (36) in~\cite{spanning}) is given where the two following $\,\, _5F_4\,$
 hypergeometric functions take place:
\begin{eqnarray}
\hspace{-0.3in}&&_5F_4\Bigl([{{5} \over {4}}, \,
 {{3} \over {2}}, \, {{ 7} \over {4}}, \, 1, \, 1], \, 
[2, \, 2, \, 2, \, 2], \, \, {{ 256 \, x^3} \over {9 \cdot \, (x+3)^4 }} \Bigr), 
\nonumber \\
\hspace{-0.3in}&&
_5F_4\Bigl([{{5} \over {4}}, \, {{3} \over {2}}, 
\, {{ 7} \over {4}}, \, 1, \, 1], \, 
[2, \, 2, \, 2, \, 2], \, \, {{ 256 \, x} \over {
9 \cdot \, (1\, + \, 3\, x)^4 }} \Bigr). 
\end{eqnarray}
The corresponding order-five linear differential operators (annihilating
these two $\, _5F_4$
 hypergeometric functions) are actually homomorphic (the intertwiners 
being order-four operators).
The relation between these two pullbacks $\, y \, = \, \, 256 \, x^3/9/(x+3)^4$
and $\, z \, = \, \, 256 \, x/9/(1\, +3\, x)^4$, remarkably {\em gives, again,
 the previous} $(y, \, z)$-{\em symmetric modular curve} (\ref{A13}). 

The  order-five linear differential operator, corresponding to the first  $\, _5F_4$
 hypergeometric function, factorizes in an order-one operator,
an order-three operator and an order-one operator, the order-three
operator being, in fact, exactly the symmetric square of an order-two 
operator:
\begin{eqnarray}
\hspace{-0.1in} L_1 \,  \,  \cdot\,  \,  Sym^2(W_2)  \,  \,  \cdot \, \,   
{\frac {{x}^{4}}{ (x\, -9)  \, (x\, +3)^{4}}} \, \,   \cdot \, \,  R_1,
\nonumber 
\end{eqnarray}
where the order-one operators read respectively
\begin{eqnarray}
\hspace{-0.95in}&& L_1  = \,  D_x \, - \, {{d} \over {dx}} \ln\Bigl(  {{x-9} \over {
(9\,x^2\,+14\,x\,+9) \cdot \, (x+3)^4 }}  \Bigr), \quad \, \, \, \,
R_1 \, = \,  \, D_x \, \, - \, {{d} \over {dx}} \ln\Bigl( {{ (x+3)^4 } \over { x^3}}\Bigr), 
\nonumber
\end{eqnarray}
and where the order-two operator $\, W_2$ reads:
\begin{eqnarray}
\hspace{-0.9in}W_2 \,\, = \, \,\,\, \, D_x^2 \,  \, \, \,
 +3\,{\frac { ( 6 \cdot \,{x}^{2} \, +7\,x \, +3)
 }{ ( 9\,{x}^{2}\, +14\,x\,+9)\cdot \,  x}}
\cdot \, D_x 
 \,\, \, + \,  {{3} \over {4}} \cdot \,
{\frac {3\,x \, +2}{ (9\,{x}^{2}\,+14\,x\,+9)\cdot \,  x}}. 
\end{eqnarray}

We have a similar result for the  order-five linear differential operator 
corresponding to the second  $\, _5F_4$ hypergeometric function.

Another solution of this order-five linear differential operator reads:
\begin{eqnarray}
\label{3F2pull}
\hspace{-0.8in}\,\,\, {{(x\, +3)^4} \over {x^3}} \cdot \, 
\int \, {{x\, -9 } \over {( x+3) \cdot \, x }} 
 \, \cdot  \,  \, 
_3F_2\Bigl( [{{1} \over {4}}, \, {{1} \over {2}}, \, {{ 3} \over {4}}], \, 
[1, \, 1], \, \, {{ 256 \, x^3} \over {9 \cdot \, (x+3)^4 }}  \Bigr)
     \, \cdot \, dx. 
\end{eqnarray}
The expansion of the $\, _3F_2$ hypergeometric function 
in (\ref{3F2pull}) is globally bounded
(change $\,\, x \, \rightarrow \, 9 \, x\,$ to get a series with integer coefficients). 

Recalling the two previous pullbacks we have, in fact, the following identity:
\begin{eqnarray}
\label{followident}
\hspace{-0.5in}&& 3 \cdot \, (1 \, + \, 3 \, x) \cdot \, \,
 _3F_2\Bigl( [{{1} \over {4}}, \, {{1} \over {2}}, \, {{ 3} \over {4}}], \, 
[1, \, 1], \, \, {{ 256 \, x^3} \over {9 \cdot \, (x+3)^4 }}  \Bigr)
\nonumber \\
\hspace{-0.5in}&& \qquad \quad  \,  \, = \,\, \,  \,  \, \,
 (x\, +\, 3) \cdot \, \,  _3F_2\Bigl( [{{1} \over {4}},
 \, {{1} \over {2}}, \, {{ 3} \over {4}}], \, 
[1, \, 1], \, \, {{ 256 \, x} \over {9 \cdot \, (1\, + \, 3\, x)^4 }}  \Bigr).
\end{eqnarray}
However this $\, _3F_2$ hypergeometric function is nothing but the
square of a $\, _2F_1$ hypergeometric function
\begin{eqnarray}
\hspace{-0.3in} _3F_2\Bigl( [{{1} \over {4}}, \, {{1} \over {2}}, \, {{ 3} \over {4}}], \, 
[1, \, 1], \, \, x \Bigr)
 \,  \,\, \,  = \, \, \, \,  \,  \,
 _2F_1\Bigl( [{{1} \over {8}}, \,  {{ 3} \over {8}}], \, 
[1], \, \, x  \Bigr)^2.
\end{eqnarray}
Thus, the previous identity (\ref{followident}) is nothing but the
 identity on a $\, _2F_1$ hypergeometric function with {\em two different
pullbacks}:
\begin{eqnarray}
\label{newident}
\hspace{-0.6in}&& (1+3\,x)^{1/2} \cdot \, \,
  _2F_1\Bigl( [{{1} \over {8}}, \,  {{ 3} \over {8}}], \, 
[1], \, \,  {{ 256 \, x^3} \over {9 \cdot \, (x+3)^4 }} 
 \Bigr) \nonumber \\
\hspace{-0.6in}&& \qquad \quad  \quad \,  \, = \, \,  \, \,   \,   \, 
\Bigl(1 \, +{{x} \over {3}}\Bigr)^{1/2} \, \cdot \, 
 _2F_1\Bigl( [{{1} \over {8}}, \,  {{ 3} \over {8}}], \, 
[1], \, \,  {{ 256 \, x} \over {9 \cdot \, (1\, + \, 3\, x)^4 }}   \Bigr).
\end{eqnarray}
The expansion of (\ref{newident}) is globally bounded. One gets a series 
with {\em positive integer}  coefficients using the simple rescaling
 $\, x \, \rightarrow \, 36 \cdot \, x$.
Note that the two pullbacks can be exchanged by the simple ``Atkin'' involution
 $\, x \, \leftrightarrow \, \, 1/x$, being related by the modular curve 
occurring for the simple cubic lattice, namely (\ref{A13}).

We have a similar result for the  other $\,\, _5F_4\,$
 hypergeometric functions popping out in~\cite{spanning}.

 For instance, for 
the diamond lattice one gets an expression (see eq. (50)
 in~\cite{spanning}) where the two following $\, _5F_4$
 hypergeometric functions take place\footnote[2]{Note a small misprint 
in eq. (50) of~\cite{spanning}: one should read $\, -27 z^2/4/(1-z^2)^3$ 
instead of $\, -27 z^4/4/(1-z^2)^3$.}:
\begin{eqnarray}
\label{diampull}
\hspace{-0.3in}&&_5F_4\Bigl([{{5} \over {3}}, \,
 {{3} \over {2}}, \, {{ 4} \over {3}}, \, 1, \, 1], \, 
[2, \, 2, \, 2, \, 2], \, \, {{ - 27 \, x^2 } \over { 4 \cdot \, (1\, - \, x^2)^3 }} \Bigr), 
\nonumber \\
\hspace{-0.3in}&&
_5F_4\Bigl([{{5} \over {3}}, \, {{3} \over {2}}, 
\, {{ 4} \over {3}}, \, 1, \, 1], \, 
[2, \, 2, \, 2, \, 2], \, \,  {{  27 \, x^4 } \over {  (4\, - \, x^2)^3 }}  \Bigr). 
\end{eqnarray}
These two pullbacks can be exchanged by the simple ``Atkin'' involution
 $\, x \, \leftrightarrow \, \, 2/x$. These two pullbacks have been seen to be
related by the (genus-zero) $(y, \, z)$-symmetric modular curve (\ref{encore2}):
\begin{eqnarray}
\label{encore3}
\hspace{-0.6in}&&4\,{y}^{3} \, {z}^{3} \, \, \,   -12\,{y}^{2} \, {z}^{2} \cdot \, (y \, +z)\,
  \,  \, \,  +3\,y\, z \left( 4\,{y}^{2}+4\,{z}^{2}-127\,y \, z \right)\,
 \nonumber \\ 
\hspace{-0.6in}&&\qquad \qquad \quad   -4 \cdot \, (y\, +z) \cdot \, 
 (y^2 \, + \, {z}^{2}+83\, y\, z) \, \, \, \,   +432\, y\, z
 \, \,\,\,\,   = \,\, \,\,\,  \, 0. 
\end{eqnarray}

Similarly to (\ref{followident}) we have an identity between two
$\, _3F_2$ hypergeometric functions 
(namely $\, _3F_2([2/3,\,1/2, \, 1/3],[1, \, 1],z)$) with 
the two pullbacks (\ref{diampull}), and these
$\, _3F_2$ hypergeometric functions being the square of 
$\, _2F_1$ hypergeometric functions, one finds that the 
``deus ex machina'' is the 
identity similar to (\ref{newident}):
 \begin{eqnarray}
\label{newidentdiam}
\hspace{-0.6in}&& (1 \,- x^2)^{1/2} \cdot \, \,
  _2F_1\Bigl( [{{1} \over {3}}, \,  {{ 1} \over {6}}], \, 
[1], \, \,   {{  27 \, x^4 } \over {  (4\, - \, x^2)^3 }} 
 \Bigr) \nonumber \\
\hspace{-0.6in}&& \qquad \quad  \quad \,  \, = \, \,  \, \,   \, \,
(1 \, -{{x^2} \over {4}})^{1/2} \cdot \, 
 _2F_1\Bigl( [{{1} \over {3}}, \,  {{ 1} \over {6}}], \, 
[1], \, \,  {{ - 27 \, x^2 } \over { 4 \cdot \, (1\, - \, x^2)^3 }}   \Bigr).
\end{eqnarray}
The series expansion of (\ref{newidentdiam}) 
is globally bounded. Rescaling the
$\, x$ variable as $\, x \,\rightarrow  \, 4 \, x$, the series expansion 
becomes a series with {\em positive integer} coefficients 
(up to the first constant term).

 For the face-centred cubic lattice one gets an expression (see eq. (52)
 in~\cite{spanning}) where the two following $\, _5F_4$
 hypergeometric functions take place\footnote[1]{There is one more misprint
in~\cite{spanning}:
the pullback $\, -x \, (x+3)/(x-1)^3$ must be changed into
 $\, x \, (x+3)/(x-1)^3$.}:
\begin{eqnarray}
\label{fcc}
\hspace{-0.3in}&&_5F_4\Bigl([{{5} \over {3}}, \,
 {{3} \over {2}}, \, {{ 4} \over {3}}, \, 1, \, 1], \, 
[2, \, 2, \, 2, \, 2], \, \, {{  x \cdot \, (x\,+3)^2} \over { (x\, -1)^3 }} \Bigr), 
\nonumber \\
\hspace{-0.3in}&&
_5F_4\Bigl([{{5} \over {3}}, \, {{3} \over {2}}, 
\, {{ 4} \over {3}}, \, 1, \, 1], \, 
[2, \, 2, \, 2, \, 2], \, \, {{  x^2 \cdot \, (x\,+3)} \over { 4 }}  \Bigr). 
\end{eqnarray}
This example is nothing but the previous diamond lattice example (\ref{diampull}) 
with the change of variable $\, x \, \rightarrow \, -3\, x^2/(x^2-4)$
in (\ref{fcc}). Therefore, the two pullbacks in (\ref{fcc}) are, again,
related by the modular curve (\ref{encore2}). The two pullbacks in (\ref{fcc}) can 
actually be seen directly in the following identity
 (equivalent to (\ref{newidentdiam})): 
\begin{eqnarray}
\hspace{-0.9in}&&\quad  _2F_1\Bigl( [{{1} \over {3}}, \,  {{ 1} \over {6}}], \, 
[1], \, \,    {{  x \cdot \, (x\,+3)^2} \over { (x\, -1)^3 }}   \Bigr)
\,  \,\, = \, \,  \, \,   \, \, 
 (1 \,- x^2)^{1/2} \cdot \, \,
  _2F_1\Bigl( [{{1} \over {3}}, \,  {{ 1} \over {6}}], \, 
[1], \, \,   {{  x^2 \cdot \, (x\,+3)} \over { 4 }} 
 \Bigr).
\nonumber 
\end{eqnarray}

Finally, the equation (17) of~\cite{spanning} on Mahler measures, 
the two following $\, _4F_3$
 hypergeometric functions take place:
\begin{eqnarray}
\label{lastpull}
\hspace{-0.3in}&&_4F_3\Bigl([{{5} \over {3}}, \,
 {{4} \over {3}},  \, 1, \, 1], \, 
[2, \, 2, \, 2], \, \, {{ 27 \, x} \over { (x \, -2)^3 }} \Bigr), 
\nonumber \\
\hspace{-0.3in}&&
_4F_3\Bigl([{{5} \over {3}}, \,
 {{4} \over {3}},  \, 1, \, 1], \, 
[2, \, 2, \, 2], \, \, {{ 27 \, x^2} \over { (x \, +4)^3 }} \Bigr). 
\end{eqnarray}
These two previous pullbacks can be exchanged by an ``Atkin'' involution
$\, x \, \leftrightarrow \, -8/x$ and are related by the (genus-zero)
$(y, \, z)$-symmetric modular curve: 
\begin{eqnarray}
\hspace{-0.6in}&&8\,{y}^{3} \, {z}^{3} \,  \,  \, 
-12\,\,{y}^{2}{z}^{2}\cdot \,  (y \, +z) \,  \, \, 
  +3\, y \, z \cdot \, ( 2\,{y}^{2} +2\,{z}^{2} \, +13\,y\, z)
  \nonumber \\
\hspace{-0.6in}&& \quad \quad \quad \quad \quad \quad \,  
\, \,  - \, (y \,+ z)  \cdot \, (y^2 \, +{z}^{2} \, +29\,y \, z) \, \,  \,  +27\,y \, z 
\,  \,  \,  \,\,\, =  \, \,   \,  \, \,  \, 0.  
\end{eqnarray}
The underlying identity on $\, _2F_1$ hypergeometric functions with
the two pullbacks (\ref{lastpull}) read:
\begin{eqnarray}
\label{newidentdiam3}
\hspace{-0.6in}&& -\, 2 \cdot \, (x \,- 2) \cdot \, \,
  _2F_1\Bigl( [{{1} \over {3}}, \,  {{ 2} \over {3}}], \, 
[1], \, \,   {{  27 \, x^2 } \over {  (x \, + \, 4)^3 }} 
 \Bigr) \nonumber \\
\hspace{-0.6in}&& \qquad \quad \quad \quad \,  \, = \, \,  \, \,   \,   \,   \, 
\, (x \, + \, 4)\cdot \, 
 _2F_1\Bigl( [{{1} \over {3}}, \,  {{ 2} \over {3}}], \, 
[1], \, \,  {{  27 \, x } \over { (x \, - \, 2)^3 }}   \Bigr).
\end{eqnarray}
The series expansion of (\ref{newidentdiam3}) is globally bounded. Rescaling the
$\, x$ variable as $\, x \,\rightarrow  \, -8 \, x$, the series expansion 
becomes a series with {\em positive integer} coefficients.

\section{Another logarithmically bounded series}
\label{locally}

Let us display other logarithmically bounded series than (\ref{logarithbounded2F1}).
The hypergeometric function $\, _2F_1([N/3,1/6],\, [7/6],\, 9 \, x) \,$ is  
{\em not globally bounded}
but is such that the order-one operator $\, 6 \, \theta\, +1$ acting on it,
is a series with integer coefficients for every integer value of $\, N$:
\begin{eqnarray}
\hspace{-0.9in}&&(6 \, \theta\, +1)\, 
[_2F_1\Bigl([{{N} \over {3}}, \, {{1} \over {6}}],
\,[{{7} \over {6}} ], \,  9 \, x  \Bigr)] \, \,\, \, \, = \, \, \,\,\,
 _1F_0\Bigl([{{N} \over {3}}],
\,[], \,  9 \, x  \Bigr) \, \, \,\, \, 
\nonumber \\
\hspace{-0.9in}&& \, \,  = \, \, \,  1+3\, x+18\, x^2
+126\, x^3+945\, x^4+7371\, x^5\, +58968\, x^6
\, + \, \cdots 
  \, \,\, \quad  \,\,   \hbox{for} \, \,  N =  \, 1, 
\nonumber \\
\hspace{-0.9in}&& \, \,  = \, \, \, 1+6\, x+45\, x^2
+360\, x^3+2970\, x^4+24948\, x^5+212058\, x^6
\, + \, \cdots 
     \,\, \,   \hbox{for} \, \,   N =  \, 2, 
\nonumber \\
\hspace{-0.9in}&& \, \,  = \, \, \, 1+9\, x+81\, x^2
+729\, x^3+6561\, x^4+59049\, x^5+531441\, x^6
\, + \, \cdots 
      \,\, \,  \hbox{for} \, \,  N =  \, 3, \,\,  \cdots 
 \nonumber
\end{eqnarray}

Similarly 
\begin{eqnarray}
U \, \,\,= \, \,\,\,\,
 _3F_2\Bigl([{{1} \over {4}}, \, {{7} \over {12}},\,{{1} \over {7}} ],
\,[{{4} \over {3}},\,{{8} \over {7}} ], \, 64\, x  \Bigr), 
\end{eqnarray}
which is such that the action of the order-one operator 
$\, 7\, \theta \, +\, 1$ changes it into a {\em globally bounded} function
\begin{eqnarray}
\hspace{-0.8in}&&(7\, \theta \, +\, 1)(U) \,\,\, = \, \,\,\, \,
 _2F_1\Bigl([{{1} \over {4}}, \, {{7} \over {12}} ],
\,[{{4} \over {3}}], \, 64\, x  \Bigr)
\,\,\, = \,\, \, \,\,\,\,\,\, 1 \,\, \,\, +7\, x\,\, +190\,x^2\,\,+7068\,x^3 
\nonumber \\
\hspace{-0.8in}&&\,\,\quad \quad  \quad  \,\, \, \, \,+303924\,x^4\,\,
+14208447\,x^5\,\, +701448594\,x^6\,\, +35983401900\,x^7 \,\,\,\, \,+ \, \cdots 
\nonumber
\end{eqnarray}

\vskip .2cm 

\section{ $\, \Phi_D^{(n)}(w)$ as diagonals }
\label{Chebi}

The family of simple integrals $\, \Phi_D^{(n)}(w)$ was
 introduced in \S4 of ~\cite{bo-ha-ma-ze-07}, as a way to simplify the study of the
singularities of the Ising integrals $\chi^{(n)}$. By definition, they are
equal to
\begin{equation}
\label{def:PhiD}
\hspace{-0.4in}\Phi_D^{(n)}(w)
 \,\, = \,  \,\, \,\,  \,  -\frac{1}{n!} \,\, \, \,  \, + \frac{2}{n!}\int_0^{2\pi} \frac{d\phi}{2\pi}
\,\, \,   
\frac{1}{
1 \, \, \,  - x^{n-1}(w,\phi) \cdot x(w,(n-1) \phi)},
\end{equation}
where
\begin{equation}
x(w,\phi) \,\,\,  \,= \,\,  \,  \, \,  \,\frac{2w}{1 \,\,  -2w  \, \cos(\phi)\, \, 
  +\sqrt{(1 \, -2w\cos(\phi))^2\, -4w^2}}.
\end{equation}
By an easy change of variables, it follows that 
\begin{eqnarray}
\label{eq:PhiD-PsiD}
\Phi_D^{(n)}(w)  \,\,\,  \, \,
=  \,\,\,\,  -\frac{1}{n!}\,  \,  \, \,\,  + \, \frac{2}{n!} \cdot \Psi_D^{(n)}(w),
\end{eqnarray}
where 
\begin{eqnarray}
\Psi_D^{(n)}(w) \, \,\,\,   = \, \, \,\, \, \,  \, 
  \frac{1}{\pi}\int_{-1}^1\, {F_n(w,t) \cdot  \, \frac{dt}{\sqrt{1-t^2}}},
\end{eqnarray}
the algebraic function $F_n(w, \, t)$ being defined by
\begin{eqnarray}
F_n(w, \, t)\,\,\,  \,  = \, \,\,\,\, \,   \frac{1}{1\,\, \,  -h(w,t)^{n-1} \cdot h(w,T_{n-1}(t))},
\end{eqnarray}
where 
\begin{eqnarray}
h(w,t) \,\, \, \,  =\,\, \, \, \,  \, \frac{2w}{1\,\, -2w \,  t\,\, +\sqrt{(1-2wt)^2\,-4w^2}},
\end{eqnarray}
and where $\, T_{m}(t)$ is the $\, m$-th Chebyshev polynomial of 
the first kind, that is, the unique polynomial of degree~$m$ such that 
$ \, \,  \cos(mt)\,=\,\,T_{m}(\cos t)$.

In order to express $ \, \Phi_D^{(n)}(w)$ as the diagonal of an 
algebraic function in two variables, it is sufficient to use the following general result:

\begin{quote}\em 
	If $F(w,t)$ is a bivariate power series, then the univariate power series
\begin{eqnarray} 
\Psi(w)\,\, \,  \,\, = \,\,\, \, \,  \, \,\,
  \frac{1}{\pi} \cdot \int_{-1}^1 \frac{F(w, \, t)}{\sqrt{1-t^2}} \cdot \,  dt
\end{eqnarray}
	is the diagonal of the generalised power 
series\footnote[3]{In the sense of~\cite{BA-JPB}.}
\begin{eqnarray}
  G(w,t)\, \, \, \, \,= \,\, \, \, \, \,  \,  \frac{F(w \, t, \, 1/t)}{\sqrt{1-t^2}}.
\end{eqnarray}
\end{quote}	
The only non-trivial point in the proof of this fact is the classical integral evaluation
\begin{eqnarray}
\hspace{-0.6in}\frac{1}{\pi} \cdot \int_{-1}^1 \frac{dt}{(1-ut) \cdot \sqrt{1-t^2}} 
\;\,\, \, = \;\, \,\,\, 
 \frac{1}{\sqrt{1-u^2}}, 
\quad \quad  \quad \text{for} \,\, \,\quad |u| < 1.
\end{eqnarray}
\vskip .1cm 

{\bf Expanded proof:} Letting 
$\, \, \,F(w,t)\, = \,\,\,  \sum_{\ell \geq 0} \, f_\ell(t) \cdot \, w^\ell$,
 the series $\Psi(w)$ is equal to
\begin{eqnarray} 
\Psi(w)\,\,\,\,\, = \,\,\,\,\,\,\,
\sum_{\ell \geq 0} \, \frac{1}{\pi} \cdot
 \int_{-1}^1 \, \frac{f_\ell(t)}{\sqrt{1-t^2}} \cdot \, w^\ell \; \, dt, 
\end{eqnarray}
 while the series $G(w, \, t)$ is equal to
\begin{eqnarray}
 G(w, \, t)\, \, \,\,  = \, \, \,  \,  \,\, \,
\sum_{\ell \geq 0}  \, \, \frac{f_\ell(1/t) \,\, t^\ell}{\sqrt{1-t^2}} \cdot \,  w^\ell.
\end{eqnarray}

It follows that the diagonal of $G$ is equal to
\begin{eqnarray} 
\Diag(G)(w)  \,  \,  \,\, \,= \, \, \, \,  \,  \,\,   \, 
 \sum_{\ell \geq 0}  \, \left[ t^0 \right]
 \frac{f_\ell(1/t)}{\sqrt{1-t^2}} \cdot w^\ell.
\end{eqnarray}
To prove that $\Psi \,=\, \Diag(G)$, it thus suffices 
to show that for any power series $f$,
\begin{eqnarray}
\hspace{-0.4in}&&\frac{1}{\pi} \cdot \int_{-1}^1 \frac{f(t)}{\sqrt{1-t^2}} \cdot \,  dt 
\,\,  \,\,   \; = \; \,\, \,\,  \,   
\left[ t^0 \right] \frac{f(1/t)}{\sqrt{1-t^2}}.
\end{eqnarray}
By linearity, it thus suffices to prove that for any non-negative integer $s$,
\begin{eqnarray}
\hspace{-0.6in}\qquad \quad &&\frac{1}{\pi} \cdot \int_{-1}^1 \frac{t^s}{\sqrt{1-t^2}} \cdot \, dt
 \;\, \, \,\, = \;\,\, \,\, \,\,
 \left[ t^s \right] \frac{1}{\sqrt{1-t^2}}.
\end{eqnarray}
This follows from the classical integral evaluation
\begin{eqnarray}
\hspace{-0.6in}\qquad \quad &&\frac{1}{\pi} \cdot \, 
\int_{-1}^1 \frac{dt}{(1-ut) \cdot \sqrt{1-t^2}}
 \,\; \, \,  = \;\, \,\, \,  \,\frac{1}{\sqrt{1-u^2}},
 \quad \quad  \quad \text{for}\,\,\,    \quad |u| < 1. \qquad\qquad 
 \end{eqnarray}

For example, when $n=\, 2$, the previous construction shows that 
\begin{eqnarray} 
\hspace{-0.5in}&&\Phi_D^{(2)}(w) \;\,\,  = \; \,\, \,\,\, \,  
 \frac14 \,\,\, \, \,  + \frac14 \cdot \, 
 _2F_1\Bigl([{{1} \over {2}}, \, {{1} \over {2}} ] \, [1], \, 16 \, w^2 \Bigr)  \\
\hspace{-0.5in}&& \qquad \quad  \quad = \, \, \,  \,   \, \, 
\frac12\,\, \, \,  \,  + w^2 \,\, \,   +9w^4\,\,   +100w^6\,\,
   +1225w^8\,\,   +15876 w^{10}
 \,\,  \,\, \,   + \,  \, \cdots   \nonumber 
\end{eqnarray}	
is equal to the diagonal of the algebraic function
\begin{eqnarray}  
 \frac{1 \,\,  -2w \, +\sqrt{(1-2w)^2 \,\,   - 4w^2t^2}}
{2 \, \cdot \, \sqrt{1-t^2} \cdot \sqrt{(1-2w)^2\, \,  - 4w^2t^2}}
\,  \, \,  \,  \,\,  \,   - \frac12 .
\end{eqnarray}

\section{Creative telescoping: computing ODEs for diagonals}
\label{telesc}

The notion of diagonal of rational function is ubiquitous in
combinatorics~\cite{Stanley99}. Its importance comes from the fact that many
operations on power series with a combinatorial relevance (Hadamard products,
constant terms, or positive parts, of Laurent power series, etc) {\em can be
encoded as diagonals}. A classical result by
Lipshitz~\cite{Lipshitz} predicts that {\em diagonals of rational
functions are D-finite}. The question is then: how to obtain algorithmically a
differential equation satisfied by the diagonal $\Diag(f)$ of a given rational
function $f(x_1, \ldots, x_n)$? The question can be reformulated 
in terms of computing a
multiple integral with parameters, over an algebraic surface 
(``vanishing cycle'' or ``\'evanescent cycle'' 
in Deligne's terminology~\cite{Deligne84}), 
and thus can be attacked from a geometric viewpoint.

A first answer to this algorithmic question is provided
 by Lipshitz's result~\cite{Lipshitz}:
 if $F$ denotes the rational function $\,F\,  = \, f(x_1, x_2/x_1, \ldots, 
x_n/x_{n-1})/(x_1 \cdots x_{n-1})$, and 
if the following equality, called the
{\em creative telescoping equation}, 
\begin{equation}
\label{eq:trivariate-rat-ct}
	L\left(x_n, \frac{\partial}{\partial x_n} \right)(F)
 \,\,\, \,  = \,\,\,\,   \,  \frac{\partial 	R_1}{\partial x_1} 
  \,\, \,\,   + \, \,\cdots \,  \, \,+ \, 
 \frac{\partial R_{n-1}}{\partial x_{n-1}}, 
\end{equation}
admits a solution $(L, \, g_1,\, \ldots, \, g_{n-1})$, where
 $\, P$ (called {\em telescoper}) is a linear 
differential operator with coefficients in $ \, \mathbb{Q}[x_n]$, and where
$ \, R_1, \, \ldots, \, R_{n-1}$ are rational functions in
$\mathbb{Q}(x_1,\ldots,x_n)$ (called \emph{certificates}), then $\, P$
 annihilates the diagonal $\Diag(f)$ of $f$.

Several algorithms exist for solving 
equation~\eqref{eq:trivariate-rat-ct}. A common weakness
 of currently known algorithms for
solving~\eqref{eq:trivariate-rat-ct} is that they are not able to compute the
telescoper~$L$ without computing the certificates $(R_1, \, \ldots, \, R_{n-1})$. This
is unfortunate, since in practice only the telescoper is really needed, while
the size of the certificates is much more important than that of
the telescoper.

Lipshitz's initial argument requires the construction of a non-zero operator
annihilating~$F$ which involves \emph{all\/} the partial derivatives
$\partial/\partial x_i$. This reduces the resolution
of~\eqref{eq:trivariate-rat-ct} to that of a linear system over~$\mathbb{Q}$.
The big practical issue with this approach is the size of the linear system, 
which is about several millions even for the simple rational function
 $\, f \, = \, 1/(1-x_1 - x_2 - x_3)$.
A much more efficient algorithm for solving equation~\eqref{eq:trivariate-rat-ct} 
is Chyzak's extension~\cite{Pech,Chyzak00} of the Zeilberger's 
celebrated {\em creative telescoping method}~\cite{Zeilberger90},
 although the computational complexity of Chyzak's algorithm 
is not yet well understood.
The most efficient implementation of Chyzak's algorithm
 is due to Koutschan~\cite{Koutschan}.

More generally, for $\, n$-fold parameterised integrals
of \emph{D-finite functions}, Chyzak's creative telescoping algorithm
delivers a system of PDEs. For two variables (anisotropic Ising model), one
will get a system of PDEs corresponding to two ``telescopers'', that can be
written in the following form:
\begin{eqnarray}
\hspace{-0.8in}P_1\Bigl(x, \, y, \,  {{\partial } \over { \partial x}} \Bigr)
 \,\, = \, \,\,  \, \sum_{n=0}^{N}\,  p_n(x, \, y) \cdot D_x^n, \qquad
P_2\Bigl(x, \, y, \,  {{\partial } \over { \partial x}} \Bigr)
 \,\, = \, \, \, \, \sum_{n=0}^{M}\,  q_n(x, \, y) \cdot D_y^n.
\nonumber 
\end{eqnarray}
where the $\, p_n$'s, and  the $\, q_n$'s, are polynomials 
of the two variables $\, x$ and $\, y$.

Note that, in practice, the down-to-earth physicist's
 {\em guessing techniques} we have used in our various
papers~\cite{ze-bo-ha-ma-04,ze-bo-ha-ma-05c,ze-bo-ha-ma-05b,bo-ha-ma-ze-07},
which amount to getting\footnote[2]{Strictly speaking, the correctness of the linear
 ODEs obtained by ``guessing'' is not mathematically guaranteed.
However, one may be convinced on the correctness of the ODE,
since, in practice, one has longer series than what is used in
the guessing. Also, some properties (as global nilpotence,
expected known structures) are retrieved.
}
the linear ODE from the series expansion of the
diagonal (or, in general, of the parameterised integral) is much more
efficient\footnote[1]{Serious programming improvements of the creative
telescoping method have been developed recently~\cite{Koutschan,Koutschan2},
and it is now possible to get the linear ODE for the isotropic Ising model 
$\, \tilde{\chi}^{(3)}$, from creative telescoping calculations.} than the
creative telescoping approach. This is moral, since time consuming computations
 are the price to pay in order to guarantee the correctness of the ODE.

\section{Christol's theorem in  more heuristic terms}
\label{onfuscation}

Let us give a sketch in  heuristic terms of how  the main theorem 
in~\cite{Christol} is proved.

\vskip .1cm 
The first step is purely algebraic-geometric.  The algebraic function 
 $F$ involved in the integral representation
\begin{eqnarray}
\label{repintc}
f(x) \, \,\,\,    = \, \,  \,\,
  \int_C  \, F(x;x_1,\ldots,x_n) \cdot \, \,dx_1 \, \cdots \,  dx_n, 
\end{eqnarray}
 lives on  a  complex $(n+1)$-manifold $V$, or, more precisely, 
 on a family of smooth complex $n$-manifolds $ \,V_x$.
One applies to it the so-called {\em embedded resolution 
of singularities}~\cite{embedded}. This
 process uses a succession of {\em blowing up} which is theoretically 
explicit but seems to be {\em inaccessible for computation}. 

Roughly speaking, we so obtain a new family of manifold $ \,\widetilde V$ with 
 $\,\widetilde{V}_x\,=\, V_x$ for $x \, \neq \,  0$, and  $\, \widetilde V_0$, a 
{\em divisor with normal crossing}, namely, a union of  ``smooth algebraic''
 $n$-manifolds $ \,D_i$ that meets ``transversally''.
In particular, if non void,  an intersection $ \, D_{i_1}\cap\cdots\cap D_{i_m}$
 of $m$ distinct $D_i$'s  is of (complex) dimension $n-m+1$. It is obvious 
that, at most, $\, n+1\,$ divisors $ \, D_i$'s can intersect 
at a given point of $\, \widetilde V_0$. 

Moreover, if there are really $\, n+1$ divisors $D_i$ intersecting at
 the point $\,P$ of $\,\widetilde V_0$, then, choosing $P$ as origin and
 equations $\,X_i= \, 0$ of $\,D_i$ ($0\leq i\leq n$) as new variables, 
the equation of $\widetilde V$ becomes, at least locally,
  $X_0...X_n = \, x$. Applying
 the (algebraic) change of variables 
$(X_1,\ldots,X_n) = \, \varphi(x_1, \, \ldots,\, x_n)$ to (\ref{repintc})
($x$ coming in the picture as a parameter%
), one gets 
\begin{eqnarray}
\label{repintb}
\hspace{-0.3in}f(x)\,\,\,\,   = \, \,\, \,\int_{\varphi(C)} \,
 F(x\,;\,X_1, \, \ldots, \, X_n) \cdot \,
 \frac{dX_1}{X_1}\,\,  \cdots \, \, \frac{dX_n}{X_n}, 
\end{eqnarray}
for an algebraic function $ \, F$. If we are lucky, the cycle
 $\,\varphi(C)$ is (homotopic to) $C_P$, 
the vanishing cycle around $P$, and  (\ref{repintb}) 
is an avatar of (\ref{4056}). 
Then  
 \begin{eqnarray}
\hspace{-0.95in}\, f \,= \,\, \Diag(\widetilde F)
 \quad \,\,  \hbox{ with} \quad \, \, \,  \, \,
 \widetilde F(X_0,\, \ldots,\, X_n)
\,\, = \,\,\, \,  F(X_0 \cdots X_n\,;\, \,X_1,\, \ldots,\, X_n),
\end{eqnarray}
 (up to a multiplicative constant) and $\, f$ is the diagonal of an 
algebraic function (in $n+1$ variables), hence, the diagonal of
 a rational function (in $\, 2n+2$ variables).

\vskip .2cm 

Actually the computation of $ \,\varphi(C)$ is inaccessible. 
To  find hypothesis under which it is possible to conclude,
we turn to  the Picard-Fuchs equation $ \,L_V$  because it does not depend
 on the cycle $ \,C$.  Moreover it is stable under birational maps
 like the blowing up used for the {\em desingularisation}. 
All reasoning we will do from now do concern
 all solutions of $ \, L_V$ and cannot
 be done by considering only, for instance, the minimal order
linear differential equation of $ \,f$.

Reverting the process, we conclude from formula \eqref{repintb}
 that integration on $\, C_P$ gives a solution of the Picard-Fuchs 
equation in the ring of diagonals of rational functions
(proposition 11 in~\cite{Christol}). 
Let us recall that the
 Picard-Fuchs linear differential equation is given by the derivation
 $ \, \frac \partial{\partial x}$ acting (through derivation under the 
integral sign) on the space $ \,H^n(\widetilde V_x)$ of $n$ differentials
 modulo exact ones and a solution of this ``differential module'' is 
a $ \,\mathbb{C}(x)$-linear application from this space to some function 
space (here the diagonals of rational function) that do commute 
with $ \, \frac \partial{\partial x}$.

It is difficult to decide, a priori, whether, or not, such a  solution is
 zero on the particular differential 
 $ \, F(x;X_1\ldots X_n)\frac{dX_1}{X_1}\cdots \frac{dX_n}{X_n}$
 we begin with. But we can assert that it is non zero for  differentials  
  $\, G(x\,;\,X_1,\ldots,X_n)\frac{dX_1}{X_1}\cdots \frac{dX_n}{X_n}$
 such that  $\widetilde G(0,\ldots,0)\neq \, 0$.
  So, we consider the $P$-residue, for 
$\, P\,\, = \, D_1\cap\cdots\cap D_n$,
 which, roughly speaking,
 associates to a given differential the coefficient of
 $\frac{dX_1}{X_1}\cdots \frac{dX_n}{X_n}$ it contains. Then the 
{\em Poincar\'e residue} map associates to a differential the family of its residues 
in all the point $P$ of $\, \widetilde V_0$ which 
are the intersection of $n+1$ divisors $D_i$ (this 
set could be void)\footnote[3]{It is rather easy to convince himself that differential
 $\, \frac{dX_1}{X_1}\cdots \frac{dX_i}{X_i^h}\cdots\frac{dX_n}{X_n}$ are exact 
ones for $h> \,1$ and then are $0$ in $\, H^n(\widetilde V_x)$. The naming logarithmic 
pole comes from this remark. It is much less obvious to prove that the
 Poincar\'e residue is actually well-defined on  $\, H^n(V_x)$ but it does.}.
 
The last step is to connect the Poincar\'e residue and   the monodromy filtration
 on  the space of solutions of this differential module. This is more or 
less contained in~\cite{Steenbrink}. Actually this paper
 shows how to compute subspaces  of differential with  logarithmic poles of 
given order by means of spaces built from the monodromy
 filtration. A by-product of this construction (cf theorem 12 in~\cite{Christol}) 
 says that a differential of $\, H^n(\widetilde V_x)$ the 
Poincar\'e residue of which  is $0$ (i.e.  it has a zero residue for each $P$)
 is in the kernel of any solution of maximal (monodromy) weight for $\, L_V$.

As a consequence, the kernel of a solution of maximal weight  for $\, L_V$ contains
 the intersection of the kernel of the solution corresponding to integrate
 on the vanishing cycles $C_P$. But the monodromy filtration is characterised by    
its ``dual'' filtration on $H^n(\widetilde V_x)$ given by corresponding kernels. So
 we conclude that the solution obtained by integrating on $C$ is in the span of
 solutions obtained by integrating on the vanishing cycles  $C_P$. Hence it takes its value
 {\em in the set of diagonals of rational functions} (in $ \, 2n+2$ variables).

When $L_V$ is MUM a simpler argument is the following : 
by hypothesis the solution associated to $C$ is of maximal weight and the differential, 
we started with, is not in its kernel because $f \, \neq \,0$. So there is, at least, one point
 $ \, P$ such that integration on $ \, C_P$ of that differential is not zero and gives 
a diagonal of rational function $\, g$.  In particular $ \, g$ is analytic (near zero) 
and one can conclude by unicity, up to a constant,  of the analytic  solution for $ \, L_V$.

\section{Other hypergeometric ``blind spots'' for Christol's conjecture}
\label{black}

Let us give a list of $\, _3F_2$ that are not {\em algebraic hypergeometric}
functions\footnote[2]{The  $\, _2F_1$ case is well-known.}, 
that are not obviously Hadamard product of algebraic functions,
 but actually correspond to
series with {\em integer} coefficients:
\begin{eqnarray}
 _3F_2\Bigl([{{N_1} \over {9}}, \,{{N_2} \over {9}}, \, {{N_3} \over {9}}, ], 
\, [{{M_1} \over {3}}, \, 1], \,\,  3^6 \, x\Bigr),
\end{eqnarray}
where the four integers $(\, N_1, \, N_2, \, N_3; \,M_1)$ read respectively:
\begin{eqnarray}
\hspace{-0.9in}&&[1, 2, 7; 2], \quad  [1, 2, 8; 2],
 \quad [1, 4, 5; 1], \quad  [1, 4, 7; 1], \quad 
[1, 4, 7; 2], \quad [1, 4, 8; 2],\quad  [1, 5, 8; 1], 
\nonumber \\
\hspace{-0.9in}&&[1, 7, 8; 1], \quad [2, 4, 5; 1],\quad [2, 4, 7; 1],
\quad [2, 5, 7; 2], \quad  [2, 5, 8; 1], \quad 
[2, 5, 8; 2],\quad  [2, 7, 8; 1], \quad
\nonumber \\
\hspace{-0.9in}&& [4, 5, 7; 2], \quad [4, 5, 8; 2]. \nonumber
\end{eqnarray}
The series expansion of the first candidate reads: 
\begin{eqnarray}
\hspace{-0.9in}&& \quad  _3F_2\Bigl([{{1} \over {9}},
 \,{{2} \over {9}}, \, {{7} \over {9}}, ], 
\, [{{2} \over {3}}, \, 1], \, \, 3^6 \, x\Bigr) 
\,\,  \,\, = \,\, \,\, \,  \, \, 1\,\,\, \,  
+21\, x\, \,\, +5544\, x^2\, +2194500\, x^3\,
\nonumber \\
\hspace{-0.9in}&&  \quad  \quad  \quad \quad  \quad
  +1032711750\, x^4\, \, 
 +535163031270\, x^5\,\,  + 294927297193620\, x^6
\nonumber \\
\hspace{-0.9in}&&   \quad \quad\,\quad \quad  \quad  \quad  \quad 
 +169625328357359160\, x^7 \,  +100668944872954458000\, x^8\, 
\,\, \, \,\, \,+ \, \cdots \nonumber 
\end{eqnarray}
Other examples read for instance:
\begin{eqnarray}
\hspace{-0.9in}&&  \, \, \,  _3F_2\Bigl([{{1} \over {7}},
 \,{{2} \over {7}}, \, {{4} \over {7}}, ], 
\, [{{1} \over {2}}, \, 1], \, \, 7^4 \, x\Bigr), \qquad 
 _3F_2\Bigl([{{1} \over {11}},
 \,{{2} \over {11}}, \, {{6} \over {11}}, ], 
\, [{{1} \over {2}}, \, 1], \, \, 11^4 \, x\Bigr).
\end{eqnarray}
Do note that, even if these various hypergeometric
functions look very much alike, the linear differential operators 
that annihilate them are {\em not equivalent} (no homomorphisms 
between these operators\footnote[1]{Associated with these various 
hypergeometric functions with the same singularities 
at $\, x \, = \, 0, \, 1, \, \infty$.} or their symmetric powers).

One can also try to find, systematically, $\, _4F_3$ hypergeometric functions
that are not {\em algebraic hypergeometric}, 
that are not obviously Hadamard product of algebraic functions, 
but actually correspond to
series with integer coefficients. Note that some of these $\, _4F_3$
are deduced from the previous $\, _3F_2$, as a consequence
 of a Hadamard product 
by an algebraic function:
\begin{eqnarray}
\hspace{-0.9in}&&  _4F_3\Bigl([{{1} \over {9}}, 
\,{{2} \over {9}}, \, {{7} \over {9}},
 \, {{1} \over {4}}], 
\, [{{2} \over {3}}, \, 1, \, 1], \, \, 3^6 \,2^3 \,  x\Bigr)  
 \\
\hspace{-0.9in}&&\quad \, \,\, \,  \, \, = \, \, \,\, 
 _3F_2\Bigl([{{1} \over {9}}, \,{{2} \over {9}}, \, {{7} \over {9}}, ], 
\, [{{2} \over {3}}, \, 1], \, \, 3^6 \, x\Bigr)
 \, \star \, (1\, -2^3 \, x)^{-1/4}
 \, \,\, = \, \,\,  \, \, \, 1\,\,\,\,   +42\, x \,\,  
+55440 \, x^2 \,
\nonumber \\
\hspace{-0.9in}&&\quad \quad \quad \quad \, \,  \, \, +131670000 \, x^3 
\,\,  +402757582500 \, x^4 \,\,  +1419252358928040 \, x^5 
 \nonumber \\
\hspace{-0.95in}&& \qquad \quad \quad \quad \quad \, \,  \, +
5475030345102361680 \, x^6 \,\,  +22492318540185824616000 \, x^7 
\,  \,\, \, + \, \cdots \nonumber
\end{eqnarray}

\section{Proof of integrality of series (\ref{contre1})}
\label{proof}

Let us sketch the proof of the integrality of series (\ref{contre1}),
namely, the integrality of coefficients (\ref{far1}). 
For each power of the integer number $\, q \, = \, \, p^n$ 
a term like $\, 4\, + 9 \, n$ is periodically divisible (period $p$) 
by $\, q$. In order to have the ratio (\ref{ratio1}) be an integer, 
one needs the numerator to be divisible by this factor $\, q$
{\em before} the denominator. The case $\, p\, = \, 3$ is an easy one.
The other prime $\, p$ do not divide $\, 9$. One needs to find 
the first case of divisibility, namely the first integer $\, n$ such that 
$\,4\,+9\,n\, =\,\, k\, q$ (this corresponds to the smallest $\, k$). 
If $\, d \, q \, = \, \, 1$, $mod.\,  9$ then 
$\, k \, = \, \, 4 \, d$, $mod.\,  9$.
In other words, the smallest $\, k$ is the rest of  $4 \, d$, $mod.\, 9$. 
Consequently, we have replaced the calculations, for every integer $\, q$,
by a {\em finite set of} calculations for 
$\, d\, = \, \, 1, \, 2, \, 4, \, 5, \, 7, \, 8$.
Let us use this approach for the ratio (\ref{ratio1}). 

\vskip .1cm
{\bf Remark:} The terms $\, n\, +1$ are always the last to be divisible by $\, q$.
Hence, one can forget its factors. However, one needs as many factors 
at the numerator than at the denominator.
For the other terms, the following table of the rest of  $\, d \cdot a$
gives the complete proof:
\begin{eqnarray}
&&. \quad 	1\quad 	2\quad 	4\quad 	5\quad 	7\quad 	8 \nonumber  \\
&&1\quad 	1\quad 	2\quad 	4\quad 	5\quad 	7\quad 	8 \nonumber \\
&&4\quad 	4\quad 	8\quad 	7\quad 	2\quad 	1\quad 	5\nonumber \\
&&5\quad 	5\quad 	1\quad 	2\quad 	7\quad 	8\quad 	4\nonumber \\
&&3\quad 	3\quad 	6\quad 	3\quad 	6\quad 	3\quad 	6\nonumber \\
\end{eqnarray}
One finds out that this is always a factor of the numerator, before the occurrence
of a factor at the denominator.

\section{Integrality of differential geometry modular form series }
\label{Goly}

\subsection{Golyshev and Stienstra examples~\cite{Golyshev}}
\label{Golyex}

{}From a differential Geometry viewpoint, 
Golyshev and Stienstra gave a set of selected
order-three linear differential operators in~\cite{Golyshev}.
The Wronskians of all these Golyshev and Stienstra examples,
displayed in~\cite{Golyshev},
are square roots of simple rational functions.
Consequently, the differential Galois groups of
the order-three  operators displayed in~\cite{Golyshev}
 will be $\, O(3, \,\mathbb{C})$
instead of $\, SO(3, \,\mathbb{C})$.

Furthermore, all these Golyshev and Stienstra order-three operators
are symmetric squares of order-two linear differential operators.

For instance for $\, G_5$, it is the symmetric square of
\begin{eqnarray}
\label{HeunG5}
\hspace{-0.8in}&&{\cal H}_5\,\,\, = \,\, \,\,\,\,\,\,
 D_x^2 \,\,\,\, \,
+\,\, {{1-66\,x-32\,x^2} \over {(1-44\,x\, -16\, x^2) \cdot x}} \cdot D_x 
\,\,\,\,\,
-\, {{3 \cdot \, (x+1)} \over {(1-44\, x\, -16\, x^2) \cdot x}}.
\end{eqnarray}

This Heunian second order operator $\, {\cal H}_5$ is 
actually solvable in $_2F_1$
hypergeometric function with a (modular) pullback
\begin{eqnarray}
\label{solHeung5}
\hspace{-0.2in}\qquad \quad \quad \kappa^{1/4}\,  \cdot \,
_2F_1\Bigl([{{1} \over {12}}, {{5} \over {12}} ], \, [1]; \, P_u\Bigr)
\end{eqnarray}
To see this, one can, for example, calculate the nome
$q$ in terms of the variable $x$ from the series solutions $y_0(x)$ 
and $y_1(x)$  in $x$ of (\ref{HeunG5}), i.e. from $q= \, \exp(y_1(x)/y_0(x))$,
then obtain a series expansion in $x$ for $P_u$ using 
the well-known expansion of the modular invariant 
$j(q)= \, 1728/P_u =  \, 1/q + 784 + 196884 q + 21493760 q^2  \, +\cdots$
 in terms of $q$,
and then recognise the algebraic equation satisfied by  $P_u$ using for example
Maple's command  $\textsf{gfun[seriestoalgeq]}$. One obtains:
\begin{eqnarray}
\hspace{-0.6in}&& (144\,x^2+216\,x+1 )^3 \cdot \, P_u^2\, \, -1728\,x \,
( 3456\,x^5+7776\,x^4-12600\,x^3
 \nonumber \\
\hspace{-0.6in}&&\quad \quad \quad \quad +1890\,x^2-80\,x +1 ) \cdot \, P_u \, \, 
+2985984\,x^6\,\,\,\,\, = \,\, \,\,\, 0, 
\end{eqnarray}
then:
\begin{eqnarray}
\label{pull}
\hspace{-0.8in}&&P_u \, \, = \, \, \, \, \,
{ { 864 \, x \cdot\, (1\,-80\, x\,+1890\,x^2\,
-12600\,x^3\,+7776\,x^4\,+3456\,x^5) }
 \over { (1+216\,x+144\,{x}^{2})^3  }}
\\
\hspace{-0.8in}&&\quad \quad \quad \quad \quad  \quad 
 +\, { {864 \cdot \, x \cdot \, (1-4\, x) \, (1\, -18\, x)\, (1\, -36\, x)   }
 \over { (1+216\,x+144\,{x}^{2})^3  }} \cdot  (1\, -44\, x\, -16\, x^2)^{1/2},
\nonumber
\end{eqnarray}
\begin{eqnarray}
\label{kappa}
\hspace{-0.6in}&&\kappa  \, \, = \, \, \, \, \,
{\frac {5 \cdot \, (36\,x-13)\,\,  
 + 60 \cdot \, (1\, -44\, x\, -16\, x^2)^{1/2}
 }{1\,+216\,x\,+144\,{x}^{2}}}.
\nonumber
\end{eqnarray}
They correspond to the following series expansions:
\begin{eqnarray}
\hspace{-0.6in}&&P_u \, \, = \, \, \, \, \, \, \,
1728 \, x \, \,
\,\, - \,  1257984  \cdot x^2 \,
+\,  575828352  \cdot x^3 \,
- \,  214274336256   \cdot x^4
 \nonumber \\
\hspace{-0.6in}&&\qquad \qquad 
+\,  70880897026368  \cdot x^5 \,
- \,  21731780729723904   \cdot x^6
\, \, \, + \, \cdots , \nonumber \\
\hspace{-0.6in}&&\kappa  \, \, = \, \, \, \,\, \,
-125 \, \, \, \,+ 28500 \, x \, \, -6123000\, x^2\, \, +1318794000 \,x^3\, \,
-283968657000\,x^4\,
\nonumber \\
\hspace{-0.6in}&&\qquad \qquad 
+ 61147607046000 \,x^5 \,
\,-13166982207738000\,x^6 \, \,
\,\, + \,\, \cdots  \nonumber
\end{eqnarray}
Note that the series for  $\, \kappa$,
 and $\,  P_u$,
have {\em integer coefficients}.

Introducing the rational parametrisation of curve
 $\, y^2 \, - \, \,(1\, -44\, x\, -16\, x^2) \, = \, \, 0$, in order to get
rid of the square root $\, (1\, -44\, x\, -16\, x^2)^{1/2}$, namely
\begin{eqnarray}
\label{param}
\hspace{-0.5in}x \,\, = \, \,\,
\, \frac{\mu}{(125\, +22\,\mu\, +\mu^2)},
\quad \quad \quad \quad
y \, = \, \,\,
\pm \, \frac {(\mu^2-125)}{(125\, +22\,\mu\, +\mu^2)},
\end{eqnarray}
the corresponding two pullbacks and $\kappa$'s  reading
\begin{eqnarray}
\hspace{-0.6in}&&P_u \, \,= \, \, \,
\frac{1728\, \mu}{(5\, +10\,\mu\, +\mu^2)^3},
\quad \quad \quad \,  \,  \, 
\kappa_u \,\, = \, \, \,
- \frac{5\, (125\, +22\,\mu\, +\mu^2)}{(5\, +10\,\mu\, +\mu^2)},
\nonumber \\
\hspace{-0.6in}&&P_v \, = \, \,
\frac{1728\, \mu^5}{(3125\, +220\,\mu\, +\mu^2)^3},
\quad \, \quad \quad
\kappa_v \,\, = \, \, \,
- \frac{125\, (125\, +22\,\mu\, +\mu^2)}{(3125\, +250\,\mu\, +\mu^2)}.
\nonumber
\end{eqnarray}
where it is straightforward to see the {\em ``Atkin'' symmetry}
$\,\,\, \mu \,\,\,   \longleftrightarrow \, \,\,  \, 125/\mu$:
\begin{eqnarray}
\hspace{-0.6in}&&P_v(\mu)   \, \,= \, \, \,
P_u\Bigl( {{125} \over {\mu }} \Bigr), \qquad \qquad
 \kappa_v(\mu) \,\, = \, \, \,\kappa_u\Bigl( {{125} \over { \mu }} \Bigr),
 \\
\hspace{-0.6in}&&x\Bigl( {{125} \over { \mu }} \Bigr)   \, \,= \, \, \,x(\mu),
\quad \quad  \,\,\,\,
 y\Bigl( {{125} \over {\mu }} \Bigr)   \, \,= \, \, -\,y(\mu),
\quad \quad  \,\,\,\,
\lambda_5\Bigl( {{125} \over { \mu }} \Bigr)  \,\, = \, \, \,
-\, \lambda_5(\mu).
\nonumber
\end{eqnarray}

The relation between the two pullbacks $\, P_u$ and
$\, P_v$ corresponds to a (rational) modular curve.
{\em One immediately recognises the modular curve}
corresponding to the elimination of $\mu$
between (the two Hauptmoduls) $P_u$, and $P_v$,
which is well-known to correspond to $\, q \, \rightarrow \, q^5$,
or $\, \tau \rightarrow \, 5 \cdot \tau$
(namely the fundamental modular curve
$\Phi(j(\tau), \, j(5\cdot \tau)) \, = \, \, 0$).

All these results for equation (\ref{HeunG5}) can be 
summarised in the following equation
\begin{eqnarray}
\hspace{-0.7in}\rho_5(a_5 \cdot \, x)^{-1} \cdot 
\hbox{Pullback}\Bigl[{\cal H}_5, x\rightarrow c_5(a_5 \cdot \,x)\Bigr]
 \cdot \rho_5(a_5 \cdot \,x)
\, \,\, \,  \,  = \,\,\, \, \,  \,   \omega_5,
\end{eqnarray}
where we denote by $\omega_5$, the order-two operator 
corresponding to the modular solution
$D_5  \, \cdot \,  _2F_1\Bigl([{{1} \over {12}}, {{5} \over {12}} ], 
\, [1]; \,   {{1728} \over {j_5}} \Bigr)$, where
$D_5$ and $j_5$ are given by Maier (see tables 4 and 12 in~\cite{Maier1}).
The expression of $\, c_5(x)$ is given in (G.4):
 $\,  c_5(x)\, = \, \, x/(x^2\,+\, 22\, x\,+\,  125)$ 
and $\, \rho_5(x)$ is given below.

This means that the order-two linear differential operator, associated 
with Golyshev and Stienstra example $G_5$, when pull-backed 
by appropriate functions $c_5(x)$, and $\rho_5(x)$, is simply
 the operator corresponding to the modular solution
$j_5$.

In a similar way, one can perform the same calculations for other examples given
in~\cite{Golyshev3}. The results are displayed in the 
following table with the same notations as before:
\begin{eqnarray}
\hspace{-0.6in}\rho_n(a_n\cdot  \, x)^{-1} \cdot 
\hbox{Pullback}\Bigl[{\cal H}_n, x\rightarrow c_n(a_n \cdot \, x)\Bigl]
 \cdot \rho_n(a_n \cdot  \,x) \,\,\,\,\,  = \,\,\, \,  \,\omega_n, 
\end{eqnarray}
\vskip 5mm
 \begin{tabular}{|l|p{3.5cm}|p{3.5cm}|p{2cm}|}
\hline
n   & $c_n(x)$ & $\rho_n(x)$ & $a_n=A_n\cdot B_n$\\ \hline
2  & $x/(x+64)^2$ & $(x+64)^{1/4}$ & $2^{12}$ \\ \hline
3  & $x/(x+27)^2$ & $(x+27)^{1/3}$ & $ 3^6$ \\ \hline
4  & $x/(x+16)^2$ & $(x+16)^{1/2}$ & $ 2^8$ \\ \hline
5  & $x/(x^2+22x+125)$ & $(x^2+22x+125)^{1/4}$ & $2^2  \, 5^3$ \\ \hline
6  & $x/(x+9)/(x+8)$ & $((x+9)(x+8))^{1/2}$ & $2^3  \, 3^2$ \\ \hline
7  & $x/(x^2+13x+49)$ & $(x^2+13x+49)^{1/3}$ & $7^2  \, 3^2$  \\ \hline
8  & $x/(x+8)/(x+4)$ & $((x+8)(x+4))^{1/2}$ & $ 2^5$ \\ \hline
9  & $x/(x^2+9x+27)$ & $(x^2+9x+27)^{1/2}$ & $ 3^3$ \\ \hline
\hline
\end{tabular}
\vskip 5mm
Operators ${\cal H}_n$ for $n\, =\, 6, \, 7, \, 8, \, 9 $ are given in the 
next subsection (for $n\, =\, 2, \, 3, \, 4$, see~\cite{Golyshev3}).  

Let us note that all these examples are associated to genus zero curves.
Another example $G_{11}$ given in~\cite{Golyshev3} 
which is associated to a genus one curve 
will be considered in detail in \ref{omegahigher}.

\vskip .3cm 

\subsection{More details}

The order-three operator $\, G_6$ is the {\em symmetric square} of 
\begin{eqnarray}
\label{Heung6}
\hspace{-0.7in}&&{\cal H}_6 \, \,\,    = \, \,\,  \,\,\, \,\, 
 D_x^2 \,\, \, \, \, 
+\,{{1\, -51\,x \, +2\,x^2 } \over  {x \cdot (1-34\,x\, +x^2)}} \cdot D_x
\, \,\,\,  + \, \, {{x-10} \over { 4 \, x \cdot (1-34\, x\, +x^2) }}.
\end{eqnarray}

Let us introduce the rational parametrisation 
of $\,\, \,y^2 \, - \, (1\,-34\,x\,+x^2 ) \, = \, \, 0$:
\begin{eqnarray}
\hspace{-0.3in}x\,\, = \, \, \, {{ \mu} \over { (\mu+9) \cdot (\mu+8) }}, 
\quad \quad \quad \,   \, 
y \, \, = \, \, \,{{\mu^2\, -72 } \over { (\mu+9) \cdot (\mu+8)  }}.
\nonumber 
\end{eqnarray}
With this new parametrisation the operator (\ref{Heung6}) becomes
\begin{eqnarray}
\label{H6mu}
\hspace{-0.7in}&&L_{\mu} \, \,\, = \, \, \, \,\, \, 
4 \cdot (\mu+9)^2 \cdot (\mu+8)^2 \cdot \theta_{\mu}^2  \, \,  
\,\,\, -10 \cdot (\mu+9) \cdot (\mu+8) \cdot \mu \, \, \,  \, \, +\mu^2, 
\end{eqnarray}
which is covariant by the ``Atkin'' involution 
$\, \mu \, \leftrightarrow \, \, 72/\mu$,
$\, x$ being invariant by this involution:
\begin{eqnarray}
\hspace{-0.2in}x\Bigl({{72} \over {\mu}} \Bigr) \, \, = \, \, \, x(\mu),
 \qquad \quad 
 y\Bigl({{72} \over {\mu}} \Bigr) \, \, = \, \, \, -\, y(\mu), 
\end{eqnarray}
It is worth recalling the rational parametrisation of the
modular curve $\, \tau  \, \rightarrow \,  \, 6 \cdot \tau$ 
namely~\cite{Maier1}:
\begin{eqnarray}
\hspace{-0.6in}&&j_6 \, \,  = \, \, \, 
 {{(\mu\, +6)^3  \cdot 
(\mu^3\,+18\, \mu^2 +\, 84 \, \mu \, +\, 24) } \over { 
\mu \cdot (\mu\, +8)^3 \cdot  (\mu\, +9)^2 }},
 \\
\hspace{-0.6in}&&j'_6 \, \,  = \, \, \,
 {{(\mu\, +12)^3  \cdot
 (\mu^3\,+252\, \mu^2 +\, 3888 \, \mu \, +\, 15552) } \over {
 \mu^6 \cdot (\mu\, +8)^2 \cdot  (\mu\, +9)^3  }}
\, \,\, \,  = \, \, \, \, \,j_6 \Bigl({{72 }  \over {\mu}}\Bigr). 
 \nonumber 
\end{eqnarray}

The solutions of (\ref{H6mu}) read:
\begin{eqnarray}
\label{solChHeunG5mu}
\Bigl( {{(\mu\, +8)^3 \, (\mu \, +9)^4 } \over { \mu}}\Bigr)^{1/12}
\, \cdot \, _2F_1\Bigl([{{1} \over {12}}, {{1} \over {12}} ], \, [{{2} \over {3}}]; \, 
 {{j_6} \over {1728}} \Bigr),
\end{eqnarray}
but can also be written as:
\begin{eqnarray}
\label{solChHeunG5mu}
\hspace{-0.9in}&&\Bigl( {{(\mu\, +8)^2 \, (\mu\, +9)^2 } \over {
(\mu+6)\, (\mu^3\,+18\, \mu^2 +\, 84 \, \mu \, +\, 24) }}  \Bigr)^{1/4}
\,  \cdot \, 
_2F_1\Bigl([{{1} \over {12}}, {{5} \over {12}} ], \, [1]; \, 
 {{1728} \over {j_6}} 
\Bigr),
 \\
\hspace{-0.9in}&&\Bigl( {{36 \cdot (x+8)^2 \, (x+9)^2 } \over {
(\mu\, +12)  \cdot (\mu^3\,+252\, \mu^2 +\, 3888 \, \mu \, +\, 15552) 
}} \Bigr)^{1/4} \,  \cdot \, 
  _2F_1\Bigl([{{1} \over {12}}, {{5} \over {12}} ], \, [1]; \, 
 {{1728} \over {j'_6}} 
\Bigr).\nonumber
\end{eqnarray}

\vskip .2cm

The third order operator $\, G_7$ is the symmetric square 
of the second order linear differential operator 
\begin{eqnarray}
\label{HeunG_7}
\hspace{-0.7in}&&{\cal H}_7\, \, \, = \,\,\,\,\,\, \,D_x^2 \,\,\,\,\,
+{{1-39\,x-54\,x^2} \over { (1-27\,x) \,(x+1)\, x }} \cdot D_x\,\,
\,\, \, -\, {{ 2 \cdot (1+3\,x) } \over { (1-27\,x) \,(x+1)\, x }}.
\end{eqnarray}
Introducing the parametrisation of the rational curve
\begin{eqnarray}
y^2 \, \, \,\, -\,\, \,  (1\, +x) \cdot (1-27\,x)
\, \, \,\, =\,\, \,\, \,0, 
\end{eqnarray}
namely
\begin{eqnarray}
\hspace{-0.3in}x\, \, = \, \, \, {{\mu } \over {\mu^2 \, +13 \, \mu \, +49 }}, 
\qquad  \quad \,  
y\, \, = \, \, \, {{\mu^2 \, -49 } \over {\mu^2 \, +13 \, \mu \, +49 }},
\end{eqnarray}
where one verifies the existence of an ``Atkin'' involution:
\begin{eqnarray}
\hspace{-0.2in}x\Bigl({{49} \over {\mu}} \Bigr) \, \, = \, \, \, x(\mu),
 \qquad \quad \quad 
y\Bigl({{49} \over {\mu}} \Bigr) \, \, = \, \, \, -\, y(\mu),
\end{eqnarray}
With this change of variables the second order differential
operator (\ref{HeunG_7}) reads:
\begin{eqnarray}
\label{H7mu}
\hspace{-0.8in}&&L_ {\mu} \, \,\, = \, \,  \, \,\,\,
(\mu^2+13\, \mu \, +49)^2 \cdot \theta_{\mu}^2 \, \,\,  \, \, \,
-2  \cdot (\mu^2+13\, \mu \, +49) \cdot \mu \, \,\,  \,\,  -6 \, \mu^2.
\end{eqnarray}

It is worth recalling the rational parametrisation of the
modular curve $\, \tau  \, \rightarrow \,  \, 7 \cdot \tau$ 
namely~\cite{Maier1}:
\begin{eqnarray}
\hspace{-0.6in}&&\quad j_7 \, \,  = \, \, \, 
 {{(\mu^2\,+13\, \mu \, +49) \cdot (\mu^2\,+5\, \mu \, +\, 1) } \over { \mu }},
 \\
\hspace{-0.6in}&&\quad j'_7 \, \,  = \, \, \,
{{(\mu^2\,+13\, \mu \, +49) \cdot (\mu^2\,+245\, \mu \,+\, 2401) 
} \over { \mu^7 }} \, \, \,  = \, \, \, \, \, 
j_7\Bigl({{49 } \over {\mu }}  \Bigr).
 \nonumber 
\end{eqnarray}

The solution of (\ref{H7mu}) reads:
\begin{eqnarray}
\label{solChHeunG7mu}
\hspace{-0.2in}&&\Bigl( {{(\mu^2\,+13\, \mu \, +49)^4} \over {\mu}} \Bigr)^{1/12}
\,   \cdot \, 
_2F_1\Bigl([{{1} \over {12}}, {{1} \over {12}} ], \, [{{2} \over {3}} ]; \, 
 {{j_7} \over {1728}} 
\Bigr),
\nonumber  \\
\hspace{-0.2in}&&\Bigl( {{ 7^6 \cdot (\mu^2\,+13\, \mu \, +49 )^4 } \over {
 \mu^7}} \Bigr)^{1/12}\,  \cdot \, 
_2F_1\Bigl([{{1} \over {12}}, {{1} \over {12}} ], \, [{{2} \over {3}} ]; \, 
 {{j'_7} \over {1728}} 
\Bigr). \nonumber 
\end{eqnarray}

\vskip .3cm
The order-three operator $\, G_8$ is the symmetric square of 
\begin{eqnarray}
\label{Heung8}
\hspace{-0.9in}&&{\cal H}_8\,  \,  \,= \,  \,\, \, \,\,  D_x^2 
\, \,\, \,\,
+\,{{ 1\, -36\,x \, +\,32\,x^2 } \over  {
x \cdot (1\, -24\, x\, +\,16\, x^2)}} \cdot D_x\,\,
\,\,\,  -  \, 2 \, \, {{1\, -2 \, x} \over { x \cdot (1\, -24\, x\, +\,16\, x^2) }}.
\end{eqnarray}
Furthermore one has an ``Atkin'' symmetry 
$\, x \, \leftrightarrow \, 1/16/x$.
Introducing the parametrisation of
 $\, y^2 \, - \, (1\, -24\, x\, +16 \, x^2) \, = \, \, 0$,
namely
\begin{eqnarray}
\hspace{-0.2in}x \, \, = \, \,  \,
 {{\mu } \over {(\mu\, +4) \cdot (\mu\, +8) }},
 \qquad  \quad 
y \, \, = \, \,  \, {{\mu^2 \, -32 } \over {(\mu\, +4) \cdot (\mu\, +8) }},
\end{eqnarray}
the linear differential operator (\ref{Heung8})
becomes
\begin{eqnarray}
\label{H8mu}
\hspace{-0.6in}&&L_{\mu} \,\,  \, = \,\, \, \, \, \, 
(\mu\, +4)^2 \cdot (\mu\,+ 8)^2\cdot \theta_ {\mu}^2\,
\, \,  \,\, -2\cdot (\mu\, +4)\cdot (\mu\,+ 8)\cdot \mu
 \, \,\,\, \, +4\cdot \mu^2,
\nonumber 
\end{eqnarray}
this operator being covariant by the ``Atkin'' involution
which leaves $\, x$ invariant: 
\begin{eqnarray}
\hspace{-0.2in}\quad x\Bigl({{32} \over {\mu}} \Bigr) \, \, = \, \, \, x(\mu), 
\qquad \quad \, 
y\Bigl({{32} \over {\mu}} \Bigr) \, \, = \, \, \, -\, y(\mu).
\end{eqnarray}

It is worth recalling the rational parametrisation of the
modular curve representing  $\, \tau  \, \rightarrow \,  \, 8 \cdot \tau$, 
namely~\cite{Maier1}:
\begin{eqnarray}
\hspace{-0.6in}&&\quad j_8 \, \,  = \, \, \, 
 {{ (\mu^4 \, + \, 16 \, \mu^3 \,+80 \, \mu^2 \,+128 \,\mu \, +16)^3 } \over { 
\mu \cdot (\mu\, +4)^2 \, (\mu \, +8) }},
 \\
\hspace{-0.6in}&&\quad j'_8 \, \,  = \, \, \,
 {{ (\mu^4 \, + \, 256 \, \mu^3 \,+5120 \, \mu^2 
\,+32768 \,\mu \, +65536)^3} \over { \mu \cdot (\mu\, +4) \, (\mu \, +8)^2 }}
\, \, \, \,   = \, \,\,  \, \,
j_8 \Bigl({{32 }  \over {\mu}}\Bigr). 
 \nonumber 
\end{eqnarray}
A solution of (\ref{H8mu}) reads
\begin{eqnarray}
\label{solChHeunG8mu}
\hspace{-0.2in}\Bigl({{ (\mu\, +\, 8)^5 
(\mu\, +\, 4)^4 } \over {\mu }}  \Bigr)^{1/12} \, 
\cdot \, 
_2F_1\Bigl([{{1} \over {12}}, {{1} \over {12}} ], \, [ {{2} \over {3}}]; \, 
 \, {{j_8} \over {1728}} 
\Bigr),
\end{eqnarray}
but can also be written
\begin{eqnarray}
\label{solChHeunG8mu2}
\hspace{-0.2in}\Bigl( (\mu \, +4) \cdot (\mu \, +8)\Bigr)^{1/2}
\,  \cdot \, 
_2F_1\Bigl([{{1} \over {2}}, {{1} \over {2}} ], \, [1]; \, 
 -\, {{\mu \cdot (\mu \, +8)} \over {16}} 
\Bigr),
\nonumber  
\end{eqnarray}
or equivalently, using Gauss-Kummer identity:
\begin{eqnarray}
\hspace{-0.6in}\Bigl( (\mu \, +4) \cdot (\mu \, +8)\Bigr)^{1/2}
\,  \cdot \, 
_2F_1\Bigl([{{1} \over {4}}, {{1} \over {4}} ], \, [1]; \, 
 -\, {{\mu \cdot (\mu \, +8)\cdot (\mu \, +4)^2} \over {64}} \Bigr).
\nonumber 
\end{eqnarray}
\vskip .1cm

Finally, the order-three operator $\, G_9$ is the symmetric square of 
\begin{eqnarray}
\label{Heung9}
\hspace{-0.95in}&&\quad {\cal H}_9\, \, = 
\,\,\,\,  \,  \, \,  D_x^2 \,\,  \, \, \,
+\,{{ 1\, -27\,x \, -\,54\,x^2 } \over  {
x \cdot (1\, -18\, x\, -\,27\, x^2)}} \cdot D_x \, \, 
\, \, \, -  \,  {{ 6 \, +27 \, x} \over {
4\,  x \cdot (1\, -18\, x\, -\,27\, x^2) }}. 
\end{eqnarray}

With the parametrisation of the rational curve
$\,\,\, y^2 \, - \, \, (1\, -18\, x\, -\,27\, x^2) \, = \, \, 0$,
namely
\begin{eqnarray}
\hspace{-0.3in}\quad x \,\, = \, \, \, {{\mu } \over { \mu^2 \, +9 \,\mu\, +27 }}, 
\qquad \quad \,\,\,
y  \,\, = \, \, \, {{\mu^2 \, -27 } \over { \mu^2 \, +9 \,\mu\, +27 }},
\end{eqnarray}
the linear differential operator (\ref{Heung9}) becomes
\begin{eqnarray}
\label{H9mu}
\hspace{-0.6in}&&L_{\mu} \, \,  \, = \, \, \, \, \,\, \, 
4 \cdot (\mu^2 \, +9 \,\mu\, +27 )^2 \cdot \theta_{\mu}^2  \, \, \, \,\,
\, -6 \cdot (\mu^2 \, +9 \,\mu\, +27 ) \cdot \mu \,\,\,\, \, \, -27 \, \mu^2,
 \nonumber 
\end{eqnarray}
which is covariant by the ``Atkin'' involution
leaving $\, x$ invariant:
\begin{eqnarray}
\hspace{-0.2in}x \Bigl({{27} \over {\mu}} \Bigr) \, \, = \, \, \, x(\mu), 
\qquad \qquad 
y\Bigl({{27} \over {\mu}} \Bigr) \, \, = \, \, \, -\, y(\mu).
\end{eqnarray}

It is worth recalling the rational parametrisation of the
modular curve $\, \tau  \, \rightarrow \,  \, 9 \cdot \tau$ 
namely~\cite{Maier1}:
\begin{eqnarray}
\hspace{-0.6in}&&\quad j_9 \, \,  = \, \, \, 
 {{(\mu\, +3)^3  \cdot (\mu^3\,+9\, \mu^2 +\, 27 \, \mu \, +\, 3) } \over { 
\mu \cdot (\mu^2 \, +9 \,\mu\, +27) }},
 \\
\hspace{-0.6in}&&\quad j'_9 \, \,  = \, \, \,
 {{(\mu\, +9)^3  \cdot (\mu^3\,+243\, \mu^2 +\, 2187 \, \mu \, +\, 6561) } \over {
 \mu^9 \cdot (\mu^2 \, +9 \,\mu\, +27) }}
\, \, \, \, = \, \, \, \, \,j_9 \Bigl({{27 }  \over {\mu}}\Bigr). 
 \nonumber 
\end{eqnarray}

A solution of (\ref{H9mu}) reads
\begin{eqnarray}
\label{solH9}
\hspace{-0.2in}\Bigl( {{(\mu^2\, +\, 9 \, \mu \, +27)^5} \over { \mu}} \Bigr)^{1/12}
\,  \cdot \, 
_2F_1\Bigl([{{1} \over {12}}, {{1} \over {12}} ], \, [ {{2} \over {3}}]; \, 
 \, {{j_9} \over {1728}} 
\Bigr),
\end{eqnarray}
but can also be written as:
\begin{eqnarray}
\label{G32}
\hspace{-0.7in}\Bigl( (\mu^2\, +\, 9 \, \mu \, +27)\Bigr)^{1/2}
\,  \cdot \, 
_2F_1\Bigl([{{1} \over {3}}, {{1} \over {3}} ], \, [1]; \, \,
 -\, {{\mu \cdot (\mu^2\, +\, 9 \, \mu \, +27)} \over {27}} 
\Bigr),
\end{eqnarray}
\begin{eqnarray}
\hspace{-0.7in}\Bigl( {{ 27 \cdot (\mu^2\, +\, 9 \, \mu \, +27)
} \over {\mu^2 }}\Bigr)^{1/2}
\,  \cdot \, 
_2F_1\Bigl([{{1} \over {3}}, {{1} \over {3}} ], \, [1]; \, \,
 -\,27 \,  {{ \mu^2\, +\, 9 \, \mu \, +27} \over {\mu^3 }} 
\Bigr).
\nonumber 
\end{eqnarray}
The relation between these two pullbacks corresponds to the modular curve
\begin{eqnarray}
\hspace{-0.9in}{x}^{3}{y}^{3}\, \,\, -270\,\cdot \, {x}^{2}{y}^{2}\, \,\,
 +972\,\cdot \, x \, y \cdot \, (x \, +y) \,\,
  -729\,\cdot \, ( {x}^{2} \,+\,xy \, +\,{y}^{2})
\, \,\,  = \, \, \,\, 0.
\end{eqnarray}

\section{Higher genus modular forms}
\label{omegahigher}

For {\em non-zero genus} modular curves, we have generalisations of these
structures associated with an ``Atkin'' involution 
of the form $\, z \, \, \rightarrow \, \, A/z$, which 
correspond to the introduction of the
so-called {\em Atkin\footnote[1]{Atkin's work we are interested in, 
cannot be found in papers but in emails~\cite{Atkin1}.} 
modular polynomials, or star-modular
 polynomials}~\cite{Morain}. 

\subsection{A genus-one curve}

Let us first consider the order-three Golyshev and Stienstra 
linear differential operator $\, G_{11}$ given in~\cite{Golyshev3} 
which is the symmetric square of 
\begin{eqnarray}
\label{Heung11}
\hspace{-0.7in}&& {\cal H}_{11}\, \,    = \, \, \, \,  \,    \,  \,  D_x^2
 \, \,\,   \,  \,  
+\frac{(625-12750\,x-30800\,{x}^{2}-3150\,{x}^{3}-4512\,{x}^{4})}{x \cdot \;
(125\, -1900\, x\, -40\, x^2 \, -188\, x^3) \cdot \, (8\, x+5)} \cdot D_x 
\nonumber \\
\hspace{-0.7in}&& \qquad \qquad \quad  \quad  \quad \,\,  \, 
 - \frac{6 \cdot \;(125+700\,x+105\,{x}^{2}+188\,{x}^{3})}
{(125\, -1900\, x\, -40\, x^2 \, -188\, x^3) \cdot \, (8\, x+5)}. 
\end{eqnarray}
Performing the same calculations as for (\ref{HeunG5}), 
one deduces that the solution of (\ref{Heung11}) can be written in terms of 
pull-backed hypergeometric function, namely
$\,_2F_1\Bigl([{{1} \over {12}}, {{5} \over {12}} ], \, [1]; \, P_u\Bigr)$,
with $\,P_u= \, 1728/J$, with $J$ satisfying the following algebraic equation:
\begin{eqnarray}
\label{modu_11}
&& \hspace{-0.95in} 244140625\,x^{12} \cdot \, {J}^{2}\, 
-5 \cdot \, (132645814272\,x^{11} \,  +372815032320\,x^{10} \,   +1405869696000\,x^{9} 
\nonumber\\
&& \hspace{-0.95in} \,\,+5229172080000\,x^{8} \, \,  -225383400000\,x^{7}\, \, 
 -5599578600000\,x^{6}  \, \,  +339591656250\,x^{5}  
 \nonumber \\
&& \hspace{-0.95in} \, \, +1103850000000\,x^{4} \,  \, -349421875000\,x^{3} \, 
 \,+42861328125\,x^{2} \,  \, -2363281250\,x 
\nonumber \\
&& \hspace{-0.95in}\,\, +48828125 )\cdot  \, x \cdot \, J \,
+ ( 78336\,x^{4}+181440\,x^{3}+561600\,x^{2}
+144000\,x+625)^{3}\,\, \,  =\,\, \,\,\,    0.
 \nonumber
\end{eqnarray}
Then, when we perform the pullback $\,x \,\rightarrow\, 5/(5\,z -3)$,
 we obtain  the  modular polynomial 
of order-eleven given\footnote[1]{See $\,\Phi^{*}_{11}(F, \, J)$
in  the last equation 
of subsection 2.3.2 of~\cite{Morain}. The variable $\, F$ in~\cite{Morain}
is $\, z$ here.} by Morain~\cite{Morain},
 or by Elkies (see eq. (49) in~\cite{Elkies}),
 associated to a {\em genus-one} modular curve:
\begin{eqnarray}
\label{mmodu_11}
&& \hspace{-0.9in}\qquad  {J}^{2}\, \, \, \,\,
- Q_1(z) \cdot \, J \, \,\,
 + ( z^{4}+228\,z^{3}
+486\,z^{2}-540\,z+225 )^{3} \, \,\, \, \, = \,\,\, \,\,  \,  \, 0.
\end{eqnarray}
where:
\begin{eqnarray}
\hspace{-0.9in}&&Q_1(z) \,\,\, \, = \, \, \,  \,\,\, \,
z^{11} \,\,\, -55\, z^{10}\, +1188\, z^9\, -12716\, z^8\,  +69630\, z^7\, -177408\, z^6\,
 \nonumber \\
\hspace{-0.7in}&&\quad \quad \quad +133056\, z^5\,\, \, 
+132066\, z^4 \, \,\,-187407\, z^3\,\, \,
+40095\, z^2\,\, \, +24300\, z\,\, \, -6750.
 \nonumber 
\end{eqnarray}
The $\, j$-invariant of the genus-one $(J, \, z)$-curve (\ref{mmodu_11})
 reads\footnote[5]{Use the command 
\textsf{algcurves[{j\_invariant}]} in Maple.}: 
\begin{eqnarray}
\label{jinvar}
\hspace{-0.1in}&&\quad \quad \, 
 j_{inv}  \,\,\, = \, \, \,  \,\,
-{{122023936} \over {161051}}\,\, \,\, = \, \, \,  \,\, -\, {{ 496^3 } \over {\, 11^5 }} .
\end{eqnarray}

\vskip .1cm 

\subsubsection{Atkin modular polynomial and modular forms for
 $\, \tau \, \rightarrow \, 11 \, \tau$ \newline \newline }

Let us focus on the previous
{\em Atkin modular polynomial} (\ref{mmodu_11}):
\begin{eqnarray}
\label{Phi}
\hspace{-0.9in}&&\Phi^{*}_{11}(z, \, j)
 \, \,\, = \,\, \, \, \,  \, j^2 \, \,\, \,  \, -Q_1(z) \cdot j \, \,\,\,
 + \, \, \, (z^{4}+228\,{z}^{3}+486\,{z}^{2}-540\,z+225)^3.
\end{eqnarray}
Note that there is no associated {\em ``Atkin'' involution} of the (rational) 
form  $\, z \, \, \rightarrow \, \, A/z$, since
the curve  $\, \Phi^{*}_{11}(z, \, j) \, = \, \,0$ is a {\em genus-one} curve.

The elimination of $\, z$, between the two solutions 
 $\, j_{+}$ and $\, j_{-}$
of $\, \Phi^{*}_{11}(z, \, j) \, = \, \,0$,
{\em yields the modular curve} 
 \begin{eqnarray}
\label{Phi11}
\hspace{-0.1in}&& \quad \quad \quad \quad \quad
 \Phi_{11}(j_{+}, \, j_{-})
 \, \,\,\,  = \, \, \,  \,\,\,0,
\end{eqnarray}
 which is a quite large (146 monomials) $\,(j_{+}, \, j_{-})$-symmetric 
polynomial of degree twelve in $\, j_{+}$ (resp.  $\, j_{-}$).
The discriminant of $\, \, \Phi_{11}(j_{+}, \, j_{-})\, $ in $\,\,  j_{-}\, $
reads (we denote here $\, j_{-}$ by $\, j$):
\begin{eqnarray}
\hspace{-0.9in}&&\qquad \qquad \quad \quad  -11^{11} \cdot \, j^8 \cdot \,  (j \, -(12)^{3})^6 
\cdot \, {\cal Q}^2 \cdot \,  {\cal P}^2, 
\end{eqnarray}
where $\, {\cal Q}$ reads:
\begin{eqnarray}
\hspace{-0.9in}&&{\cal Q} \, \,\, \, \,   = \,  \,  \,  \, \,\,
\, (j\,+(15)^3) \cdot  \,  (j\, -(20)^3) \cdot  \,  (j\, +(96)^3) \cdot 
(j\, +(960)^3) \cdot  \,  (j\, -(255)^3)  \nonumber \\
\hspace{-0.9in}&&\quad \,\, \times \, 
(j^2 \, -425692800 \, \, j\, +9103145472000)^2 \cdot \, 
(j^2\, +117964800 \, \, j\, -134217728000)^2, \nonumber
\end{eqnarray}
and where $\, {\cal P}$ is a polynomial that factors into 
the product of seven polynomials of degree two, fourteen  
polynomials of degree four, one polynomial of degree six and 
four polynomials of degree eight:
\begin{eqnarray}
\hspace{-0.9in}&&\qquad \quad  \quad {\cal P}  
 \,  \, \,\,  = \,\, \, \,  \,  \, \,
 P_2^{(1)} \,  \cdots  \,\,  P_2^{(7)} 
\cdot \, 
P_4^{(1)} \,  \cdots  \,\,  P_4^{(14)} \cdot \, P_6^{(1)} \,
\cdot \, 
P_8^{(1)} \,  \cdots  \,\,  P_8^{(4)}. 
\end{eqnarray}

In the $j_{+} = \, j_{-} \, = \, j$ limit it becomes
a polynomial of degree 22, which simply factors as follows:
\begin{eqnarray}
\hspace{-0.7in}&&\quad \quad 
  \Phi_{11}(j, \, j) \,\, \, \,  = \,\,\,   \, \, -\, (j \, +(32)^{3}) 
\cdot \, {\cal Q}^2 
\nonumber \\
\hspace{-0.7in}&&\quad \qquad \quad \, \,\,   \, \, \quad  \times  \,
 (j^3\, -1122662608 \, \, j^2\,+270413882112 \, \, j\, -653249011576832).
 \nonumber 
\end{eqnarray}
where, besides the {\em Complex Multiplication} value~\cite{Miller}
 $\,j \, = \, \,  (255)^{3} \, = \, \, 16581375$,  one recognises a large
 set of {\em Heegner numbers}~\cite{bo-ha-ma-ze-07b},
corresponding to the following integer values of the  $\, j$-invariant:
 $\, (20)^{3}$, $\, -(15)^{3}$,  $\, -(32)^{3}$,  $\, -(96)^{3}$,  
$\, -(960)^{3}$.
Note that the  $\, j$-invariant of the genus-one $(j_{+}, \, j_{-})$
modular curve also reads the {\em same $\, j$-invariant}
 as the one for the genus-one curve (\ref{mmodu_11}), namely
 $\, -122023936/161051$ (see (\ref{jinvar})). 

Recalling the well-known expansion of the $\, j$-function as a function
of $\, q$:
\begin{eqnarray}
\label{jq}
\hspace{-0.7in}&&j(q) \, \, = 
\, \, \, \,\, \,{{1} \over {q}} \, \,\,  \,+ 744 \,\, \, \,
+196884 \,q \, \, \,+21493760 \,q^2 \,\,  +864299970 \,q^3 \,   \\
\hspace{-0.7in}&&\quad \quad \quad \quad  \quad 
+20245856256 \,q^4 \, +333202640600 \,q^5 \, +4252023300096 \,q^6 
\,\,\, \, + \,  \, \cdots \nonumber
\end{eqnarray}
one verifies immediately that {\em it does
 correspond} to $\, \tau \, \rightarrow   \, \, 11 \, \tau$: 
\begin{eqnarray}
\hspace{-0.1in}\quad \quad \quad 
  \Phi_{11}\Bigl(j(q), \, j(q^{ 11})\Bigr) \, \, \,= \, \,\,  \, \,0.
\end{eqnarray}
Recalling the modular discriminant of Weierstrass,
$\, \Delta(q)$ (i.e. the 24-th power of the Dedekind eta 
function up to $\, (2\pi)^{12}$ factor), it reads:
\begin{eqnarray}
\hspace{-0.3in}&&\quad \quad  \quad  \quad \, \Delta(q) \, \, = \, \, \, \, \, 
  q \cdot \prod_{n=1}^{\infty} \, (1\, - \, q^n)^{24}.
\nonumber 
\end{eqnarray}
Let us introduce the 
{\em Eisenstein series}\footnote[2]{Quasi-modular form~\cite{Quasi}: 
$\, G_2((a\, \tau\, +b)/(c\, \tau\, +d)) \, $
$= \, \, (c\, \tau\, +d)^2 \cdot \, G_2(\tau) \, - \, (c\, \tau\, +d)\, c/4/\pi/i$.} 
$\,E_2(q)$ and the Euler-like function $\,\rho(q)$:
\begin{eqnarray}
\hspace{-0.8in}&&E_2(q) \,\,  = \, \, \,
 {{q} \over {24}} \cdot  {{d \ln(G(q))} \over {d \, q}}, 
 \,  \, \qquad 
\hbox{where:} \qquad \quad  \, 
G(q) \,\,  = \, \, \, {{ \Delta(q^{11}) } \over {  \Delta(q)}}, 
\nonumber \\
\hspace{-0.8in}&&\qquad \qquad  \rho(q)  \, \,  \,= \, \, \,  \,\,
 q \cdot \prod_{n=1}^{\infty} \, (1\, - \, q^n)^2 \cdot 
\prod_{n=1}^{\infty} \, (1\, - \, q^{11\, n})^2.
\end{eqnarray}
and, finally $\, z(q)$ (denoted $\, F$ in Morain~\cite{Morain}):
\begin{eqnarray}
\hspace{-0.5in}&&z(q)  \,  \, = \,\,\, \, \, {{3} \over {5}} \,
+ \, \,  {{12} \over {5}} \cdot  {{E_2(q)} \over {\rho(q)}} 
\,  \, \,\, \,  = \, \, \, \, \, \,
{{1} \over {q}} \,  \,\, \, +5 \,\, \,+17 \,q \, +46\, q^2\,
 +116\, q^3 \,  +252\, q^4  
\nonumber \\
\hspace{-0.5in}&& \qquad \qquad \quad \quad   +533\, q^5\,\, \, 
+1034\, q^6\,\, \,  +1961\, q^7\,\, \,\,\, + \, \,  \cdots 
\end{eqnarray}
One verifies that the two relations on the {\em Atkin modular
 polynomial} (\ref{Phi})
are satisfied: 
\begin{eqnarray}
\hspace{-0.4in} \Phi^{*}_{11}(z(q), \, j(q)) \,\,  = \,\,  \,0, 
\qquad \hbox{and:} \qquad  \, \, 
\Phi^{*}_{11}(z(q), \, j(q^{11})) \,\,  = \,\,  \,0.
\end{eqnarray}

Introducing $\, x(q) \, = \, \,  1/z(q)$ one has the following series
expansion with integer coefficients:
\begin{eqnarray}
\hspace{-0.9in} x(q) \,\, = \, \, \, \,  \,\, q \, \, \,   \,-5 \,q^2 \, 
\, +8 \,q^3 \, -q^4  \,-17\,q^5 
 \,+62\,q^6 \,-176\,q^7  \,+339\,q^8 \,-386 \,q^9
 \,\,\,\,   + \,\,  \,  \cdots
 \nonumber 
\end{eqnarray}

Let us now show an identity characteristic of modular forms:
\begin{eqnarray}
\hspace{-0.6in} A(x) \cdot \, _2F_1\Bigl([{{ 1} \over {12}}, 
\, {{5 } \over {12 }}], \,[1];
   {{1728} \over {j(q^{11})}}  \Bigr) 
\, \,\, \,   = \, \,  \, \, \, 
 _2F_1\Bigl([{{ 1} \over {12}}, \, {{5 } \over {12 }}], \,[1]; 
  {{1728} \over {j(q)}}  \Bigr), 
\end{eqnarray}
where $\, A(x)$ is the algebraic function  
\begin{eqnarray}
\hspace{-0.9in}\quad {{A(x)^4} \over {11^2}} \, + \, \,  {{11^2} \over {A(x)^4}} 
 \,\,  \,\,  = \, \,\, \,  \,\,  {{2} \over {11^2}} \cdot \, \, 
{{7321-87612\,x+73206\,x^2+21060\,x^3-23175\,x^4 } \over {
1+228\,x+486\,x^2-540\,x^3+225\,x^4 }}.
\nonumber 
\end{eqnarray}

\subsubsection{A change of variables \newline \newline }

Let us rewrite (\ref{Phi}) in this variable $\, x \, = \, 1/z$,
 and in the Hauptmodul $\, H \, = \, \, 1728/j$,
introducing a new star-modular polynomial
 $\,P^{*}_{11}(x, \, H) = \,$
$ H^2 \cdot \, z^{12} \cdot \, \Phi^{*}_{11}(1/x, \, 1728/H)$:
\begin{eqnarray}
\label{Phibis}
\hspace{-0.9in}&&P^{*}_{11}(x, \, H)
\, \,\, = \, \, \,\, \,\,
 (1\, +228\, x \, +486\, x^2 \, -540\, x^3 +225\, x^4)^3 \cdot \, H^2
\\
\hspace{-0.9in}&& \qquad \qquad \quad  \quad  \quad \quad 
\,  -1728 \cdot \, {\cal Q}_1(x) \cdot x \, \cdot \, H \,\,\,\, \,\, +1728^2 \, x^{12}, 
\qquad \quad 
\hbox{with:} 
\nonumber
\end{eqnarray}
\begin{eqnarray}
\hspace{-0.9in}&&{\cal Q}_1(x) \,  \, \,  =  \, \,   \, \,  \,  \, \,\,
1 \, \, \,\,\,  -55\, x\,\,\, +1188\, x^2 \,\,
 -12716\, x^3 \,\, +69630\, x^4\, \, -177408\, x^5\,
+133056\, x^6 \nonumber \\
\hspace{-0.9in}&& \quad \quad \quad \quad \quad 
+132066\, x^7 \, -187407\, x^8\,
+40095\, x^9\, +24300\, x^{10}\, -6750\, x^{11}. 
\nonumber
\end{eqnarray}
The two Hauptmodul solutions of polynomial $\,P^{*}_{11}(x, \, H)$
expand respectively as:
\begin{eqnarray}
\hspace{-0.95in}&&{{1}\over{j(q)}}\,=\,\,{{H_1} \over {1728}} \,\, = \, \, \,  \,
x  \, \,  \, -739\, x^2  \, +349254\, x^3 \,
  -135092042\, x^4 \,  +46600204623\, x^5 \,\,   + \, \cdots,  
\nonumber \\
\hspace{-0.95in}&&{{1}\over{j(q^{11})}}\, =\,\,  {{H_2} \over {1728}}
 \,\, = \, \,\,\,\,
  x^{11} \, \, \, +55\, x^{12}\,  +1837\, x^{13} \,
 +48411\, x^{14}\,  +1109999\, x^{15}\, \, \\
\hspace{-0.95in}&&\quad \quad \quad \qquad  \quad \quad \quad 
 +23244727\, x^{16}  \, \,\, + \, \, \cdots 
\nonumber
\end{eqnarray}
The corresponding expansions of
 $\, _2F_1([1/12, \, 5/12], \,[1];\, H_1)$
and $\, _2F_1([1/12, \, 5/12], \,[1];\, H_2)$
read respectively two series with {\em integer} coefficients:
\begin{eqnarray}
\hspace{-0.9in}&&_2F_1([1/12, \, 5/12], \,[1];\, H_1) \, = \,\,   \, \, \,
1\,\,\,\, +60\,x\, -4560\,{x}^{2}
\,+614400\,{x}^{3} \, -95660400\,{x}^{4} 
\nonumber \\
\hspace{-0.9in}&&\quad \quad \quad \, \, +16231863060\,{x}^{5}\, 
 -2905028387700\,{x}^{6}\, \, 
+ \, \, \,  \cdots \, \, \,   \,  +20000242239261022140\,{x}^{9}\, 
\nonumber  \\
\hspace{-0.9in}&&\quad \quad \quad \, \, -3953288123422938241560\,{x}^{10}\,
+\,791518845663517087144740 \, {x}^{11}\, \, + \, \, \cdots 
\nonumber 
\end{eqnarray}
\begin{eqnarray}
\label{twoseries}
\hspace{-0.9in}&&_2F_1([1/12, \, 5/12], \,[1];\, H_2)
\, \,\,\,\, = \, \, \,\,\,\,
 1\,\,\,  \, +60\, x^{11}\,\, +3300\, x^{12} \,
 +110220\, x^{13}\,\, +2904660\, x^{14}
\nonumber \\
\hspace{-0.9in}&& \qquad \quad \quad\, \,
+66599940\, x^{15} \, \, +1394683620 \, x^{16} \, \, +27425371380\, x^{17}
\,\, \,\, \,   + \, \,  \cdots 
\end{eqnarray}
Defining $A(x)$ as the ratio of these two series expansions:
\begin{eqnarray}
\label{defining}
\hspace{-0.6in}\quad  _2F_1\Bigl([{{ 1} \over {12}}, 
\, {{5 } \over {12 }}], \,[1];
   \, H_1  \Bigr)\, \,   \,\,\,
 = \,  \,  \,  \, \, A(x) \cdot \,  _2F_1\Bigl([{{ 1} \over {12}}, 
\, {{5 } \over {12 }}], \,[1];
   \, H_2  \Bigr),
\end{eqnarray}
one can easily see that this ratio $\, A(x)$ is solution 
of the {\em genus-one} algebraic equation:
\begin{eqnarray}
\label{direaussi}
\hspace{-0.6in}\qquad \quad {{A(x)^4} \over {11^2}} \,\, 
 + \, \,  {{11^2} \over {A(x)^4}} \,\,\,\, \,\,=  \,\, \,\,\,\,
{{2} \over {11^2}} \cdot \, \, 
{{\gamma(x)} \over {\delta(x) }}.
\end{eqnarray}

It is worth noting that this {\em genus-one} algebraic
curve (\ref{direaussi}), in the two variables  $\, x$  and 
$\, y \, = \, A(x)^4$, has the {\em same $\, j$-invariant}, namely
 $\, -122023936/161051\, $
(see (\ref{jinvar})), as the 
 {\em genus-one} $(x, \, y)$-curve (\ref{mmodu_11}) corresponding to the
 {\em Atkin-modular polynomial}. 

The ratio $\, A(x)$ then reads:
\begin{eqnarray}
\hspace{-0.9in}\quad \quad   A(x)^4 \, \, = \, \, \,
 {{ \gamma(x) \, -\, \, 120 \cdot \, \alpha(x)  \cdot \, 
\beta(x)^{1/2} 
} \over {  \delta(x)   }},
 \qquad \quad \quad \hbox{where:}
\end{eqnarray}
\begin{eqnarray}
\hspace{-0.9in}&&\,\, \alpha(x) \,\, = \,  \, \,
 45\, x^2\, -246\,x \, +61, \qquad \quad 
 \beta(x) \,\, = \, \,  \,  (1\, +x) \cdot \, (1\, -17\, x\, +19\, x^2\, -7\, x^3  ),
 \nonumber \\
\hspace{-0.9in}&&\,\, \gamma(x) \,\, = \,\, \,  \, \, 7321\, \, \,-87612\, x\, \, \,
+73206\, x^2\, \, +21060\, x^3\,  \, -23175\, x^4, 
\nonumber \\
\hspace{-0.9in}&&\,\, \delta(x) \,\, = \,\, \,  \,\, \,
 1\,\, \, +228\, x\,\, \, +486\,x^2\,\, \, -540\,x^3\,\, \, +225\,x^4. 
\end{eqnarray}
Its series expansion is a series with integer coefficients:
\begin{eqnarray}
\hspace{-0.9in}&&A(x) \, = \,\,  \, \,   \, \, \,1\,\,\,\, +60\,x\,
-4560\,{x}^{2}
\,+614400\,{x}^{3} \, -95660400\,{x}^{4} \, +16231863060\,{x}^{5}
\nonumber \\
\hspace{-0.9in}&&\quad \quad \quad \quad -2905028387700\,{x}^{6}
\,\, \,  \quad 
+ \, \, \,  \,  \cdots \, \, \, \,   \,  +20000242239261022140\,{x}^{9}\, 
 \\
\hspace{-0.9in}&&\quad \quad \quad \quad -3953288123422938241560\,{x}^{10}\,
+791518845663517087144680\,{x}^{11}\,\, \, + \, \, \cdots 
\nonumber 
\end{eqnarray}

Introducing the polynomial $\, \zeta(x) \, = \,\, 
1\, +8\,x \,-9\,x^2\,+10 \,x^3\, -6\, x^4$, 
one has the following relations on  $\, A(x)$:  
\begin{eqnarray}
\label{dlnA}
\hspace{-0.1in}&&\qquad \,\,\, {{ d \ln A(x)}\over{dx}}
 \,\,\,\,\, = \, \,\,\,\,\,\, 
{{60 \cdot \zeta(x)}\over{  \beta(x)^{1/2} \cdot \delta(x) }},
\end{eqnarray}

\subsubsection{Order-four operator \newline \newline } 

Let us introduce the order-four linear differential  operator 
$\, L_4$ annihilating  $\, \,\,  _2F_1\Bigl([1/12, \, 5/12], \,[1];\, H_1)$.
Quite remarkably this linear differential  operator $\, L_4$  {\em also annihilates} 
$\,\,  _2F_1\Bigl([1/12, \, 5/12], \,[1];\, H_2)$.
Therefore, this order-four operator is not a MUM operator
(it has two series-solutions (\ref{twoseries}), analytic at $\, x\, = \, 0$).

The symmetric square of $\, L_4$ is of {\em order nine}, instead 
of the order ten one can expect generically. Furthermore, the exterior square of 
$\, L_4$ is of order six, but, 
remarkably, this exterior square
factorizes into a {\em direct sum of an order-one operator}, 
$\,M_1$, {\em  an order-two operator}, $\,M_2^{(1)}$, and an 
{\em order-three operator}, which is the symmetric square of 
{\em  another order-two operator}, $\,M_2^{(2)}$:
\begin{eqnarray}
\label{extdecomp}
\hspace{-0.1in}&&\quad ext^2(L_4) \,\, \,   \, = \,  \,  \, \,  \, \, 
M_1 \, \oplus \, M_2^{(1)} \, \oplus \, Sym^2(M_2^{(2)}),
\end{eqnarray}
where
\begin{eqnarray}
\label{mu}
\hspace{-0.9in}&&\quad \, \,  M_1 \, \,   = \,  \,  \, \,  \, 
 D_x \, - \,{{1} \over {4}}\cdot \,  {{d \ln(\mu(x))} \over {dx}}, 
\qquad \hbox{with:} \qquad \mu(x)  \, = \,  \,  \,
 {{ \delta(x)} \over { x^4 \cdot \, \beta(x)^2 }}
\end{eqnarray}
the Wronskian of  the two order-two operators $\,M_2^{(1)} $
 and  $\,M_2^{(2)} $ reading respectively:
\begin{eqnarray}
\hspace{-0.9in}&&\,  \, \, \, Wr(M_2^{(1)})^2   \, \,   = \,  \,  \, \,  \,
 {{\zeta(x)^2} \over {x^4 \cdot \, \beta(x)^3 \cdot \, \delta(x)}},
 \quad \, \, 
Wr(M_2^{(2)})^4   \, \,   = \,  \,  \, \,  \, 
{{\zeta(x)^4} \over {x^4 \cdot \, \beta(x) \cdot \, \delta(x)^3}}.
\end{eqnarray}

This factorisation is a consequence of relation (\ref{defining}) 
and indicates that  $\, L_4$ is very special. 

The Hauptmoduls $\, H_1$ or $\, H_2$ are {\em not rational functions}
 of $\, x$, they are  {\em algebraic}.
Therefore $\, \,\,  _2F_1\Bigl([1/12, \, 5/12], \,[1];\, H_1)\, \, $ 
has no reason to be solution of a second-order operator. As far as 
factorisation of linear differential operators in operators with 
polynomial coefficients, $\, L_4$ is
 irreducible\footnote[1]{The command \textsf{DEtools[DFactor]} in Maple.}. Let us show 
that such a reduction to a second order linear differential operator actually exists.

\subsubsection{Order-two operator: operator $\omega_{11}$ \newline \newline } 

Introducing the algebraic function $\, {\cal A}(x)$ 
\begin{eqnarray}
\label{defcalA}
\hspace{-0.9in}&&\quad \quad  {\cal A}(x) \,\,\,\, =\, \,\,\,\,\,
 {{1}\over {\delta(x)^{1/8} \cdot A(x)^{1/2}}} \,\,\, \, = \, \,\,\, \, 
 \Bigl({{1}\over {\gamma(x) \, -\, \, 120 \cdot \, \alpha(x)  \cdot \, 
\beta(x)^{1/2}  }}\Bigr)^{1/8}
\nonumber \\
\hspace{-0.9in}&&\qquad \qquad \qquad \quad  \,\,\, \, = \, \,\,\, \, 
 { { \beta(x)^{1/4} \cdot \,\delta(x)^{3/8} } \over
{ \zeta(x)^{1/2}  }}  \cdot \, 
\Bigl({{-1} \over {\, \, 60}} \cdot \,  {{d \,(1/A(x)) } \over {d x}} \Bigr)^{1/2},
\end{eqnarray}
which has the series expansion 
\begin{eqnarray}
\hspace{-0.9in}&& \quad  {\cal A}(x) \,\, \, = \, \,\,\, \,  \, \,\, 
1\, \, \, \,\,\,   - {{117 } \over {2 }} \,x\,\,  \, + {{64635 } \over {8}} \,x^2\,\, \, 
 - {{21853425} \over {16 }}\,x^3\,\, \,  +  {{32050683795} \over {128}}\,x^4\, 
\nonumber \\
\hspace{-0.9in}&& \quad \quad \quad \quad  \quad  \quad \, 
  - {{12299248285371} \over {256 }}\,x^5\, \,
+ {{9718868161850799 } \over {1024 }} \,x^6\,\,\,\,  \,   +\, \,   \cdots,   
\end{eqnarray}
one finds that 
\begin{eqnarray}
\label{seromeg11}
\hspace{-0.9in}&& \quad \quad {\cal A}(x) \cdot \,  _2F_1\Bigl([{{ 1} \over {12}}, 
\, {{5 } \over {12 }}], \,[1];   \, H_1  \Bigr)\, \, \, \,\, = \, \, \, \, \, \,\,  \,\,
 1 \,\, \,\, \, + {{3} \over {2}}\, x \,\,\, + {{75} \over {8}}\, x^2 \,\,\,
+ {{1335} \over {16}}\, x^3\, 
\nonumber \\
\hspace{-0.9in}&& \quad \quad \quad  \quad \quad \quad  \quad \quad 
+ {{111795} \over {128}}\, x^4 \,\,  + {{2559789} \over {256}}\, x^5\,  \, 
+ {{124177119} \over {1024}}\, x^6 \,
\,  \, \,\, + \, \, \cdots 
\end{eqnarray}
is actually solution of an {\em order-two} linear differential operator:
\begin{eqnarray}
\label{omeg11}
\hspace{-0.9in}&&\quad \quad  \quad  \quad {\tilde \omega}_{11} 
\,\,\,\,  = \,\, \,\,   \,\,\,
 D_x^2 \, \,  \, \,\, 
+ \, {{1-24\,x+4\,{x}^{2}+30\,{x}^{3}-21\,{x}^{4} } \over { 
 x  \, (x+1)  \, 
 (1\, -17\, x \,+19\, x^2\, -7\, x^3)}}
\cdot \, D_x  \nonumber \\
\hspace{-0.9in}&&\qquad \quad  \quad  \quad \quad
 \quad \quad \quad \quad \quad \quad \, 
-   {{3\, \, (x-1)  \, (7\,{x}^{2}-x-2)
 } \over {
 4 \, x  \cdot \, (x+1)  \, 
 (1\, -17\, x \,+19\, x^2\, -7\, x^3) }}.
\end{eqnarray}
Note that
\begin{eqnarray}
\label{seealso}
\hspace{-0.9in}&&\,\, M_2^{(2)}\,\,\,  = \,\, \,\,   \, 
{{1} \over {\lambda(x)}} \cdot \, {\tilde \omega}_{11} 
 \cdot \,\lambda(x), \quad \,  \hbox{with:} \qquad
 \lambda(x) \,\,\,  = \,\, \,\,   \,
 {{ \delta(x)^{3/8} \cdot \, \beta(x)^{1/4} } \over {\zeta(x)^{1/2}}}.
\end{eqnarray}

The series (\ref{seromeg11}) is {\em globally bounded}, the rescaling
 $\,\,\, x \, \rightarrow \, 4 \, x\,\,\,$
changing this series into a series with {\em integer} coefficients,
\begin{eqnarray}
\hspace{-0.7in}&& \quad 1\,\,\,\,+6\,x\,\,\,  +150\,x^2\,+5340\,x^3\,\, +223590\,x^4\,\,
+10239156\,x^5\,\,+496708476\,x^6\,\,
\nonumber \\
\hspace{-0.7in}&&\qquad \quad \quad \quad +25083657720\,x^7\,\,\, +1304819854470\,x^8
\,\, \,\,\, + \, \, \cdots,  
\end{eqnarray}
solution of the {\em order-two} operator (pullback of (\ref{omeg11})
by $\,\, x \, \rightarrow \, 4 \, x\,\,$):
\begin{eqnarray}
\label{omeg11integ}
\hspace{-0.9in}&&\quad  \quad  \omega_{11} \,\,\,  = \,\, \, \,\, \,\,\,
D_x^2 \, \, \,\,\,  \,  
 +\,\, {{1-96\,x+64\,{x}^{2}+1920\,{x}^{3}-5376\,{x}^{4}}\over{
x\, \cdot \, ( 1+4\,x ) \, \cdot \, ( 1-68\,x+304\,{x}^{2}-448\,{x}^{3} )}} \cdot \,  D_x
\nonumber \\
\hspace{-0.9in}&&\quad \quad \qquad \quad  \quad  \quad  \quad \quad  \,  
  -  {{6\, ( 1-4\,x) \, ( 1+2\,x-56\,{x}^{2}) }\over{
x\, \cdot \, ( 1+4\,x ) \, \cdot \, ( 1-68\,x+304\,{x}^{2}-448\,{x}^{3} )}} . 
\end{eqnarray}

\subsubsection{{}From order-two operator to order-four operator \newline \newline } 

This result can be revisited as follows. The function
$\, \,\,  _2F_1\Bigl([1/12, \, 5/12], \,[1];\, H_1)$, known to be solution of the
order-four operator $\, L_4$ is also solution of the order-two operator:
\begin{eqnarray}
\label{opH1}
\hspace{-0.95in}&&{{1} \over {{\cal A}(x)}}  \cdot \,  {\tilde \omega}_{11}  \cdot \, {\cal A}(x)
\,\,   = \,\,\, 
 A(x)^{1/2}  \cdot \, {\hat \omega}_{11}  \cdot \, {{1} \over { A(x)^{1/2}}} 
\, \, \,   = \,\,\, \,\,      
{\hat \omega}_{11}\,   + \,  {{1} \over {4}} \cdot \, \Bigl({{ d \ln A(x)}\over{dx}}\Bigr)^2 
\, - \,   {{ \, \,\, \Omega_1} \over {2}}, 
\nonumber \\
\hspace{-0.95in}&&\quad \quad  \quad  \,  \hbox{where:}
 \,  \quad \quad \quad \quad {\hat \omega}_{11} 
 \,\, \,\,   = \,\, \, \,\, \,\,\,  
  \delta(x)^{1/8} \cdot \,  {\tilde \omega}_{11}  \cdot \,  {{1} \over {\delta(x)^{1/8}}}
 \\
\hspace{-0.95in}&&\qquad \qquad \quad \quad \quad \quad   \quad  \quad 
\,\, \,\,   = \,\, \, \,\, \,  D_x^2 \, \,\, 
+ \, {{ p_1(x)} \over { x \cdot \, \delta(x) \cdot \, \beta(x) }} \cdot D_x \,\,
 + \, \,  {{p_0(x) } \over { x \cdot \, \delta(x)^2 \cdot \, \beta(x) }}
 \nonumber \\
\hspace{-0.95in}&&\qquad \qquad \quad \quad \quad  \quad  \quad  \quad 
\,\, \,\,   = \,\, \, \,\, \,  D_x^2 \, \,\, 
 - \,{{1} \over {4}}\cdot \,  {{d \ln(\mu(x))} \over {dx}} \cdot D_x \,\,
 + \, \,  {{p_0(x) } \over { x \cdot \, \delta(x)^2 \cdot \, \beta(x) }}, 
 \nonumber
\end{eqnarray}
 with $\, \mu(x)$ is given in (\ref{mu}) and 
\begin{eqnarray}
\hspace{-0.9in}&&\quad p_1(x)  \,\,   = \,\, \, \,\,\, 
1 \,\,  +147\,x\,-4313\,{x}^{2}\,-7083\,{x}^{3}\,+14073\,{x}^{4}\, +4125\,{x}^{5}\,
-19395\,{x}^{6}
\nonumber \\
\hspace{-0.9in}&&\quad \quad \qquad \qquad 
 \,+12555\,{x}^{7}\,-3150\,{x}^{8}, 
  \\
\hspace{-0.9in}&&\quad p_0(x)  \,\,   = \,\, \, \,\,\, 
1\, \,  -19\,x\, +1389\,{x}^{2}\, -6497\,{x}^{3}\, +27603\,{x}^{4} \, 
-37155\,{x}^{5}  -18369\,{x}^{6}
\nonumber \\
\hspace{-0.9in}&&\quad \quad \qquad \qquad 
 \, +45477\,{x}^{7}\, -17280\,{x}^{8}\, -270\,{x}^{9},  
\end{eqnarray}
and where the order-one operator $\, \Omega_1$ reads
(using (\ref{dlnA})): 
\begin{eqnarray}
\hspace{-0.9in}&&\Omega_1  \,\,\,   = \,\, \, \,\,\,
{{ d \ln A(x)}\over{dx}} \cdot \, 
\Bigl(2 \, D_x \, + \,{{ p_1(x)} \over {
 x \cdot \, \delta(x) \cdot \, \beta(x) }} \Bigr)
 \,\,\, \,+ \, \, {{ d^2  \ln A(x)}\over{dx^2}} \\
\hspace{-0.9in}&&\quad  \,\,   = \,\, \, \,\, 
{{60 \cdot \, \beta(x)^{1/2} \cdot \zeta(x)}\over{  \beta(x) \cdot \delta(x) }} \cdot \, 
\Bigl(2 \, D_x \,  - \,{{1} \over {4}}\cdot \,  {{d \ln(\kappa(x))} \over {dx}} \Bigr),
 \quad \hbox{where:} \quad  \, \, \, 
\kappa(x) \,\,  = \, \, \, \, {{\delta(x)^5 } \over { x^4 \cdot \, \zeta(x)^4}}.
 \nonumber 
\end{eqnarray}

\vskip .2cm

The function $\, \,\,  _2F_1\Bigl([1/12, \, 5/12], \,[1];\, H_2)$, 
known to be solution of the
order-four operator $\, L_4$ is also solution of the order-two operator:
\begin{eqnarray}
\label{opH2}
\hspace{-0.9in}&&\quad {{1} \over {A(x) \cdot \,{\cal A}(x)  }} 
 \cdot \,  {\tilde \omega}_{11}  \cdot \, {\cal A}(x) \cdot \, A(x)
\,\, \, \, \,   = \,\,\,\,  \,\,  
 {{1} \over { A(x)^{1/2}}}   \cdot \, {\hat \omega}_{11}  \cdot \,   A(x)^{1/2}
 \nonumber  \\
\hspace{-0.5in}&&\quad \qquad \qquad  \qquad \quad \quad 
\, \,  \,    = \,\,\,  \, \,     
{\hat \omega}_{11}    \,   \,  \, \,    
+ \, \,  {{1} \over {4}} \cdot \, \Bigl({{ d \ln A(x)}\over{dx}}\Bigr)^2 
\, \,  + \,  \,   {{ \, \,\, \Omega_1} \over {2}}. 
\end{eqnarray}

These last two order-two operators (\ref{opH1}) and (\ref{opH2}) 
can be written as  $\, \Omega_{11} \, \pm \Omega_1/2$
where:
\begin{eqnarray}
\hspace{-0.6in}&&\quad \quad \Omega_{11} \,\,   = \,\,\,  \, \, 
{\hat \omega}_{11}   \,  \, \, 
+ \,  {{1} \over {4}} \cdot \, \Bigl({{ d \ln A(x)}\over{dx}}\Bigr)^2
 \nonumber  \\
\hspace{-0.6in}&&\quad  \qquad \quad \quad 
 \,\,   = \,\,\,  \,\,    D_x^2 \, \,\, \,  
+ \, {{ p_1(x)} \over { x \cdot \, \delta(x) \cdot \, \beta(x) }} \cdot D_x
 \,\, \,  - \, \, 30 \cdot \,   {{q_0(x) } \over {
 x \cdot \, \delta(x)^2 \cdot \, \beta(x) }},
 \nonumber
\end{eqnarray}
 with
\begin{eqnarray}
\hspace{-0.7in}&&q_0(x)  \,\, \,   = \,\, \, \,\,\, \, \, 
1\,\, \,\,   -49\,x\,\,\,  +909\,{x}^{2}\,-7877\,{x}^{3}\,
+31323\,{x}^{4}\,-44025\,{x}^{5}\,-10089\,{x}^{6}\, 
 \nonumber  \\
\hspace{-0.7in}&&\qquad \quad \quad \quad  \quad 
\, +39237\,{x}^{7} \, -13680\,{x}^{8} \, -1350\,{x}^{9}.
\end{eqnarray}
These  last two order-two operators (\ref{opH1}) and (\ref{opH2}) 
 are not linear differential operators 
with rational coefficients, but with {\em algebraic} coefficients: there 
are $\, \beta(x)^{1/2}$ terms.

One can verify directly that
the order-four operator $\, L_4$ {\em can actually be seen as the 
direct sum of these last two order-two operators} with algebraic coefficients 
(\ref{opH1}) and (\ref{opH2}):
\begin{eqnarray}
\label{generalresult}
\hspace{-0.7in}&&  \,  L_4  \,\, \,\,   = \,\, \, \,\, 
\Bigl( A(x)^{1/2}  \cdot \, {\hat \omega}_{11}  \cdot \, {{1} \over { A(x)^{1/2}}}\Bigr) 
 \,  \, \oplus \,  \,
 \Bigl( {{1} \over { A(x)^{1/2}}}  \cdot \, {\hat \omega}_{11}  \cdot \, A(x)^{1/2}  \Bigr) 
 \nonumber \\
\hspace{-0.7in}&&  \,\, \qquad  \quad   \,\, \,\,   = \,\, \, \,\,\,  
 ( \Omega_{11} \, - {{\Omega_1} \over {2}})  \,  \, \oplus \,  \, 
  ( \Omega_{11} \, + {{\Omega_1} \over {2}}).  
\end{eqnarray}
Note that this result is a particular case 
of a more general result. Let us consider 
the direct-sum of the two order-two operators with algebraic coefficients 
depending on one parameter $\, u$
 \begin{eqnarray}
\hspace{-0.6in}&& \qquad  \quad L_4(u)  \,\, \,\,\,   = \,\, \, \,\,\, \, 
 ( \Omega_{11} \, + u \cdot \, {{\Omega_1} \over {2}})  \,  \, \oplus \,  \, 
  ( \Omega_{11} \, - u \cdot \,  {{\Omega_1} \over {2}}).  
\end{eqnarray}
The order-four operator is actually a linear differential operator with
rational coefficients for any $\, u$. It is of the form
\begin{eqnarray}
\hspace{-0.6in}&& \qquad  \quad L_4(u)  \,\, \,\,\,   = \,\, \, \,\,\, \, 
M_2 \cdot  \, \Omega_{11} \,\,\, + \, u^2 \cdot \, M_1 \cdot N_1, 
\end{eqnarray}
where $\, M_2$ and $\, M_1$ are respectively  order-two and order-one operators with
rational coefficients and the order-one operator $\, N_1$ reads:
\begin{eqnarray}
\hspace{-0.65in}&&\,\,  N_1 \,\, \,   = \,\, \,   \, 
D_x \, \, -\, {{1} \over {8}} \cdot \, {{d \ln(\rho(x))} \over {dx}}, 
\quad \quad  \hbox{with:} \quad \quad \quad 
\rho(x)\,\,   = \,\, \,  \, {{\delta(x)^5} \over { x^4 \cdot \, \zeta(x) }}.
\end{eqnarray}

We have the following general result: the exterior square of 
a direct sum of the form (\ref{generalresult}) is a direct sum
of three order-one operators and the symmetric square 
of an order-two linear differential operator. Let 
us denote $\, W^{+}$, $\, W^{-}$ and  $\, W$
the Wronskian of respectively
\begin{eqnarray}
\hspace{-0.8in}&& \,\,\, {\hat \omega}_{11}^{+} \,\, = \, \,\,
  A(x)^{1/2}  \cdot \, {\hat \omega}_{11}  \cdot \, {{1} \over { A(x)^{1/2}}},
 \,\, \, \, \,\quad  \,
{\hat \omega}_{11}^{-} \,\, = \, \, \,
 {{1} \over { A(x)^{1/2}}}  \cdot \, {\hat \omega}_{11}  \cdot \, A(x)^{1/2},
  \,\, \,\, \, \quad  \,
{\hat \omega}_{11}. \nonumber 
\end{eqnarray}
One has the following direct sum decomposition for an arbitrary order-two 
linear differential  operator $\, {\hat \omega}_{11}$: 
\begin{eqnarray}
\label{general}
\hspace{-0.95in}&&\quad \quad ext^2\Bigl({\hat \omega}_{11}^{+} 
 \,  \, \oplus \,  \, {\hat \omega}_{11}^{-}  \Bigr) 
 \, \,  \, \,  \, \,   \, =  \\
\hspace{-0.95in}&&  \quad \quad \quad \quad \, \, \,  
\Bigl(D_x \, -{{d \ln(W^{+})} \over {dx}} \Bigr)  \, \oplus
 \,\Bigl(D_x \, -{{d \ln(W^{-})} \over {dx}} \Bigr) \, \oplus
 \,\Bigl(D_x \, -{{d \ln(W)} \over {dx}} \Bigr)   \, \oplus \, M_2^{(3)},
 \nonumber 
\end{eqnarray}
where the order-three operator $\,M_2^{(3)}$ can, for instance, be written
as a conjugation of a symmetric square (see also (\ref{seealso})):
\begin{eqnarray}
\hspace{-0.6in}\qquad \quad M_2^{(3)}  \, \,  \, \,  = \, \,   \, \,   \,
 Sym^2\Bigl(  {{d A(x)} \over {dx}}^{1/2}  {\hat \omega}_{11}^{-} 
 \cdot \, \Bigl({{d A(x)} \over {dx}} \Bigr)^{-1/2} \Bigr).
\end{eqnarray}
The order-one operator $\, D_x \, -{{d \ln(W)} \over {dx}}$ in (\ref{general})
 actually corresponds to 
(\ref{opH1}) together with (\ref{mu}). However we see from (\ref{general}) 
that the irreducible order-two operator $\, M_2^{(1)}$, we found in 
the direct sum decomposition (\ref{extdecomp}), 
can in fact be decomposed in a direct sum of order-one operators with {\em algebraic} 
coefficients.

\subsubsection{Back to Golyshev and Stienstra operator \newline }

Do note that the order-two operator ${\cal H}_{11}$, previously encountered
with a genus-one situation (see (\ref{Heung11})), is conjugated to a
pullback of (\ref{omeg11}):
\begin{eqnarray}
\hspace{-0.3in}&&\, {\cal H}_{11}  \,\,\,\,  = \,\, \, \,\, \, 
 {{1}\over{\sqrt{5+3x}}} \cdot \, \, 
{\tilde \omega}_{11}\Bigl(x \, \rightarrow \, \,
 {{5 \, x}  \over {5\, +\, 3\, x}} \Bigr) \cdot \, \,   \sqrt{5+3x}.
\end{eqnarray}
The pullback of ${\cal H}_{11}$ by  $\,\, x \, \rightarrow \, 5 \, x\,\,$
reads:
\begin{eqnarray}
\label{better11}
\hspace{-0.9in}&& {\cal H}_{11}\Bigl(x \, \rightarrow \, 5 \, x\Bigr) 
\,  \,\, = \, \, \,\, \,\,\, D_x^2 \, \, \,\,  \,\,
+{{ 4512\, x^4+630\, x^3+1232\, x^2+102\, x-1} \over {
 (188\, x^3+8\, x^2+76\, x-1) \cdot \, (8\, x+1) \cdot \, x}} \cdot \, D_x \,
 \nonumber \\
\hspace{-0.9in}&&\qquad \quad   \quad \quad  \quad  \quad \quad \quad
 +{{6\, (188\,x^3+21\,x^2+28\,x+1) } \over {
(188\, x^3+8\, x^2+76\, x-1) \cdot \, (8\, x+1) \cdot \, x }},
\end{eqnarray}
which has as a solution the series with {\em integer} coefficients:
\begin{eqnarray}
\hspace{-0.9in}&&\quad \quad 1\,\, \,  \, +6\, x\, \,  \, +204\, x^2\, +8790\, x^3
\,\,  +445170\, x^4\, +24577236\, x^5\,\, 
+1436107596\, x^6\, \nonumber \\
\hspace{-0.9in}&&\qquad \quad \quad  \quad  \,
 +87310665684\, x^7\, +5466252149820\, x^8
\,\,\, \, \, \,   + \, \cdots 
\end{eqnarray}

\subsubsection{Hadamard products \newline }

The Hadamard square of $\, \omega_{11}$ is a linear differential operator of order ten.
The Hadamard square of ${\cal H}_{11}$ (or its pullback
by  $\,\, x \, \rightarrow \, 5 \, x$) is also a linear differential
operator of order ten. This order-ten operator is not MUM.

Its head polynomial is
\begin{eqnarray}
\hspace{-0.9in}&&x^6 \cdot \, (1\, -64 \, x) \cdot \,
(96256\, x^3-512\, x^2+608\, x+1) \cdot \,
(35344\, x^3+28512\, x^2+5792\, x-1)
 \,  \nonumber \\
\hspace{-0.9in}&&\qquad \quad \quad \times \,  \,
 (35344\, x^3-14288\, x^2-8\, x-1) \cdot \, P_{29}, 
\nonumber
\end{eqnarray}
where $\,  P_{29}$ is a polynomial of degree 29.
One verifies easily that Hadamard's theorem~\cite{Hadamard}
 on the location of the singularities is 
verified. If one denotes $\,x_1$, $\,x_2$, $\,x_3$, the
 three roots of polynomial $\,188\, x^3+8\, x^2+76\, x-1 $
(see (\ref{better11})), one sees that the three roots of  polynomial
 $\,96256\, x^3-512\, x^2+608\, x+1$
are nothing but $\,-x_1/8$, $\,-x_2/8$, $\,-x_3/8$, the three roots of  polynomial
$\, 35344\, x^3+28512\, x^2+5792\, x-1$ are nothing but  $\,x_1^2$, $\,x_2^2$, $\,x_3^2$,
 the three roots of  polynomial
 $\,35344\, x^3-14288\, x^2-8\, x-1 $ are nothing but 
 $\,x_1\, x_2$, $\,x_2 \, x_3$, $x_1 \,x_3$,
and of course, besides $\, x= \, 0\,$ and $\, x\, = \, \infty$, 
 $\,\, 1/64\, =  \, (-1/8)^2$.  

\vskip .1cm

\subsection{Other higher genus modular forms}
\label{omega23}

Similar calculations can be performed for the 
modular forms associated with genus-one 
modular curves, corresponding, for instance to $\, \tau \, \rightarrow \, N \, \tau$,
for $\, N \, = \, \, 17, \, 19$. For  $\, N \, = \, \,23$
 the modular curve is a genus-two
curve~\cite{Fricke,Fricke2}. 
These detailed analysis and calculations will be given in a forthcoming publication. 

The $\, N \, = \, \,23 \, $ genus-two case requires to introduce 
the order-two operator
\footnote{where one notes that the polynomial $\, 1\, -x^2 \, +x^3$ has the root
$\, - \,1/P$, where $\,P\, = \, 1.324717958\, ...$ is the smallest 
{\em Pisot number}~\cite{Bertin}.}:
\begin{eqnarray}
\hspace{-0.8in}&&{\tilde \omega_{23}}(x) \,\,\,\, = \,\, \,\, \, \,\,
D_x^2 \, \, \, \,\,\,
+\, {{1\, -12\, x \, +4\, x^2 \, +5\, x^3 \, -33\, x^4 \, +35\, x^5 \, -28\, x^6
} \over {
x \cdot \, (1\, -8\, x \, +3\, x^2 \, -7\, x^3) \cdot \, (1\, -x^2 \, +x^3) }}
 \cdot \, D_x 
\nonumber \\
\hspace{-0.8in}&&\qquad \qquad   \quad \quad \quad \quad
\, - \, \,  {{1\, -\, x \, -\, x^2 \, +12 \, x^3 \, -15\, x^4 \, +14\, x^5 
} \over {
 x \cdot \, (1\, -8\, x \, +3\, x^2 \, -7\, x^3) \cdot \, (1\, -x^2 \, +x^3) }},
 \nonumber
\end{eqnarray}
This operator has the analytic solution with integer coefficients: 
\begin{eqnarray}
\hspace{-0.9in}&&1\,\, \,  \, +x\,\, \,  +3\, x^2\, \,  +13\, x^3\, +67\, x^4\,
 +375\, x^5\, +2223\, x^6\,
 +13713\, x^7\,+87123\, x^8\,  \,\,\, + \, \, \cdots
 \nonumber
\end{eqnarray}

The pullback of $\, {\tilde \omega_{23}}(x)$ by $\, x \, \rightarrow \, 1/x$
gives an order-two operator with two analytic solutions (no logarithm):
\begin{eqnarray}
\hspace{-0.8in}&&\quad x\,\, \, \, \, 
+ {{ 5} \over {14 }} \, x^2\,\,\,  +{{11 } \over {196 }} \, x^3\,\, 
- {{85 } \over {1372 }} \, x^4\,\,  - {{3499 } \over {57624 }} \, x^5\, 
- {{2041 } \over {57624 }} \, x^6\,  \,-{{18317 } \over {672280 }} \, x^7\,\,
 \nonumber \\
\hspace{-0.8in}&&\quad \quad  \quad  \quad  \,
 -{{332455 } \over {19765032 }} \, x^8\, \, 
+ {{21994361 } \over {1383552240 }}\, x^9\, 
\,\, \,\,  + \, \, \cdots, 
\end{eqnarray}
and 
\begin{eqnarray}
\hspace{-0.6in}&&x^2\,\, + {{ 5} \over {14 }}\, x^3\,\,
 - {{ 3} \over {98 }}\, x^4\,\, 
- \, {{ 251} \over {1372 }} x^5\, \, - {{ 137} \over {1372 }} \, x^6\, \,
 -{{507} \over {9604}} \, x^7\,  -{{ 24007} \over {470596 }} \, x^8\,
\nonumber \\
\hspace{-0.6in}&&\qquad \qquad  - {{ 144083} \over {6588344 }} \, x^9
 \,\, \,  \,\, + \, \, \cdots 
\end{eqnarray}
These two series are {\em not globally bounded}
 (but a linear combination of these
two series  {\em may be} globally bounded ...).
The fact that the two previous solutions have 
no logarithmic terms {\em excludes any relation
like} (\ref{omeg5}). We encounter the same situation 
with the order-two operators
$\,{\tilde \omega_{11}}(x)$, $\,{\tilde \omega_{17}}(x)$, 
$\,{\tilde \omega_{19}}(x)$,
 $\,{\tilde \omega_{29}}(x)$,
 $\,{\tilde \omega_{31}}(x)$,  $\,{\tilde \omega_{41}}(x)$,
  $\,{\tilde \omega_{47}}(x)$,
 $\,{\tilde \omega_{59}}(x)$, and  $\,{\tilde \omega_{71}}(x)$,
the corresponding two series at $\, x \,= \, \,  \infty$ 
having no logarithmic terms
 yielding the same obstruction for a relation like (\ref{omeg5}).
These various  order-two operators correspond 
to higher order genus modular curves~\cite{Hibino},
namely  {\em genus-one}
 for $\,{\tilde \omega_{11}}(x)$, $\,{\tilde \omega_{17}}(x)$, 
$\,{\tilde \omega_{19}}(x)$,
{\em genus-two} for $\,{\tilde \omega_{29}}(x)$,
 $\,{\tilde \omega_{31}}(x)$,
{\em genus-three} for $\,{\tilde \omega_{41}}(x)$,  {\em genus-four }
for $\,{\tilde \omega_{47}}(x)$,
{\em genus-five} for  $\,{\tilde \omega_{59}}(x)$, and {\em genus-six} 
 for $\,{\tilde \omega_{71}}(x)$.

Note that all these higher-genus $\,{\tilde \omega_{n}}(x)$'s are simply 
homomorphic to their adjoint. They are such that
\begin{eqnarray}
\hspace{-0.3in}R(x)^{1/M} \cdot \, adjoint({\tilde\omega_{n}}(x)) 
 \, \,\, \, \,  = \,\,  \, \, \, \, \tilde\omega_{n}(x) \cdot \, R(x)^{1/M},  
\end{eqnarray}
where $\, R(x)$ is a rational function, and 
where $\, M\, = \, \, 2$, except for 
$\, n\, = \, \, 19$ where $\, M\, = \, \, 6$.

\vskip .1cm

\subsection{Order-two operators associated with modular Atkin equations}

Let us give the explicit expressions of some $\, \omega_{n}$'s for
 $\, n\, = \, \, 17, \, 19, \,  23, \, 31, \, 39, \,41, \, 47, \, 59, \, 71$.
The higher genus  $\, \omega_{n}$'s read
\begin{eqnarray}
\hspace{-0.6in} \qquad \qquad \omega_{i} \,\,\,\,= \, \,\, \,\,\,\,
D_x^2 \,\,\, +\frac{A_{i}}{C_{i}} D_x \,\,\,\, +\frac{B_{i}}{C_{i}}, 
  \quad
\nonumber
\end{eqnarray}
with:
\begin{eqnarray}
\hspace{-0.6in}&&A_{11}\, =\, \, 84\,x^{4}-120\,x^{3}-16\,x^{2}+96\,x-4, 
\nonumber \\
\hspace{-0.6in}&&B_{11}\, =\,\,  3\, (x-1) \,  (7\,x^{2}-x-2), 
\nonumber \\
\hspace{-0.6in}&&C_{11}\, =\, \, 4\, x \cdot \, (7\,x^{3}-19\,x^{2}+17\,x-1)\,  (x+1), 
\nonumber
\end{eqnarray}
\begin{eqnarray}
\hspace{-0.6in}&&A_{17}\, =\, \, 448\,x^{5}-576\,x^{4}+280\,x^{3}+224\,x^{2}-264\,x+16, 
\nonumber \\
\hspace{-0.6in}&&B_{17}\, =\, \, 168\,x^{4}-180\,x^{3}+67\,x^{2}+41\,x-20, 
\nonumber \\
\hspace{-0.6in}&&C_{17}\, =\, \, 16\, x ( x-1 )  ( 8\,x^{4}-4\,x^{3}+3\,x^{2}+10\,x-1), 
\nonumber
\end{eqnarray}
\begin{eqnarray}
\hspace{-0.6in}&&A_{19}\, =\, \, 264\,x^{5}-354\,x^{4}-258\,x^{3}+300\,x^{2}+93\,x-9, 
\nonumber \\
\hspace{-0.6in}&&B_{19}\, =\, \, 110\,x^{4}-130\,x^{3}-69\,x^{2}+56\,x+6, 
\nonumber \\
\hspace{-0.6in}&&C_{19}\, =\, \, 9\, x  \cdot \, (2\,x+1)\, 
 (x+1) \,(4\,x^{3}-12\,{x}^{2}+10\,x-1), 
\nonumber
\end{eqnarray}
\begin{eqnarray}
\hspace{-0.6in}&&A_{23}\, =\, \, 28\,x^{6}-35\,x^{5}+33\,x^{4}-5\,x^{3}-4\,x^{2}+12\,x-1, 
\nonumber \\
\hspace{-0.6in}&&B_{23}\, =\, \, 14\,x^{5}-15\,x^{4}+12\,x^{3}-x^{2}-x+1, 
\nonumber \\
\hspace{-0.6in}&&C_{23}\, =\, \, x \cdot \, (7\,x^{3}-3\,x^{2}+8\,x-1)  (x^{3}-x^{2}+1), 
\nonumber
\end{eqnarray}
\begin{eqnarray}
\hspace{-0.6in}&&A_{29}\, =\, \, 504\,x^{7}-64\,x^{6}-896\,x^{5}
-480\,x^{4}+400\,x^{3}+512\,{x}^{2}+72\,x -16, 
\nonumber \\
\hspace{-0.6in}&&B_{29}\, =\, \, 315\,x^{6}-35\,x^{5}-441\,x^{4}
-213\,x^{3}+109\,x^{2}+103\,x+4, 
\nonumber \\
\hspace{-0.6in}&&C_{29}\, =\, \, 16\,  x \cdot \, (x+1)  
(7\,x^{6}-8\,x^{5}-8\,x^{4}-2\,{x}^{3}+12\,x^{2}+4\,x-1), 
\nonumber
\end{eqnarray}
\begin{eqnarray}
\hspace{-0.6in}&&A_{31}\, =\, \, 108\,x^{6}+343\,x^{5}+477\,x^{4}
+235\,x^{3}+28\,x^{2}-6\,x-1, 
\nonumber \\
\hspace{-0.6in}&&B_{31}\, =\, \, 60\,x^{5}+161\,x^{4}
+180\,x^{3}+69\,x^{2}+5\,x-1, 
\nonumber \\
\hspace{-0.6in}&&C_{31}\, =\, \, x  \cdot \, 
(x^{3}+3\,x^{2}+4\,x+1 )  ( 27\,x^{3}+17\,x^{2}-1 )
\nonumber
\end{eqnarray}
\begin{eqnarray}
\hspace{-0.6in}&&A_{41}\, =\, \, 704\,x^{9}+960\,x^{8}-360\,x^{7}
-1472\,x^{6}-672\,x^{5}+480\,x^{4}
\nonumber \\ 
\hspace{-0.6in}&&\quad \quad +720\,x^{3}-128\,x^{2}-120\,x+16, 
\nonumber \\
\hspace{-0.6in}&&B_{41}\, =\, \, 616\,x^{8}+756\,x^{7}-289\,x^{6}
-969\,x^{5}-353\,x^{4}+265\,x^{3}
\nonumber \\ 
\hspace{-0.6in}&&\quad \quad +261\,x^{2}-39\,x-12, 
\nonumber \\
\hspace{-0.6in}&&C_{41}\, =\, \, 16\,  x \cdot \, (x-1)  
(8\,x^{8}+20\,x^{7}+15\,x^{6}-8\,x^{5}-20\,x^{4}-10\,x^{3}
\nonumber \\ 
\hspace{-0.6in}&&\quad \quad +8\,x^{2}+4\,x-1), 
\nonumber
\end{eqnarray}
\begin{eqnarray}
\hspace{-0.6in}&&A_{47}\, =\, \, 66\,x^{10}-154\,x^{9}+190\,x^{8}
-135\,x^{7}+52\,x^{6}+56\,{x}^{5}-57\,x^{4}
\nonumber \\ 
\hspace{-0.6in}&&\quad \quad +60\,x^{3}-22\,x^{2}+9\,x-1, 
\nonumber \\
\hspace{-0.6in}&&B_{47}\, =\, \, 66\,x^{9}-140\,x^{8}+158\,x^{7}
-103\,x^{6}+35\,x^{5}+33\,x^{4}
\nonumber \\ 
\hspace{-0.6in}&&\quad \quad -28\,x^{3}+24\,x^{2}-6\,x+1, 
\nonumber \\
\hspace{-0.6in}&&C_{47}\, =\, \, x \cdot \, (11\,x^{5}-6\,x^{4}
+15\,x^{3}-5\,x^{2}+5\,x-1) \,
 (x^{5}-2\,x^{4}+x^{3}
\nonumber \\
\hspace{-0.6in}&&\quad \quad +x^{2}-x+1), 
\nonumber
\end{eqnarray}
\begin{eqnarray}
\hspace{-0.6in}&&A_{59}\, =\, \, 308\,x^{12}+624\,x^{11}+1632\,x^{10}
+2640\,x^{9}+3520\,x^{8}+3816\,x^{7}
\nonumber \\
\hspace{-0.6in}&&\quad \quad  +3232\,x^{6}+2128\,x^{5}+1008\,x^{4}+280\,x^{3}-24\,x-4, 
\nonumber \\
\hspace{-0.6in}&&B_{59}\, =\, \, 385\,x^{11}+720\,x^{10}+1748\,x^{9}
+2600\,x^{8}+3128\,x^{7}+3024\,x^{6}
\nonumber \\
\hspace{-0.6in}&&\quad\quad   +2217\,x^{5}+1216\,x^{4}+448\,x^{3}+86\,x^{2}-4\,x-4, 
\nonumber \\
\hspace{-0.6in}&&C_{59}\, =\, \, 4\,  x \cdot \,  (x^{3}+2\,x+1)\,  
(11\,x^{9}+24\,x^{8}+46\,x^{7}+61\,x^{6}+60\,x^{5}
\nonumber \\
\hspace{-0.6in}&& \quad\quad  +44\,x^{4}+21\,x^{3}+4\,x^{2}-2\,x-1), 
\nonumber
\end{eqnarray}
\begin{eqnarray}
\hspace{-0.6in}&&A_{71}\, =\, \, 88\,x^{14}-30\,x^{13}-280\,x^{12}
-195\,x^{11}+420\,x^{10}+671\,x^{9}-5\,x^{8}
\nonumber \\
\hspace{-0.6in}&& \quad \quad  -666\,x^{7}-444\,x^{6}+91\,x^{5}+231\,x^{4}
+95\,x^{3}+4\,x^{2}-6\,x-1, 
\nonumber \\
\hspace{-0.6in}&&B_{71}\, =\, \, 132\,x^{13}-42\,x^{12}-372\,x^{11}
-248\,x^{10}+475\,x^{9}+717\,x^{8}
\nonumber \\
\hspace{-0.6in}&&\quad  +25\,x^{7}-563\,x^{6}-362\,x^{5}+29\,x^{4}+112\,{x}^{3}+38\,x^{2}+x-1, 
\nonumber \\
\hspace{-0.6in}&&C_{71}\, =\, \, x \cdot \,
 (11\,x^{7}-4\,x^{6}-18\,x^{5}-5\,x^{4}+11\,x^{3}+7\,{x}^{2}-1)
\,   ( x^{7}-2\,x^{5}
\nonumber \\
\hspace{-0.6in}&& \quad \quad  -3\,x^{4}+x^{3}+5\,x^{2}+4\,x+1). 
\nonumber 
\end{eqnarray}

\vskip .1cm

\section{Yukawa coupling as ratio of determinants}
\label{Yukawaratio}

Consider an order-four MUM linear differential operator.
Let us introduce the determinantal variables $\, W_m \, = \, \, \det(M_m)$ 
which are the determinants\footnote[2]{For an order-four 
operator the Wronskian is $\, W_4$.}
of  the following $\, m \times m$ matrices
 $\, M_m$, $\, m\,= \,  \,  1, \, \cdots, \, 4$, 
with entries expressed in terms of derivatives of the four solutions
$y_0(x)$, $y_1(x)$, $y_2(x)$ and $y_3(x)$ 
of the MUM linear differential operator 
(see Section  (\ref{plan}) for the definitions). 
One takes
$W_1 (x) \,\, = \, \, \,   y_0 (x)\, \, $ and:
\begin{eqnarray}
\label{filtration}
\hspace{-0.9in}&&M_2 \, = \, \,  
 \left[ \begin {array}{cc} 
y_0  &y_1
 \\ \noalign{\medskip} y_0'  &y_1' 
 \end {array}
 \right], \quad 
M_3 \, = \, \,  
\left[ \begin {array}{ccc} y_0  &y_1
  &y_2  \\ 
\noalign{\medskip} y_0'  &y_1' &y_2' \\ 
\noalign{\medskip}y_0'' &y_1''  &y_2'' 
 \end {array} \right],   
 \quad M_4 \, = \, \,  
 \left[ \begin {array}{cccc} 
y_0  &y_1  &y_2  &y_3  \\ 
\noalign{\medskip}y_0'  &y_1'  &y_2' &y_3' \\ 
\noalign{\medskip}y_0''  &y_1''  &y_2'' 
 &y_3''  \\ 
\noalign{\medskip}y_0'''  & y_1'''  &y_2'''  &y_3'''
  \end {array} \right],
  \nonumber \\
\hspace{-0.9in}&&\, \,  \hbox{where:} \qquad  \quad  
y_i' \, = \, \, {\frac {d}{dx}}{y_i}, 
\quad \quad \quad 
y_i'' \, = \, \, {\frac {d^2}{dx^2}}{y_i}, 
\quad \quad  \quad 
y_i''' \, = \, \, {\frac {d^3}{dx^3}}{y_i}. 
\end{eqnarray}

Since $\,  \, q \, = \, \, exp(y_1/y_0)$, and hence, 
\begin{eqnarray}
\label{derivq}
 q \cdot {{d} \over {dq }} 
 \, \, \, = \, \, \,  \, \, 
{{W_1^2} \over {W_2}} \cdot  {{d} \over {dx }} \, \, \, = \, \, \,  \, \, 
{{y_0^2} \over {W_2}} \cdot  {{d} \over {dx }},
\end{eqnarray}
and thus
\begin{eqnarray}
\hspace{-0.5in}\Bigl(q \cdot {{d} \over {dq }}  \Bigr)^2
\, \,  \, \, = \, \, \, \,  \, \,  \, 
{{y_0^4} \over {W_2^2}}\,  \cdot \,  {{d^2} \over {dx^2 }} \,\,  
\, + \, \, 2 \, {{y_0^3} \over {W_2^2}}\,  
 {{d y_0} \over {dx }} \cdot \,  {{d} \over {dx }} \, \, 
\,  - {{y_0^4} \over {W_2^3}}\,  {{d W_2} \over {dx }}
  \cdot \,  {{d} \over {dx }}, 
\end{eqnarray}
we deduce, after some simple algebra,
 an alternative definition for the 
{\em Yukawa coupling}: 
\begin{eqnarray}
\label{Yukawa}
\hspace{-0.2in}K(q) \,\, = \, \, \, \,
 \Bigl( q \cdot {{d} \over {dq }} \Bigr)^2
 \Bigl(  {{y_2} \over {y_0}}\Bigr)
 \, \,\, = \, \,\,\, \,\, 
 {{W_1^3 \cdot W_3 } \over {W_2^3 }}
\, \,\, = \, \,\,\, \,\, 
 {{y_0^3 \cdot W_3 } \over {W_2^3 }}. 
\end{eqnarray}
to be compared with the other previous 
alternative expression previously given (\ref{otherYuk}) 
for the {\em Yukawa coupling} 
\begin{eqnarray}
\label{otherYuk2}
\hspace{-0.65in}K(q) \, \, \, = \, \, \, \,
  {{ x(q)^3 \, \cdot W_4^{1/2}} \over { y_0^2}} \cdot
 \, \Bigl({{q} \over {x(q)}} \cdot {{d x(q)} \over {dq}}\Bigr)^3
\, \, = \, \, \, \, 
  {{ W_4^{1/2}} \over { y_0^2}} \cdot
 \, \Bigl( q \cdot {{d x(q)} \over {dq}}\Bigr)^3.
\end{eqnarray}

In fact from (\ref{derivq}) we deduce 
\begin{eqnarray}
 \Bigl(q \cdot {{d x(q)} \over {dq }} \Bigr)^3
 \, \, \, = \, \, \,  \, \, 
{{W_1^6} \over {W_2^3}}  \, \, \, = \, \, \,  \, \, 
{{y_0^6} \over {W_2^3}},
\end{eqnarray}
and, so, (\ref{otherYuk2}) is compatible with (\ref{Yukawa})
{\em if the following identity is verified}:
\begin{eqnarray}
  W_3^2  
\, \, \, = \, \, \, \,\,   W_4 \cdot \, y_0^2 
\,\,  \, = \, \, \, \,\,   W_4 \cdot \, W_1^2. 
\end{eqnarray}
This identity is in fact specific of {\em order-four operators 
conjugated to their adjoint} (see below (\ref{cond})). 
Therefore we prefer to use definition (\ref{Yukawa}) 
for the Yukawa coupling, instead of the more restricted 
definition (\ref{otherYuk2}).

\vskip .1cm 

Let us assume that the pullback $\, p(x)$ 
has a series expansion of the form
\begin{eqnarray}
\label{formpull}
p(x) \, \,  \, = \, \, \, \,\,   \lambda \cdot \, x^r \cdot \, A(x), 
\end{eqnarray}
where the exponent $\, r$ is an integer, where $\,\lambda $
is a constant, and where $\, A(x)$ is a function analytic at
 $\, x \, = \, \, 0$ with the series expansion:
\begin{eqnarray}
\hspace{-0.1in}A(x) \, \, = \, \, \, \,\, \, \,  
 1\,\, \,\,   \, + \alpha_1 \cdot \, x\,  \,  \,
+ \alpha_2 \cdot \, x^2\, \, \,  \, +\, \cdots 
\nonumber 
\end{eqnarray}
The determinantal variables $\, W_m$'s transform 
very nicely under pullbacks $\, p(x)$ of the
form (\ref{formpull}):
\begin{eqnarray}
\label{covar}
\hspace{-0.98in}&&(W_1(x), \, W_2(x), \, W_3(x), \, W_4(x)) 
 \, \,\, \, \quad  \longrightarrow  \\
\hspace{-0.98in}&&\qquad 
\Bigl(W_1(p(x)),\, \,{{p'} \over {r}}  \cdot \,W_2(p(x)), \,\, 
 {{p'^3} \over {r^3}}  \cdot \, W_3(p(x)), \,\, 
 {{p'^6} \over {r^6}} \cdot \, W_4(p(x))\Bigr),
 \quad \, \, \, \,    
p' \, = \, \, \, {{d p(x)} \over {d x}}. \nonumber 
\end{eqnarray}

One can show that the nome (\ref{nome}) of an order-$\, N$
operator transforms under a pullback $\, p(x)$:
\begin{eqnarray}
\label{previous}
\hspace{-0.2in}q(x) \, \, \longrightarrow  \, \, Q(x)
 \qquad \, \,   \hbox{with:} \qquad \quad  \,  \, 
 \lambda \cdot \, Q(x)^r \, \, = \, \, \, q(p(x)). 
\end{eqnarray}

{}From the covariance property (\ref{covar}), and 
from the previous transformation
$\, q \, \rightarrow \, \lambda \cdot \,q^r$ for the nome,   
one easily gets the transformation of the Yukawa coupling seen
as a function of the nome
$\,\, K(q) \, \, \rightarrow \, \, K(\lambda \cdot \, q^r)$:
\begin{eqnarray}
\label{Yukawachange}
\hspace{-0.9in}&&K(q(x)) \,
  \, \,\, = \, \,\,\, \,\, 
 {{W_1(x)^3 \cdot W_3(x) } \over {W_2(x)^3 }} \\
\hspace{-0.9in}&& \quad \quad \quad \quad \quad  \,\,\,  
 \, \, \longrightarrow   \quad \quad  \, \, 
 {{W_1(p(x))^3 \cdot W_3(p(x)) } \over {W_2(p(x))^3 }}
\,\,\,  \,  = \, \, \,  \, \,K(q(p(x)))
\,\, = \, \, \, K(\lambda \cdot \, Q(x)^r).
 \nonumber 
\end{eqnarray}

For $\, \lambda \, = \, \, 1$ and $\, r \, = \, \, 1$ (i.e. when
 the pullback is a deformation of the 
identity transformation), one recovers the known 
invariance of the Yukawa coupling 
by pullbacks (see Proposition 3 in~\cite{Prop3}).

\vskip .1cm 

One finds {\em another}
pullback invariant ratio, namely:
\begin{eqnarray}
\label{Kstar}
 K^{\star} \, \,\,  = \, \, \, \,\, \,\,  
{{W_1 \cdot W_3^3} \over {W_4 \cdot W_2^3 }},
\end{eqnarray}
which is, in fact, 
{\em nothing but the Yukawa coupling for the adjoint} 
of the original operator.
\vskip .1cm
Another invariance property is worth noting. Let us consider 
two linear differential operators $\, \Omega_1$ and $\, \Omega_2$ 
of order $\, N$ that are equivalent, in 
the sense of the equivalence of linear differential operators. This 
means that there exists  linear differential operators intertwiners $\, I_1$,
 $\, I_2$,  $\, J_1$, $\, J_2$, of order at most $\, N-1$  
such that 
\begin{eqnarray}
\hspace{-0.2in}\Omega_1 \cdot I_1 \,\,\, = \, \,\, \, I_2 \cdot \Omega_2, 
\quad \, \quad\hbox{and:}   \quad \,\, \quad \,
J_1 \cdot \Omega_1 \,\,\, = \, \,\,\,  \Omega_2 \cdot J_2.
\end{eqnarray}
Let us assume that one of these intertwiners is a linear 
differential operator
of order zero (a function), then the previous homomorphism between
 operators amounts to saying that
the two operators are conjugated by a function:
\begin{eqnarray}
\label{conju}
\hspace{-0.1in}\Omega_2 \,\,\,\, = \, \,\,\, \,\, 
  \rho(x)  \cdot \Omega_1 \cdot \rho(x)^{-1},
\end{eqnarray}
which correspond to changing the four solutions
as follows: 
$\, y_i \, \, \rightarrow \, \, \rho(x) \cdot \, y_i$.
In such a case (quite frequent as will be seen in forthcoming publication)
the previous determinant variables transform, again,
very nicely under the ``gauge'' function $\, \rho(x)$:
\begin{eqnarray}
\label{I11}
\hspace{-0.9in}(W_1, \, W_2, \, W_3, \, W_4) 
 \,\,   \, \,\, \rightarrow \, \,\,\,\,    \, 
(\rho(x) \cdot W_1,\, \, \rho(x)^2 \cdot  W_2, \,\,
  \rho(x)^3 \cdot  W_3, \,\,  \rho(x)^4 \cdot  W_4). 
\end{eqnarray}

It is straightforward to see that the Yukawa coupling 
and the ``dual Yukawa'', $\,K^{\star}$, are 
{\em invariant by such a transformation}\footnote[5]{ $\,K$
and $\,K^{\star}$ (and their combinations) are the only monomials 
$\,\, W_1^{n_1} \, W_2^{n_2} \, W_3^{n_3} \, W_4^{n_4}$
to be invariant by (\ref{covar}) and (\ref{I11}). }.
Two  conjugated operators (\ref{conju}) 
automatically have the same Yukawa coupling.

Do note that the Yukawa couplings for two operators, which are 
non trivially homomorphic to each other
 (intertwiners of order one, two, ...), 
are actually different.
{\em The (pullback invariant) Yukawa coupling is not preserved by 
operator equivalence} (see subsection (\ref{operatornontriv})).
\vskip .1cm
\vskip .1cm
{\bf Remark:} The definition of these determinantal variables
$\, W_i$'s heavily relies on the MUM structure of the 
operator\footnote[1]{Or recalling the powerful result 
of Steenbrink~\cite{Steenbrink}
that the filtration by the logs is the same as the Hodge filtration,
the definition relies on this filtration.}. The four solutions 
are not on the same footing: the log filtration
 {\em imposes a natural order}
between the four solutions the definition of $\, W_i$'s relies on. 
It is worth noting that if one permutes the four solution $\, y_i$,
one would get 24 other sets of $(W_1, \,W_2,  \,W_3,  \,W_4)$
 which are actually {\em also nicely covariant by pullbacks}, 
thus yielding a finite set of other ``Yukawa couplings''
 or  adjoint Yukawa coupling $\, K^{\star}$ {\em also invariant by pullbacks}.
In fact these ``Yukawa couplings'', and other  $\, K^{\star}$, 
can even be defined when the linear differential 
operator is not MUM, and they are still
invariant by pullbacks. 

\subsection{Pullback-invariants for higher order ODEs}
\label{invar}

These simple calculations can straightforwardly be generalised to higher order
linear differential equations. We give here 
the invariants for higher order linear differential operators.

Let us give, for the $\, n$-th order linear differential
operator the list of the $\, K_n$  invariants by pullback transformations:
\begin{eqnarray}
\hspace{-0.9in}&&K_3 \, = \, \, {{W_1^3 \cdot W_3} \over { W_2^3}}, \, \quad\, \,  
K_4 \, = \, \, {{W_1^8 \cdot W_4} \over { W_2^6}}, \, \quad \, \, 
K_5 \, = \, \, {{W_1^{15} \cdot W_5} \over { W_2^{10}}}, \,  \quad \, \, 
K_6 \, = \, \, {{W_1^{24} \cdot W_6} \over { W_2^{15}}}, \,\,  \, \cdots 
 \nonumber \\
\hspace{-0.9in}&&K_n \, = \, \, 
{{W_1^{a_n} \cdot W_n} \over { W_2^{b_n}}},\, \, 
 \quad \hbox{with:}  \, \, \quad \quad
 a_n \, = \, \,  n \cdot (n-2), \qquad 
b_n \, = \, \, {{ n \cdot (n-1)} \over { 2}}.
\end{eqnarray}
A $\, n$-th order linear differential operator has  $\, K_n$ as
an invariant by pullback transformation, as well as all the  $\, K_m$ 
with $\, m \, \le n$.
$K_3$ is the Yukawa coupling, and one remarks, for the order-four
operators, that the other pullback invariant
 $\, K^{\star}$ (see (\ref{Kstar})),
 which is actually also the Yukawa coupling of the adjoint
 operator, is nothing but $\, K_3^3/K_4$.

\vskip .3cm 
For  order-four operators  conjugated to their adjoint
 (see (\ref{conju})) (i.e. operators 
homomorphic to their adjoint, 
the intertwiner being
an order zero differential operator, a function), one has the equality 
\begin{eqnarray}
\label{cond}
\hspace{-0.8in}K_4 \, = \, \, K_3^2, \qquad  \hbox{i.e. } 
\qquad K_3 \, = \, \, K^{\star}, \qquad  \hbox{or} \qquad 
W_3^2      \, = \, \, W_1^2 \cdot W_4, 
\end{eqnarray}
to be compared with the equality in Almkvist et al. 
(see Proposition 2 in~\cite{Almkvist})
\begin{eqnarray}
\hspace{-0.1in}&&\quad y_0 \, y_3' \, - \, \, y_3 \, y_0'
\,\,\,\, = \,\, \,\,\,\,\,
y_1 \, y_2' \, - \, \, y_2 \, y_1',
\end{eqnarray}
which is satisfied when the Calabi-Yau condition that
the exterior square is of order five is satisfied.

If a linear differential operator $\, \Omega_4$
 verifies condition (\ref{cond}),
 its conjugate by a function, 
$\,\rho(x) \cdot \, \Omega_4 \cdot \rho(x)^{-1} $,
also verifies condition (\ref{cond}) (their Yukawa couplings
are equal). 

The condition (\ref{cond}) is not satisfied for linear differential 
operators homomorphic to their adjoint with {\em non-trivial} intertwiner
(of order greater than zero). For instance the order-four operator
(\ref{calB2}) does not satisfy condition (\ref{cond}).

\vskip .1cm

{\bf Remark:} Concerning order-three operators and hypergeometric 
functions.  It is worth noting the four examples 
\begin{eqnarray}
\hspace{-0.95in}\quad \quad 
_3F_2\Bigl([{{1} \over {2}},\, {{1} \over {2}},\,{{1} \over {2}} ],
 \, [1, \, 1 ], \, 64 \, x \Bigr), \, \quad  \quad \quad
 _3F_2\Bigl([{{1} \over {2}},\, {{1} \over {3}},\,{{2} \over {3}} ], 
\, [1, \, 1 ], \, 108 \, x \Bigr), 
\nonumber\\
\,  \, \, 
_3F_2\Bigl([{{1} \over {2}},\, {{1} \over {4}},\,{{3} \over {4}} ],
 \, [1, \, 1 ], \, 256 \, x \Bigr), \, \quad  \quad 
 _3F_2\Bigl([{{1} \over {2}},\, {{1} \over {6}},\,{{5} \over {6}} ], 
\, [1, \, 1 ], \, 1728 \, x \Bigr),
\nonumber 
\end{eqnarray}
which are such that their series expansions, as well as their associated nome, 
mirror map (compositional inverse of the nome), are 
series with {\em integer} coefficients,  the previous invariant
 $\, K_3$ being {\em the constant $\, 1$}.

This however comes as no surprise since
the four corresponding operators are all symmetric squares\footnote[1]{
Recall that 
$\,\,\, _3F_2([1/2, \, \alpha, \, 1\, -\alpha], \, [1, \, 1], \, x) \,\, = \,\, \,$ 
$\,_2F_1([\alpha/2, \, 1/2\, -\alpha/2], \, [1], \, x)^2 $. }
 of order-two 
operators. Their solutions of the form 
$ \, y_0(x)\,=\,\, u(x)^2$, $\, y_1(x)=\,u(x) \cdot \, v(x)$, and
 $\,y_2(x)\,=\,\,v(x)^2/2$,  give automatically $\,K_3 \, =\,\, 1$.

\section{Quasi-Calabi-Yau ODE associated to the 
Hadamard product of two HeunG functions}
\label{Un}

\vskip .1cm
The operator having the Hadamard product of the
two HeunG functions $\, HeunG(a, \, q, \, 1, \, 1, \,  1, \, 1; \, x)$
and $\, HeunG(A, \, Q, \, 1, \, 1, \,  1, \, 1; \, x)$ as a solution
reads:
\begin{eqnarray}
\label{abABapp}
\hspace{-0.7in}&&(x-1)  \, (x-a) 
 \, (x\,-A)  \, (x\,-A\,a)  \, ( A\,a \, -{x}^{2})^{2}
\cdot \,x^3 \cdot \,  D_x^{4}  \\
\hspace{-0.7in}&& \qquad \quad +2\, \, (x ^2 \,-A\,a) \cdot U_3 
 \cdot \,{x}^{2}\cdot \,  D_x^{3}
\, \,\, \,\, \, - U_2 \cdot \, x \cdot \,  D_x^{2} \, \, \, \, \,\,
- U_1 \cdot \,  D_x \,\, \,\, \, \, +U_0, 
 \nonumber
\end{eqnarray}
where:
\begin{eqnarray}
\hspace{-0.9in}&& U_3 \, \, = \, \, \, 
 5\,{x}^{6}\, -4\, \left( a+1 \right)  \left( A+1 \right) \cdot \,  {x}^{5}
+ \left( 3\,a{A}^{2}+3\,{a}^{2}A-Aa+3\,A+3\,a \right) \cdot \,  {x}^{4}\, 
 \nonumber \\
\hspace{-0.9in}&& \qquad 
+4\,Aa \left( a+1 \right)  \left( A+1 \right) \cdot \,  {x}^{3}\, 
-Aa \left( 5\,{a}^{2}A+5\,a{A}^{2}+9\,Aa+5\,a+5\,A \right)  \cdot \,  {x}^{2}\, 
 \nonumber \\
\hspace{-0.9in}&& \qquad 
+4\,{A}^{2}{a}^{2} \left( a+1 \right)  \left( A+1 \right) \cdot \,  x
 \,\, \,\,
-3\,{a}^{3}{A}^{3},
 \nonumber 
\end{eqnarray}
\begin{eqnarray}
\hspace{-0.9in}&& U_2 \, \, = \, \, \, 
 -25\,{x}^{8}+ \left( 14\,Aa+ \left( 14+Q \right) a+ \left( 14+q
 \right) A+14+q+Q \right)  \cdot \,   {x}^{7}
 \nonumber \\
\hspace{-0.9in}&& \qquad -2\, \left( 3\,Aa \left( a+A \right)
 +Aa \left( Q+q-29 \right) 
\, +3\,A+qA+3\,a+Qa \right)  \cdot \,  {x}^{6}\,
 \nonumber \\
\hspace{-0.9in}&& \qquad 
-Aa \left( 42\,Aa+42\,A+qA+Qa+42\,a+Q+q+42 \right) \cdot \,  {x}^{5}\,  
\nonumber \\
\hspace{-0.9in}&&\qquad 
+2\,Aa \left( 11\,Aa \left( a+A \right) +2\,Aa \left( q+Q-
2 \right) +11\,a+11\,A+2\,Qa+2\,qA \right) \cdot \,   {x}^{4}
 \nonumber \\
\hspace{-0.9in}&& \qquad \, 
+{A}^{2}\, {a}^{2} \, (30\,Aa+30\,a+30\,A-Qa-qA-Q-q+30) \cdot \,  {x}^{3}
\nonumber \\
\hspace{-0.9in}&& \qquad 
-2\,{A}^{2}{a}^{2}\, \left( 12\,A\, a \, (a+A)
 +A\,a \,(Q+q+25) +Qa+qA+12\,a+12\,A \right) \cdot \,  {x}^{2} 
\nonumber \\
\hspace{-0.9in}&& \qquad 
+{A}^{3}{a}^{3} \, 
( 14\,A\, a\, +14\,a+14\,A\, +Qa\, +qA\, +Q+q+14)  \cdot \,  x 
\,  \,\,
-7\,{A}^{4}{a}^{4},
 \nonumber 
\end{eqnarray}
\begin{eqnarray}
\hspace{-0.9in}&& U_1 \, \, = \, \, \,
-15\,{x}^{8}\, +2\, (2\,Aa+2\,A+2\,a+Qa+qA+Q+q+2) \cdot \,  {x}^{7} \,
 \nonumber \\
\hspace{-0.9in}&&  \qquad 
-2\, \left( Aa \left( Q+q-23 \right) +qA+Qa \right) \cdot \,  {x}^{6}\, 
 \nonumber \\
\hspace{-0.9in}&&\qquad 
-6\,A\,a \, (2\,Aa+2\,A+2\,a+Qa+qA+Q+q+2) \cdot \,  {x}^{5}\,
\nonumber \\
\hspace{-0.9in}&& \qquad 
-4\,A\,a \,\left( A\,a \, (a+A)
 -2\,A\,a\, (Q+q-8)
 -2\,Qa -2\,qA+a+A \right) \cdot \,  {x}^{4}\,  
\nonumber \\
\hspace{-0.9in}&& \qquad 
+2\,{A}^{2}{a}^{2} \,
 (18\,A\,a\,+Qa+qA +18\,a+18\,A +Q+q+18) \cdot \,  {x}^{3}\,
 \nonumber \\
\hspace{-0.9in}&& \qquad 
-6\,{A}^{2}{a}^{2}
 \left( 2\,A\,a \, (a+A)
 +A\,a\, (Q+q+5)\, +2\,A+2\,a+Qa+qA \right) \cdot \,  {x}^{2}\,
 \nonumber \\
\hspace{-0.9in}&& \qquad 
+2\,{A}^{3}{a}^{3} \, (2\,Aa+2\,A+2\,a+Qa+qA+Q+q+2) \cdot \,  x
\, \,\,
-{A}^{4}{a}^{4},
 \nonumber 
\end{eqnarray}
\begin{eqnarray}
\hspace{-0.9in}&& U_0 \, \, = \, \, \,
{x}^{7}\,-q\,Q\,{x}^{6} \, 
- \, (-{q}^{2}A \,+3\,Aa \,-a{Q}^{2})  \cdot \,  {x}^{5}\, 
 \nonumber \\
\hspace{-0.9in}&& \qquad 
-a\, A \, (2\,qA\, +2\,q\, +2\,Qa\, +2\,Q\, -qQ) \cdot \,  {x}^{4}\,
 \nonumber \\
\hspace{-0.9in}&&\qquad  
+\, A\, a \left( 3\,A\,a \,(2\,Q+2\,q+1)\,
 +6\,Qa \, +6\,qA-2\,a{Q}^{2}-2\,{q}^{2}A \right) \cdot \,  {x}^{3}\,
  \nonumber \\
\hspace{-0.9in}&&\qquad 
-{A}^{2}{a}^{2} \, (6\,qA\,+6\,q\,+6\,Qa\,+6\,Q\,-qQ) \cdot \,  {x}^{2}\, 
 \nonumber \\
\hspace{-0.9in}&&\qquad 
+{A}^{2}{a}^{2} \left(A \, a \, (2\,Q+2\,q-1) \, 
+2\,Qa+2\,qA \,+a{Q}^{2} \,+{q}^{2}A \right) \cdot \,  x 
\,  \,\,
-q \,Q \, {a}^{3}{A}^{3}. \nonumber 
\end{eqnarray}
This order-four operator satisfying the 
Calabi-Yau condition (its exterior square is of order five).
The solution of this order-four operator, analytic at $\, x \, = \, 0$
reads:
\begin{eqnarray}
\hspace{-0.95in}\, y_0(x) \,  = \, \,  1 \,\,\, 
 +{\frac { Q \,q  }{Aa}} \cdot \, x 
\,\, \,
+\,\,{\frac { \left( 2\,qa+2\,q+{q}^{2}-a \right)\,
  \left( 2\,QA+2\,Q+{Q}^{2}-A \right) }{16 \, {a}^{2}{A}^{2}}} \,\cdot x^2 
\,  \,  + \, \,  \cdots, 
\nonumber 
\end{eqnarray}
the nome reads
\begin{eqnarray}
\hspace{-0.3in}\qquad    x \,\, \, \,\, 
 +{\frac { (Qa \, +qA \, +Q \, +q \, -4\,qQ) }{Aa}} \cdot \, x^2
\,\,  \, \,  \, + \, \cdots, 
\end{eqnarray}
and the first terms of its Yukawa coupling $\, K(x)$ read:
\begin{eqnarray}
\hspace{-0.95in}K(x) \, = \, \,   \, 1  \,  \,
+{\frac { (1 \,  -3\,Q-3\,q  \,+a+A \, +10\,qQ+Aa-3\,qA-3\,Qa) }{Aa}} \cdot \, x 
\,  +  \, \cdots 
\end{eqnarray}

 Even inside this restricted set of  HeunG functions 
solutions  of the form $\, HeunG(a, \, q, \, 1, \, 1, \,  1, \, 1; \, x)$
it is hard to find exhaustively the values of the two parameters
$\, a$, and of the accessory parameter $\, q$,
such that the series $\, HeunG(a, \, q, \, 1, \, 1, \,  1, \, 1; \, x)$
is globally bounded. The HeunG function 
 $\, HeunG(a, \, q, \, 1, \, 1, \,  1, \, 1; \, b \, x)$
becomes a series $\, 1 \, +  N_1 \, x \, + N_2 \, x^2 \, + \, \cdots \, $
with integer coefficients $\, N_1$, $\, N_2$, ..., for 
\begin{eqnarray}
\hspace{-0.3in}a \, = \, q \cdot \, {{(4 \, N_2 \, -N_1^2) \cdot \, q \, -2 \, N_1^2
 } \over { N_1^2 \cdot \, (2\, q \, -1)}},
 \quad \quad \,  \,\, \,\, \, b \, = \, \, N_1 \cdot \, {{a} \over {q}}.
\end{eqnarray}
These are necessary conditions. One can find some other necessary conditions
on the integer coefficients of the series. Besides $\, q= \, 1/2$,
 $\, a \, = \, \, 2 \, q$,  $\, a \, = \, \,1/3\,q\,(13\,q \,-6)/(2\,q\,-1)$
(i.e. $\, N_2/N_1^2 \, = \, \, 5/4$ or $\, N_2/N_1^2 \, = \, \, 4/3$)  
that require some specific analysis, one finds that one has necessarily 
the following relation among the first four coefficients  
$\, N_1$, $\, N_2$, $\, N_3$, $\, N_4$, namely:
\begin{eqnarray}
\label{N1N2N3N4}
\hspace{-0.6in}&& 27 \cdot \, N_3^2\, \, \,  \,\, 
  +10 \, \cdot \,   (N_1^2 -9\, N_2) \cdot  \, N_1 \cdot  \, N_3
 \,\,  \,  \, \, 
+8 \cdot \, (4\, N_1^2 -3\, N_2) \cdot \, N_4 \, 
  \nonumber \\
\hspace{-0.6in}&& \qquad \quad  \quad \, 
 -9 \cdot \, N_2^2 \cdot \, (N_1^2 -6\, N_2)
 \, \,\,  \, \,  = \, \,\, \, \,   \, 0. 
\end{eqnarray}
One verifies easily that (\ref{N1N2N3N4}) is actually verified for 
(\ref{112}) and (\ref{11253}).  

Note that the ratio $\, N_p/N_1^p$ are rational expressions of $\, a$ and $\, q$,
invariant by $\, (a, \, q) \, \rightarrow \, (1/a, \, q/a)$,  
of the form $ \, P(a,\, q)/q^p$, where $ \, P(a,\, q)$ is a polynomial. 

However, finding the values of the
 accessory parameter $\, q$, such that 
$\, HeunG(a, \, q, \, 1, \, 1, \,  1, \, 1; \, x)$
is globally bounded remains difficult. Is it possible to finds
such values of the accessory parameter $\, q$ for any given integers
 $\, N_1$ and  $\, N_2$ ? 
For instance for  $\, N_1 \, = \, 2$, $\, N_2 \, = \, 10$
the series with integer coefficients
 $\, HeunG(-8, \, q, \, 1, \, 1, \,  1, \, 1; \, 8 \, x)$ 
is actually a series with integer coefficients for
 $\, q \, = \, -2$. 

Furthermore it is difficult to find  the values of the two parameters
$\, a$ and $\, q$ such that the order-two operator having 
$\, HeunG(a, \, q, \, 1, \, 1, \,  1, \, 1; \, x)$ 
as a solution is globally nilpotent. 
For instance, if one restricts to $\, a \, = \, 9/8$, it is hard to 
show that the only rational number value of the accessory parameter $\, q$
yielding global nilpotence is $\, q= \,3/4$. For $\, q$ a rational number 
one gets an infinite set of 
divisibility conditions. The prime $\, 5$ must divide the numerator
of $\,(q+3)\, (q+1)\, (q+2)\, (q^2+4\,q+1)$, the prime $\, 7$ must
 divide the numerator
of $\,(q^2+6\,q+6)\,(q+6)\,(q+1)\,(q+4)\,(q^2+4\,q+1)$,  etc. 
With $\, q \, = \, N/D$ where $\, N$ and $\, D$ integers $\, < 1000$
and with the  conditions emerging from the $p$-curvature calculations 
for the  first primes $\, \le 42$. One finds that $\, q= \,3/4$
is the only value of the accessory parameter corresponding to 
global nilpotence.

Restricting to  HeunG functions 
solutions  of the form 
$\, HeunG(a, \, q, \, 1, \, 1, \,  1, \, 1; \, 60 \, a \, x)$
for integer values of $\, a$ and $\, q$ it is hard to find 
the integer values of $\, a$ and $\, q$ such that the corresponding
 series are globally bounded.

\section{Yukawa coupling of Calabi-Yau ODEs}
\label{YukawaHmn}

Let us give the expansion of the Yukawa coupling 
for a set of other $\, H_{m,n}$ that are Calabi-Yau:
 in particular
they are MUM and their exterior squares are of order {\em five}
(the ``Calabi-Yau condition'').

For $\, H_{4,8}$ the Yukawa coupling reads: 
\begin{eqnarray}
\label{YukuM48}
\hspace{-0.9in}&&K(q) \, \,\, \,= \,\, \,K^{\star}(q) 
\, \, \,= \,\, \, \, \,\,  \,1 \,\,\, \, + \, 16 \cdot \,q\,\,  
 +  352 \cdot \, q^2 \,\, +33280 \cdot \, q^3 \,\,  + \,2058528 \cdot \, q^4 \,\,  
   \,\,  \nonumber \\
\hspace{-0.9in}&& \quad  \quad  \quad   + \, 123766016 \cdot \, q^5 \,\, 
 + \,7347718144 \cdot \, q^6 \,
+ \, 439489011712 \cdot \, q^7 \,\,  
 \nonumber \\
\hspace{-0.9in}&&  \qquad  \quad  \quad  \quad 
+ \, 26579639900960 \cdot \, q^8 
\,\,  + \, 1616513123552128\cdot \, q^9
 \,\,\,\,   + \, \, \cdots,  
\end{eqnarray}
which is Number 36 in Almkvist et al. large tables of 
Calabi-Yau ODEs~\cite{TablesCalabi}.

For $\, H_{4,9}$  the Yukawa coupling reads: 
\begin{eqnarray}
\label{YukuM49}
\hspace{-0.9in}&&K(q)\, \,\, \,= \,\, \,K^{\star}(q) 
 \, \,= \,\, \,\, \,\, \,1 \,\,\, \, + \, 12 \cdot \,q\,  \,
 -  324 \cdot \, q^2 \, \,-29544 \cdot \, q^3 \,\,\,  -1314756 \, \cdot \, q^4 \,\,  
    \nonumber \\
\hspace{-0.9in}&&\quad  \quad \quad  \,-12971988  \cdot \, q^5 \,\,
 + \,2033927928 \cdot \, q^6 \,\, + \, 146587697352 \cdot \, q^7 \,\, 
   \nonumber \\
\hspace{-0.9in}&& \qquad  \quad  \quad \quad  \, + \, 4172739566652 \cdot \, q^8
 \,\,  \,-77469253445544 \cdot \, q^9 \,\, 
\,\, \,  + \, \, \cdots, 
\end{eqnarray}
which is Number 133 in tables~\cite{TablesCalabi}.

For $\, H_{6,6}$ it reads: 
\begin{eqnarray}
\label{YukuM66}
\hspace{-0.9in}&&K(q)\, \,\, \,= \,\, \,K^{\star}(q)
  \, \,= \,\, \, \,\, \,1 \,\,\, \, + \, 37 \cdot \,q\,  
 -4523 \cdot \, q^2 \, + 412327 \cdot \, q^3 \,\,  -33924139 \, \cdot \, q^4 \,\,  
 \,\,    \nonumber \\
\hspace{-0.9in}&&\qquad \quad \, + 2662557912 \cdot \, q^5
 -203154013049 \, \cdot \, q^6 \,\,
 + \, 15217617773948 \cdot \, q^7 \,\,  
 \nonumber \\
\hspace{-0.9in}&& \qquad \quad  \qquad  \, -1125153432893483 \,  \cdot \, q^8
 \, \, + 82390368380951296 \cdot \, q^9 
\,\, \, \,  + \, \, \cdots,  
\end{eqnarray}
which is Number 144 in tables~\cite{TablesCalabi}.

For $\, H_{6,8}$ it reads: 
\begin{eqnarray}
\label{YukuM68}
\hspace{-0.9in}&&K(q)\, \,\, \,= \,\, \,K^{\star}(q)
  \, \,= \,\, \, \,\, \,1 \,\,\, \,+ \, 24 \cdot \,q\,  
 -2012\cdot \, q^2 \, + 139056 \cdot \, q^3 \,\,  -8227932 \, \cdot \, q^4 \,\,   
  \nonumber \\
\hspace{-0.9in}&& \qquad\,\quad  + 468328024 \cdot \, q^5
 \,\,   -25856580632 \, \cdot \, q^6 \,\, 
+ \,1402012096656  \cdot \, q^7 \,\,  
 \nonumber \\
\hspace{-0.9in}&& \qquad \quad  \qquad  \, -74994891745116 \,  \cdot \, q^8 
\,\,  \, + 3972880128014736 \cdot \, q^9 
\,\,\,   + \, \, \cdots,  
\end{eqnarray}
which is Number 176 in tables~\cite{TablesCalabi}.

For $\, H_{6,9}$ it reads: 
\begin{eqnarray}
\label{YukuM69}
\hspace{-0.9in}&&K(q)\, \,\, \,= \,\, \,K^{\star}(q) 
 \, \,= \,\, \, \,\, \,1 \,\,\, + \, 18 \cdot \,q\,  
 -1674 \cdot \, q^2 \, +  88209 \cdot \, q^3 \,\,  -4801770 \, \cdot \, q^4 \,\,   
 \nonumber \\
\hspace{-0.9in}&& \qquad\,\quad  + 239279643 \cdot \, q^5 \,\,   
 -11680323039 \, \cdot \, q^6 \,\,
 + \,558685593414  \cdot \, q^7 \,\,   
 \nonumber \\
\hspace{-0.9in}&& \qquad \quad  \qquad  \, -26379917556714 \,  \cdot \, q^8 
\,\,  \, + 1233104626297710 \cdot \, q^9
 \,\, \,\,   + \, \, \cdots,  
\end{eqnarray}
which is Number 178 in tables~\cite{TablesCalabi}.

For $\, H_{8,8}$: 
\begin{eqnarray}
\label{YukuM88}
\hspace{-0.9in}&&K(q)\, \,\, \,= \,\, \,K^{\star}(q) 
 \, \,= \,\, \, \,\, \,1 \,\,\, + \, 16 \cdot \,q\,  
 -864 \cdot \, q^2 \, + 47104 \cdot \, q^3 \,\,  -1890528 \, \cdot \, q^4 \,\,    
  \nonumber \\
\hspace{-0.9in}&& \qquad \quad  \, +  80502016 \cdot \, q^5 \,\, 
  -3118639104 \, \cdot \, q^6 \,\,
 + \,123287486464  \cdot \, q^7 \,\,  
  \nonumber \\
\hspace{-0.9in}&& \qquad\quad  \qquad  -4691784791264 \,  \cdot \, q^8
 \,\,   \, + 179585946086272 \cdot \, q^9
 \,\, \,  \,  + \, \, \cdots,  
\end{eqnarray}
which is Number 107 in tables~\cite{TablesCalabi}.

For $\, H_{8,9}$ the Yukawa coupling reads: 
\begin{eqnarray}
\label{YukuM89}
\hspace{-0.9in}&&K(q)\, \,\, \,= \,\, \,K^{\star}(q) 
 \, \,= \,\, \, \,\, \,1 \,\,\, + \,  12 \cdot \,q\,  
 -756 \cdot \, q^2 \, + 27192 \cdot \, q^3 \,\,  -1144644 \, \cdot \, q^4 \,\,   
  \nonumber \\
\hspace{-0.9in}&& \qquad \quad \quad  \, +  39948012 \cdot \, q^5 \,\,  
-1377082728 \, \cdot \, q^6 \,\, + \, 47882164776 \cdot \, q^7 \,\,   
 \nonumber \\
\hspace{-0.9in}&& \qquad \quad  \qquad  \, -1608623259588 \,  \cdot \, q^8 \,\,   \,  
+ 53732432848152 \cdot \, q^9
 \,\,\, \,   + \, \, \cdots,  
\end{eqnarray}
which is Number 163 in tables~\cite{TablesCalabi},
and for $\, H_{9,9}$ it reads: 
\begin{eqnarray}
\label{YukuM99}
\hspace{-0.9in}&&K(q)\, \,\, \,= \,\, \,K^{\star}(q) 
 \, \,= \,\, \,\, \,\, \,1 \,\,\,\, + \,  9 \cdot \,q\,  
 -567 \cdot \, q^2 \, + 20205  \cdot \, q^3 \,\,  -615735 \, \cdot \, q^4 \,\,   
  \nonumber \\
\hspace{-0.9in}&& \qquad \quad \quad  \, + 19431009 \cdot \, q^5 \,\,  
-608213043 \, \cdot \, q^6 \,\, + \,18406651167  \cdot \, q^7 \,\,  
  \nonumber \\
\hspace{-0.9in}&&  \qquad \quad  \qquad  \,-542566460727 \,  \cdot \, q^8 
\,\,\, + 15865350996861 \cdot \, q^9 \, \,
 \,\,\,  + \, \, \cdots, 
\end{eqnarray}
which is Number 165 in tables~\cite{TablesCalabi}.

\vskip .1cm

\section{Modular form character of $\, _2F_1([1/6,\,1/6],\, [1], \, x)$ }
\label{Special}

The modular form character of (\ref{forminv}) is clear
on the remarkable (and intriguing) identity (58) (or (59)) 
in Maier~\cite{Maier1}:
\begin{eqnarray}
\hspace{-0.6in}&&_2F_1\Bigl([{{1} \over {6}}, \, {{1} \over {6}}],
\, [1], \, {\cal P}_1\Bigr) \\
\hspace{-0.6in}&& \qquad \quad \quad   = \,\, \,  \, \, 2 \cdot \,
 \Bigl({{(x+60)\, (x+80)\, (x+96) } \over {(x+48)\, (x+120)^2 }} \Bigr)^{-1/6}
\cdot \, 
_2F_1\Bigl([{{1} \over {6}}, \, {{1} \over {6}}],\, [1], \, {\cal P}_2\Bigr), 
\nonumber 
\end{eqnarray}
where the two pullbacks  read 
respectively ($\circ$ denotes the composition of functions)
\begin{eqnarray}
\hspace{-0.6in}&&{\cal P}_1 \, \, = \, \,\, \,\,   {{-1} \over {432}} \cdot \, 
{{ x \, (x + 60)^2 \, (x + 72)^2 \, (x + 96) } \over {
(x + 48) (x + 80) (x + 120)^2 }} \\
\hspace{-0.6in}&&\qquad \quad  \,  = \,\,   \, \,\,  
{{ \, x} \over {x-1}}  \circ  {{1728 \, x} \over {(x\, +16)^3}}  \circ  
 G \,\,\, \,  \,\,    = \,  \,\, \, \,
{{ x} \over {x-1}}  \circ   {{1728 \,  x^2} \over {(x\, +256)^3}}   \, \circ \,
 {{64^2} \over {G}}, 
\nonumber \\
\hspace{-0.6in}&&{\cal P}_2 \, \, \,\,  = \,\,\, \,    
 {{-1} \over {432}} \cdot \, 
{{ x^2  (x + 48)^2  (x + 72) (x + 120)
 } \over {
 (x + 60) \, (x + 80)^2 \,  (x + 96)^2}} \\
\hspace{-0.6in}&&\qquad \quad \,  = \, \,  \,\, \, \, 
{{\, x} \over {x-1}}  \circ 
 {{1728 \, x} \over {(x\, +16)^3}}  \circ  
  {{ 64^2} \over {H}} \,\,\, \,   = \,\, \,\, \, \, 
{{ x} \over {x-1}}  \circ   {{1728 \, x^2} \over {(x\, +256)^3}}   \, \circ \,  H, 
\nonumber
\end{eqnarray}
where
\begin{eqnarray}
\hspace{-0.5in}G \,\, = \, \,  \, 
 {{8\, x \cdot \, (x+96)} \over {(x + 72) (x + 60) }}, 
\quad \quad \quad \,
 H \, = \, \,  \,  {{-64 \,x  \cdot  \, (x+48)} \over { (x+72) \, (x+120)}},
\end{eqnarray}
The algebraic relation between these two pullbacks
 $\, (u, \, v) \, = \, \, ({\cal P}_1, \, {\cal P}_2)$,
corresponds to the
(genus zero) {\em modular curve}:
\begin{eqnarray}
\hspace{-0.8in}&&
6912\,\, u\, v \cdot \, (u^4\, v^4+1) \,\,\,\,\,
 -3\, u\, v\cdot \,
 \Bigl(80\, (u^2+\, v^2) \, +529999\, u\, v\Bigr) \cdot \, (v^2\, u^2+1)
\nonumber  \\
\hspace{-0.8in}&& \quad \quad 
+16\, (u+v)\cdot \, (u\, v+1) \cdot \,
 (u^2+v^2 \, -241\, u\, v) \cdot\, (v^2\, u^2 \, -u\, v \, +1)
\nonumber \\
\hspace{-0.8in}&&  \quad \quad 
-6\, u\, v\cdot \, (u+v)\cdot \, (u\, v+1)\cdot \,
 \Bigl(640\, (u^2+v^2) \, -652959\, u\, v\Bigr)
\nonumber \\
\hspace{-0.8in}&&  \quad \quad 
+u\, v \cdot \,
 \Bigl(6912\, (u^4+v^4)\, -1589997\, (u^3\, v + v^3\, u)
 \, +21958300\, v^2\, u^2\Bigr)
\,\, \,\,  = \, \, \,\,\,  \, 0. \nonumber
\end{eqnarray}
Other identities are worth noting on these pullbacks, for instance:
\begin{eqnarray}
\hspace{-0.7in}&&{{-4 \, x} \over {(1-x)^2}} \, \circ \, {\cal P}_1 
\, \, = \,\,  \, \, \,
 {{1728 x} \over {(x\, +16)^3}}  \, \circ \, F 
\,\,  \,\,  = \,\,  \, \, \,\,   
 {{1728 x^2} \over {(x\, +256)^3}}   \, \circ \, {{4096} \over {F}}, 
\quad \quad  \quad  \hbox{and:}
\nonumber \\
\hspace{-0.7in}&&{{-4 \, x} \over {(1-x)^2}} \, \circ \, {\cal P}_2
 \, \, = \, \, \, \, \, 
 {{1728 x^2} \over {(x\, +256)^3}}   \, \circ \, F
 \,\,\,   \, = \, \, \, \,\, \, 
  {{1728 x} \over {(x\, +16)^3}}  \, \circ \,  {{4096} \over {F}}, 
\quad \quad \, \,\, \hbox{where:}
\nonumber \\
\hspace{-0.7in}&& \quad \quad \quad \quad  F \,\,  \, = \, \, \,  \,  \,
 {{x \cdot \, (x+80)\, (x+48) \, (x+96) } \over {(x+120) \, (x+72) \, (x+60) }}.
\nonumber
\end{eqnarray}

It can also be
illustrated through the identity:
\begin{eqnarray}
\label{forminvbis}
\hspace{-0.8in}&&_2F_1\Bigl([{{1} \over {6}},\, {{1} \over {6}}],
\, [1], \, \, {{27 \, x } \over { \, (1\, + \, 8\, x)^2 \, (1\, -x) }} \Bigr) \\
\hspace{-0.8in}&& \quad  \, \,  \, \, \,\,  \, \, \, \, 
= \, \,\,  \, \, 
 {{(1\, - \, x)^{1/6} \, (1\, + \, 8\, x)^{1/3} } \over {
(1\, + \, 7\, x \, + x^2)^{1/4}}}   \cdot \, 
_2F_1\Bigl([{{1} \over {12}},\, {{5} \over {12}}],\, [1], \, 
 {{27 \, x^2 \,(1\, -x)^2 \, (1\, + \, 8\, x)  } \over { 
4 \, (1\, + \, 7\, x \, + x^2)^3 }}\Bigr) \nonumber \\
\hspace{-0.8in}&& \quad  \, \,  \, \, \,\,  \, 
\, \, \, = \, \,\,\,   \,
 {{(1\, - \, x)^{1/6} \, (1\, + \, 8\, x)^{1/3} } \over {
(1\, - \, 4\, x)^{1/2}}}   \cdot \, 
_2F_1\Bigl([{{1} \over {12}},\, {{5} \over {12}}],\, [1], \, 
 {{-108 \, x \,(1\, -x) \, (1\, + \, 8\, x)^2  } \over { 
(1\, - \, 4\, x)^6 }}\Bigr),  \nonumber 
\end{eqnarray}
where the last two pullbacks are related 
by the fundamental modular 
curve~\cite{bo-ha-ma-ze-07b,CalabiYauIsing1,Maier1} (corresponding 
to $\, \tau \, \rightarrow \, 2 \, \tau$),
as can be deduced from the identities on these two pullbacks:
\begin{eqnarray}
\hspace{-0.7in}&&\quad {{27 \, x^2 \cdot \,(1\, -x)^2 \, (1\, + \, 8\, x)  } \over { 
4 \, (1\, + \, 7\, x \, + x^2)^3 }} \, \, \, \, \, = \,  \, \, \,\, \, 
{{1728 x^2} \over {(x+256)^3 }} \, \circ \, {{16 \, (1+8\, x)} \over { (x-1) \, x}},
\\
\hspace{-0.7in}&&  -\, {{108  \, x \cdot \,(1\, -x) \, (1\, + \, 8\, x)^2  } \over { 
(1\, - \, 4\, x)^6 }}\, \, \, \, = \, \, \, \, \, \, 
 {{1728 x} \over {(x+16)^3 }} \, \circ \,
 {{16 \, (1+8\, x)} \over { (x-1) \, x}}.
\end{eqnarray}

\vskip .1cm 
The relation between the hypergeometric 
function $_2F_1([1/6,1/6],[1],x)$ 
and  $_2F_1([1/12,5/12],[1],x)$, can also be 
understood from the Kummers's quadratic relation on 
$\, _2F_1([1/6,5/6],[1],x)$:
\begin{eqnarray}
\hspace{-0.4in}_2F_1\Bigl([{{1} \over {6}},\, {{5} \over {6}}],
\, [1], \, \,x\Bigr) \, \, \,  \, = \, \,  \,  \, \, \,
 _2F_1\Bigl([{{1} \over {12}},\, {{5} \over {12}}],
\, [1], \, \,4 \, x \cdot \, (1-x)\Bigr). 
\nonumber  
\end{eqnarray}

\vskip .1cm

\section{Calabi-Yau condition versus integrality: 
a Saalschutzian hypergeometric family of operators}
\label{CalabiYaucond}

The Calabi-Yau condition that the exterior square of an order-four 
linear differential operator is of order five is
a fundamental condition defining 
Calabi-Yau ODEs~\cite{Almkvist,Batyrev,TablesCalabi}.
Let us consider the following {\em Saalschutzian hypergeometric function} 
\begin{eqnarray}
\label{Saal}
\hspace{-0.1in}&&_4F_3([b  \,- c  \,+ d,\, b,\,  c,\,  d ],
 \, [e,\, b  \,+ d, \, 1  \,+ b +  \,d  \,- e ], \, x), 
\end{eqnarray}
and, let us introduce the order-four linear differential operator
$\, M_4(b, \, c, \, d, \, e)$  which annihilates
this hypergeometric function.
It is a straightforward exercise to verify that this linear differential operator 
satisfies the Calabi-Yau condition: {\em its  exterior square 
of is actually of order five  
for any values of the parameters} $\, b, \, c, \, d, \, e $. This 
 order-four operator is almost self-adjoint\footnote[1]{If one denotes by $\, N$ 
the denominator of the rational number
 $\,b + d -1$ , the symmetric $\, N$-th power 
of  $\, M_4(b, \, c, \, d, \, e)$ is up to a normalisation, self-adjoint.}:
\begin{eqnarray}
\label{selfad}
\hspace{-0.9in}&&\quad \quad   \, \, \, x^{b + d -1} \cdot \, 
 \, M_4(b, \, c, \, d, \, e)
\, \, \, = \, \, \, \, adjoint(M_4(b, \, c, \, d, \, e))
  \cdot \,  x^{b + d -1}. 
\end{eqnarray}
For generic rational parameters, the series expansion of the 
{\em Saalschutzian} hypergeometric function (\ref{Saal}) 
is {\em not globally bounded}.

Let us restrict to the condition $\,\, d \,\, = \,\, 1 \, -b$:
\begin{eqnarray}
\label{Saal2}
\hspace{-0.1in}&&\qquad _4F_3([1  \,- c,\, b,\,  c,\,  1 \, -b ], 
\, [e,\, 1, \,  \,2   \,- e ], \, x). 
\end{eqnarray}
The order-four operator is still such that  its  exterior square
 is  of order five,
but it is now a {\em self-adjoint operator} (see (\ref{selfad})). Again, 
one sees that,
for generic rational parameters, the series expansion of the 
Saalschutzian hypergeometric function (\ref{Saal2}) 
is {\em not globally bounded}.
Actually the four solutions are not of a MUM form. Besides the 
analytic at $\, x \, = \, \, 0$, non globally bounded,
 solution (\ref{Saal2})
\begin{eqnarray}
\hspace{-0.4in}&&\quad \quad 1 \, \, \,  \, \, \, \,
+{\frac {b\, c\, d \cdot \, (b-c+d)}{e \cdot \,
 (b\, +d)  \cdot \, (1\, +b \, +d -e) }} \cdot \, x
 \, \,\, \, \,  \, \, + \, \, \,  \cdots,  
\end{eqnarray}
the three other solutions have {\em no logarithm} and are, for generic 
rational parameters, {\em Puiseux series}:
\begin{eqnarray}
\hspace{-0.9in}&&\quad x^{1-e}\cdot \Bigl(1 \,\,\,\, -{\frac {(1\, +d-e)  \, ( 1 \, +c-e)
  \, (1\,+b -e)  \, (1 \, +b-c+d-e) }{ (e-2)  \, (1 \, +b+d -e) 
 \, (b +d +2 -2\,e) }} \cdot x \,\,\, + \, \cdots \Bigr), 
\nonumber \\
\hspace{-0.9in}&&\quad \quad  \quad \quad \quad x^{1-b-d}\cdot (1 \, \, + \, \cdots),
 \qquad 
\quad \quad x^{e-b-d}\cdot (1 \, + \, \cdots). \nonumber 
\end{eqnarray}
When the exponents $\, 1-e$, $\,1-b-d$, $\, e-b-d$ are integer values
one can recover solutions with logarithms for the operators.

\section{Integrality of a one-parameter Saalschutzian hypergeometric 
family of operators}
\label{appendO}

Let us consider
 the order-four linear differential  operator ($\theta \, = \, \, x \cdot D_x$):
\begin{eqnarray}
\label{theexample}
\hspace{-0.6in}&&\quad \, \,\, M_4(\mu) \, \,\, \, = \, \, \,\, \, \,
 16 \cdot \theta^2 \cdot \, (\theta\, -1)^2
\,\, \,\,\,\, \\
\hspace{-0.6in}&& \qquad \qquad \qquad \quad -\, x \cdot \, (2\, \theta\, +1)^2 
\cdot \, (2\, \theta\, -1 \, +\mu) 
\cdot \, (2\, \theta\, -1\, -\mu). \nonumber 
\end{eqnarray}
We have a one-parameter family of operators depending on $\, \mu^2$.

The Wronskian of $\, M_4(\mu)$ is independent of $\, \mu$,
 and reads $\, 1/(x-1)^2/x^4$.
This order-four operator is, {\em for any rational value of } $\, \mu$,
non-trivially homomorphic to its adjoint 
(with order-two intertwiners):
\begin{eqnarray}
\label{defL2}
\hspace{-0.9in}&&\,\qquad   adjoint(M_4(\mu)) \cdot (x-1) \, \cdot \, L_2(\mu) 
\, \,\,  = \, \,\,  \, \, L_2(\mu) \cdot \, (x-1) \cdot \, M_4(\mu),
\end{eqnarray}
\begin{eqnarray}
\label{defM2}
\hspace{-0.9in}&&\, \qquad  M_4(\mu) \cdot \, (x-1) \, \cdot \, M_2(\mu) 
\, \,\,  = \, \,\,  \,
  \, M_2(\mu)  \cdot \, (x-1) \, \cdot \, adjoint(M_4(\mu)),
\end{eqnarray}
where $\, L_2(\mu)$ and  $\, M_2(\mu)$ are two {\em self-adjoint order-two}
linear differential  operators:
\begin{eqnarray}
\hspace{-0.9in}&& \quad \quad \quad  L_2(\mu) \, \,\,\, \, = \, \, \,\,\,\, 
 \,D_x^2 \, \, \, \,\, 
- \, {{1 \, - \, \mu^2} \over {4 \, \mu^2}}
 \cdot \,  {{1 \, -\mu^2 \, x} \over {  (x-1) \cdot \, x^2}}, 
 \\
\hspace{-0.9in}&& \quad  \quad  \quad M_2(\mu) \,\,\, \, = \, \, \,\,\, \, 
 16 \cdot \, x^4 \,  \cdot \,D_x^2 \, \,\,\,\,\,
 + 64 \cdot \, x^3 \,  \cdot \,D_x \,\,\,\, 
+ \,  \, 36  \, x ^2 \, + \,  \,{{4 \, x^2 } \over { (x-1) \cdot \, \mu^2 }}. 
\nonumber
\end{eqnarray}
The order-two operator  $\, M_2(\mu)$  has simple hypergeometric solutions:
\begin{eqnarray}
\hspace{-0.9in}&& \quad \quad \qquad (1\, -x)  \, \cdot \, \, x^{-(3 \, \mu \, +1)/2/\mu} \cdot \, \,
 _2F_1\Bigl([{{2\, \mu \, -1} \over {2\, \mu }}, \,{{2\, \mu \, -1} \over {2\, \mu }}], \,
 [{{\mu \, -1} \over {\mu }} ], \, x\Bigr), 
 \nonumber \\
\hspace{-0.9in}&& \quad \qquad \quad  (1\, -x) \,  \cdot \, x^{-(3 \, \mu \, -1)/2/\mu} \cdot \, \,
 _2F_1\Bigl([{{2\, \mu \, +1} \over {2\, \mu }}, \,{{2\, \mu \, +1} \over {2\, \mu }}], \,
 [{{\mu \, +1} \over {\mu }} ], \, x\Bigr).
\nonumber 
\end{eqnarray}
note that, generically, the hypergeometric function
$ \, _2F_1([{{2\, \mu \, -1} \over {2\, \mu }}, \,{{2\, \mu \, -1} \over {2\, \mu }}], \,
 [{{\mu \, -1} \over {\mu }} ], \, x)$ 
{\em does not corresponds to a globally bounded series},
as can be seen,  for instance,  for $\, \mu \, = \, -5/11$.

The two intertwining relations (\ref{defL2}) and (\ref{defM2})
 can, in fact, be seen as a straight consequence of the fact that the 
order-four operator $\,M_4(\mu)$ can remarkably be  
written in terms of these two self-dual 
operators:
\begin{eqnarray}
\label{forthcoming}
 \hspace{-0.95in}&& \quad  \quad \qquad  M_4(\mu) 
\, \,\,\, \, = \, \, \,\,\,\, 
  M_2(\mu) \cdot \, (1\, -x)  \cdot \,
 L_2(\mu)  \, \, \, \,+ \, \, \,\,\,  \,
 {{ (\mu^2 \, -1)^2 } \over {\mu^4 \cdot \, (1\, -x)  }}.
\end{eqnarray}

This order-four operator (\ref{theexample}), or (\ref{forthcoming}),
 {\em does not satisfy the Calabi-Yau condition}. 
Its exterior square $\, {\cal M}_6(\mu)$ is of {\em order six}, 
with a, not only {\em rational function} solution, 
but a {\em constant solution}.
It is actually the direct sum of $\, D_x$ 
and of an order-five linear differential operator $\, {\cal M}_5(\mu)$:
\begin{eqnarray}
\label{1plus5}
\hspace{-0.9in}&& \quad \quad \quad \quad  {\cal M}_6(\mu) 
\,  \,\,\, \, \,  = \, \, \,\,\, \,  \,
  D_x \, \oplus \, {\cal M}_5(\mu), 
 \qquad \quad \quad  \hbox{where:} \\
\hspace{-0.95in}&& 
{\cal M}_5(\mu)\,\, \, \,  = \, \, \,\,\, \, 
{\cal M}_5(0) \, + \, \mu^2 \cdot \, {\cal M}_5^{(2)}
 \, + \, \mu^4 \cdot \, {\cal M}_5^{(4)} 
 \, + \, \mu^6 \cdot \, {\cal M}_2^{(6)} \,\cdot \, D_x \,  
+ \, \mu^8 \cdot \, x^3 \cdot \, D_x.
 \nonumber 
\end{eqnarray}
where the $\,{\cal M}_n^{(m)}$'s are  linear differential 
operators of order $\, n$.
For $\, \mu^2 \, = \, 1$  the order-five operator $\,{\cal M}_5(\mu)$ 
becomes the product\footnote[2]{It is also 
a direct sum  $\, {\hat M}_2  \, \oplus ( {\hat N}_2 \cdot \, D_x)$ 
where $\, {\hat M}_2$ and $\, {\hat N}_2$ 
are two order-two operators.} 
of two order-two operators and of $\, D_x$:
\begin{eqnarray}
\label{reca}
\hspace{-0.9in}&& \quad \quad \quad \quad \qquad \quad 
{\cal M}_5(\pm  \, 1) \,\, \, \,  = \, \, \,\,\, \,
 {\cal N}_2 \cdot \,  {\cal M}_2 \cdot \, D_x. 
\end{eqnarray}
For the other odd values of $\, \mu$ 
 ($\, \mu \, = \, \, \pm \,3, \, \pm \, 5, \, \cdots $)
the order-five operator $\, {\cal M}_5(\mu)$ factorizes 
into the product of an order-two, 
order-one and an order-two operator.

As will be seen in a forthcoming publication, the fact that the 
exterior square has a {\em rational solution} is also 
a consequence of the decomposition (\ref{forthcoming}).
We have the following general result. Any order-four linear differential 
operator of the form
\begin{eqnarray}
\label{forthcoming2}
 \hspace{-0.95in}&& \quad \quad \qquad \qquad 
 M_4 \, \,  \,\,\, \, = \, \, \,\,\, \, 
  M_2 \cdot \, c_0(x)  \cdot \,
 L_2  \, \, \, \,  \,+ \, \, \,\,\,  \,
 {{ \lambda } \over {c_0(x) }},
\end{eqnarray}
where $\, L_2$ and $\, M_2$ are two (general) self-adjoint operators
\begin{eqnarray}
 \hspace{-0.95in}&& \quad \quad  \qquad \quad 
 L_2  \, \, \,  = \, \, \,   \,  \, \, a_2(x) \cdot \, D_x^2 \,  \, 
+ \, \, {{d \, a_2(x)} \over {dx}}  \cdot \, D_x
\,\,  + \, \, a_0(x), \quad \\
 \hspace{-0.95in}&& \quad \quad \qquad  \quad 
  M_2  \, \,  \, = \, \, \, \,   \,  \, b_2(x) \cdot \, D_x^2 \, \, 
 + \, \,  {{d\, b_2(x)} \over {dx}} \cdot \, D_x
\,\,  + \, \, b_0(x),
\end{eqnarray}
is such that its exterior square has $\, A/a_2(x)$ ($A$ is any constant) 
as a solution. In the case (\ref{forthcoming}), the solution is 
the constant solution:  $\, 1/a_2(x) \, = \, \, 1$. Instead of $\, M_4$, 
we can introduce 
\begin{eqnarray}
\label{forthcoming3}
 \hspace{-0.95in}&& \quad \quad  \qquad  
\tilde{M}_4\,\, = \, \,\,\, c_0(x) \cdot \,  M_4
 \, \,  \,\,\, \, = \, \, \,\,\, \, 
 c_0(x)  \cdot \,  M_2 \cdot \, c_0(x)  \cdot \,
 L_2  \, \, \, \,  \,+ \, \, \,\,\,  \,
 \lambda, 
\end{eqnarray}
which annihilates the same solutions as $\, M_4$.
It is worth noting  that a decomposition like (\ref{forthcoming2}), provides, in fact,
interesting results on the spectrum of $\, \tilde{M}_4$.  For simplicity let us  
consider  the example (\ref{theexample}) with its decomposition (\ref{forthcoming}),
or on  $\, \tilde{M}_4(\mu)$, the decomposition 
\begin{eqnarray}
\label{forthcoming4}
 \hspace{-0.95in}&&\qquad   \qquad
\tilde{M}_4(\mu)\,\, = \, \,\,\,\, \,    (1\, -x) 
\cdot \,  M_4 \, \,  \,\,\, \, = \, \, \,\,\, \, 
  \hat{M}_4(\mu) \, \, \, \,  \,+ \, \, \,\,\,  \,
  {{ (\mu^2 \, -1)^2 } \over {\mu^4 }}, 
\end{eqnarray} 
where the order-four operator $\, \hat{M}_4(\mu)$ factors:
\begin{eqnarray}
\label{decomprod}
\hspace{-0.95in}&&  \qquad \qquad
\hat{M}_4(\mu) \,  \,\,\, \, = \, \, \,\,\,\, \, 
(1\, -x)  \cdot \,  M_2(\mu) \cdot \, (1\, -x)  \cdot \, L_2(\mu).
\end{eqnarray}
This order-four operator $\, \hat{M}_4(\mu)$ is, also, (non-trivially)
 {\em homomorphic to its adjoint}:
\begin{eqnarray}
\hspace{-0.9in}&&  \qquad \qquad adjoint(\hat{M}_4(\mu)) \cdot \, L_2(\mu)
 \, \, \,\,\, = \, \, \,\,\, L_2(\mu) \cdot \, \hat{M}_4(\mu).
\end{eqnarray}
It is clear that $\, \mu \, = \,\pm \, 1$ needs to be analysed separately.
For instance, for $\, \mu \, = \, \, 1$, the order-four operator  $\, \hat{M}_4(\mu)$ 
factors in a direct sum $\,  D_x \, \oplus \, M_3$
where the order-three operator $\, M_3$ has the three solutions:
\begin{eqnarray}
\label{decomsum}
\hspace{-0.9in}&&\qquad  9 \, x^2 \cdot \,
 _4F_3\Bigl([1, \, 1, \, {{5} \over{2}}, \, {{5} \over{2}}], \, [ 2, 3, 3], x \Bigr)
 \, \, + \, 16 \, x \cdot \, \ln(x) \,\,  \, -64, 
  \\
\hspace{-0.9in}&&\qquad \qquad \qquad x \cdot \, \int \, {{\pi} \over {2}} \cdot \,
 _2F_1\Bigl([{{1} \over{2}},{{1} \over{2}}], \, [1], \,  1 \, -x \Bigr)
 \cdot \, {{dx} \over{x^2}}, 
\qquad \,  \hbox{and:} \qquad \, \,  x.  \nonumber
\end{eqnarray}

Let us consider the solutions of $\, L_2(\mu)$ which can actually be expressed 
in terms of hypergeometric functions:
\begin{eqnarray}
\hspace{-0.95in}&&  \Psi_1 \, \, = \, \, \,  
 (1\, -x)  \cdot  \, x^{(2 \, \mu \, -\rho)/4/\mu}   \cdot \, 
 _2F_1\Bigl([{{2 \,\mu^2 \, +4\, \mu \, -\rho} \over {4 \, \mu }}, \,
{{2 \,\mu^2 \, -4\, \mu \, +\rho} \over {4 \, \mu }}], \,
 [{{ 2\, \mu \, -\rho} \over { 2 \, \mu }} ], \, x\Bigr), 
 \nonumber \\
\hspace{-0.95in}&&  \Psi_2 \, \, = \, \, \,  
(1\, -x)   \cdot  \, x^{(2 \, \mu \, +\rho)/4/\mu}  \cdot \, 
 _2F_1\Bigl([{{2 \,\mu^2 \, +4\, \mu \, +\rho} \over {4 \, \mu }}, \,
{{-2 \,\mu^2 \, +4\, \mu \, -\rho} \over {4 \, \mu }}], \,
 [{{ 2\, \mu \, +\rho} \over { 2 \, \mu }} ], \, x\Bigr),
\nonumber 
\end{eqnarray}
where $\, \rho$ reads 
$\,\,\, \rho \,  \,\,\, \, = \, \, \,\,\, 2 \cdot \, (2 \, \mu^2 \, -1)^{1/2}$.
One immediately deduces from the decomposition (\ref{forthcoming4}), 
that  $\, \Psi_1$
and  $\, \Psi_1$ are {\em two eigenfunctions\footnote[1]{See the  spectral theory of 
ordinary linear differential equations, in particular 
for self-adjoint operators.} of the order-four operator} $\, \tilde{M}_4(\mu)$,
with the {\em same eigenvalue} $\,(\mu^2 \, -1)^2/\mu^4$: 
\begin{eqnarray}
\label{spec}
\hspace{-0.9in}&& \qquad \qquad  \quad \,\,\,  \tilde{M}_4(\mu) \cdot \Psi_i
  \,\, \,\, = \, \,\,\,\,   \,\,\,  \,
  {{ (\mu^2 \, -1)^2 } \over {\mu^4 }} \cdot \Psi_i ,
 \quad \qquad \,\, i \, = \, 1, \, 2. 
\end{eqnarray}
Introducing a rational parametrisation for  $\, \mu$ and $\, \rho$, namely 
\begin{eqnarray}
\hspace{-0.9in}&& \qquad \qquad \quad 
\mu \, = \, \, {\frac {{u}^{2} \, -2\,u \, +2}{{u}^{2}-2}},
\qquad \quad 
\rho \, = \, \, - \, 2 \cdot \, {\frac {{u}^{2} -4\,u +2}{{u}^{2} \, -2}},
\end{eqnarray}
one finds that, even for rational values of the parameter $\, u$, that forces
$\, \mu$ and $\, \rho$ {\em to be rational numbers} as well, the hypergeometric functions
in the two previous eigenfunctions  $\, \Psi_i$ {\em do not correspond} 
(generically) to {\em globally bounded series}.
More generally, for rational values of the parameter $\, u$,
the four solutions of $\, \hat{M}_4(\mu)$
are Puiseux series of the form $\, x^r \cdot A(x)$, where $\, r$ is 
a rational number and $\, A(x)$ are series analytic at $\, x \, = \, 0$. 
None of the four $\, A(x)$ corresponds to a globally bounded series. 

\subsection{Saalschutzian hypergeometric solution}
\label{hadamard5}

In fact the order-four linear differential 
operator $\, M_4(\mu)$, or $\, \tilde{M}_4(\mu)$, 
has the $\, _4F_3$ (Saalschutzian) hypergeometric solution
\begin{eqnarray}
\,\,\, x \cdot \, _4F_3\Bigl([ {{1-\mu } \over { 2}}, 
\, {{1+\mu } \over { 2}}, \, {{3 } \over { 2}},
 \,{{3 } \over { 2}}], \, [1,2,2],\, x\Bigr),
\end{eqnarray}
 which expands as:
\begin{eqnarray}
\label{thisseries}
\hspace{-0.95in}&&x \, \, \,\,\, \,\, 
-{\frac {9}{64}}\, \left( \mu-1 \right) 
 \left( \mu+1 \right)\cdot \,  {x}^{2}\, \, \,\,\,
+{\frac {25}{4096}}\, \left( \mu-1 \right)  \left( \mu+1 \right)  \left( \mu-3 \right) 
 \left( \mu+3 \right)\cdot \,  {x}^{3} 
\nonumber \\
\hspace{-0.95in}&&-{\frac {1225}{9437184}}\, \left( \mu-1 \right)  \left( \mu+1 \right) 
 \left( \mu-3 \right)  \left( \mu+3 \right) 
 \left( \mu-5 \right)  \left( \mu+5 \right)\cdot \,  {x}^{4}
 \\
\hspace{-0.95in}&& \, 
+{\frac {441}{268435456}}\, \left( \mu-1 \right)  \left( \mu+1 \right)
  \left( \mu-3 \right)  \left( 
\mu+3 \right)  \left( \mu-5 \right)  \left( \mu+5 \right) 
 \left( \mu-7 \right)  \left( \mu+7 \right)\cdot \,  {x}^{5}
\, \, + \, \, \cdots
 \nonumber
\end{eqnarray}
On this expansion it is clear that the hypergeometric function truncates 
into a polynomial for $\, \mu$ {\em any odd integer} 
(positive or negative). Furthermore, this series (\ref{thisseries})
is globally bounded for {\em any rational number} $\, \mu$.

Let us take a simple rational value for $ \, \mu$, namely$ \, \mu \, = \, \, 1/3$.
The series  expansion (\ref{thisseries}) reads 
\begin{eqnarray}
\hspace{-0.9in}&&x\, \,  +{{1} \over {8}} \,{x}^{2}\, \, 
 +{\frac {125}{2592}}\,{x}^{3} \,  \, 
+{\frac {42875}{1679616}}\,{x}^{4} \, +{\frac {94325}{5971968}}\,{x}^{5} \, 
+{\frac {41544503}{3869835264}}\,{x}^{6} \, 
\, \, + \, \, \, \cdots 
\end{eqnarray}
The series expansion of the nome (as well as the mirror map) is 
{\em not globally bounded}:
\begin{eqnarray}
\hspace{-0.95in}&&\quad  q \,\,\, = \, \,\,\,  \, \,
x \,\,\,  +{\frac {17}{48}}\,{x}^{2}\, \,\, 
+{\frac {22195}{124416}}\,{x}^{3}\,\,\,  +{\frac {1913687}{17915904}}\,{x}^{4}\, \,\, 
+{\frac {2195016283}{30958682112}}\,{x}^{5}\,\,  \, \,  + \, \cdots 
\nonumber 
\end{eqnarray}
In contrast, the series expansion of the Yukawa coupling $\, K(x)$ is a 
{\em globally bounded series}
\begin{eqnarray}
\hspace{-0.9in}&&\quad \quad K(x)\,\, = \, \,\,\,  \,{{1} \over {x}} \, 
-{\frac {43}{144}} \, -{\frac {11}{288}}\,x \, 
-{\frac {31517}{3359232}}\,{x}^{2} \, -{\frac {522821}{3869835264}}\,{x}^{3} \, 
\,\, + \, \, \, \cdots 
\nonumber 
\end{eqnarray}
Actually the rescaling $\, x \, \rightarrow \, \,2^4\, \cdot 3^3 \cdot \,x $ 
changes $\,  2^4\, \cdot 3^3 \cdot \,  \, K(x)$ into
a series with {\em integer coefficients}:
\begin{eqnarray}
\label{firstex}
\hspace{-0.95in}&&\quad 2^4\, \cdot 3^3 \cdot \, K( 2^4\, \cdot 3^3 \cdot \,x)
\,\, = \, \,\,\,  \,  \, 
{{1} \over {x}} \, \,   \, -129\, \,  \,   -7128\,x\, \,   -756408\,{x}^{2}\,  \, 
-4705389\,{x}^{3}\, \nonumber \\
\hspace{-0.95in}&&\quad \quad \quad  +58331013489\,{x}^{4}\,  +38259799407522\,{x}^{5}\, 
+19576957591348938\,{x}^{6}\,
 \\
\hspace{-0.95in}&&\quad \quad \quad \quad 
  +9193736880930978297\,{x}^{7}\, 
+4149261387452007788523\,{x}^{8}\, \, \,  \, + \, \, \cdots   \nonumber
\end{eqnarray}

\subsection{A non-trivially equivalent operator for $\, \mu \, = \, \, \pm \, 1$}
\label{nontriv}

Let us introduce the order-four operator 
\begin{eqnarray}
\label{nontrivop}
\hspace{-0.6in}&&\quad \quad \, \,\, N_4 \, \,\, \, = \, \, \,\, \, \,
 16 \cdot \theta \cdot \, (\theta\, -2) \cdot \, (\theta\, -1)^2
\,\, \,\,\,\, \\
\hspace{-0.6in}&& \qquad \qquad \qquad \quad \quad 
-\, 4 \cdot \,x \cdot \,\theta \cdot \, (\theta\, -1) 
\cdot \, (2\, \theta\, -1)^2. \nonumber 
\end{eqnarray}
which is nothing but $\,{\cal M}_2 \cdot \, D_x^2 \, $   (see (\ref{reca})). 
 
This operator is non-trivially homomorphic to $\, M_4(\pm \, 1)$, that is
 (\ref{theexample}) for $\mu \, = \, \pm \, 1$:
\begin{eqnarray}
\label{homoM4N4}
\hspace{-0.8in}&& \quad \quad  N_4  \cdot  A_2 
 \, \,\, \,\, \, = \,\, \, \,\, \, \,\Bigl({{x} \over {1-x}}\Bigr)  \cdot \,
 A_2 \cdot \, \Bigl({{1-x} \over {x}}\Bigr)  \cdot \, M_4(\pm \, 1),
  \\
\hspace{-0.8in}&& \qquad \qquad \quad\quad  \hbox{where:}
 \quad   \quad  \quad \qquad  
 A_2 \, \,\, \,\, \, = \,\, \, \,\, \, \,
 (1-x) \cdot \,  \theta \cdot \, (\theta\, -1).
 \nonumber 
\end{eqnarray}
Operator $\, N_4$  {\em does verify the Calabi-Yau condition}:
 its exterior square is of {\em order five}.

The operator  $\, N_4$  can also be written as a direct sum: 
\begin{eqnarray}
\label{dirsum1}
\hspace{-0.8in}&&  \qquad \qquad \qquad  D_x^2 \, \,  \oplus \,  \, 
\Bigl(D_x^2 \,    \,-\,{\frac {1}{ 4 \, x  \cdot \, (1-x) }}
    \Bigr).
\end{eqnarray}
or, in terms of $\, \theta \, = \, \, x \cdot \, D_x$,  the direct sum: 
\begin{eqnarray}
\label{dirsum2}
 \hspace{-0.9in}&&  \qquad \qquad \quad \theta \, \oplus \, (\theta \, -1) \,
  \oplus \,  \, 
\Bigl(16 \cdot \, \theta \cdot \, (\theta \, -1) \, 
- \, 4 \, x \cdot \, (2 \theta \, -1)^2 \Bigr). 
\end{eqnarray}
Thus, besides the constant solution and  $\, y(x) \, = x$, 
its solutions can simply be written in terms of 
hypergeometric functions, for instance 
\begin{eqnarray}
\label{solN4}
\hspace{-0.8in}&& \,  \,  \, \, \,  \quad 
 x \cdot \, (1\, -x) \, \cdot \, 
 _2F_1\Bigl([{{3} \over {2}},  {{3} \over {2}}], \,[2 ];  \, x\Bigr) \\
\hspace{-0.8in}&& \,   \,  \quad \quad  \qquad \, \, \, \, 
  \,  \, \, = \, \,  \, \, 4 \,  x \cdot \, (1\, -x) \, \cdot \,
 {{d} \over {dx}} \Bigl( \, _2F_1\Bigl([{{1} \over {2}},  {{1} \over {2}}],
 \,[1];  \, x\Bigr)  \Bigr) 
 \nonumber 
 \end{eqnarray}
\begin{eqnarray}
\hspace{-0.8in}&& \,   \,  \quad \quad \quad  \qquad = \, \,  \, \, \, \, \, 
 x \cdot \, (1\, -x) \, (1 \, -2 \, x)^{-3/2} \cdot \, 
 _2F_1\Bigl([ {{3} \over {4}},  {{5} \over {4}} ], \,[2]; \,  \,
 - \, {{ 4 \, x \cdot \, (1 \, -x)} \over {(1 \, -2 \, x)^2 }}\Bigr). 
\nonumber 
\end{eqnarray}
This series is {\em globally bounded}. Changing $\,\, x \, \rightarrow \, \, 16 \,\, x$
turns this series into a series with {\em integer coefficients}:
\begin{eqnarray}
\hspace{-0.7in}&&  \quad  \,16\,x \,\,  \, +32\,{x}^{2}\, \, +192\,{x}^{3}\, 
+1600\,{x}^{4}\,  +15680\,{x}^{5}\,  +169344\,{x}^{6}\, +1951488\,{x}^{7}\,
 \nonumber  \\
\hspace{-0.7in}&& \qquad \qquad  \quad  \, +23557248\,{x}^{8}\,
 +294465600\,{x}^{9}\,\, \,  + \, \, \, \cdots
\end{eqnarray}

\subsection{A family of non-trivially equivalent operators for odd integer $\, \mu $}
\label{nontriv2}

It is tempting to try to generalise (\ref{dirsum1}),
or  (\ref{dirsum2}), with a simple $\, \mu^2$-ansatz:
\begin{eqnarray}
\label{directsummu}
\hspace{-0.8in}&&  \quad \quad  D_x^2 \, \,  \oplus \,  \, {\cal M}_2(\mu) 
\qquad \hbox{where:} \quad \quad 
 {\cal M}_2(\mu) \, \, = \, \, \, \, D_x^2 \,    \,-\,{\frac {\mu^2}{ 4 \, x  \cdot \, (1-x) }},
\end{eqnarray}
or, in terms of $\, \theta \, = \, \, x \cdot \, D_x$,  the direct sum: 
\begin{eqnarray}
 \hspace{-0.9in}&&  \quad \quad \quad \theta \, \oplus \, (\theta \, -1) \,
  \oplus \,  \, 
\Bigl(16 \cdot \, \theta \cdot \, (\theta \, -1) \, 
- \, 4 \, x \cdot \, (4 \theta^2 \,   -4 \theta \, + \, \mu^2) \Bigr). 
\end{eqnarray}
In order to compare this ansatz operator with the initial
 operator (\ref{theexample}), 
we slightly rewrite it as
\begin{eqnarray}
\label{theexample2}
\hspace{-0.6in}&&\quad \quad 
   16 \cdot \theta^2 \cdot \, (\theta\, -1)^2\,\, \, 
 -\, x \cdot \, (2\, \theta\, +1)^2 
\cdot \, (4 \theta^2 \,   -4 \theta \, +1 \, - \, \mu^2), 
\nonumber 
\end{eqnarray}
and we, also,  rewrite this ansatz operator
as follows:
\begin{eqnarray}
\label{oddresult}
\hspace{-0.9in}&&  \, \,\,\quad \quad \quad   {\cal C}_{odd}(\mu) \,\,\,\, = \, \, \,  \,\,\,
   16 \cdot \theta \cdot \, (\theta\, -1)^2 \cdot \, (\theta\, -2)
\,\,  \, \, \\
\hspace{-0.9in}&&  \, \,\,\quad \quad \quad    \quad \qquad   \qquad \qquad   
-\, 4 \, \cdot \, x \cdot \,\theta \cdot \, (\theta\, -1) \cdot 
 \, (4 \theta^2 \,   -4 \theta \, + \, \mu^2). \nonumber 
\end{eqnarray}

One finds that such an ansatz is actually 
 non-trivially\footnote[1]{The intertwiners are 
linear differential operators of {\em order three}.} 
{\em homomorphic to operator} (\ref{theexample}) for {\em any odd integer values}
 (positive or negative) of $\, \mu$,   
and that  its exterior square
is actually of {\em order-five} for
 {\em any value} of $\, \mu$ (Calabi-Yau condition).

For $\, \mu \, = \, \, 2\,t/(1+t^2)$,  with $\, t$
 a {\em rational number} (hence $\, |\mu| \, < 1$, the solutions 
of (\ref{oddresult}) read (besides the constant solution and $\, y(x) \, = \, x$)
simple hypergeometric functions, for instance:
\begin{eqnarray}
\label{globnilp1}
\hspace{-0.9in}&&\quad \quad \quad \quad \quad  
 {\cal S}_0(x) \,\, \, = \, \, \, \,\,\,  x \, \cdot  \,
 _2F_1\Bigl([{{ t^2} \over {1+t^2}}, \, {{1} \over {1+t^2}}], \, [2], \, x\Bigr),  
\end{eqnarray}
together with the solution\footnote[3]{This solution
can also be written $\, MeijerG([[],[(2+t^2)/(1+t^2), (2\,t^2+1)/(1+t^2)]],[[0, 1],[]],x)$, 
i.e. as a MeijerG function.}
 $\, {\cal S}_1(x) \, = \, \, {\cal S}_0(x) \, \cdot \ln(x) 
 \, + \, \tilde{{\cal S}_1}(x)$ 
where $\, \tilde{{\cal S}_1}(x)$ is analytic at $\, x \, = \, \, 0$, and solution of 
an order-four operator $\, {\cal N}_4$, product 
of two order-two operators, 
 $\,\,{\cal N}_4\,= \, \, \,  {\cal N}_2(\mu)  \cdot \,  {\cal M}_2(\mu)$, 
where  $\, {\cal N}_2(\mu)$ is an order-two operator homomorphic
 to  $\, {\cal M}_2(\mu)$ (see (\ref{directsummu})):
 \begin{eqnarray}
\hspace{-0.9in}&&\qquad {\cal N}_2(\mu) \, \cdot \,  {{1} \over {x^2}}
 \cdot \, (2 \, \theta \, -1) \, \, \, \, = \, \, \, \, \, \,
  {{2} \over {x}} \, \cdot  \,\Bigl(D_x \, -{{d \ln(\rho(x))} \over {dx}} \Bigr) 
 \cdot \,  {\cal M}_2(\mu),  
 \\
\hspace{-0.9in}&&\qquad \quad \quad \quad
 \hbox{where:} \quad \quad \quad  \quad 
\rho(x) \,\, \, \,  = \, \, \, \,\, \, 
 {\frac {1 \,-x \, +{\mu}^{2} \, x}{ (1 -x) \cdot \,  {x}^{3/2}}}. \nonumber 
\end{eqnarray}

The series expansion of (\ref{globnilp1}) is {\em globally bounded 
for any rational value}\footnote[5]{In contrast the series 
 $\, \tilde{{\cal S}_1}(x)$ is {\em not globally bounded}
for the rational values of $\, t$.} of the parameter $\, t$:
\begin{eqnarray}
\hspace{-0.9in}&&\quad x \,\,\,\,  
+ {{1} \over {2}}\,{\frac {{t}^{2}}{ (1+{t}^{2})^{2}}} \, \cdot \, {x}^{2}\, \,\,
\, +{{1} \over {12}}\,{\frac {{t}^{2} \, (2\,{t}^{2}+1)  \, ({t}^{2}+2) }{
 (1+{t}^{2})^{4}}} \, \cdot \, {x}^{3}
 \nonumber \\
\hspace{-0.9in}&&\qquad \qquad   +{\frac {1}{144}}\,{\frac {{t}^{2} \, (3\,{t}^{2}+2) 
 \, (2\,{t}^{2}+3)  \, (2\,{t}^{2}+1)  \, ({t}^{2}+2) }{
 (1+{t}^{2})^{6}}} \, \cdot \, {x}^{4}
 \, \, \,\, + \, \, \, \cdots 
\end{eqnarray}

For instance for $\, \mu \, = \, 4/5$ the hypergeometric solution 
 (\ref{globnilp1}) reads
  $\, {\cal S}_0(x) \, \,=  \,\, $
$x \, \cdot  \, _2F_1([1/5, \, 4/5], \, [2], \, x)$,  corresponding to 
the {\em globally bounded} solution series:
\begin{eqnarray}
\hspace{-0.9in}&&\quad \quad x\, \,+{\frac {2}{25}}\,{x}^{2}\,\,
 +{\frac {18}{625}}\,{x}^{3} \,\,
 +{\frac {231}{15625}}\,{x}^{4}\, \,
+{\frac {17556}{1953125}}\,{x}^{5}\,\, +{\frac {1474704}{244140625}}\,{x}^{6}\,\,
\, + \, \, \cdots 
\nonumber 
\end{eqnarray}
The rescaling $\, x \,\, \rightarrow \,\, 5^3 \, x$ turns this series 
into a series with {\em integer coefficients}.  

Remark that the two solutions $\, {\cal S}_0(x)$
 and $\, {\cal S}_1(x)$, for $\, \mu \, = \, \, 4/5$, 
can be replaced by the two solutions well-suited for $\, x$ large ($\, z \, = \,1/x$):
\begin{eqnarray}
\hspace{-0.95in}&&\,\,\,  z^{-4/5}\, \cdot  \,(1-z) \, \cdot  \,
 _2F_1\Bigl([{{ 1} \over {5}}, \, {{6} \over {5}}], \, [{{ 2} \over {5}}], \, z\Bigr),\, \,\,
\quad  z^{-1/5}\, \cdot  \,(1-z) \, \cdot  \,
 _2F_1\Bigl([{{ 4} \over {5}}, \, {{9} \over {5}}], \, [{{ 8} \over {5}}], \, z\Bigr).
\nonumber 
\end{eqnarray}
The two hypergeometric functions $\,  _2F_1([1/5, \, 6/5], \, [2/5], \, z)$
and $\,   _2F_1([4/5, \, /5], \, [8/5], \, z)$ 
do not correspond to globally bounded series. We have a similar result
 for the other rational values of $\, t$.

Do note, however, that for the other rational values of $\, \mu$ 
we have a drastically different situation: the operator  (\ref{oddresult})
 is {\em no longer globally nilpotent}\footnote[2]{For instance one sees 
explicitly on (\ref{globnilpsol}) that its exponents 
are {\em not rational numbers}, therefore operator (\ref{oddresult}) {\em cannot be 
a globally nilpotent operator}.}, as can be seen on the solution of 
$\, D_x^2 \,    \,-\, \mu^2/ 4 /x  / (1-x) \, $ in (\ref{directsummu})
\begin{eqnarray}
\label{globnilpsol}
\hspace{-0.9in}&&\quad \quad  \quad  x \cdot \, (1-x) \cdot  \,
 _2F_1\Bigl(\Bigl[{{ 3 \, + \, (1 \, -\mu^2)^{1/2}} \over {2}},
 \, {{ 3 \, - \, (1 \, -\mu^2)^{1/2}} \over {2}}\Bigr], \, [2], \, x\Bigr),
\end{eqnarray}
which has the series expansion
\begin{eqnarray}
\label{solmu}
\hspace{-0.9in}&&x \,\,\, + {{1} \over {8}}\, \,{\mu}^{2} \cdot \, {x}^{2}\,\,  
+{\frac {1}{192}}\,\,{\mu}^{2} \cdot \, ({\mu}^{2}\, +8) \cdot \, {x}^{3}\, 
 \, +{\frac {1}{9216}}\, \, {\mu}^{2} \cdot \, ({\mu}^{2} +8) ({\mu}^{2}+24)  \cdot \,   {x}^{4}
\nonumber \\
\hspace{-0.9in}&&\qquad \qquad \quad 
+ {\frac {1}{737280}}\,{\mu}^{2} \cdot \, ({\mu}^{2}+8)   
 \, ({\mu}^{2}+24) \, ({\mu}^{2}+48)  \, \cdot \,  {x}^{5}
 \,\,\,  + \,\, \cdots 
\end{eqnarray}
For rational values of $\, \mu$ that are not of the form $\, \mu \, = \, \, 2\,t/(1+t^2)$,
 the solution-series (\ref{solmu}) of the operator (\ref{oddresult})
is {\em not globally bounded} as can be checked with the two
 rational values of $\, \mu$ such that  $\, |\mu| \, <1$ and  $\, |\mu| \, > 1$
respectively. 
One can verify that  the solution-series (\ref{solmu}) of (\ref{oddresult})
for $\, \mu \, = \, \, 1/3$
\begin{eqnarray}
\hspace{-0.95in}&&\,\,\,    x \, \, +{\frac {1}{72}}\,{x}^{2} \,
 +{\frac {73}{15552}}\,{x}^{3} \, +{\frac {15841}{6718464}}\,{x}^{4}
+{\frac {6859153}{4837294080}}\,{x}^{5} \, 
+{\frac {4945449313}{5224277606400}}\,{x}^{6} \, 
\, + \, \, \cdots
 \nonumber 
\end{eqnarray}
and for $\, \mu \, = \, \,3$:
\begin{eqnarray}
\hspace{-0.95in}&&\quad x\,\,  +{\frac {9}{8}}\,{x}^{2}\,\,  +{\frac {51}{64}}\,{x}^{3} \,\, 
 +{\frac {561}{1024}}\,{x}^{4}\,\,  +{\frac {31977}{81920}}\,{x}^{5}\, \, 
+{\frac {948651}{3276800}}\,{x}^{6}\, \,   +{\frac {40791993}{183500800}}\,{x}^{7}
\,\,\,  + \, \, \cdots \nonumber 
\end{eqnarray}
are {\em not globally bounded}.

\vskip .1cm 

\vskip .1cm 

\subsection{Seeking for equivalent operators for other values of $\, \mu $}
\label{nontriv3}

The operator (\ref{oddresult}), which is non-trivially homomorphic to (\ref{theexample}), 
(or (\ref{theexample2})) for {\em odd integer values}
 of the parameter $\, \mu$, is {\em not valid for
even  integer values} of  $\, \mu$. For example, 
 for $\, \mu \, = \, 0$,  operator (\ref{theexample}) 
 actually verifies the 
Calabi-Yau condition, its exterior square being nothing but
$\, {\cal M}_5(\mu)$ for $\, \mu \, = \, 0$ (see (\ref{1plus5})). 
Operator $\, {\cal M}_5(0)$ reads
\begin{eqnarray}
\label{theexample5}
\hspace{-0.6in}&&\quad 
 \qquad  16 \cdot \theta^2 \cdot \, (\theta\, -1)^2 \,\,\,  -\, x \cdot \,  (2 \theta \, \, +1)^2 
 \cdot \,  (2 \theta \, \, -1)^2.
\end{eqnarray}
It is not of the form (\ref{oddresult}), which
cannot encapsulate all the operators homomorphic to (\ref{theexample2})
satisfying the Calabi-Yau condition.
As previously 
remarked, the series expansion
of the solution 
$\, x \cdot \, _4F_3\Bigl([ {{1} \over { 2}}, 
\, {{1} \over { 2}}, \, {{3 } \over { 2}},
 \,{{3 } \over { 2}}], \, [1,2,2],\, x)$
is globally bounded, and can be turned into a series with integer coefficients with
the rescaling $\, x \, \rightarrow \, 256 \, x$. 

Note that (\ref{theexample5}) can be used as a ``seed''
 to get a family of $\, \mu^2$-dependent
operator verifying the Calabi-Yau condition. The linear differential operator 
\begin{eqnarray}
\label{theexample7}
\hspace{-0.9in}&&\quad \quad   {\cal C}(\mu) \,\,\, = \, \,  \,\,\, 
 16 \cdot \theta^2 \cdot \, (\theta\, -1)^2 \,\, 
\\
\hspace{-0.9in}&&\quad \quad \quad \qquad 
 -\, x \cdot  \,  ( 2 \theta \, \, +1\, -\mu) \cdot \,  ( 2 \theta \, \, +1\, +\mu) 
\cdot \, ( 2 \theta \, \, -1\, -\mu) \cdot \, ( 2 \theta \, \, -1\, +\mu), \nonumber 
\end{eqnarray}
is such that {\em its exterior square is actually of order five}. Very fortunately, 
operator (\ref{theexample7}) is {\em non-trivially homomorphic}
 to (\ref{theexample}) for {\em any integer value} 
(positive or negative, even or odd) of 
parameter $\, \mu$. Again the intertwiners are of order-three. 

For instance, for $\, \mu \, = \, \, 2$ one has the intertwining relation
between operator (\ref{theexample}) and  (\ref{theexample7}) 
\begin{eqnarray}
\label{intertheexample7}
\hspace{-0.9in}&&\, \,  \,   {\cal C}(2)\, \cdot \,
  \Bigl((8 \, \theta^2 \cdot \, (2\, \theta \, -3) \, + \, 4 \, \theta \,+1) \,  \, 
- \, 2 \, x \, \cdot \, (2\, \theta \, -3) \, (2\, \theta \, +1)^2 \Bigr) 
\\
\hspace{-0.9in}&&\quad    \, \,   \,\,\, = \, \,  \,
 \Bigl((8 \, \theta^2  \cdot \, (2\, \theta \, -3) \, + \, 4 \, \theta \,+1) \,  \, 
- \, 2 \, x \, \cdot \, (2\, \theta \, -3) \, (2\, \theta \, +1)
 \, (2\, \theta \, +3)  \Bigr)  \, \cdot \, M_4(2). \nonumber
\end{eqnarray}
For larger values of the integer $\, \mu$ the intertwiners become more and more involved.
They are still of degree three in $\, \theta$ but of higher degree in $\, x$.

Operator (\ref{theexample7}) has simple hypergeometric solutions 
for {\em any value} of $\, \mu$:
\begin{eqnarray}
\label{other4F3}
\hspace{-0.9in}&& 
 \quad \quad \quad \quad 
 x \cdot \, _4F_3\Bigl([{{\mu \, +3} \over {2}}, \, {{-\mu \, +3} \over {2}}, \,
  {{\mu \, +1} \over {2}},\, {{-\mu \, +2} \over {2}}],
  \, [1, \,  2, \, 2]; \,\,  x) 
\end{eqnarray}

\vskip .1cm 

For {\em any rational value} of $\, \mu$, the series expansion 
of the solution of (\ref{theexample7}) analytic at $\, x \, = \, 0$, 
namely (\ref{other4F3})
\begin{eqnarray}
\label{terminating}
\hspace{-0.9in}&&\qquad x \, \, \,
 +{\frac {1}{64}}\,   \, (\mu-1)  \, (\mu+1)  \, (\mu-3) \cdot \,(\mu+3)  \,   {x}^{2} 
\\ 
\hspace{-0.9in}&&\quad \quad \qquad
\,\, +{\frac {1}{36864}}\,  \, (\mu-1)  \, (\mu+1) 
 \, (\mu-3)^{2} \,(\mu+3)^{2} \, (\mu-5)\, (\mu+5) \,  \cdot \, {x}^{3}\,  \, \, + \, \cdots 
\nonumber  
\end{eqnarray}
is {\em globally bounded}\footnote[1]{Operator (\ref{theexample7})
 is thus globally nilpotent for any rational value of $\, \mu$.}. 

For instance, for $\, \mu \, = \, 1/3$, the series (\ref{terminating})
\begin{eqnarray}
\hspace{-0.95in}&&x \, +{\frac {10}{81}}\,{x}^{2} \,
 +{\frac {2800}{59049}}\,{x}^{3} \,
 +{\frac {1078000}{43046721}}\,{x}^{4}
\, +{\frac {53953900}{3486784401}}\,{x}^{5} \,
 +{\frac {26709338656}{2541865828329}}\,{x}^{6} \, 
 \, + \, \, \cdots 
\nonumber 
\end{eqnarray}
can be turned into a series with integer coefficients 
after the rescaling $\, x \, \rightarrow \, \, 3^6 \, x$.
However, the series expansion of the {\em nome}
 for  (\ref{theexample7})
for $\, \mu \, = \, 1/3$ is {\em not globally bounded}:
\begin{eqnarray}
\hspace{-0.9in}&&\quad \quad  x \, \,+{\frac {19}{54}}\,{x}^{2} \,\,
 +{\frac {250403}{1417176}}\,{x}^{3}\, \,
+{\frac {218211473}{2066242608}}\,{x}^{4} \, \,
+{\frac {281241377443}{4016775629952}}\,{x}^{5} 
 \nonumber \\
\hspace{-0.9in}&& \qquad \qquad \qquad \quad 
+{\frac {1456188325082179}{29282294342350080}}\,{x}^{6} \, 
\, +{\frac {3167628271177596809}{85387170302292833280}}\,{x}^{7}
\,\, + \, \,\, \cdots  \nonumber 
\end{eqnarray}
In contrast, the Yukawa coupling $\, K(x)$ for $\, \mu \, = \, 1/3$ is
 {\em actually globally bounded}
\begin{eqnarray}
\hspace{-0.9in}&&\quad K(x)\,  \, = \, \, \,
 {{1} \over {x}} \,\,  -{\frac {49}{162}} \, \, 
-{\frac {2179}{59049}}\,x \,\,  -{\frac {1508129}{172186884}}\,{x}^{2} \, 
+{\frac {47590097}{167365651248}}\,{x}^{3} \,
\, + \, \, \cdots \nonumber 
\end{eqnarray}
Actually the rescaling
  $\, x \, \rightarrow \, \, 2 \cdot  \, 3^6 \,  x\,\,$
 on  $\,\, 2 \cdot  \, 3^6 \, \, K(x)$
turns the previous series expansion into a series with
 {\em integer coefficients}:
\begin{eqnarray}
\hspace{-0.9in}&&\, \, \,  2 \cdot  \, 3^6 \,  \cdot \,K(2 \cdot  \, 3^6 \,x)
 \,\, = \, \,\, \, {{1} \over {x}} \, \, -441 \, -78444\,x \,  \,
-27146322\,{x}^{2} \, \, +1284932619\,{x}^{3} \, \,
 \nonumber \\
\hspace{-0.9in}&& \quad \quad +27674475754905\,{x}^{4} \,
 +59119113109746798\,{x}^{5} \, 
+100896041483693939736\,{x}^{6} \nonumber \\
\hspace{-0.9in}&& \quad \quad  \, +158984355721045048019613\,{x}^{7} \,\,
 +241323001023828827752150059\,{x}^{8}
 \,\,\,  + \, \, \cdots \nonumber 
\end{eqnarray}
to be compared with (\ref{firstex}) for operator (\ref{theexample}).
We have similar result for $\, \mu \, = \, 4/5$ (the series (\ref{terminating})
can be turned into a series with integer coefficients 
after the rescaling
 $\, x \, \rightarrow \, \,  2^8\cdot \, 5^5 \, x$, and 
the series expansion of the nome is not
globally bounded).
 The Yukawa coupling $\, K(x)$ for $\, \mu \, = \, 4/5$ 
can be changed into a series with {\em integer coefficients}:
\begin{eqnarray}
\hspace{-0.9in}&&\quad \quad  2^8 \cdot  \, 3 \cdot \, 5^5   \,
 K(2^8 \cdot  \, 3 \cdot  \, 5^5  \, x) 
 \,\, = \, \, \,\, \, {{1} \over {x}} \, \, \, - 1057980 \, \, - 92574954000 \, x \,
\\
\hspace{-0.9in}&& \quad \quad \quad \qquad  \quad
 - 51733629839745000 \, x^2 \, + 74509092036686778685920 \, x^3
  \, + \, \cdots \nonumber
\end{eqnarray}

Note that, for {\em any odd integer}, (positive or negative), 
(\ref{terminating}) is a {\em terminating series},
 reducing to a {\em polynomial}.

\vskip .1cm 

As a byproduct, we see, for {\em any odd integer} value of $\, \mu$, that 
operator (\ref{theexample}) is (non-trivially) homomorphic to {\em two different}
operators (\ref{oddresult}) and (\ref{theexample7}), such that their 
exterior square is of order five. As it should these two operators 
(\ref{oddresult}) and (\ref{theexample7}) are, for {\em any odd integer value} of $\, \mu$, 
 homomorphic\footnote[1]{Note that for even integer values of $\, \mu$, 
 operator (\ref{theexample7}) is irreducible when (\ref{oddresult}) is reducible.
Therefore operators (\ref{oddresult}) and (\ref{theexample7})
 {\em cannot be homomorphic} for 
 {\em even integer} values of $\, \mu$.} (with intertwiners of order three). 
For $\, \mu\, = \, \, 3$ 
one gets the homomorphism:
\begin{eqnarray}
\label{interCC}
 \hspace{-0.9in}&&\quad {\cal C}_{odd}(3) \cdot \, \Bigl(4 \, \theta \cdot \, (\theta -1)^2  \,
 -4 \, x \cdot \, (\theta -1) \cdot \,  (\theta -2) \cdot \, (\theta +2)\Bigr) 
  \\
 \hspace{-0.9in}&& \qquad \quad \quad \quad \,    \, \, \, = \, \, \, \,  \, \, \, 
 \Bigl(4 \, (\theta -2) \cdot \, (\theta -1)^2   \,
 - \, x \cdot \, (\theta -1) \cdot \, (4 \, \theta^2 \, -4 \, \theta  \, + \, 9) \Bigr) 
 \cdot \, {\cal C}(3).  \nonumber 
\end{eqnarray}
For other values of $\, \mu$ one gets slightly more involved 
intertwiners that are no longer of degree one in $\, x$.

 The homomorphisms, for {\em odd integer values} of $\, \mu$,
between a non globally nilpotent operator (\ref{oddresult}) and the globally nilpotent
 operator (\ref{theexample7}), or between operators 
with non globally bounded and globally bounded solutions, may seem misleading.
It is important to note that, for {\em odd integer values} of $\, \mu$,
the  operator (\ref{theexample7}) is 
{\em non longer irreducible}\footnote[2]{For any odd integer value of $\, \mu$
 the operator (\ref{theexample7})
is the product of four order-one operators such that their wronskian
is a rational function. Operator (\ref{theexample7}) 
is thus globally nilpotent.}, that the intertwiners between (\ref{oddresult})
and (\ref{theexample7}) (see (\ref{interCC}))  are not globally nilpotent,
and, furthermore, that the globally bounded
 infinite series (\ref{terminating}), reduces 
to a polynomial. The intertwining relation between  (\ref{oddresult})
 and (\ref{theexample7}) then matches this terminating hypergeometric
 series  (\ref{terminating}) with the $\, y(x) \, = \, x$ solution of (\ref{oddresult}).

\vskip .1cm 

\vskip .3cm 

Do note that, given an order-four operator such that its {\em exterior square has a 
 rational solution} (``extended Calabi-Yau condition''), 
finding an order-four operator (non-trivially) 
homomorphic to the first operator such that its {\em exterior square is of order five}
(Calabi-Yau condition) is an extremely difficult task\footnote[5]{In particular
 because, as we have seen, this reduction to an operator satisfying the Calabi-Yau 
condition is not unique.
In contrast, starting from an operator satisfying the Calabi-Yau condition, like (\ref{theexample7}),
it is straightforward to get operators that are non-trivially homomorphic 
to this operator, and are such that their exterior square has a 
 rational solution (just perform the LCLM of this operator with any order-three
linear differential operator). This will be explained
 in a forthcoming publication.}, even if one assumes
decompositions like (\ref{forthcoming2}), (\ref{forthcoming3}).
We will address this difficult question of the reduction of the 
``extended Calabi-Yau condition'' to the ``Calabi-Yau condition'' 
in a forthcoming publication.

\section{Modular forms and selected $\, _2F_1$ 
hypergeometric functions with two pullbacks}
\label{Miscell}
We display here a (non exhaustive) list of miscellaneous 
identities (between modular forms
and their representations as $\, _2F_1$ hypergeometric functions with two pullbacks)
that we often encountered in 
our studies of the Ising model, lattice Green functions, or Calabi-Yau ODEs:
\begin{eqnarray}
\label{identity2}
\hspace{-0.7in}&&
 _2F_1\Bigl([{{1} \over {6}}, \, {{1} \over {3}}],[1];\, 
 108\cdot x^2 \cdot(1+4\, x)\Bigr) \\
\hspace{-0.7in}&& \quad  \quad \, \, = \, \, \,\, 
(1-12\, x)^{-1/2} \, \cdot \, _2F_1\Bigl([{{1} \over {6}}, \, {{1} \over {3}}],[1];\,\, 
 -\, {{ 108 \cdot x \cdot(1+4\, x)^2} \over {(1-12\, x)^3 }}\Bigr).
 \nonumber \\
\hspace{-0.7in}&& \quad  \quad \quad \, \, = \, \, \,\, \,\, \, 
1\,\,\,\,   +6\, x^2\,\,  +24\, x^3\,\,  +252\, x^4\,\,  +2016\, x^5\,\,  +19320\, x^6\,
 +183456\, x^7\,
\nonumber \\
\hspace{-0.7in}&& \quad  \quad  \quad  \quad  \quad  \quad  \quad  \quad 
 +1823094\, x^8\,\,  +18406752\, x^9\,\,  +189532980\, x^{10}
\,   \,\,\,\,  + \, \, \cdots 
\nonumber
\end{eqnarray}

\begin{eqnarray}
\label{identity3}
\hspace{-0.7in}&&\quad (1\, +2\, x) \, \cdot \,
 _2F_1\Bigl([{{1} \over {6}}, \, {{1} \over {3}}],[1];\, \,
 {{27\, x^2 \, (1\, +x)^2 } \over { 4\, (1\, +\, x\, +x^2)^3}}\Bigr)
 \\
\hspace{-0.7in}&& \qquad \quad \quad \, \, = \, \, \,\,  \,
(1\, +\, x\, +x^2)^{1/2} \, \cdot \,
 _2F_1\Bigl([{{1} \over {6}}, \, {{1} \over {3}}],[1];\,\, 
{{27\, x \, (1\, +x) } \over { 4\, (1\, +2\, x)^6}}\Bigr).
 \nonumber 
\end{eqnarray}
The series expansion of (\ref{identity3}) are globally bounded 
and can be changed into a series with integer coefficients
 after the rescaling $\, x \, \rightarrow \, 4\, x$. This identity 
is nothing but  identity (\ref{identity2}) after a change of variable.
Another example corresponds to HeunG functions of the form 
$\, HeunG(a, \, q,  \, 1, \, 1, \, 1, \, 1; \, \, x)$, such 
that the two parameters $\, a$ and $\, q$
are associated with  fixed points of the symmetries 
(\ref{HeunGidentity}) of these HeunG functions : 
\begin{eqnarray}
\label{HeunGhyp}
\hspace{-0.9in}&&HeunG({{1} \over {2}}, \,{{1} \over {2}}, 1, 1, 1, 1, 4\, x)
 \, \, \, \, = \, \, \, \, \, \,
HeunG(2, \, 1, 1, 1, 1, 1, 8\, x) 
 \nonumber \\
\hspace{-0.9in}&& \quad  \,  \, \, \,
 \, \, \, = \, \, \,  \, \,
{{1} \over{1\, -4\, x}} \cdot \,
 HeunG\Bigl(-1, \, 0, 1, 1, 1, 1, \, -{{4\, x} \over {1\, -4\, x}}\Bigr)
\nonumber 
\end{eqnarray}
\begin{eqnarray}
\hspace{-0.9in}&& \quad  \quad    \,  = \,\, \,   
{{1} \over {1\, -4\, x}} \cdot \,
 _2F_1([{{1} \over {4}}, \, {{1} \over {4}}], \, [1], \, 
{{64 \, x^2 \cdot \, (1\, -8\, x)} \over {(1\, -4\, x)^4}}\Bigr) 
 \nonumber 
\end{eqnarray}
\begin{eqnarray}
\hspace{-0.9in}&& \quad \quad    \,  \, = \,\, \, \,
  _2F_1([{{1} \over {4}}, \, {{1} \over {4}}], \, [1], \, 
64 \, x \cdot (1\, -4\, x) \cdot (1\, -8\, x)^2\Bigr)\,   
\end{eqnarray}
\begin{eqnarray}
\hspace{-0.9in}&& \quad \quad     \, \, = \, \, \,\, 
 (1\,-256\, x\,+5120\,x^2\,-32768\,x^3\,+65536\,x^4)^{-1/4}
   \nonumber \\
\hspace{-0.9in}&& \quad  \quad \quad \times \,  
_2F_1\Bigl([{{1} \over {12}}, \,{{5} \over {12}}],\, [1],\, 
 -\, 1728 \cdot {{x \cdot (1\,-4\, x) \cdot (1\,-8\, x)^2
} \over  { (1\,-256\, x\,+5120\,x^2\,-32768\,x^3\,+65536\,x^4)^3 }}\Bigr)
 \nonumber 
\end{eqnarray}
\begin{eqnarray}
\hspace{-0.9in}&& \quad \quad   \, \, = \, \, \,\, 
 (1\,-16\, x\,+80\,x^2\,-128\,x^3\,+256\,x^4)^{-1/4} 
  \nonumber \\
\hspace{-0.9in}&& \qquad  \quad   \quad \times \,   
 _2F_1\Bigl([{{1} \over {12}}, \,{{5} \over {12}}],\, [1],\, 
 1728 \cdot {{x^4 \cdot (1\,-4\, x)^4  \cdot (1\,-8\, x)^2 } \over  {
(1\,-16\, x\,+80\,x^2\,-128\,x^3\,+256\,x^4)^3 }}\Bigr)
 \nonumber 
\end{eqnarray}
\begin{eqnarray}
\hspace{-0.9in}&& \quad \quad \quad   \,  \, = \,\, \,\,  \, \,
1\,\, \,  \, +4 \, x  \,\,+20 \,x^2 \,
 \,+112 \,x^3 \, \,+676 \,x^4 \,+4304 \,x^5 \, \,
+28496 \,x^6 \,+194240 \,x^7 
 \nonumber \\
\hspace{-0.9in}&& \quad \quad \quad   \quad  \quad  \quad    \quad   \quad  
 \,+1353508 \,x^8 \,\,+9593104 \,x^9 \,\, \, \, + \,\, \cdots \nonumber 
\end{eqnarray}
The previous HeunG function can also be written 
\begin{eqnarray}
\label{alternative}
\hspace{-0.9in}&&  \quad 
_2F_1\Bigl( [{{1} \over {2}}, \, {{1} \over {2}}],\, [1], \,
 16 \, x \cdot (1\,-4\,x) \Bigr)\, \, \, = \, \, \,  \, \,
{{1} \over {1\, -4 \, x}} \cdot \,
_2F_1\Bigl( [{{1} \over {2}}, \, {{1} \over {2}}],\, [1], \,
 {{16 \, x^2} \over {(1\, -4 \, x)^2}}\Bigr)  
\nonumber \\
\hspace{-0.9in}&& \quad \quad\quad   \, = \, \,  \,  \,
 (1\, -4 \, x) \cdot \, (1\, -8 \, x) \cdot \, 
 _2F_1\Bigl( [{{3} \over {2}}, \, {{3} \over {2}}],\, [2], \,
 16 \, x \cdot (1\,-4\,x) \Bigr) 
 \\
\hspace{-0.9in}&& \quad \quad \quad \quad \quad \qquad \qquad  \quad  \quad   \, \, 
- {{2 \, x} \over {(1-4\,x)^3}}   \cdot \,
  _2F_1\Bigl( [{{3} \over {2}}, \, {{3} \over {2}}],\, [2], \, 
{{16 \, x^2} \over {(1\, -4 \, x)^2}}\Bigr)  
\nonumber 
\end{eqnarray}
Using the identity 
\begin{eqnarray}
\hspace{-0.95in}&& \qquad  \Bigl( 16 \, x \cdot (1\,-4\,x)\Bigr) \circ \,  
\Bigl({\frac { 9 \, {x}^{2}}{1+4\,x+40\,{x}^{2}}} \Bigr)
 \,\,  \,  = \, \,  \,  \,   \,   \, 
144\,{\frac {{x}^{2} \cdot \, (1 \, + 2\,x)^{2}}{ (1+4\,x+40\,{x}^{2})^{2}}}
 \nonumber \\
\hspace{-0.95in}&&\qquad \qquad \qquad  \quad  \quad  \quad \,\, = \, \,  \,  \,   \,   \,   \, 
 \Bigl({{16 \, x^2} \over {(1\, -4 \, x)^2}}   \Bigr)
\circ \,  
\Bigl(\,{\frac { -3 \, x \cdot \, (1 \, + 2\,x) }{ (1 \, -4\,x)^{2}}} \Bigr)
\end{eqnarray}
one easily deduces from (\ref{alternative}) a {\em linear functional identity} between 
the {\em same} hypergeometric function with  {\em three different rational pullbacks}:
\begin{eqnarray}
\label{equFunctional}
\hspace{-0.95in}&&\qquad 18 \,{x}^{2} \cdot \, (1+4\,x+40\,{x}^{2})^{3}
 \cdot \, (1 \, - 4\,x)^{6}
\cdot \,  _2F_1\Bigl( [{{3} \over {2}}, \, {{3} \over {2}}],\, [2], \,  \, p_1(x) \Bigr)
\nonumber \\
\hspace{-0.95in}&&
 \quad \qquad
 \, - \, (1 \, + 2\,x) ^{7} \cdot \, (1 \, - 4\,x)^{7}
\cdot \,  _2F_1\Bigl( [{{3} \over {2}}, \, {{3} \over {2}}],\, [2], \,  \, p_2(x) \Bigr)
 \\
\hspace{-0.95in}&&
 \quad \qquad \,\,
+ \, (1\, +8\,x)^{2} \,  \cdot \, (1+4\,x+40\,{x}^{2})^{3} \cdot \, (1 \,+  2\,x)^{6}
\cdot \,  _2F_1\Bigl( [{{3} \over {2}}, \, {{3} \over {2}}],\, [2], \,  \, p_3(x) \Bigr)
\, \,\,\, \, = \, \,\, \, \,\, 0, 
\nonumber 
\end{eqnarray}
where the three pullbacks read respectively:
\begin{eqnarray}
\hspace{-0.95in}&&
 \quad
\quad p_1(x) \,\, = \, \, \,\,\, 1296\,\cdot \,{\frac {{x}^{4}}{(1 \, + 2\,x)^{4}}}
\,\,\,\, = \, \, \, \,\,\,
\Bigl( {{16 \, x^2} \over {(1\, -4 \, x)^2}}\Bigr) \circ \,
 \Bigl({\frac { 9 \, {x}^{2}}{1+4\,x+40\,{x}^{2}}} \Bigr), 
\nonumber \\
\hspace{-0.95in}&&
\quad \quad p_2(x) \,\, = \, \, \, 
144\,\cdot  \,{\frac {{x}^{2} \cdot \, 
(1 \, +2\,x)^{2}}{(1 \, +4\,x \, +40\,{x}^{2})^{2}}}
\nonumber \\
\hspace{-0.95in}&&
\quad \quad \qquad  \qquad  \qquad 
\, \,\, = \, \, \, \, \,\,  \Bigl( 16 \, x \cdot (1\,-4\,x)\Bigr) \circ \, 
\Bigl({\frac { 9 \, {x}^{2}}{1+4\,x+40\,{x}^{2}}} \Bigr), 
\nonumber \\
\hspace{-0.95in}&&
 \quad \quad p_3(x)\, \, = \, \, \, \,\,
 -48 \,\cdot  \, {\frac { x \cdot \, (1 \, + 2\,x) 
 \cdot \, (1+4\,x+40\,{x}^{2}) }{ (1 \, - 4\,x)^{4}}} 
 \\
\hspace{-0.95in}&&
\quad \quad \qquad  \qquad  \qquad 
\,\, = \, \, \,\, \, \Bigl( 16 \, x \cdot (1\,-4\,x)\Bigr) \circ \,
\Bigl(\,{\frac { -3 \, x \cdot \, (1 \, + 2\,x) }{ (1 \, -4\,x)^{2}}} \Bigr).
\nonumber 
\end{eqnarray}
Note that each of the three terms in (\ref{equFunctional}) corresponds to 
series with {\em integer coefficients}, their
series reading respectively: 
\begin{eqnarray}
\hspace{-0.9in}&&\quad \, \,  18\,{x}^{2}\, -216\,{x}^{3}\, +2160\,{x}^{4}\,-25344\,{x}^{5}
\, +223236\,{x}^{6}\,
-1810512\,{x}^{7}  \,\,  + \cdots, \nonumber \\
\hspace{-0.9in}&&\quad -1 \, +14\,x \, -190\,{x}^{2} \, +2524\,{x}^{3}
 \, -26732\,{x}^{4} \, +270184\,{x}^{5} \, 
-2650164\,{x}^{6} 
 \,\,  + \cdots, \nonumber \\
\hspace{-0.9in}&&\quad \, \,  1 \, -14\,x \, +172\,{x}^{2} \, -2308\,{x}^{3} \, +24572\,{x}^{4} \, 
-244840\,{x}^{5} \,  +2426928\,{x}^{6}
 \, \, + \cdots \nonumber 
\end{eqnarray}

Another $\, HeunG(a, \, q,  \, 1, \, 1, \, 1, \, 1; \, \, x)$ example, such 
that the two parameters $\, a$ and $\, q$
correspond to fixed points of the symmetries 
(\ref{HeunGidentity}) of these particular  HeunG functions,
 read: 
\begin{eqnarray}
\hspace{-0.9in}&& HeunG\Bigl({{1\, + i \, 3^{1/2}} \over {2}},
 \, {{3\, + i \, 3^{1/2}} \over {6}},  \, 1, \, 1, \, 1, \, 1; \, \, \,
9 \cdot \, {{3\, + \, i \, 3^{1/2} } \over {2 }} \cdot \,  x\Bigr) 
\nonumber \\
\hspace{-0.9in}&&\quad \,\, = \, \, \,\,\,\, 
_2F_1\Bigl([{{1} \over{3}}, \, {{1} \over{3}}], \, [1]; \, \, \,
 81 \cdot \, x \cdot \, (1\, -27\,x\, + 243\,{x}^{2})   \Bigr)   
\nonumber \\
\hspace{-0.9in}&&\quad \,\, = \, \, \,\,\,\, {{1} \over {x}} \cdot \, 
_2F_1\Bigl([{{1} \over{3}}, \, {{1} \over{3}}], \, [1]; \, \, \,
 {{1\, -27\,x\, + 243\,{x}^{2}} \over { 729 \, x^3}}   \Bigr)   
\nonumber \\
\hspace{-0.9in}&&\quad \,\, = \, \, \,\,\,\, 
 \, 1\,\,\, \, +9\,x\,\,\,  +81\,{x}^{2}\,\,+567\,{x}^{3}\,
+729\,{x}^{4}\,-72171\,{x}^{5}\,-1764909\,{x}^{6}\,-28284471\,{x}^{7}
\nonumber \\
\hspace{-0.9in}&& \qquad \quad \quad \quad
\,-343842327\,{x}^{8}\,-2859802119\,{x}^{9}\,
-2072088459\,{x}^{10}\,+523309421259\,{x}^{11}
\nonumber \\
\hspace{-0.9in}&& \qquad\,\quad \quad  \quad \quad \quad +13407709577211\,{x}^{12}\,
+226522478442087\,{x}^{13}
\,\, \,  \,\,  + \, \,  \cdots,  
\end{eqnarray}
where the coefficients are actually integers, their sign having  a period six. 
This example is nothing but revisiting (\ref{G32})
with a $\, x \rightarrow \, -81 \, x$ change of variable. 

Among the fixed points of the symmetries 
(\ref{HeunGidentity}) the case $\, a \, = \, 0$ for {\em any value}
 of $\, q$,
corresponds to a special limit. Let us write $\, q$ as 
$\, q \, = \, (1-r^2)/4$, one has
\begin{eqnarray}
\hspace{-0.9in}lim_{a \rightarrow \, 0} \, 
HeunG(a,\,  {{1\, -r^2} \over {4}},\, 1,\, 1,\, 1,\, 1; \, \, 2^4 \, a  \, x)
\, \,\, \,  = \, \,  \,\,\, \, _2F_1\Bigl([{{1\, -r} \over {2}}, \,
 {{1\, +r} \over {2}}], \, [1], \, \, 2^4\, x), 
\nonumber  
\end{eqnarray}
which is, of course, a series with integer coefficients
 for any even integer values of $\, r$
(it is a simple polynomial expression for odd integer values of $\, r$).

\vskip .3cm

More identities\footnote[1]{Not to be confused with Goursat-type identities
on hypergeometric functions (see for instance section 4 in~\cite{Vidunas2}). 
Here, the hypergeometric 
functions have the {\em same} parameters but different pullbacks.} read: 
\begin{eqnarray}
\label{identitybis}
\hspace{-0.9in}&&HeunG(-3,\,  0,\, 1/2,\, 1,\, 1,\, 1/2; \, 12\cdot x)
\, \,  \, \, = \, \, \, \, \, 
 _2F_1\Bigl([{{1} \over {6}}, \, {{1} \over {3}}],[1];\, 
 108\cdot x^2 \cdot(1+4\, x)\Bigr)
  \nonumber \\
\hspace{-0.9in}&& \qquad \, \,  \, = \, \, \,  \, \,
(1-12\, x)^{-1/2} \, \cdot \, _2F_1\Bigl([{{1} \over {6}}, \, {{1} \over {3}}],[1];\, 
 -\, {{ 108 \cdot x \cdot(1+4\, x)^2} \over {(1-12\, x)^3 }}\Bigr)
 \nonumber \\
\hspace{-0.9in}&& \qquad  \,  \, \, = \,\, \, \, \,  \,\,  \,
1\,\,\,\,  \,  +12\, x^2\,\,  +48\, x^3\, \,
 +540\, x^4\, \, +4320\, x^5\, \, +42240\, x^6\, \,
 +403200\, x^7\,
\nonumber \\ 
\hspace{-0.9in}&& \qquad   \qquad   \qquad  \quad  \quad  
 +4038300\, x^8 \,\, \, \, \, + \, \,  \cdots,
\nonumber 
\end{eqnarray}
\begin{eqnarray}
\label{twopullrecall}
\hspace{-0.9in}&&HeunG(4, {{1} \over {2}}, {{1} \over {2}}, {{1} \over {2}},
 1, {{1} \over {2}}; \,  64\, x)
 \, \, \, \, = 
\nonumber \\
\hspace{-0.9in}&&\quad 
\,  \, = \, \,  \,\,   (1\, -16 \, x)^{-1/2} \, \cdot \,
 _2F_1\Bigl([{{1} \over {6}},\, {{1} \over {3}}], \, [1]; \,
 {{1728 \, x^2 } \over {(1\, -16 \, x)^3 }} \Bigr) 
 \nonumber \\
\hspace{-0.9in}&&\quad 
\,  \,\,  \,= \, \,  \,\,   (1\, -64 \, x)^{-1/2} \, \cdot \,
 _2F_1\Bigl([{{1} \over {6}},\, {{1} \over {3}}], \, [1]; \,
 -\, {{432 \, x } \over {(1\, -64 \, x)^3 }} \Bigr)
 \\
\hspace{-0.9in}&&\quad \,  \,\, \,  \, = \, \,  \, \,\,  \,\,  
1\,\,\,  \, \,  +8\, x\,\,  +192\, x^2\, +6656\, x^3\, +275968\, x^4\, +12644352\, x^5\,
 +616562688\, x^6\, 
 \nonumber \\
\hspace{-0.9in}&&\quad \quad  \quad  \quad  \quad \, \, \,  \, +31366053888\, x^7
 \,\,  +1645521666048\, x^8
 \,   \, +88371818921984\, x^9\, 
 \,\, \,  + \, \, \cdots,  \nonumber
\end{eqnarray}

\begin{eqnarray}
\hspace{-0.8in}&&_3F_2\Bigl([{{1} \over {3}}, \,{{1} \over {2}}, 
\, {{2} \over {3}}], \, [1, \, 1],
 \, -\, {{108 \, x} \over { (1\,-16\, x)^3}}\Bigr)
 \,\,  \,    \\
\hspace{-0.8in}&& \quad \qquad \,\,  \, = \, \,  \,\, \,\, 
 {{1\, -16\, x} \over {1\, -4\, x}} \cdot \, 
_3F_2\Bigl([{{1} \over {3}}, \,{{1} \over {2}}, \, {{2} \over {3}}], \, [1, \, 1],
 \, {{108 \, x^2} \over { (1\,-4\, x)^3}}\Bigr)
  \nonumber \\
\hspace{-0.8in}&& \quad \qquad \,\,  \, = \, \,  \,\, \,\, \, 
1\,\,\,\,   -12\, x\,\, \, -36\, x^2\,\,  -192\, x^3\,\, 
 -1380\, x^4\,\,  -11952\, x^5\, 
-116928\, x^6\, 
 \nonumber \\
\hspace{-0.8in}&& \quad \qquad \quad  \quad  \quad  \quad  \quad 
\,\, \, -1242624\, x^7\,\, 
 -14006628\, x^8\,\,  -164954640\, x^9\,\, \, 
\, \,  + \, \, \cdots,   \nonumber
\end{eqnarray}

\begin{eqnarray}
\hspace{-0.9in}&&\quad  _2F_1\Bigl([{{1} \over {2}}, \,{{1} \over {2}}],\, [1],\, 
 x \Bigr)\, \,  \, \, \, = \, \, \, \, \,\, 
(1\, -16\,x\, +16\, x^2)^{-1/4} \,  
 \\
\hspace{-0.9in}&&\quad  \quad \quad \quad \quad \quad \quad \quad \quad \quad
 \times \,  _2F_1\Bigl([{{1} \over {12}}, \,{{5} \over {12}}],\, [1],\, \, 
 -108 \cdot {{x \cdot (1-x)} \over  { (1\, -16\,x\, +16\, x^2)^3}}\Bigr). 
\nonumber
\end{eqnarray}
This globally bounded series becomes a series with integer coefficients 
after the rescaling $\, x \, \rightarrow  \, 16 \, x$.

\begin{eqnarray}
\hspace{-0.8in}&&\quad  _2F_1\Bigl([{{1} \over {4}}, \,{{3} \over {4}}],\, [1],\, 
 x \Bigr)\, \, 
 \\
\hspace{-0.8in}&&\quad  \quad \quad \quad \quad \quad  \, \, = \, \, \, \, \, 
(1\, +3 \,x)^{-1/4} \, \cdot \,   
 _2F_1\Bigl([{{1} \over {12}}, \,{{5} \over {12}}],\, [1],\,\,  
 27 \cdot {{x \cdot (1-x)^2} \over  { (1\, +3\,x)^3}}\Bigr). \nonumber
\end{eqnarray}
This globally bounded series becomes a series with integer coefficients 
after the rescaling $\, x \, \rightarrow  \, 64 \, x$.

\begin{eqnarray}
\hspace{-0.7in}&&\quad  _2F_1\Bigl([{{1} \over {3}}, \,{{2} \over {3}}],\, [1],\, 
 x \Bigr)\, \,  \\
\hspace{-0.7in}&& \quad \quad \quad \quad \quad \quad \, \, = \, \, \,  \,\, 
\Bigl({{9} \over {9 \, -8 \, x }} \Bigr)^{1/4} \, \cdot \, 
 _2F_1\Bigl([{{1} \over {12}}, \,{{5} \over {12}}],\, [1],\, \, 
 64 \cdot {{x^3 \cdot (1-x)} \over  { (9 \, -8 \, x)^3}}\Bigr).
\nonumber 
\end{eqnarray}
This globally bounded series becomes a series with integer coefficients 
after the rescaling  $\, \, x \, \rightarrow  \, 27 \, x$.

\vskip .1cm 

A few hypergeometric functions occur in the analysis of the Yang-Baxter
integrable hard-hexagon~\cite{hard-hexagon0,hard-hexagon,hard-hexagon2}:
\begin{eqnarray}
\hspace{-0.8in}&&_2F_1\Bigl([{{1} \over {12}}, \,{{5} \over {12}}], \, [1]; \,
-1728\,{\frac {x \left( 1+11\,x-{x}^{2} \right)^{5}}{
 \left( 1-228\,x+494\,{x}^{2}+228\,{x}^{3}+{x}^{4} \right)^{3}}}
\Bigr)   \\
\hspace{-0.8in}&& \qquad  \, \, = \, \,  \,\, \,  \,
 \Bigl( {\frac {1-228\,x+494\,{x}^{2}+228\,{x}^{3} +{x}^{4}}{{x}^{4}
-12\,{x}^{3}+14\,{x}^{2}+12\,x+1}}
  \Bigr)^{1/4}  \nonumber \\
\hspace{-0.8in}&& \qquad  \qquad   \quad  
\times  \, _2F_1\Bigl([{{1} \over {12}}, \,{{5} \over {12}}], \, [1]; \,
-1728\,{\frac {{x}^{5} \left( 1+11\,x-{x}^{2} \right) }
{ \left( {x}^{4}-12\,{x}^{3}+14\,{x}^{2}+12\,x+1 \right) ^{3}}}
\Bigr),  \nonumber 
\end{eqnarray}
and the globally bounded algebraic hypergeometric functions 
 $\, _2F_1([1/6, \, 2/3], \, [1/2], \, x)$
 $\, _2F_1([1/4, \, 3/4], \, [2/3], \, x)$: 
\begin{eqnarray}
\hspace{-0.8in}&&_2F_1([{{1} \over {6}}, \,{{2} \over {3}}], 
\, [{{1} \over {2}}]; \, \, 27 \, x)
 \, \, = \, \,  \,  \, \,
1\, \,\,  +6\,x \, +105\,{x}^{2}\, 
+2184\,{x}^{3}\, +48906\,{x}^{4}\, +1141140\,{x}^{5}\,
 \nonumber \\
\hspace{-0.8in}&& \qquad  \qquad  \quad  +27335490\,{x}^{6}\, +666865800\,{x}^{7}\, 
+16488256905\,{x}^{8}\, \,  \,  + \, \, \cdots  
\end{eqnarray}
\begin{eqnarray}
\hspace{-0.8in}&&_2F_1([{{1} \over {4}}, \,{{3} \over {4}}], \, [{{2} \over {3}}]; \, \, 64 \, x)
 \, \, = \, \,  \,  \, \,
1\, \,\, +18\,x\,+756\,{x}^{2}\,+37422\,{x}^{3}\,+1990170\,{x}^{4}\,
\nonumber \\
\hspace{-0.8in}&& \qquad \quad  \quad  +110198556\,{x}^{5}  \,+6261870888\,{x}^{6} 
\,+362293958520\,{x}^{7}\,\, 
\, + \, \, \cdots 
\end{eqnarray}
Note that the two second-order operators 
\begin{eqnarray}
\hspace{-0.3in}&&D_x^2 \, \, \,
 + \, {{1} \over {6}} \cdot \,\,{\frac { (11\,x-3) }{x \,\cdot \, (x-1) }}
\, \cdot D_x  \, \,
\, +\, {{1} \over {9}}\,{\frac {1}{x \cdot \, (x-1) }}, 
 \nonumber \\
\hspace{-0.3in}&&
D_x^2 \,  \, \, +{{2} \over {3}} \cdot \,{\frac { (3\,x-1)}{x \,\cdot \, (x-1) }}\, \cdot D_x 
 \, \, \,
+ {{3} \over {16}} \,{\frac {1}{x \,\cdot \, (x-1) }}, 
\end{eqnarray}
annihilating respectively the two algebraic functions
 $\, _2F_1([1/6, \, 2/3], \, [1/2], \, x)$
 $\, _2F_1([1/4, \, 3/4], \, [2/3], \, x)$, are such that 
their symmetric sixth power are (non-trivially) homomorphic, 
the intertwiners being order-six operators.

 \pagebreak 

\section*{References}

\providecommand{\newblock}{}


\begin{thebibliography}{100}
\expandafter\ifx\csname url\endcsname\relax
  \def\url#1{{\tt #1}}\fi
\expandafter\ifx\csname urlprefix\endcsname\relax\def\urlprefix{URL }\fi
\providecommand{\eprint}[2][]{\url{#2}}

\bibitem{Butera}
Butera P and Comi M 2002 An on-line library of extended high-temperature
  expansions of basic observables for the spin-{$S$} {I}sing models on two- and
  three-dimensional lattices {\em J. Statist. Phys.\/} {\bf 109} 311--315
  \urlprefix\url{http://arxiv.org/abs/hep-lat/0204007}

\bibitem{Vicari}
Campostrini M, Pelissetto A, Rossi P and Vicari E 2002 25th-order
  high-temperature expansion results for three-dimensional {I}sing-like systems
  on the simple-cubic lattice {\em Phys. Rev. E\/} {\bf 65} 066127
  \urlprefix\url{http://arxiv.org/abs/cond-mat/0201180}

\bibitem{Fujiwara}
Arisue H and Fujiwara T 2003 Algorithm of the finite-lattice method for
  high-temperature expansion of the {I}sing model in three dimensions {\em
  Phys. Rev. E\/} {\bf 67} 066109
  \urlprefix\url{http://arxiv.org/abs/hep-lat/0209002}

\bibitem{Bessis}
Bessis J~D, Drouffe J~M and Moussa P 1976 Positivity constraints for the
  {I}sing ferromagnetic model {\em J. Phys. A\/} {\bf 9} 2105
  \urlprefix\url{http://stacks.iop.org/0305-4470/9/i=12/a=015}

\bibitem{Mahler}
Lalin M~N and Rogers M~D 2007 Functional equations for {M}ahler measures of
  genus-one curves {\em Algebra Number Theory\/} {\bf 1} 87--117
  \urlprefix\url{http://arxiv.org/abs/math/0612007v3}

\bibitem{Eisenstein}
Eisenstein G 1852 \"{U}ber eine allgemeine {E}igenschaft der
  {R}eihen-{E}ntwicklungen aller algebraischen {F}unktionen {\em Bericht
  K\"onigl. Preuss. Akad. Wiss. Berlin\/}  441--443

\bibitem{Heine}
Heine E 1854 \"{U}ber die {E}ntwickelung von {W}urzeln algebraischer
  {G}leichungen in {P}otenzreihen {\em J. Reine Angew. Math.\/}  267--275
  \urlprefix\url{http://dx.doi.org/10.1515/crll.1854.48.267}

\bibitem{Fatou}
Fatou P 1906 S\'eries trigonom\'etriques et s\'eries de {T}aylor {\em Acta
  Math.\/} {\bf 30} 335--400
  \urlprefix\url{http://dx.doi.org/10.1007/BF02418579}

\bibitem{Polya}
P{\'o}lya G and Szeg{\H{o}} G 1998 {\em Problems and theorems in analysis.
  {II}\/} Classics in Mathematics (Berlin: Springer-Verlag) ISBN 3-540-63686-2
  {T}heory of functions, zeros, polynomials, determinants, number theory,
  geometry

\bibitem{Polya2}
P{\'o}lya G 1916 \"{U}ber {P}otenzreihen mit ganzzahligen {K}oeffizienten {\em
  Math. Ann.\/} {\bf 77} 497--513
  \urlprefix\url{http://dx.doi.org/10.1007/BF01456965}

\bibitem{Flajolet}
Flajolet P, Gerhold S and Salvy B 2005 On the non-holonomic character of
  logarithms, powers, and the {$n$}th prime function {\em Electron. J.
  Combin.\/} {\bf 11} Research Paper A2, 16 pp. (electronic)
  \urlprefix\url{http://www.combinatorics.org/ojs/index.php/eljc/article/view/v11i2a2}

\bibitem{Carlson}
Carlson F 1921 \"{U}ber {P}otenzreihen mit ganzzahligen {K}oeffizienten {\em
  Math. Z.\/} {\bf 9} 1--13
  \urlprefix\url{http://dx.doi.org/10.1007/BF01378331}

\bibitem{Polya3}
P{\'o}lya G 1923 Sur {L}es {S}eries {E}nti\`eres \`a {C}oefficients {E}ntiers
  {\em Proc. London Math. Soc.\/} {\bf 21} 22--38
  \urlprefix\url{http://dx.doi.org/10.1112/plms/s2-21.1.22}

\bibitem{ze-bo-ha-ma-04}
Zenine N, Boukraa S, Hassani S and Maillard {J-M} 2004 The {F}uchsian
  differential equation of the square lattice {I}sing model {$\chi^{(3)}$}
  susceptibility {\em J. Phys. A\/} {\bf 37} 9651--9668
  \urlprefix\url{http://arxiv.org/abs/math-ph/0407060}

\bibitem{Khi6}
Boukraa S, Hassani S, Jensen I, Maillard {J-M} and Zenine N 2010 High-order
  {F}uchsian equations for the square lattice {I}sing model: {$\chi^{(6)}$}
  {\em J. Phys. A\/} {\bf 43} 115201, 22
  \urlprefix\url{http://arxiv.org/abs/0912.4968}

\bibitem{ze-bo-ha-ma-05c}
Zenine N, Boukraa S, Hassani S and Maillard {J-M} 2005 Square lattice {I}sing
  model susceptibility: connection matrices and singular behaviour of
  {$\chi^{(3)}$} and {$\chi^{(4)}$} {\em J. Phys. A\/} {\bf 38} 9439--9474
  \urlprefix\url{http://arxiv.org/abs/math-ph/0506065}

\bibitem{bo-ha-ma-ze-07b}
Boukraa S, Hassani S, Maillard {J-M} and Zenine N 2007 Singularities of
  {$n$}-fold integrals of the {I}sing class and the theory of elliptic curves
  {\em J. Phys. A\/} {\bf 40} 11713--11748
  \urlprefix\url{http://arxiv.org/abs/0706.3367}

\bibitem{mccoy3}
Boukraa S, Hassani S, Maillard {J-M}, McCoy B~M and Zenine N 2007 The diagonal
  {I}sing susceptibility {\em J. Phys. A\/} {\bf 40} 8219--8236
  \urlprefix\url{http://arxiv.org/abs/math-ph/0703009}

\bibitem{bo-gu-ha-je-ma-ni-ze-08}
Boukraa S, Guttmann A~J, Hassani S, Jensen I, Maillard {J-M}, Nickel B and Zenine
  N 2008 Experimental mathematics on the magnetic susceptibility of the square
  lattice {I}sing model {\em J. Phys. A\/} {\bf 41} 455202, 51
  \urlprefix\url{http://arxiv.org/abs/0808.0763}

\bibitem{Stanley80}
Stanley R~P 1980 Differentiably finite power series {\em European J. Combin.\/}
  {\bf 1} 175--188

\bibitem{Lipshitz89}
Lipshitz L 1989 {$D$}-finite power series {\em J. Algebra\/} {\bf 122} 353--373
  \urlprefix\url{http://dx.doi.org/10.1016/0021-8693(89)90222-6}

\bibitem{Bernstein}
Bern{\v{s}}te{\u\i}n I~N 1972 Analytic continuation of generalized functions
  with respect to a parameter {\em Funkcional. Anal. i Prilo\v zen.\/} {\bf 6}
  26--40
  \urlprefix\url{http://www.math.tau.ac.il/~bernstei/Publication_list/publication_texts/Bern-a-cont-FAN.pdf}

\bibitem{Kashiwara}
Kashiwara M 1978 On the holonomic systems of linear differential equations.
  {II} {\em Invent. Math.\/} {\bf 49} 121--135
  \urlprefix\url{http://dx.doi.org/10.1007/BF01403082}

\bibitem{Takayama}
Takayama N 1992 An approach to the zero recognition problem by {B}uchberger
  algorithm {\em J. Symbolic Comput.\/} {\bf 14} 265--282
  \urlprefix\url{http://dx.doi.org/10.1016/0747-7171(92)90039-7}

\bibitem{wu-mc-tr-ba-76}
Wu T~T, McCoy B~M, Tracy C~A and Barouch E 1976 Spin-spin correlation functions
  for the two-dimensional {I}sing model: {E}xact theory in the scaling region
  {\em Phys. Rev. B\/} {\bf 13} 316--374
  \urlprefix\url{http://dx.doi.org/10.1103/PhysRevB.13.316}

\bibitem{ze-bo-ha-ma-05b}
Zenine N, Boukraa S, Hassani S and Maillard {J-M} 2005 Ising model
  susceptibility: the {F}uchsian differential equation for {$\chi^{(4)}$} and
  its factorization properties {\em J. Phys. A\/} {\bf 38} 4149--4173
  \urlprefix\url{http://arxiv.org/abs/cond-mat/0502155}

\bibitem{High}
Bostan A, Boukraa S, Guttmann A~J, Hassani S, Jensen I, Maillard {J-M} and Zenine
  N 2009 High order {F}uchsian equations for the square lattice {I}sing model:
  {$\tilde\chi{}^{(5)}$} {\em J. Phys. A\/} {\bf 42} 275209, 32
  \urlprefix\url{http://arxiv.org/abs/0904.1601}

\bibitem{bo-bo-ha-ma-we-ze-09}
Bostan A, Boukraa S, Hassani S, Maillard {J-M}, Weil {J-A} and Zenine N 2009
  Globally nilpotent differential operators and the square {I}sing model {\em
  J. Phys. A\/} {\bf 42} 125206, 50
  \urlprefix\url{http://arxiv.org/abs/0812.4931}

\bibitem{Andre5}
Andr{\'e} Y 2003 Arithmetic {G}evrey series and transcendence. {A} survey {\em
  J. Th\'eor. Nombres Bordeaux\/} {\bf 15} 1--10 {L}es XXII{\`e}mes
  Journ{\'e}es Arithmetiques (Lille, 2001)
  \urlprefix\url{http://jtnb.cedram.org/item?id=JTNB_2003__15_1_1_0}

\bibitem{Andre6}
Andr{\'e} Y 1989 {\em {$G$}-functions and geometry\/} Aspects of Mathematics,
  E13 (Braunschweig: Friedr. Vieweg \& Sohn) ISBN 3-528-06317-3

\bibitem{Kratten}
Krattenthaler C and Rivoal T 2010 On the integrality of the {T}aylor
  coefficients of mirror maps {\em Duke Math. J.\/} {\bf 151} 175--218
  \urlprefix\url{http://arxiv.org/abs/0907.2577}

\bibitem{CalabiYauIsing1}
Bostan A, Boukraa S, Hassani S, van Hoeij M, Maillard {J-M}, Weil {J-A} and Zenine
  N 2011 The {I}sing model: from elliptic curves to modular forms and
  {C}alabi-{Y}au equations {\em J. Phys. A\/} {\bf 44} 045204, 44
  \urlprefix\url{http://arxiv.org/abs/1007.0535}

\bibitem{CalabiYauIsing}
Assis M, Boukraa S, Hassani S, van Hoeij M, Maillard {J-M} and McCoy B~M 2012
  Diagonal {I}sing susceptibility: elliptic integrals, modular forms and
  {C}alabi-{Y}au equations {\em J. Phys. A\/} {\bf 45} 075205, 32
  \urlprefix\url{http://arxiv.org/abs/1110.1705}

\bibitem{SP4}
Yang Y and Zudilin W 2010 On {${\rm Sp}_4$} modularity of {P}icard-{F}uchs
  differential equations for {C}alabi-{Y}au threefolds {\em Gems in
  experimental mathematics\/} ({\em Contemp. Math.\/} vol 517) (Providence, RI:
  Amer. Math. Soc.) pp 381--413 \urlprefix\url{http://arxiv.org/abs/0803.3322}

\bibitem{LianYau}
Lian B~H and Yau S~T 1996 Mirror maps, modular relations and hypergeometric
  series. {II} {\em Nuclear Phys. B Proc. Suppl.\/} {\bf 46} 248--262
  {$S$}-duality and mirror symmetry (Trieste, 1995)
  \urlprefix\url{http://arxiv.org/abs/hep-th/9507153}

\bibitem{Ford}
Ford D, McKay J and Norton S 1994 More on replicable functions {\em Comm.
  Algebra\/} {\bf 22} 5175--5193
  \urlprefix\url{http://dx.doi.org/10.1080/00927879408825127}

\bibitem{McKay}
McKay J and Sebbar A 2007 Replicable functions: an introduction {\em Frontiers
  in number theory, physics, and geometry. {II}\/} (Berlin: Springer) pp
  373--386 \urlprefix\url{http://dx.doi.org/10.1007/978-3-540-30308-4_10}

\bibitem{Basraoui}
{El Basraoui} A 2005 {\em Modular Functions and Replicable Functions\/}
  Master's thesis University of Ottawa Canada
  \urlprefix\url{http://www.ruor.uottawa.ca/fr/bitstream/handle/10393/10771/MR16975.PDF?sequence=1}

\bibitem{Livne}
Livn{\'e} R, Sch{\"u}tt M and Yui N 2010 The modularity of {$K3$} surfaces with
  non-symplectic group actions {\em Math. Ann.\/} {\bf 348} 333--355
  \urlprefix\url{http://arxiv.org/abs/0904.1922}

\bibitem{Schutt}
Sch{\"u}tt M 2004 New examples of modular rigid {C}alabi-{Y}au threefolds {\em
  Collect. Math.\/} {\bf 55} 219--228
  \urlprefix\url{http://arxiv.org/abs/math/0311106}

\bibitem{Gouvea}
Gouv{\^e}a F~Q and Yui N 2011 Rigid {C}alabi-{Y}au threefolds over {$\Bbb Q$}
  are modular {\em Expo. Math.\/} {\bf 29} 142--149
  \urlprefix\url{http://arxiv.org/abs/0902.1466}

\bibitem{Fontaine}
Fontaine J~M and Mazur B 1995 Geometric {G}alois representations {\em Elliptic
  curves, modular forms, \& {F}ermat's last theorem ({H}ong {K}ong, 1993)\/}
  Ser. Number Theory, I (Int. Press, Cambridge, MA) pp 41--78
  \urlprefix\url{https://www.dpmms.cam.ac.uk/~ty245/2010_GalRep_L24/Fontaine-Mazur.pdf}

\bibitem{Saito}
Saito M~H and Yui N 2001 The modularity conjecture for rigid {C}alabi-{Y}au
  threefolds over {$Q$} {\em J. Math. Kyoto Univ.\/} {\bf 41} 403--419
  \urlprefix\url{http://arxiv.org/abs/math/0009041}

\bibitem{Apery}
van~der Poorten A 1978/79 A proof that {E}uler missed{$\ldots $}{A}p\'ery's
  proof of the irrationality of {$\zeta (3)$} {\em Math. Intelligencer\/} {\bf
  1} 195--203 {A}n informal report
  \urlprefix\url{http://dx.doi.org/10.1007/BF03028234}

\bibitem{GlasserGuttmann}
Glasser M~L and Guttmann A~J 1994 Lattice {G}reen function (at {$0$}) for the
  {$4$}{D} hypercubic lattice {\em J. Phys. A\/} {\bf 27} 7011--7014
  \urlprefix\url{http://arxiv.org/abs/cond-mat/9408097}

\bibitem{Prell}
Guttmann A~J and Prellberg T 1993 Staircase polygons, elliptic integrals,
  {H}eun functions, and lattice {G}reen functions {\em Physical Review E\/}
  {\bf 47} 2233--2236
  \urlprefix\url{http://www.maths.qmul.ac.uk/~tp/papers/pub004.pdf}

\bibitem{Zucker}
Joyce G~S, Delves R~T and Zucker I~J 2003 Exact evaluation of the {G}reen
  functions for the anisotropic face-centred and simple cubic lattices {\em J.
  Phys. A\/} {\bf 36} 8661--8672
  \urlprefix\url{http://dx.doi.org/10.1088/0305-4470/36/32/307}

\bibitem{GoodGuttmann}
Guttmann A~J 2010 Lattice {G}reen's functions in all dimensions {\em J. Phys.
  A\/} {\bf 43} 305205, 26 \urlprefix\url{http://arxiv.org/abs/1004.1435}

\bibitem{Christol}
Christol G 1990 Globally bounded solutions of differential equations {\em
  Analytic number theory ({T}okyo, 1988)\/} ({\em Lecture Notes in Math.\/} vol
  1434) (Berlin: Springer) pp 45--64
  \urlprefix\url{http://dx.doi.org/10.1007/BFb0097124}

\bibitem{WuJPA}
Boukraa S, Hassani S and Maillard {J-M} 2012 Holonomic functions of several
  complex variables and singularities of anisotropic {I}sing $n$-fold integrals
  {\em J. Phys. A\/} {\bf 45} 33 pp.
  \urlprefix\url{http://arxiv.org/abs/1207.1784}

\bibitem{Barouch}
Barouch E 1980 On the {I}sing model in the presence of magnetic field {\em
  Phys. D\/} {\bf 1} 333--337
  \urlprefix\url{http://dx.doi.org/10.1016/0167-2789(80)90031-7}

\bibitem{Guttmann}
Guttmann A~J 2009 Lattice {G}reen functions and {C}alabi-{Y}au differential
  equations {\em J. Phys. A\/} {\bf 42} 232001, 6
  \urlprefix\url{http://dx.doi.org/10.1088/1751-8113/42/23/232001}

\bibitem{Arnold}
Arnold W~I 2001 On teaching mathematics {\em Wiadom. Mat.\/} {\bf 37} 17--26
  {T}ranslated from the Russian by Danuta {\'S}ledziewska-B{\l}ocka

\bibitem{Arnold2}
Arnold V~I 1998 Sur l'\'education math\'ematique {\em Gaz. Math.\/}  19--29
  \urlprefix\url{http://smf4.emath.fr/Publications/Gazette/1998/78/smf_gazette_78_19-29.pdf}

\bibitem{Pochekutov}
Pochekutov D~Y 2009 Diagonals of the {L}aurent series of rational functions
  {\em Sibirsk. Mat. Zh.\/} {\bf 50} 1370--1383
  \urlprefix\url{http://dx.doi.org/10.1007/s11202-009-0119-z}

\bibitem{Purdue}
Lipshitz L and van~der Poorten A~J 1990 Rational functions, diagonals, automata
  and arithmetic {\em Number theory ({B}anff, {AB}, 1988)\/} (Berlin: de
  Gruyter) pp 339--358

\bibitem{Lipshitz}
Lipshitz L 1988 The diagonal of a {$D$}-finite power series is {$D$}-finite
  {\em J. Algebra\/} {\bf 113} 373--378
  \urlprefix\url{http://dx.doi.org/10.1016/0021-8693(88)90166-4}

\bibitem{BA-JPB}
Adamczewski B and Bell J~P Diagonalization and {R}ationalization of algebraic
  {L}aurent series {P}reprint \url{http://arxiv.org/pdf/1205.4090.pdf}

\bibitem{legacy2}
Christol G 1986-1987 Fonctions hyperg\'eom\'etriques born\'ees {\em Groupe de
  travail d'analyse ultram\'etrique\/} {\bf 14} 1--16
  \urlprefix\url{http://archive.numdam.org/article/GAU_1986-1987__14__A4_0.pdf}

\bibitem{Polya21}
P\'olya G 1921--1922 Sur les s\'eries enti\`eres, dont la somme est une
  fonction alg\'ebrique {\em Enseignement Math.\/} {\bf 22} 38--47
  \urlprefix\url{http://retro.seals.ch/digbib/view?rid=ensmat-001:1921-1922:22::220}

\bibitem{CaMa38}
Cameron R~H and Martin W~T 1938 Analytic continuation of diagonals and
  {H}adamard compositions of multiple power series {\em Trans. Amer. Math.
  Soc.\/} {\bf 44} 1--7 \urlprefix\url{http://dx.doi.org/10.2307/1990100}

\bibitem{Hadamard}
Hadamard J 1892 Essai sur l'\'etude des fonctions donn\'ees par leur
  d\'eveloppement de {T}aylor {\em J. Math. Pures Appl.\/} {\bf 4} 101--186
  \urlprefix\url{http://portail.mathdoc.fr/JMPA/PDF/JMPA_1892_4_8_A4_0.pdf}

\bibitem{FlSe09}
Flajolet P and Sedgewick R 2009 {\em Analytic combinatorics\/} (Cambridge:
  Cambridge University Press) ISBN 978-0-521-89806-5
  \urlprefix\url{http://algo.inria.fr/flajolet/Publications/AnaCombi/anacombi}

\bibitem{Deligne84}
Deligne P 1984 Int\'egration sur un cycle \'evanescent {\em Invent. Math.\/}
  {\bf 76} 129--143 \urlprefix\url{http://dx.doi.org/10.1007/BF01388496}

\bibitem{Pech}
Bostan A, Chyzak F, van Hoeij M and Pech L 2011 Explicit formula for the
  generating series of diagonal {3D} rook paths {\em S\'eminaire Lotharingien
  de Combinatoire\/} {\bf B66a}
  \urlprefix\url{http://www.emis.de/journals/SLC/wpapers/s66bochhope.html}

\bibitem{Christol84}
Christol G 1984 Diagonales de fractions rationnelles et \'equations
  diff\'erentielles {\em Study group on ultrametric analysis, 10th year:
  1982/83, {N}o. 2, Exp. No. 18\/} (Paris: Inst. Henri Poincar\'e) pp 1--10
  \urlprefix\url{http://archive.numdam.org/article/GAU_1982-1983__10_2_A4_0.pdf}

\bibitem{Christol85}
Christol G 1985 Diagonales de fractions rationnelles et \'equations de
  {P}icard-{F}uchs {\em Study group on ultrametric analysis, 12th year,
  1984/85, {N}o.\ 1, Exp.\ No.\ 13\/} (Paris: Secr\'etariat Math.) pp 1--12
  \urlprefix\url{http://archive.numdam.org/article/GAU_1984-1985__12_1_A8_0.pdf}

\bibitem{Christol369}
Christol G 1988 Diagonales de fractions rationnelles {\em S\'eminaire de
  {T}h\'eorie des {N}ombres, {P}aris 1986--87\/} ({\em Progr. Math.\/} vol~75)
  (Boston, MA: Birkh\"auser Boston) pp 65--90

\bibitem{Fu}
Furstenberg H 1967 Algebraic functions over finite fields {\em J. Algebra\/}
  {\bf 7} 271--277
  \urlprefix\url{http://dx.doi.org/10.1016/0021-8693(67)90061-0}

\bibitem{Ihara}
Ihara Y 1974 Schwarzian equations {\em J. Fac. Sci. Univ. Tokyo Sect. IA
  Math.\/} {\bf 21} 97--118
  \urlprefix\url{http://repository.dl.itc.u-tokyo.ac.jp/dspace/bitstream/2261/6446/1/jfs210106.pdf}

\bibitem{Honda}
Honda T 1981 Algebraic differential equations {\em Symposia {M}athematica,
  {V}ol. {XXIV} ({S}ympos., {INDAM}, {R}ome, 1979)\/} (London: Academic Press)
  pp 169--204

\bibitem{Koike99}
Koike M 1999 Hypergeometric polynomials over finite fields {\em Tohoku Math. J.
  (2)\/} {\bf 51} 75--79
  \urlprefix\url{http://dx.doi.org/10.2748/tmj/1178224854}

\bibitem{Deuring}
Deuring M 1941 Die {T}ypen der {M}ultiplikatorenringe elliptischer
  {F}unktionenk\"orper {\em Abh. Math. Sem. Hansischen Univ.\/} {\bf 14}
  197--272 \urlprefix\url{http://dx.doi.org/10.1007/BF02940746}

\bibitem{Igusa}
Igusa J~i 1958 Class number of a definite quaternion with prime discriminant
  {\em Proc. Nat. Acad. Sci. U.S.A.\/} {\bf 44} 312--314
  \urlprefix\url{http://dx.doi.org/10.1073/pnas.44.4.312}

\bibitem{Necer}
Necer A 1997 S\'eries formelles et produit de {H}adamard {\em J. Th\'eor.
  Nombres Bordeaux\/} {\bf 9} 319--335
  \urlprefix\url{http://archive.numdam.org/article/JTNB_1997__9_2_319_0.pdf}

\bibitem{Fliess74}
Fliess M 1974 Sur divers produits de s\'eries formelles {\em Bull. Soc. Math.
  France\/} {\bf 102} 181--191
  \urlprefix\url{http://archive.numdam.org/article/BSMF_1974__102__181_0.pdf}

\bibitem{Hurwitz1898}
Hurwitz A 1898 Ueber die {E}ntwickelungscoefficienten der lemniscatischen
  {F}unctionen {\em Math. Ann.\/} {\bf 51} 196--226
  \urlprefix\url{http://dx.doi.org/10.1007/BF01453637}

\bibitem{Hurwitz1899}
Hurwitz A 1898 Sur un th\'eor\`eme de {M}. {H}adamard {\em C. R. Math. Acad.
  Sci. Paris\/} {\bf 128} 350--353

\bibitem{Lamperti}
Lamperti J 1958 On the coefficients of reciprocal power series {\em Amer. Math.
  Monthly\/} {\bf 65} 90--94
  \urlprefix\url{http://www.jstor.org/stable/2308880}

\bibitem{Trjitzinsky}
Trjitzinsky W~J 1930 On composition of singularities {\em Trans. Amer. Math.
  Soc.\/} {\bf 32} 196--215 \urlprefix\url{http://dx.doi.org/10.2307/1989490}

\bibitem{Denef}
Denef J and Lipshitz L 1987 Algebraic power series and diagonals {\em J. Number
  Theory\/} {\bf 26} 46--67
  \urlprefix\url{http://dx.doi.org/10.1016/0022-314X(87)90095-3}

\bibitem{bo-ha-ma-ze-07}
Boukraa S, Hassani S, Maillard {J-M} and Zenine N 2007 Landau singularities and
  singularities of holonomic integrals of the {I}sing class {\em J. Phys. A\/}
  {\bf 40} 2583--2614 \urlprefix\url{http://arxiv.org/abs/math-ph/0701016}

\bibitem{Sathaye}
Sathaye A 1983 Generalized {N}ewton-{P}uiseux expansion and {A}bhyankar-{M}oh
  semigroup theorem {\em Invent. Math.\/} {\bf 74} 149--157
  \urlprefix\url{http://dx.doi.org/10.1007/BF01388535}

\bibitem{Kauers}
Aparicio-Monforte A and Kauers M 2012 Formal {L}aurent series in several
  variables {\em Expo. Math.\/} {T}o appear
  \urlprefix\url{http://www.risc.jku.at/people/mkauers/publications/aparicio12.pdf}

\bibitem{nickel-99}
Nickel B 1999 On the singularity structure of the 2{D} {I}sing model
  susceptibility {\em J. Phys. A\/} {\bf 32} 3889--3906
  \urlprefix\url{http://dx.doi.org/10.1088/0305-4470/32/21/303}

\bibitem{nickel-00}
Nickel B 2000 Addendum to: ``{O}n the singularity structure of the 2{D} {I}sing
  model susceptibility'' [{J}. {P}hys. {A} {\bf 32} (1999), 3889--3906] {\em J.
  Phys. A\/} {\bf 33} 1693--1711
  \urlprefix\url{http://dx.doi.org/10.1088/0305-4470/33/8/313}

\bibitem{2005-chi3-method}
Zenine N, Boukraa S, Hassani S and Maillard {J-M} 2005 Square lattice {I}sing
  model susceptibility: series expansion method and differential equation for
  {$\chi^{(3)}$} {\em J. Phys. A\/} {\bf 38} 1875--1899
  \urlprefix\url{http://arxiv.org/abs/hep-th/0411051}

\bibitem{Almkvist}
Almkvist G and Zudilin W 2006 Differential equations, mirror maps and zeta
  values {\em Mirror symmetry. {V}\/} ({\em AMS/IP Stud. Adv. Math.\/} vol~38)
  (Providence, RI: Amer. Math. Soc.) pp 481--515
  \urlprefix\url{http://arxiv.org/abs/math/0402386}

\bibitem{Morrison}
Morrison D~R 1993 Mirror symmetry and rational curves on quintic threefolds: a
  guide for mathematicians {\em J. Amer. Math. Soc.\/} {\bf 6} 223--247
  \urlprefix\url{http://arxiv.org/abs/alg-geom/9202004}

\bibitem{Batyrev}
Batyrev V~V and van Straten D 1995 Generalized hypergeometric functions and
  rational curves on {C}alabi-{Y}au complete intersections in toric varieties
  {\em Comm. Math. Phys.\/} {\bf 168} 493--533
  \urlprefix\url{http://arxiv.org/abs/alg-geom/9307010}

\bibitem{TablesCalabi}
Almkvist G, van Enckevort C, van Straten D and Zudilin W Tables of
  {C}alabi--{Y}au equations {P}reprint \url{http://arxiv.org/pdf/math/0507430}
  and database \url{http://enriques.mathematik.uni-mainz.de/CYequations/}

\bibitem{chen-yang-yui-08}
Chen Y~H, Yang Y and Yui N 2008 Monodromy of {P}icard-{F}uchs differential
  equations for {C}alabi-{Y}au threefolds {\em J. Reine Angew. Math.\/} {\bf
  616} 167--203 {W}ith an appendix by Cord Erdenberger
  \urlprefix\url{http://arxiv.org/abs/math/0605675}

\bibitem{Andre7}
Andr{\'e} Y 2004 Sur la conjecture des {$p$}-courbures de {G}rothendieck-{K}atz
  et un probl\`eme de {D}work {\em Geometric aspects of {D}work theory. {V}ol.
  {I}, {II}\/} (Walter de Gruyter GmbH \& Co. KG, Berlin) pp 55--112

\bibitem{Almkvist1}
Almkvist G 2006 {C}alabi-{Y}au differential equations of degree 2 and 3 and
  {Y}ifan {Y}ang's pullback {P}reprint \url{http://arxiv.org/abs/math/0612215}

\bibitem{Almkvist2}
Almkvist G 2007 Binomial identities related to {C}alabi--{Y}au differential
  equations {P}reprint \url{http://arxiv.org/abs/math/0703255}

\bibitem{Zagier}
Zagier D 2009 Integral solutions of {A}p\'ery-like recurrence equations {\em
  Groups and symmetries\/} ({\em CRM Proc. Lecture Notes\/} vol~47)
  (Providence, RI: Amer. Math. Soc.) pp 349--366
  \urlprefix\url{http://people.mpim-bonn.mpg.de/zagier/files/tex/AperylikeRecEqs/fulltext.pdf}

\bibitem{Egorychev}
Egorychev G~P 1984 {\em Integral representation and the computation of
  combinatorial sums\/} ({\em Translations of Mathematical Monographs\/}
  vol~59) (Providence, RI: American Mathematical Society) ISBN 0-8218-4512-8
  {T}ranslated from the Russian by H. H. McFadden, Translation edited by Lev J.
  Leifman

\bibitem{Letterto}
Deligne P 1973 Deligne's letter to {P}iatetski-{S}hapiro
  \urlprefix\url{http://www.math.ias.edu/~jaredw/DeligneLetterToPiatetskiShapiro.pdf}

\bibitem{Strehl}
Strehl V 1994 Binomial identities---combinatorial and algorithmic aspects {\em
  Discrete Math.\/} {\bf 136} 309--346 {T}rends in discrete mathematics
  \urlprefix\url{http://dx.doi.org/10.1016/0012-365X(94)00118-3}

\bibitem{Schmidt}
Schmidt A~L 1995 Legendre transforms and {A}p\'ery's sequences {\em J. Austral.
  Math. Soc. Ser. A\/} {\bf 58} 358--375
  \urlprefix\url{http://dx.doi.org/10.1017/S1446788700038350}

\bibitem{Zudilin}
Zudilin W 2004 On a combinatorial problem of {A}smus {S}chmidt {\em Electron.
  J. Combin.\/} {\bf 11} Research Paper 22, 8 pp. (electronic)
  \urlprefix\url{http://www.combinatorics.org/Volume_11/Abstracts/v11i1r22.html}

\bibitem{KZ}
Kontsevich M and Zagier D 2001 Periods {\em Mathematics unlimited---2001 and
  beyond\/} (Berlin: Springer) pp 771--808
  \urlprefix\url{http://people.mpim-bonn.mpg.de/zagier/files/periods/fulltext.pdf}

\bibitem{JA}
Ayoub J 2012 Une version relative de la conjecture des p\'eriodes de
  {K}ontsevich-{Z}agier {\em Ann. of Math.\/} {T}o appear
  \urlprefix\url{http://www.math.uiuc.edu/K-theory/1010/rel-KZ.pdf}

\bibitem{evanescent}
Perron B 1989 Les cycles \'evanescents sont d\'enou\'es {\em Ann. Sci. \'Ecole
  Norm. Sup. (4)\/} {\bf 22} 227--253
  \urlprefix\url{archive.numdam.org/article/ASENS_1989_4_22_2_227_0.pdf}

\bibitem{Hattori}
Hattori A and Kimura T 1974 On the {E}uler integral representations of
  hypergeometric functions in several variables {\em J. Math. Soc. Japan\/}
  {\bf 26} 1--16
  \urlprefix\url{http://projecteuclid.org/euclid.jmsj/1240435361}

\bibitem{Driver}
Driver K~A and Johnston S~J 2006 An integral representation of some
  hypergeometric functions {\em Electron. Trans. Numer. Anal.\/} {\bf 25}
  115--120

\bibitem{GR1}
Grothendieck A 1966 On the de {R}ham cohomology of algebraic varieties {\em
  Inst. Hautes \'Etudes Sci. Publ. Math.\/}  95--103

\bibitem{Hodge}
Katz N~M 1972 Algebraic solutions of differential equations ({$p$}-curvature
  and the {H}odge filtration) {\em Invent. Math.\/} {\bf 18} 1--118
  \urlprefix\url{http://dx.doi.org/10.1007/BF01389714}

\bibitem{bernie2010}
Nickel B, Jensen I, Boukraa S, Guttmann A~J, Hassani S, Maillard {J-M} and Zenine
  N 2010 Square lattice {I}sing model {$\tilde\chi{}^{(5)}$} {ODE} in exact
  arithmetic {\em J. Phys. A\/} {\bf 43} 195205, 24
  \urlprefix\url{http://arxiv.org/abs/1002.0161}

\bibitem{DGS}
Dwork B, Gerotto G and Sullivan F~J 1994 {\em An introduction to
  {$G$}-functions\/} ({\em Annals of Mathematics Studies\/} vol 133)
  (Princeton, NJ: Princeton University Press) ISBN 0-691-03681-0

\bibitem{BeHe89}
Beukers F and Heckman G 1989 Monodromy for the hypergeometric function
  {$_nF_{n-1}$} {\em Invent. Math.\/} {\bf 95} 325--354
  \urlprefix\url{http://dx.doi.org/10.1007/BF01393900}

\bibitem{JAW}
van Hoeij M and Weil {J-A} 2005 Solving second order linear differential
  equations with {K}lein's theorem {\em Proceedings of the 2005 international
  symposium on Symbolic and algebraic computation\/} ISSAC'05 (New York, NY,
  USA: ACM) pp 340--347 ISBN 1-59593-095-7
  \urlprefix\url{http://doi.acm.org/10.1145/1073884.1073931}

\bibitem{Renorm}
Bostan A, Boukraa S, Hassani S, Maillard {J-M}, Weil {J-A}, Zenine N and Abarenkova
  N 2010 Renormalization, isogenies, and rational symmetries of differential
  equations {\em Adv. Math. Phys.\/}  Art. ID 941560, 44
  \urlprefix\url{http://arxiv.org/abs/0911.5466}

\bibitem{AtkinMorain}
Atkin A~O~L and Morain F 1993 Elliptic curves and primality proving {\em Math.
  Comp.\/} {\bf 61} 29--68 \urlprefix\url{http://dx.doi.org/10.2307/2152935}

\bibitem{Chan}
Chan H~H 1998 On {R}amanujan's cubic transformation formula for {${}_2F_1(\frac
  13,\frac 23;1;z)$} {\em Math. Proc. Cambridge Philos. Soc.\/} {\bf 124}
  193--204 \urlprefix\url{http://dx.doi.org/10.1017/S0305004198002643}

\bibitem{Maier}
Maier R~S 2005 On reducing the {H}eun equation to the hypergeometric equation
  {\em J. Differential Equations\/} {\bf 213} 171--203
  \urlprefix\url{http://arxiv.org/abs/math/0203264}

\bibitem{Golyshev}
Golyshev V and Stienstra J 2007 Fuchsian equations of type {DN} {\em Commun.
  Number Theory Phys.\/} {\bf 1} 323--346
  \urlprefix\url{http://arxiv.org/abs/math/0701936}

\bibitem{Huse}
Husem{\"o}ller D 2004 {\em Elliptic curves\/} 2nd ed ({\em Graduate Texts in
  Mathematics\/} vol 111) (New York: Springer-Verlag) ISBN 0-387-95490-2 {W}ith
  appendices by Otto Forster, Ruth Lawrence and Stefan Theisen

\bibitem{Bouw}
Bouw I~I and M{\"o}ller M 2010 Differential equations associated with
  nonarithmetic {F}uchsian groups {\em J. Lond. Math. Soc. (2)\/} {\bf 81}
  65--90 \urlprefix\url{http://arxiv.org/abs/0710.5277}

\bibitem{Dettweiler}
Dettweiler M and Reiter S 2010 On globally nilpotent differential equations
  {\em J. Differential Equations\/} {\bf 248} 2736--2745
  \urlprefix\url{http://arxiv.org/abs/math/0605383}

\bibitem{Manin}
Manin Y~I 1999 {\em Frobenius manifolds, quantum cohomology, and moduli
  spaces\/} ({\em American Mathematical Society Colloquium Publications\/}
  vol~47) (Providence, RI: American Mathematical Society) ISBN 0-8218-1917-8

\bibitem{mirror}
Lian B~H and Yau S~T 1996 Arithmetic properties of mirror map and quantum
  coupling {\em Comm. Math. Phys.\/} {\bf 176} 163--191
  \urlprefix\url{http://arxiv.org/abs/hep-th/9411234}

\bibitem{Maier1}
Maier R~S 2009 On rationally parametrized modular equations {\em J. Ramanujan
  Math. Soc.\/} {\bf 24} 1--73
  \urlprefix\url{http://arxiv.org/abs/math/0611041}

\bibitem{Maier7}
Maier R~S 2007 The 192 solutions of the {H}eun equation {\em Math. Comp.\/}
  {\bf 76} 811--843 (electronic)
  \urlprefix\url{http://arxiv.org/abs/math/0408317}

\bibitem{Cooper}
Cooper S 2009 Inversion formulas for elliptic functions {\em Proc. Lond. Math.
  Soc. (3)\/} {\bf 99} 461--483
  \urlprefix\url{http://dx.doi.org/10.1112/plms/pdp007}

\bibitem{Varilly}
V{\'a}rilly-Alvarado A and Zywina D 2009 Arithmetic {$E_8$} lattices with
  maximal {G}alois action {\em LMS J. Comput. Math.\/} {\bf 12} 144--165
  \urlprefix\url{http://arxiv.org/abs/0803.3063}

\bibitem{Villegas}
Rodriguez-Villegas F Integral ratios of factorials and algebraic hypergeometric
  functions {P}reprint
  \url{http://www.math.utexas.edu/users/villegas/publications/oberwolfach-05.pdf}
  (Summary of talk at Oberwolfach 2005)

\bibitem{Chebyshev}
Chebyshev P~L 1852 M\'emoire sur les nombres premiers {\em J. de Math. Pures
  Appl.\/} {\bf 17} 366--390
  \urlprefix\url{http://archive.numdam.org/article/ASENS_1920_3_37__1_0.pdf}

\bibitem{Delaygue}
Delaygue E 2011 {\em Propri\'et\'es arithm\'etiques des applications miroir\/}
  Ph.D. thesis Universit\'e de Grenoble
  \urlprefix\url{http://tel.archives-ouvertes.fr/tel-00628016/}

\bibitem{Bober}
Bober J~W 2009 Factorial ratios, hypergeometric series, and a family of step
  functions {\em J. Lond. Math. Soc. (2)\/} {\bf 79} 422--444
  \urlprefix\url{http://arxiv.org/abs/0709.1977}

\bibitem{Katz}
Katz N~M 1990 {\em Exponential sums and differential equations\/} ({\em Annals
  of Mathematics Studies\/} vol 124) (Princeton, NJ: Princeton University
  Press) ISBN 0-691-08598-6; 0-691-08599-4

\bibitem{Kean}
McKean H and Moll V 1997 {\em Elliptic curves\/} (Cambridge: Cambridge
  University Press) ISBN 0-521-58228-8; 0-521-65817-9 {F}unction theory,
  geometry, arithmetic

\bibitem{Vidunas}
Vid{\=u}nas R and Filipuk G 2009 Parametric transformations between the {H}eun
  and {G}auss hypergeometric functions {P}reprint
  \url{http://arxiv.org/abs/0910.3087}

\bibitem{Valent}
Valent G 2007 Heun functions versus elliptic functions {\em Difference
  equations, special functions and orthogonal polynomials\/} (World Sci. Publ.,
  Hackensack, NJ) pp 664--686
  \urlprefix\url{http://arxiv.org/abs/math-ph/0512006}

\bibitem{Verdier}
Treibich A and Verdier {J-L} 1990 Solitons elliptiques {\em The {G}rothendieck
  {F}estschrift, {V}ol.\ {III}\/} ({\em Progr. Math.\/} vol~88) (Boston, MA:
  Birkh\"auser Boston) pp 437--480 {W}ith an appendix by J. Oesterl{\'e}
  \urlprefix\url{http://dx.doi.org/10.1007/978-0-8176-4576-2_11}

\bibitem{spanning}
Guttmann A~J and Rogers M~D 2012 Spanning tree generating functions and
  {M}ahler measures {P}reprint \url{http://arxiv.org/abs/1207.2815}

\bibitem{Stanley99}
Stanley R~P 1999 {\em Enumerative combinatorics. {V}ol. 2\/} ({\em Cambridge
  Studies in Advanced Mathematics\/} vol~62) (Cambridge: Cambridge University
  Press) ISBN 0-521-56069-1; 0-521-78987-7 {W}ith a foreword by Gian-Carlo Rota
  and appendix 1 by Sergey Fomin

\bibitem{Chyzak00}
Chyzak F 2000 An extension of {Z}eilberger's fast algorithm to general
  holonomic functions {\em Discrete Math.\/} {\bf 217} 115--134
  \urlprefix\url{http://dx.doi.org/10.1016/S0012-365X(99)00259-9}

\bibitem{Zeilberger90}
Zeilberger D 1991 The method of creative telescoping {\em J. Symbolic
  Comput.\/} {\bf 11} 195--204
  \urlprefix\url{http://dx.doi.org/10.1016/S0747-7171(08)80044-2}

\bibitem{Koutschan}
Koutschan C 2010 A fast approach to creative telescoping {\em Math. Comput.
  Sci.\/} {\bf 4} 259--266 \urlprefix\url{http://arxiv.org/abs/1004.3314}

\bibitem{Koutschan2}
Koutschan C 2011 Lattice {G}reen's {F}unctions of the {H}igher-{D}imensional
  {F}ace-{C}entered {C}ubic {L}attices {P}reprint
  \url{http://arxiv.org/abs/1108.2164}

\bibitem{embedded}
Schoutens H 1999 Embedded resolution of singularities in rigid analytic
  geometry {\em Ann. Fac. Sci. Toulouse Math. (6)\/} {\bf 8} 297--330
  \urlprefix\url{http://archive.numdam.org/article/AFST_1999_6_8_2_297_0.pdf}

\bibitem{Steenbrink}
Steenbrink J and Zucker S 1985 Variation of mixed {H}odge structure. {I} {\em
  Invent. Math.\/} {\bf 80} 489--542
  \urlprefix\url{http://dx.doi.org/10.1007/BF01388729}

\bibitem{Golyshev3}
Golyshev V~V 2007 Classification problems and mirror duality {\em Surveys in
  geometry and number theory: reports on contemporary {R}ussian mathematics\/}
  ({\em London Math. Soc. Lecture Note Ser.\/} vol 338) (Cambridge: Cambridge
  Univ. Press) pp 88--121 \urlprefix\url{http://arxiv.org/abs/math/0510287}

\bibitem{Atkin1}
Atkin A~O~L 1992 The number of points on an elliptic curve (modulo a prime) {E}-mails to the

\urlprefix\url{https://listserv.nodak.edu/cgi-bin/wa.exe?A2=NMBRTHRY;d054c463.9202} and
\urlprefix\url{https://listserv.nodak.edu/cgi-bin/wa.exe?A2=NMBRTHRY;d6b18300.9207}

\bibitem{Morain}
Morain F 1995 Calcul du nombre de points sur une courbe elliptique dans un
  corps fini: aspects algorithmiques {\em J. Th\'eor. Nombres Bordeaux\/} {\bf
  7} 255--282 {L}es Dix-huiti{\`e}mes Journ{\'e}es Arithm{\'e}tiques (Bordeaux,
  1993)
  \urlprefix\url{http://archive.numdam.org/article/JTNB_1995__7_1_255_0.pdf}

\bibitem{Elkies}
Elkies N~D 1998 Elliptic and modular curves over finite fields and related
  computational issues {\em Computational perspectives on number theory
  ({C}hicago, {IL}, 1995)\/} ({\em AMS/IP Stud. Adv. Math.\/} vol~7)
  (Providence, RI: Amer. Math. Soc.) pp 21--76
  \urlprefix\url{http://www.math.harvard.edu/~elkies/modular.pdf}

\bibitem{Miller}
Miller R~L 2011 Proving the {B}irch and {S}winnerton-{D}yer conjecture for
  specific elliptic curves of analytic rank zero and one {\em LMS J. Comput.
  Math.\/} {\bf 14} 327--350 \urlprefix\url{http://arxiv.org/abs/1010.2431}

\bibitem{Quasi}
Kaneko M and Zagier D 1995 A generalized {J}acobi theta function and
  quasimodular forms {\em The moduli space of curves ({T}exel {I}sland,
  1994)\/} ({\em Progr. Math.\/} vol 129) (Boston, MA: Birkh\"auser Boston) pp
  165--172
  \urlprefix\url{http://www2.math.kyushu-u.ac.jp/~mkaneko/papers/genjacobi.pdf}

\bibitem{Fricke}
Cohn H 1988 Fricke's two-valued modular equations {\em Math. Comp.\/} {\bf 51}
  787--807 \urlprefix\url{http://dx.doi.org/10.2307/2008779}

\bibitem{Fricke2}
Fricke R 1915 {\em Die elliptischen {F}unktionen und {I}hre {A}nwendungen.
  {E}rster {T}eil.\/} (Leipzig: B. G. Teubner)

\bibitem{Bertin}
Bertin {M-J}, Decomps-Guilloux A, Grandet-Hugot M, Pathiaux-Delefosse M and
  Schreiber {J-P} 1992 {\em Pisot and {S}alem numbers\/} (Basel: Birkh\"auser
  Verlag) ISBN 3-7643-2648-4 {W}ith a preface by David W. Boyd
  \urlprefix\url{http://dx.doi.org/10.1007/978-3-0348-8632-1}

\bibitem{Hibino}
Hibino T and Murabayashi N 1997 Modular equations of hyperelliptic {$X_0(N)$}
  and an application {\em Acta Arith.\/} {\bf 82} 279--291
  \urlprefix\url{http://citeseerx.ist.psu.edu/viewdoc/download?doi=10.1.1.113.3530&rep=rep1&type=pdf}

\bibitem{Prop3}
Almkvist G, van Straten D and Zudilin W 2011 Generalizations of {C}lausen's
  formula and algebraic transformations of {C}alabi-{Y}au differential
  equations {\em Proc. Edinb. Math. Soc. (2)\/} {\bf 54} 273--295
  \urlprefix\url{http://dx.doi.org/10.1017/S0013091509000959}

\bibitem{Vidunas2}
Vid{\=u}nas R 2005 Transformations of some {G}auss hypergeometric functions
  {\em J. Comput. Appl. Math.\/} {\bf 178} 473--487
  \urlprefix\url{http://http://arxiv.org/abs/math/0310436}

\bibitem{hard-hexagon0}
Baxter R~J 1980 Hard hexagons: exact solution {\em J. Phys. A\/} {\bf 13}
  L61--L70 \urlprefix\url{http://stacks.iop.org/0305-4470/13/L61}

\bibitem{hard-hexagon}
Joyce G~S 1989 On the icosahedral equation and the locus of zeros for the grand
  partition function of the hard-hexagon model {\em J. Phys. A\/} {\bf 22}
  L237--L242 \urlprefix\url{http://stacks.iop.org/0305-4470/22/L237}

\bibitem{hard-hexagon2}
Joyce G~S 1988 On the hard-hexagon model and the theory of modular functions
  {\em Philos. Trans. Roy. Soc. London Ser. A\/} {\bf 325} 643--702
  \urlprefix\url{http://dx.doi.org/10.1098/rsta.1988.0077}

\end{thebibliography}
\end{document}